\documentclass[oneside,letterpaper,12pt]{book}
\usepackage{wuthesisx}
\usepackage[margin=1 in]{geometry}
\usepackage{amssymb}
\usepackage{amsfonts}
\usepackage{wrapfig}
\usepackage{setspace}
\usepackage[toc,page]{appendix}
\usepackage{color}

\usepackage{cite}
\usepackage{fancyhdr}
\usepackage{graphicx}
\graphicspath{{graphs/}}

\usepackage{bm, amsmath, amssymb}
\numberwithin{equation}{section}
\usepackage{xcolor}
\usepackage{color}
\usepackage{soul}
\usepackage{pdfpages}
\usepackage{amstext}
\usepackage[perpage]{footmisc}
\usepackage{bbold}
\usepackage{esint}
\usepackage{caption}
\usepackage{floatrow}

\newcommand{\ad}[1]{\textcolor{blue}{#1}}

\def\be{\begin{equation}}
\def\ee{\end{equation}}
\def\bea{\begin{eqnarray}}
\def\eea{\end{eqnarray}}
\def\bse{\begin{subequations}}
\def\ese{\end{subequations}}

\defheaders{\FANCY}

\renewcommand{\title}{Introduction to Experimental Quantum Measurement with Superconducting Qubits}
\renewcommand{\author}{Mahdi Naghiloo, Murch Lab}
\renewcommand{\month}{May}
\renewcommand{\year}{2019}
\renewcommand{\date}{\month\ \year}
\newcommand{\dept}{Physics}
\newcommand{\chair}{Kater Murch}

\begin{document}
\begin{frontmatter}

\thispagestyle{empty}
\begin{abstract}[1]{\title }{\author}{\dept }{\year}{\chair}
\textbf{Abstract}---Quantum technology has been rapidly growing due to its potential revolutionary applications. In particular, superconducting qubits provide a strong light-matter interaction as required for quantum computation and in principle can be scaled up to a high level of complexity. However, obtaining the full benefit of quantum mechanics in superconducting circuits requires a deep understanding of quantum physics in such systems in all aspects. One of the most crucial aspects is the concept of measurement and the dynamics of the quantum systems under the measurement process. This document is intended to be a pedagogical introduction to the concept of quantum measurement from an experimental perspective. We study the dynamics of a single superconducting qubit under continuous monitoring. We demonstrate that weak measurement is a versatile tool to investigate fundamental questions in quantum dynamics and quantum thermodynamics for open quantum systems.
\end{abstract}

\tables

\newpage

\end{frontmatter}

\begin{main}

\chapter{Introduction\label{ch1}} 
Quantum mechanics has revolutionized our understanding of nature since its development in the 20th century. Its prescription for the workings of nature is full of unexpected rules that remain counterintuitive even after over a century of confirmations. In the past decades, we have witnessed enormous progress in technology and control over quantum systems. These technologies aim to use the counterintuitive properties of quantum mechanics for real-life applications such as secure communication~\cite{gisin2007quantum}, high-precision sensing~\cite{degen2017quantum}, and information processing~\cite{wendin2017quantum}.
These ambitious and revolutionary goals have driven a tremendous effort in the implementation of quantum devices in a variety of platforms ranging between, photonics, atomic systems, nano-mechanical structures, and superconducting circuits. Each platform offers a unique capability over others; photons are suited for transmitting quantum information, while atoms can serve as long-lived quantum memories. In this regard, superconducting circuits have gained a lot of attention for quantum computation owing to the strong light-matter interaction achievable in these circuits. 

Apart from computational goals, the superconducting circuit architecture is a powerful technology to explore quantum physics and can serve as a testbed for fundamental questions in science. In part, this is because the characteristics of quantum systems made of artificial atoms are rather easy to manipulate which opens possibilities to explore non-trivial quantum systems by the versatile design and engineering of superconducting circuits. The pronounced interplay between science and engineering in the superconducting circuit technology brings on active research from different perspectives ranging from fundamental studies to practical applications. In particular, understanding the physics of open quantum systems and the concept of measurement is considered a core problem in modern physics~\cite{rotter2015review}.

Open systems appear in many disciplines in science, from environmental science to social science and atomic physics to biophysics.  With recent progress in quantum technology and its applications, a deeper understanding of open quantum systems is required to face practical challenges. However, the importance of open quantum systems is not limited to practical applications. From the fundamental point of view, many questions tie into open quantum systems in some ways---questions such as how classical laws emerge from underlying quantum laws, the classical-quantum boundary~\cite{zurek1992environment,modi2012classical,tan2016quantum}, the arrow of time~\cite{dressel2017arrow,manikandan2018fluctuation,harrington2018characterizing}, and exploring quantum thermodynamics~\cite{vinjanampathy2016quantum}. 

The dynamics of open quantum systems cannot be described by the Schr\"odinger equation due to the interaction with the environment. This interaction results in dissipation and decoherence in quantum systems. Superconducting circuits naturally tend to interact with all available degrees of freedom which makes them highly controllable systems yet presents a challenge to preserve quantum coherence. Therefore, one of the most active areas of research in quantum circuit technology is directed toward understanding and controlling decoherence channels and encoding quantum information in states that are protected from decoherence~\cite{houck2008controlling,gladchenko2009superconducting,gambetta2011superconducting,kockum2018decoherence,ningyuan2015time,dempster2014understanding}. Another approach to cope with dissipation and decoherence is to come up with clever designs and protocols out of imperfect parts that enable to correct for imperfections and perform a perfect tasks~\cite{kelly2015state,ofek2016extending,corcoles2015demonstration,reed2012realization}.  

From the quantum measurement point of view, if we are able to monitor dissipation of a quantum system, we could then maintain its coherence~\cite{korotkov2010decoherence,kim2012protecting}. In fact, measurement on quantum system can be used as a resource for feedback to control dynamics~\cite{gillett2010experimental,vijay2012stabilizing}, to herald non-trivial states~\cite{sayrin2011real}, and to prepare entangled states~\cite{sorensen2003measurement,ruskov2003entanglement,roch2014observation}. Therefore the concept of measurement in open quantum systems is important in many ways.

In particular, weak measurement enables one to continuously monitor a quantum system without destroying its quantum coherence~\cite{murch2013observing}. This provides a powerful tool to explore quantum dynamics in its most fundamental level~\cite{hacohen2016quantum, foroozani2016correlations, naghiloo2016mapping, weber2014mapping,naghiloo2017quantum}. Understanding the dynamics of continuously monitored systems in turn opens new ways for novel applications such as sensing~\cite{cujia2018watching,naghiloo2017achieving} and parameter estimation~\cite{kiilerich2016bayesian}.

Also, superconducting circuits and quantum measurement techniques have a lot to offer to the newly emerging field of quantum thermodynamics~\cite{vinjanampathy2016quantum}. The hope is that understanding the dynamics of quantum systems lead us to an understanding of underlying thermodynamic law in the quantum regime. In this context, the quantum system (e.g. a qubit) is in contact with the environment as a reservoir. By continuous monitoring of the reservoir, we can learn about energy exchange between the system and the reservoir. These observations would be helpful to understand the underlying thermodynamical laws and fluctuations in the system. This raises many new questions about the relevant thermodynamics parameters in the quantum regime like heat, work, and entropy~\cite{naghiloo2017thermodynamics}, the validity of the classical thermodynamics laws for quantum systems~\cite{brandao2015second, toyabe2010experimental}, the emergence of thermalization and irreversibility~\cite{gogolin2016equilibration,dressel2017arrow} from quantum mechanical principles, and the energy-information connection~\cite{parrondo2015thermodynamics,naghiloo2018information}. Many of these questions can be addressed by a deep understanding of open quantum system dynamics given from quantum measurement techniques.

Finally, the superconducting quantum systems can be engineered to realize non-trivial systems such as hybrid systems~\cite{kurizki2015quantum}, ``giant" atoms~\cite{kockum2018decoherence}, engineered baths~\cite{murch2012cavity,harrington2018bath}, and non-Hermitian systems~\cite{el2018non,peng2014parity, chen2017exceptional} where each of these hybrid systems opens new opportunities to explore unprecedented areas in physics. In particular, non-Hermitian systems which obey Parity-Time (PT) symmetry have gained a lot of attention both from theoretical \cite{bender2016pt,bender2017behavior,bender2016comment,bender2018series,bender2016analytic,bender2018p,bender2018pt} and experimental \cite{el2018non,naghiloo2019quantum} perspective owing to their topological and nonreciprocal properties.

\section{Overview}
This document is intended to be a pedagogical introduction to quantum measurement with a focus on experiments in the superconducting qubit platform. A goal of this document is to provide a clear and simple picture of quantum measurement in superconducting qubit circuits for those who are new to the field. To this end, I will try to address questions I encountered when beginning this research and cover questions I have received from other students during my PhD studies.
Chapter~2 provides a basic theoretical discussion about the light-matter interaction and preliminary theory for measurement and characterization of superconducting circuits.
Chapter~3 provides basic experimental knowledge about quantum measurement and superconducting circuits in close connection with the theoretical discussions of Chapter~2.
Chapter~4 provides a pedagogical discussion of generalized measurements and continuous monitoring of a qubit and provides experimental procedures for two types of continuous measurements corresponding to measurement operators $\sigma_z$ and $\sigma_-$.
Chapter~5 and 6 discuss two experiments in close connection with the pedagogical discussions of previous chapters.

In Chapter~5, we will study how measurement affects the dynamics of quantum systems. In particular, I discuss the situation where the spontaneous emission of a quantum emitter is measured by homodyne detection. Typically, spontaneous emission is associated with the sudden jump of an atom or molecule from an excited state to lower energy state by emission of a photon. Spontaneous jump dynamics occur because most of detectors are sensitive to energy quanta. However, light has both wave and particle nature, and here we explore how the spontaneous emission process is altered if we detect the wave rather than the particle nature of light. To do this, we interfere the spontaneously emitted light from a quantum emitter with another electromagnetic wave, measuring a specific amplitude of the emission. The dynamics of the quantum emitter under such a detection scheme are drastically different than what is observed when photons are detected, for the state of the quantum emitter can no longer simply jump between energy levels. Rather, the emitter's state takes on diffusive dynamics and follows a continuous quantum trajectory between its excited and ground state.

\begin{figure}[ht]
  \begin{center}
    \includegraphics[width=0.7\textwidth]{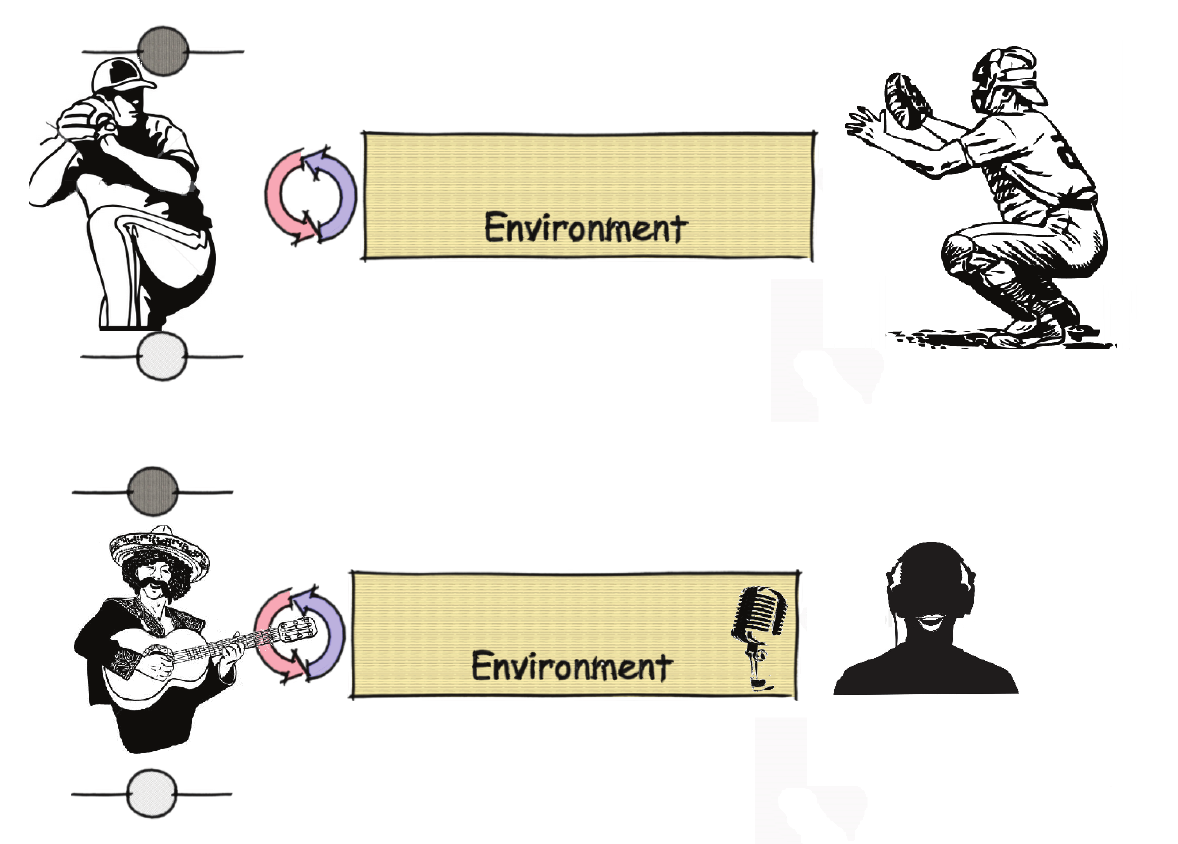}
\caption[Photon detection vs. homodyne detection]{ \footnotesize \textbf{Photon detection vs. homodyne detection (discussed in Chapter~5):} The behavior of a quantum emitter depends on how we detect its emission. If we hire a catcher as a detector which is sensitive to the energy quanta (addresses the particle notion of light), the emitter behaves like a pitcher (spontaneous jump behavior). However, if we instead ``listen" to the emitter, it behaves according to the wave nature of its emitted energy (diffusive behavior).} \label{catcher}
\end{center}
\end{figure}

Chapter~6 discusses quantum thermodynamics under the guise of Maxwell's demon. The thought experiment of Maxwell demon, whereby knowing the position and velocity of the molecules, a demon can sort hot and cold particle in a box was in apparent violation of 2$^\text{nd}$ law of thermodynamics. This thought experiment revealed a profound connection between energy and information in thermodynamics and has driven a lot of theoretical and experimental studies to understand this connection in many different platforms. In Chapter~6, we study the experimental realization of Maxwell's demon in a quantum system using continuous monitoring. We show that the second law of thermodynamics can be violated by a quantum Maxwell's demon unless we consider the demon's information. In our case, this information is quantum information which is susceptible to decoherence.

\begin{figure}[ht]
  \begin{center}
    \includegraphics[width=0.4\textwidth]{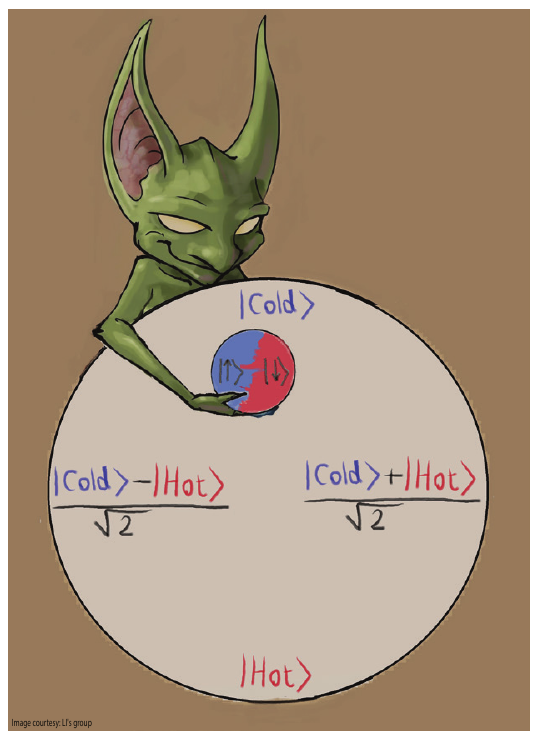}
\caption[Quantum Maxwell's demon]{ \footnotesize \textbf{ Quantum Maxwell's demon (discussed in Chapter~6):} We experimentally study a quantum version of Maxwell's demon who sorts particles that are in a quantum superposition of both hot and cold. We will see that the information obtained by the demon can be lost due to the decoherence and inefficient detection. Image adopted from Li's group.}\label{fig:test2}
\end{center}
\end{figure}

\chapter{The Light-Matter Interaction\label{ch2}} 

This chapter provides the basic theoretical concepts of the light-matter interaction. The aim of this chapter is to pedagogically introduce concepts related to the rest of this document, especially Chapter~3 where we experimentally discuss qubit-cavity characterization.

We consider the simplest example\footnote{One would think that the simplest situation is a qubit in free space. However free space supports infinite continuum of modes. In this regard, the free space situation is not the simplest situation.} of the light-matter interaction where a two-level quantum system (a qubit) interacts only with a single mode of light\footnote{A mode of light contains photons all of the same frequency, polarization, and spatial distribution.}. In practice, this situation can be achieved by placing the qubit inside a cavity that supports a discrete set of  modes. By a proper choice of qubit and cavity frequencies, cavity mode geometry, qubit placement and orientation, the qubit can effectively interact with only one of the modes of the cavity\footnote{However this assumption works fine for many practical situations, it may not be accurate enough in general. In fact, this is an issue of fundamental importance see for example see~\cite{malek17cutoff,gely17converg}}.\\

\section{One-dimensional cavity modes} 
The electromagnetic mode of a cavity can be described by Maxwell's equations in classical electrodynamics. In the next section, we discuss the proper description of an electromagnetic field in quantum mechanics. Here we focus on a one-dimensional (1D) cavity but we will see that the result can be simply extended to higher dimensions.

Here, I follow the conventional quantization found in quantum optics textbooks (e.g. Ref.~\cite{gerr05,walls2007quantum}) and discuss the quantization of electromagnetic field of an actual cavity (a volume bounded by perfect conductors) which is relevant to the three-dimensional (3D) architecture of cavity quantum electrodynamics (cQED)\footnote{Often in quantum circuit literature, this discussion is introduced by quantization of an $LC$ circuit; we will discuss this when we study the qubit. In this chapter we will see theoretically why a cavity bounded by superconducting walls is an $LC$ circuit, and later study this physically in Chapter~\ref{ch3} (See Fig.~\ref{fig:rect_cav}).}.

In order to quantize the electromagnetic field, we may solve Maxwell's equations for a given set of boundary conditions and identify a corresponding canonical position $q$ and canonical momentum $p$. Then we transition to the quantum case by promoting $q$ and $p$ to operators\footnote{This is a convenient way to quantized photons, massless particles. For ``massive'' particles (e.g. electron in a box) one can solve the Schr\"odinger equation.}.

\begin{figure}[ht]
\centering
\includegraphics[width = 0.7\textwidth]{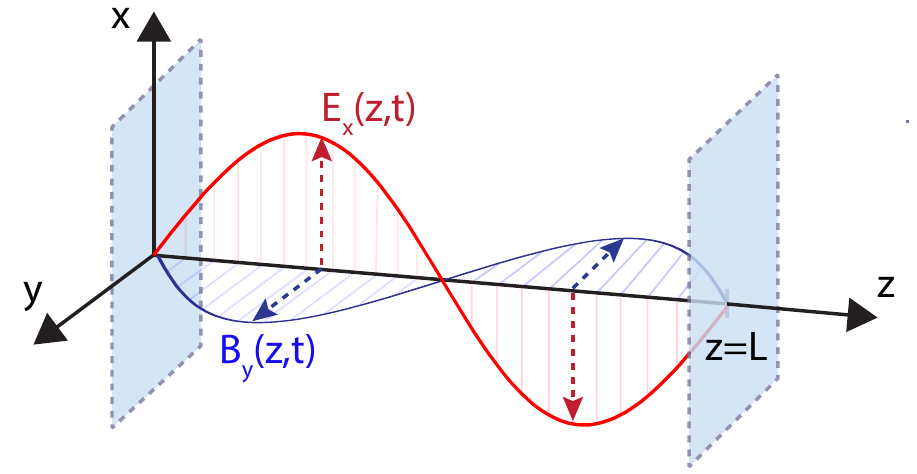}
\caption[One dimensional cavity]{ {\footnotesize \textbf{One dimensional cavity:} Two infinite superconducting walls separated by distance $L$ form a cavity that supports a discrete number of electromagnetic modes in the $z$-dimension (the second mode is shown). Due to the translational symmetry in $x$ and $y$ directions, the electromagnetic fields are only functions of $z$. For simplicity, we assume the electric field (red lines) has polarization along $x$ axis and consequently the magnetic field (blue lines) is along $y$ axis.}} 
\label{fig:1Dcavity}
\end{figure}

For a one-dimensional cavity, consider a pair of infinite perfect conducting walls separated by the distance $L$ along the $z$-direction as depicted in Figure~\ref{fig:1Dcavity}. This configuration can be considered as one-dimensional because we have a continuous translational invariance in $x$ and $y$ dimensions. Therefore the electric and magnetic fields only depend on the $z$-coordinate. For simplicity, we assume that the electric field is polarized along $x$-axis which implies that the magnetic field is only along the $y$-axis. This is an empty cavity with no external current or charge source, therefore for Maxwell's equations we have,
\begin{subequations}
\begin{eqnarray}
\triangledown \times \vec{E} = - \frac{\partial \vec{B}}{\partial t }  & \rightarrow & \frac{\partial E_x(z,t)}{\partial z } = - \frac{\partial B_y(z,t)}{\partial t },\\
\triangledown \times \vec{B} = \varepsilon_0 \mu_0 \frac{\partial \vec{E}}{\partial t }  & \rightarrow & - \frac{\partial B_y(z,t)}{\partial z } = \varepsilon_0 \mu_0 \frac{\partial E_x(z,t)}{\partial t },\\
\triangledown \cdot \vec{E} = 0  & \rightarrow & \frac{\partial E_x(z,t)}{\partial x } = 0,\\
\triangledown \cdot  \vec{B} = 0  & \rightarrow & \frac{\partial B_y(z,t)}{\partial y } = 0.
\end{eqnarray} \label{eq:maxwells}
\end{subequations}
Given perfect conducting walls, the electric field is required to vanish at the boundaries; $E_x(z=0,t)=0$ and $E_x(z=L,t)=0 $. One can show that the solution for electric and magnetic field inside the cavity are,
\begin{subequations}\label{eq:EandB}
\begin{eqnarray}
E_x(z,t) &=& \mathcal{E} \ q(t) \sin(k z), \label{eq:Ex}\\
B_y(z,t) &=& \mathcal{E} \  \frac{\mu_0 \varepsilon_0}{k}  \dot{q}(t) \cdot \cos(k z). \label{eq:By0}
\end{eqnarray}
\end{subequations}
The normalization constant $\mathcal{E}$ is conveniently set to be $\mathcal{E} =\sqrt{\frac{2 \omega_c^2}{V \epsilon_0 }}$ where $V$ is the effective volume of the cavity\footnote{Here the constant $\mathcal{E}$ is defined in a way that the total energy in the cavity finds a compact form in Equation~\eqref{eq:H_singmode} which conveniently ensures that $\hat{q}$ and $\hat{p}$ obey the canonical commutation relation $[ \hat{q}, \hat{p}]=i \hbar$.}. The parameter, $k=m \pi/L,\  m=1, 2, ... $ is wave number corresponding to the frequency $\omega_c=\frac{k}{\sqrt{\mu_0 \epsilon_0}}$. The function $q(t)$ describes the time-evolution for modes and has a dimension of length\footnote{The actual form for $q$ is $q(t)=\sin(\omega t +\phi)$. But for now, we rather to implicitly represent it by $q(t)$ and we will see that it acts as the canonical position.}. Each integer value $m$ corresponds to one mode of the cavity. Figure~\ref{fig:1Dcavity} shows the electric and magnetic field for the second mode of the cavity ($m=2$). The total electromagnetic energy (per unit of volume) stored in one mode can be written as,
\begin{eqnarray}
H = \frac{1}{V} \int dV \left( \ \ \frac{\epsilon_0}{2} |  E_x(z,t)|^2 + \frac{1}{2 \mu_0} |B_y(z,t)|^2 \ \  \right).
\label{eq:H_singlemode}
\end{eqnarray}
By substituting Equation~\eqref{eq:EandB} in \eqref{eq:H_singlemode}, one can show that total energy is equal to,
\begin{eqnarray}
H = \frac{1}{2} \left[ p^2(t) + \omega^2_c q^2(t) \right],
\label{eq:H_singmode}
\end{eqnarray}
where $p(t)=\dot{q}(t)$. From Eq.~\eqref{eq:H_singmode}, it is apparent that the energy of an electromagnetic mode is analogous to the energy of a classical harmonic oscillator if we consider $q(t)$ and $p(t)$ as the canonical position and momentum. Having canonical position and momentum identified, the Hamiltonian may be treated quantum mechanically by promoting the canonical parameters to be operators ($p,q \longrightarrow \hat{p}, \hat{q}$). This results in a \emph{quantum} Hamiltonian for a harmonic oscillator:
\begin{eqnarray}
\hat{H} = \frac{1}{2} \left[ \hat{p}^2(t) + \omega^2_c \hat{q}^2(t)   \right].
\label{eq:H_singmodeq}
\end{eqnarray}
Therefore, we may conclude that each mode of the cavity acts as a quantum harmonic oscillator\footnote{In this transition, we may keep/drop the time-dependence to work in Heisenberg/Schr\"odinger picture.}. Note that in the classical description of Equation~\eqref{eq:EandB}, we already found that the cavity has discrete modes. However in that picture, each mode could have continuous amount of energy. The transition to a quantum mechanical description happens in Equation~\eqref{eq:H_singmode} $\to$~\eqref{eq:H_singmodeq} which results in quantization of the energy spectrum for each mode. To see this, it is convenient to define non-Hermitian operators\footnote{Here $\hbar$ is introduced indicating we enter quantum world. However, we set $\hbar=1$ throughout this thesis except for few confusing situations.}
\begin{subequations}\label{eq:aadag}
\begin{eqnarray}
\hat{a} &=& \frac{1}{\sqrt{2  \omega_c}} ( \omega_c \hat{q} + i \hat{p}), \\
\hat{a}^{\dagger} &=& \frac{1}{\sqrt{2  \omega_c}} ( \omega_c \hat{q} - i \hat{p}),
\end{eqnarray}
\end{subequations}
which are annihilation and creation operators for a photon in the corresponding mode of the cavity and obey the commutation relation $[\hat{a} ,\hat{a}^{\dagger} ]=1$. The electric and magnetic fields, which are now operators, can be represented by $\hat{a}$ and $\hat{a}^{\dagger}$ as, 
\begin{subequations}\label{eq:EBaadag}
\begin{eqnarray}
\hat{E}_x(z,t) &=& \mathcal{E}_0 (\hat{a}  + \hat{a}^{\dagger} )  \sin(k z), \label{eq:Eaadag}
\\
\hat{B}_y(z,t) &=& i \mathcal{B}_0 (\hat{a}  - \hat{a}^{\dagger} )  \cos(k z).
\label{eq:Baadag}
\end{eqnarray}
\end{subequations}
The Hamiltonian Eq.~\eqref{eq:H_singmodeq} also takes a compact representation in terms of  $\hat{a}$ and $\hat{a}^{\dagger}$,
\begin{eqnarray}
\hat{H}=  \omega_c (\hat{a}^{\dagger}\hat{a}  + \frac{1}{2} ) =  \omega_c (\hat{n} + \frac{1}{2} ),
\label{eq:Haadag}
\end{eqnarray}
where the operator $\hat{n}=\hat{a}^{\dagger}\hat{a} $ is the \textit{number} operator. Knowing the Hamiltonian for the electromagnetic field of a single cavity mode, we can describe the state of the cavity by solving the corresponding eigenvalue problem. Considering Hamiltonian \eqref{eq:Haadag} we have, 
\begin{eqnarray}
\hat{H} |n\rangle = E_n |n\rangle , \hspace{0.5cm} n=0,1,2,...
\label{eq:hnen}
\end{eqnarray}
where $\{ |n \rangle \}$ are \emph{photon number states}  or \emph{Fock states} representing the energy eigenstate for the single mode cavity field with the corresponding energy $E_n= \omega_c (n+\frac{1}{2})$. The photon-number states $\{|n\rangle \}$ form a complete basis to describe any arbitrary state of the cavity. That means at any given time, the cavity is either in one of states  $|n\rangle$ or in some linear superposition of them, $\sum_n c_n|n \rangle$. However, In general, the cavity state can be in a mixed state, an incoherent superposition of Fock states, like thermal states, which are conveniently represented by the density matrix $\rho=\sum_n P_n |n \rangle \langle n|$.

\subsection{How to visualize the state of light}
We may describe the quantum state of the light inside the cavity by a wave function $|\psi\rangle$ which can be represented in any arbitrary basis e.g. photon-number basis, $|\psi\rangle=\sum_n c_n|n \rangle$. Now the question is what is the best way to characterize and visualize $|\psi\rangle$. One way to do this is by looking at the expectation value of the electric and magnetic fields, $\langle \psi|E|\psi \rangle$ and  $\langle \psi|B|\psi \rangle$ and their fluctuations. In the previous section we learned that electric and magnetic fields are quantum objects and are described by operators, Eq.~\eqref{eq:EBaadag}. From these equations, it is apparent that the electric and magnetic operators are directly related to the canonical position and momentum respectively.
\begin{subequations}\label{eq:EB_prop_xp}
\begin{eqnarray}
\hat{E} \propto (\hat{a}  + \hat{a}^{\dagger} ) \propto \hat{q} \label{eq:E_prop_x}
\\
\hat{B} \propto (\hat{a}  - \hat{a}^{\dagger} ) \propto \hat{p} \label{eq:B_prop_p}
\end{eqnarray}
\end{subequations}
The only difference between operators $\{\hat{E},\hat{B}\}$ and $\{\hat{q},\hat{p}\}$ is an extra spatial dependence term in the electromagnetic operators which depends on the details of the geometry in the system\footnote{The spatial dependence has to do with the geometry of the problem which sets the spatial property for all photons in the same way. Let me explain this by a question; What is the difference between a Fock state, let's say $|1\rangle$, of a cylindrical cavity and a rectangular cavity? The is no difference. They both represent having a photon in a cavity. But if you were asked about the spatial probability distribution of that photon inside that cavity, then the answer indeed depends on the geometry of each cavity. Later when we discuss the qubit placement inside the cavity, we will see this spatial dependence comes into play implicitly in the coupling between cavity and qubit.}. Therefore, a general way to visualize the state of light, regardless of the geometry of the cavity, is to look at the probability distribution of photons in phase space, $W(q,p)$. This ``quasi-probability" distribution\footnote{It is called ``quasi-probability'' because unlike a normal probability distribution, the Wigner function may be negative for a non-classical light.} is known as the \emph{Wigner function}. There are a bunch of different representations for the Wigner function in different bases. For example, in the canonical position basis $\{|q\rangle\}$, the Wigner function has the following definition for a given pure state $|\psi\rangle$,
\begin{eqnarray}
W(q,p) =\frac{1}{2\pi   } \int_{-\infty}^{+\infty}  \langle q+x/2| \psi\rangle \langle \psi | q-x/2 \rangle e^{+ipx} dx.
\label{eq:wigner}
\end{eqnarray}
In this section, we will see that Wigner function has an intuitive distribution for classical light (e.g. coherent light, thermal light) but it is somewhat nonintuitive for non-classical light (e.g. single photon state). 
Now we briefly discuss a few common states of light for a single mode of a cavity.\\


\emph{\textbf{Fock state}}-- As we introduced earlier, Fock states, or photon-number states, are eigenstates of the quantum harmonic oscillator. Thus they have the simplest representation in the photon-number basis\footnote{They are simple in terms of representation but experimentally, the preparation of a cavity in a Fock state is not simple~\cite{houck2007generating}.} and describing the situation where exactly $n$ photons exist in the cavity,
\begin{eqnarray}
|\psi\rangle = |n\rangle \hspace{0.5cm}& &\hspace{0.5cm} \mbox{(Fock state)}
\label{eq:fockstate}
\end{eqnarray}
including \textit{vacuum} state $|0\rangle$ where there is no photon in a cavity\footnote{There is no clear spatial visualization of photon-number states inside a cavity. But for our purpose one may have some sort of visualization by combining both notions of light; wave and particle. In that sense, one can imagine that each photon is a packet of energy that extended inside the cavity so that its spatial probability distribution follows the distribution of the energy on that mode.  A conventional way to characterize the state of the light is by calculating its \textit{Wigner function} which is somehow a probability distribution as a function of canonical position and momentum but it doesn't give any visualization in real space.}. Photon-number states are orthogonal to each other, $\langle n | m \rangle = \delta_{n,m}$, which means, experimentally, one should be able to distinguish $|n\rangle$ from $|m\rangle$ without any ambiguity. Considering this orthogonality, it is easy to check that the expectation values of the electromagnetic field operators (Eq.~\ref{eq:EBaadag}) for photon-number states are zero regardless of the number of photons. But the expectation value for the electromagnetic field squared (e.g. $\langle E^2 \rangle$) and electromagnetic fluctuations (e.g. $\Delta E= \sqrt{\langle E^2 \rangle - \langle E \rangle^2}$) are nonzero even for a vacuum state\footnote{However, $\langle E \rangle$ and $\langle B \rangle$ for a superposition of two or more Fock states can be non-zero.}.\\

\noindent\fbox{\parbox{\textwidth}{
\textbf{Exercise~1:} Show that  $\langle E \rangle=0$, $\langle B \rangle=0$ for a Fock state $|n\rangle$, but $\langle E^2 \rangle\neq0$, $\langle B^2 \rangle\neq0$. What is the electric field uncertainty $\Delta E$ for the vacuum state?
}} \vspace{0.25cm}

For example the Wigner function for state $|0\rangle$ and $|1\rangle$ has following form,
\begin{eqnarray}
|0\rangle & \to &  W_0(q,p) =\frac{1}{2\pi  } e^{-(q^2+p^2)} \label{eq:W0}\\
|1\rangle & \to & W_1(q,p) =\frac{1}{2\pi  } (2q^2+2p^2-1) e^{-(q^2+p^2)}.
\label{eq:W1}
\end{eqnarray}
It is somewhat easy to find some classical interpretation for a Wigner function when it is not negative. For that just consider that $q$ and $p$ are related to the electric and magnetic field\ad{s}. For example the vacuum state (Eq.~\eqref{eq:W0}) depicted in Fig.~\ref{fig:wigner_fock}a shows the probability is maximum for  $q=0, p=0$, corresponding to zero electric and magnetic field. But there is some probability for a non-zero electromagnetic field around zero which comes from vacuum fluctuations and accounts for a vacuum energy $\omega_c/2$. So even an empty cavity has some amount of energy and electric and magnetic field fluctuate around zero\footnote{This makes the vacuum state non-classical.}.
\begin{figure}[ht]
\centering
\includegraphics[width = 0.9\textwidth]{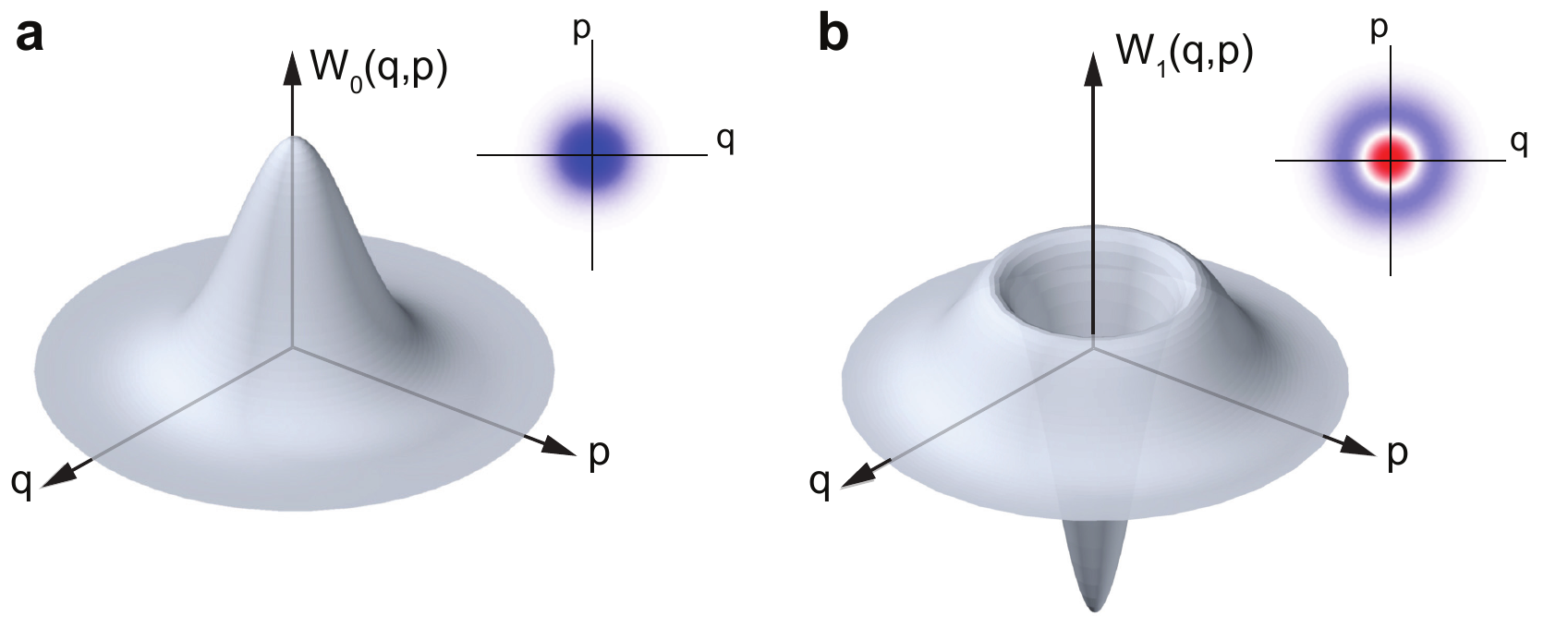}
\caption[Wigner distribution for photon-number states]{ {\footnotesize \textbf{Wigner distribution for photon-number states:} \textbf{a}, The vacuum state $|0\rangle$ has a Gaussian distribution centered at the origin of the phase space. \textbf{b},  The single photon state $|1\rangle$ exhibits negative probabilities around the origin.}} 
\label{fig:wigner_fock}
\end{figure}

However, the Wigner distribution is not very intuitive for photon-number states other than the vacuum state. For example, as it is apparent from  Equation~\eqref{eq:W1} (also depicted in Fig~\ref{fig:wigner_fock}b), the Wigner function is negative in some region for the state $|1\rangle$. It is hard to interpret the negative probability density, thus states with negative Wigner functions are called non-classical states.

Note that the photon-number states are eigenstates of the harmonic oscillator Hamiltonian (Eq.~\ref{eq:Haadag}), thus the Fock state Wigner functions are stationary and do not evolve in time. 

It is worth here to mention some common operational relations for Fock states:
\begin{subequations}\label{eq:aadag_fock}
\begin{eqnarray}
\hat{a} |n\rangle &=& \sqrt{n} |n-1\rangle \\
\hat{a}^{\dagger} |n\rangle &=& \sqrt{n+1} |n+1\rangle \\
\hat{n} |n\rangle &=& \hat{a}^{\dagger} \hat{a} |n\rangle =  n |n\rangle
\end{eqnarray}
\end{subequations}
Where $\hat{a} (\hat{a}^{\dagger})$ annihilates (creates) a photon and $\hat{n}$ leaves the state intact and gives the number of photons. With that, let's finish the discussion of Fock states by a ``counterintuitive" question.\\

\noindent\fbox{\parbox{\textwidth}{
\textbf{Exercise~2:} Consider a situation where the single cavity mode contains superposition of two Fock states described by $|\psi\rangle = \sqrt{0.99} |0\rangle + \sqrt{0.01} |100\rangle$. What is the average number of photons in the cavity? If you annihilate a photon by acting annihilation operator $\hat{a}$ on this state, then how many photons remain in the cavity? Interpret the result.
}} \vspace{0.25cm}

\emph{\textbf{Coherent state}}-- One of the most common types of light is \textit{coherent} light which is also known as classical light. In fact, the output of a laser or a signal generator is coherent light. Experimentally, one can simply send the output of a signal generator at the right frequency to a cavity to produce a coherent state in the cavity. The coherent state can be represented in the photon-number basis as,
\begin{eqnarray}
|\psi\rangle = |\alpha\rangle = \sum_n c_n |n\rangle \hspace{0.2cm},\hspace{0.2cm} c_n=e^{-|\alpha|^2/2} \frac{\alpha^n}{\sqrt{n!}},
\label{eq:coherent}
\end{eqnarray}
where $c_n$ indicates the contribution of each photon-number state in the coherent state. The parameter $\alpha$ is a constant\footnote{Note, $\alpha=|\alpha|e^{i\phi}$ can be any complex number. We will later see that the phase $\phi$ has a very simple meaning (the phase of the oscillations) when we discuss the coherent state in analogy with a classical oscillator.} whose magnitude is related to the average number of photons, $\langle \hat{n} \rangle =|\alpha|^2$, of the coherent state $|\alpha\rangle$. In Figure~\ref{fig:coh_c_n} we plot $c_n$ versus $n$ for two different values of $\alpha$.
\begin{figure}[ht]
\centering
\includegraphics[width = 0.8\textwidth]{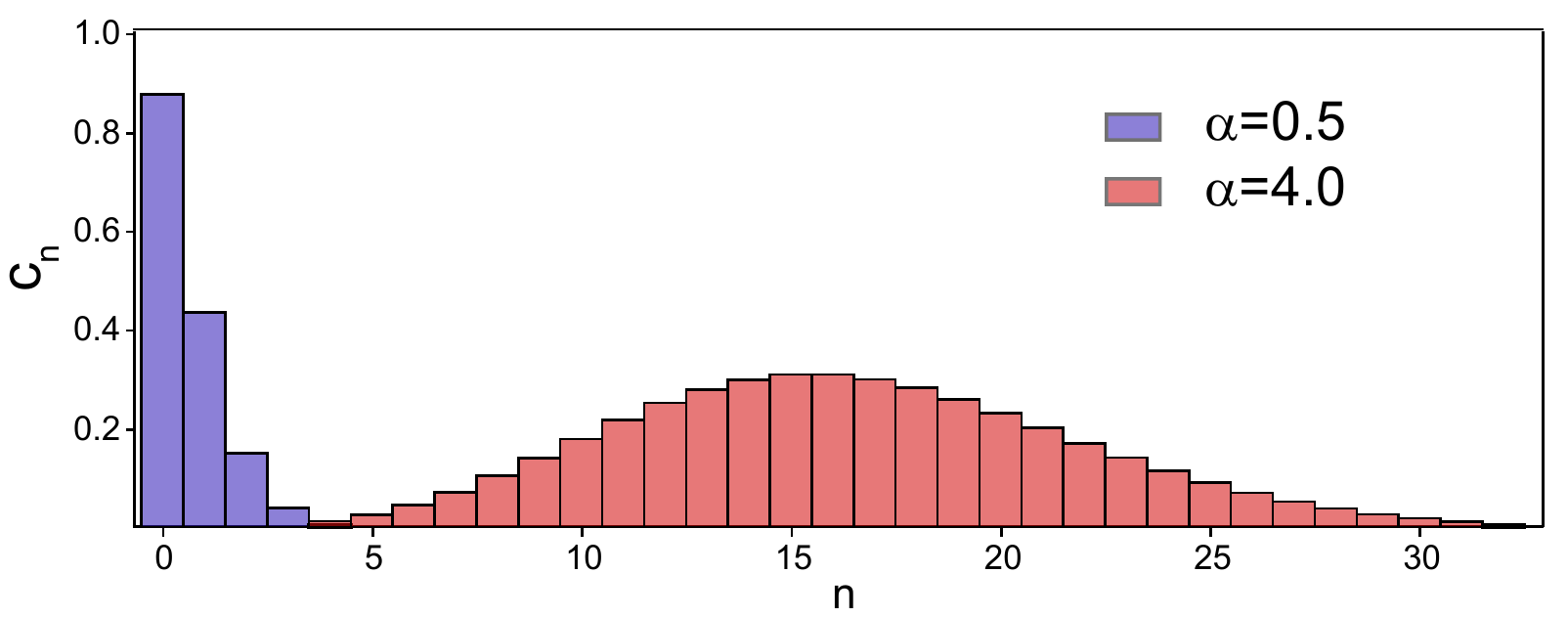}
\caption[Photon number distributions for coherent states]{ {\footnotesize \textbf{Photon number distributions for coherent states:} The blue (red) distribution shows the photon number distribution for a coherent state which has an average photon number $\bar{n}=1/4\  (\bar{n}=16)$. The photon number distribution for the higher average number of photons is more like a Gaussian distribution.}} 
\label{fig:coh_c_n}
\end{figure}\\
The blue distribution is for $\alpha=1/2$ which corresponds to the average number of photons $\bar{n}= \langle n \rangle =1/4$. That means if we measure the number of photons in the cavity, we mostly ($c_0^2=0.88^2 \sim 0.78$) find zero photons but on average we get 1/4 photon. The red distribution shows the distribution of photon-states for the coherent state that has 16 photons on average\footnote{It is important to distinguish coherent light with other incoherent mixed distributions of photons. It is possible that an incoherent light gives the same distribution of photons as a coherent light does, but a coherent state requires a certain relative phase between Fock states. For example, a qubit evolves totally different interacting with coherent light versus incoherent light even if they have a same photon number distribution.}. The fact that the distribution for the higher average number of photons is more like a Gaussian distribution, follows from the \textit{central limit theorem} for a Poisson distribution.\\

\noindent\fbox{\parbox{\textwidth}{
\textbf{Exercise~3:} Show that in the limit $\alpha\gg 1$ the photon distribution $c_n$ approaches to a Gaussian distribution centered at $|\alpha|^2$ and variance of $|\alpha|^2$.
}} \vspace{0.25cm}

It is easy to show that coherent state is the eigenstate of annihilation operator,
\begin{eqnarray}
\hat{a} | \alpha \rangle =\alpha  | \alpha \rangle.
\label{eq:alpha_alpha_a_alpha}
\end{eqnarray}
However, since $\hat{a}$ is a non-Hermitian operator, the corresponding eigenstates $\{|\alpha\rangle\}$ do not form a orthogonal basis\footnote{Two coherent states $|\alpha\rangle$ and $|\beta\rangle$ are orthogonal only in the limit of $|\alpha - \beta| \gg1$.}. Unlike the photon-number state, the coherent state is not an eigenstate of the Hamiltonian \eqref{eq:Haadag}, therefore it has time evolution\footnote{Here we assume that we are in Schr\"odinger picture which is more intuitive and convenient to discuss the evolution of the system. However, calculating the expectation values are often more straightforward in Heisenberg picture.}. But it turns out that the time evolution of a coherent state is simply a rotation in phase space.\\ 

\noindent\fbox{\parbox{\textwidth}{
\textbf{Exercise~4:} Show that a coherent state remains a coherent state under the time evolution but $\alpha$ acquires a phase: $|\alpha\rangle (t)= |\alpha_t\rangle$, where $\alpha_t=e^{-i \omega_c t} \alpha$.
}} \vspace{0.25cm}

Now it is the time to discuss why coherent light often is considered classical light. As we see in Equation~\eqref{eq:coherent}, a coherent state is indeed a superposition of quantized photon number states. But it turns out that most of its characteristics can be understood in a close analogy with a classical light. In other words, when a cavity is populated with coherent light, the behavior of the cavity corresponds to classical oscillatory motion. For example, by considering Equation~\eqref{eq:alpha_alpha_a_alpha} and the fact that $|\alpha\rangle (t)= |e^{-i \omega_c t} \alpha \rangle$, it is easy to show that the expectation value of the electromagnetic field (Eq.~\ref{eq:EBaadag}) for a coherent state is non-zero and oscillatory in time,
\begin{subequations}
\begin{eqnarray}
\langle \alpha_t| E |\alpha_t\rangle &=& 2 \mathcal{R}[\alpha_t]  \mathcal{E}_0 \ \sin(k z) \ \cos(\omega_c t ) \nonumber\\
&=& 2|\alpha|  \mathcal{E}_0 \ \sin(k z) \ \cos(\omega_c t+\phi),
\label{eq:E_exp_coherent}\\
\langle \alpha_t| B |\alpha_t\rangle &=& 2 \mathcal{I}[\alpha_t] \mathcal{B}_0 \ \cos(k z)\ \sin(\omega_c t) \nonumber \\
&=& 2|\alpha| \mathcal{B}_0 \ \cos(k z)\ \sin(\omega_c t-\phi), \label{eq:B_exp_coherent}
\end{eqnarray}
\end{subequations}
where we used the fact that $\alpha_t=e^{-i\omega_c t} \alpha$ and $\alpha$ itself is a complex number $\alpha=|\alpha|e^{i\phi}$. You may notice that the expectation value for the electric and magnetic fields are similar to the classical solutions of Maxwell's equation (Eq.~\ref{eq:EandB}). Therefore, the quantum description of the coherent state is consistent with our classical understanding of the oscillating electric and magnetic modes of a harmonic oscillator.

In addition, one can show that the coherent state has minimum quantum fluctuations equal to the vacuum fluctuations. This is a minimum uncertainty allowed by the Heisenberg uncertainty principle, assuming no squeezing. These fluctuations can be considered as an intrinsic uncertainty related to determining both the amplitude and phase of the electromagnetic field.\\

\noindent\fbox{\parbox{\textwidth}{
\textbf{Exercise~5:} Show that coherent state has minimum fluctuations (like a vacuum state) in each quadrature, $\langle ( \Delta I)^2 \rangle=\langle ( \Delta Q)^2 \rangle= 1/4$, where $I=(\hat{a} + \hat{a}^{\dagger})/2$, $Q=(\hat{a} - \hat{a}^{\dagger})/2i$. 
}} \vspace{0.25cm}

Thus the Wigner function for a coherent state is a vacuum Wigner function displaced in phase space (Fig.~\ref{fig:wigner_coherent}a) by an amount $\alpha$ which can be written in this form,
\begin{eqnarray}
|\alpha \rangle & \to &  W_{\alpha}(q,p) =\frac{1}{2\pi   } \exp( -(q-\mathcal{R}[\alpha])^2-(p-\mathcal{I}[\alpha])^2),
\label{eq:Wc}
\end{eqnarray}
where $\mathcal{R}[\alpha]$ ($\mathcal{I}[\alpha]$) is the real (imaginary) part of $\alpha$.
\begin{figure}[ht]
\centering
\includegraphics[width = 0.9\textwidth]{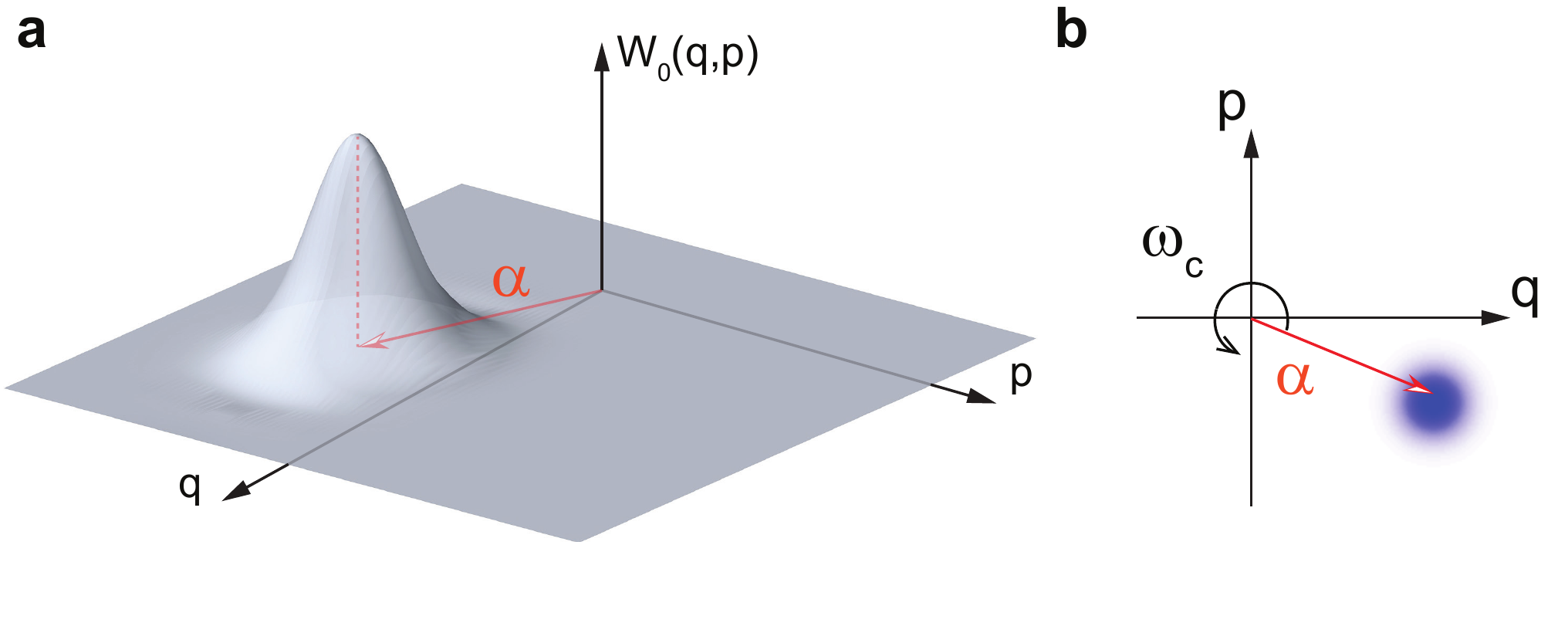}
\caption[Winger function for a coherent state]{ {\footnotesize \textbf{Winger function for a coherent state:} \textbf{a}, The Wigner function for a coherent state is a Gaussian distribution displaced from the origin by amount of $\alpha$. The coherent state has minimum uncertainty in each quadrature like a vacuum state. \textbf{b}, The evolution of coherent state under harmonic oscillator Hamiltonian  is simply a rotation around the origin.}} 
\label{fig:wigner_coherent}
\end{figure}

As illustrated in Figure~\ref{fig:wigner_coherent}b, the coherent state evolves around the origin of phase space by frequency $\omega_c$. That means the energy swings back-and-forth from electric (potential) to magnetic (kinetic). Therefore one may realize that the coherent state's Wigner function is very similar to the classical ``phasor diagram" of a noisy signal. The difference is that when considering classical signals, we assume one can in principle reduce the noise and make it arbitrarily small. But for the coherent state the ``noise" in each quadrature is quantum noise, originating from vacuum fluctuations as described by the Heisenberg uncertainty principle. In the limit of large average photon number, the noise (either classical or quantum) is negligible compared to the actual signal. Therefore, the classical picture and quantum picture completely overlap in that limit.

It worth mentioning here that the coherent state has a very important role in quantum measurement. In particular, a precise measurement of the phase of a coherent signal is an essential component for most quantum measurement experiments. Usually, we are not interested in the natural oscillation frequency of a coherent signal. Therefore we go to a frame that exactly rotates with that frequency. In that \emph{rotating frame}, the coherent state doesn't rotate anymore in phase space. The coherent state Wigner distribution is either along $q$ or $p$ or somewhere between and remains a steady-state. So in the rotating frame we freeze the time evolution for the oscillator. For simplicity let us assume the oscillator state is along the $q$ axis, which means all the energy is potential (like a stretched spring or a pendulum at its turning point). In the rotating frame, the coherent state is stationary and the phase is fixed unless, for any reason, the coherent state experiences an external phase shift (or a kick) on top of its normal phase evolution due to a perturbative interaction. In such case, the coherent state rotates to a new place in phase space. We can easily detect that displacement in the rotating frame\footnote{Rotating frames are useful in many ways; both in theory and experiment. Theoretically, sometimes it is easier to solve a problem in a rotating frame or it is more clear to see the dynamics of a system. Experimentally, as we will see in the next chapter, it is very natural and easy to work in a rotating frame. Otherwise, it wouldn't possible to precisely measure the phase shifts in a rapidly rotating signal (typically $\omega_c/2\pi \sim 5$ GHz).}(see Figure~\ref{fig:sig_kick_rf}). We will see in the next chapter that this type of phase detection is the essence of qubit readout measurement.

\begin{figure}[ht]
\centering
\includegraphics[width = 0.9\textwidth]{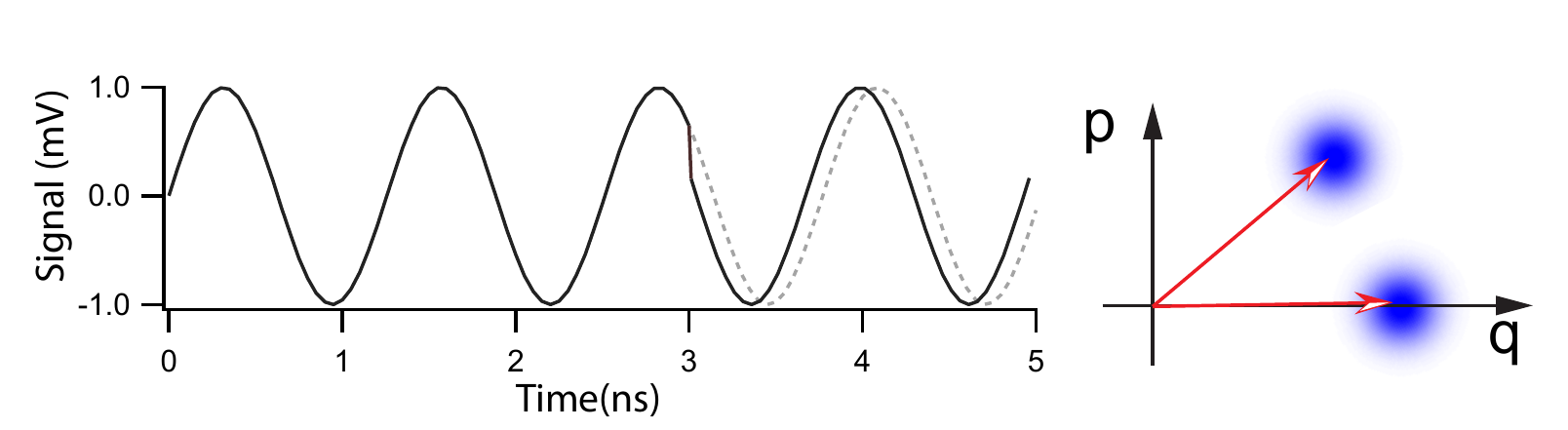}
\caption[Phase shifts for coherent state in the rotating frame]{ {\footnotesize \textbf{Phase shifts for coherent state in the rotating frame:} The phase shift of a coherent signal is easily detectable in the rotating frame.}} 
\label{fig:sig_kick_rf}
\end{figure}


\section{Qubit}
Experimentally there are many ways to realize a qubit. Here we discuss theoretically how to realize a qubit with a superconducting circuit. In our circuit toolbox we have only three elements to work with: capacitors, inductors, and Josephson junctions (JJ)\footnote{Of course we wanted to avoid resistors in our toolbox but this is something that comes for free. Even in superconducting circuits, there are various ways that energy can dissipate. e.g. photon emission/radiation, coupling to phonons.} (Fig.~\ref{fig:QED_toolbox}). 
\begin{figure}[ht]
\centering
\includegraphics[width = 0.58\textwidth]{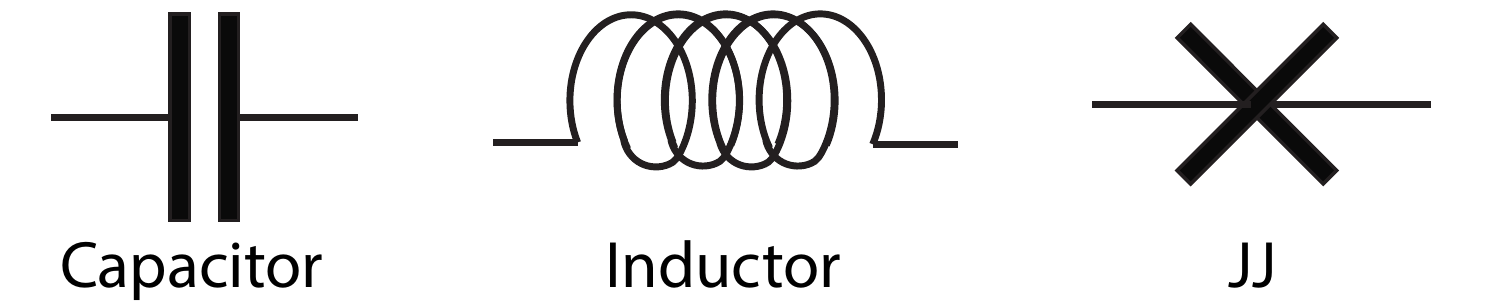}
\caption[Circuit QED toolbox]{ {\footnotesize \textbf{Circuit QED toolbox:} Quantum circuit technology relies on these three elements. The required nonlinearity comes from the JJ which is basically a dissipationless nonlinear inductor.}} 
\label{fig:QED_toolbox}
\end{figure}
The most important element is the Josephson junction which introduces a circuit nonlinearity necessary to form a qubit. In order to realize a qubit, the idea is to make a nonlinear (anharmonic) oscillator out of Josephson junction and use only the two lowest energy states as a qubit\footnote{Normally in circuit QED literature the transmon discussion is introduced by a circuit called Cooper pair box. The transmon is a Cooper pair box in a limit of a large shunt capacitance---see a nice discussion in Ref~\cite{schu07thesis}. Here, I approach the discussion of the qubit by starting from transmon as a nonlinear oscillator.}.
\subsection{Josephson junctions}
The Josephson junction (JJ) comprises of a thin ($\sim 1$ nm) layer of an insulator sandwiched between two superconducting slabs (Fig~\ref{fig:jj}). The superconducting leads consist of many atoms, but due to their superconducting state they can be described by a single complex number, $\Psi_{1,2}= \sqrt{n_{1,2}} e^{i \theta_{1,2}}$, where $n_{1,2}$ and $\theta_{1,2}$ indicate the number of Cooper pairs and the phase of the superconducting order parameter on each side of the junction. 
\begin{figure}[ht]
\centering
\includegraphics[width = 0.4\textwidth]{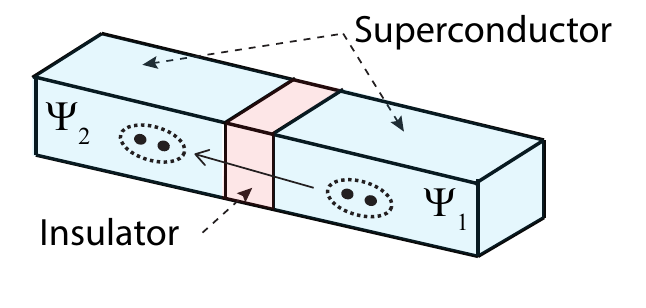}
\caption[Josephson junction]{ {\footnotesize \textbf{Josephson junction:} The JJ consists of two superconductors separated by a thin layer of insulator. The Cooper pairs on each side can tunnel through the insulator and create a super-current $I$. Remarkably, the current can be non-zero even when $V=0$. The highly nonlinear $I$-$V$ characteristics of the JJ can be exploited for quantum circuits.}} 
\label{fig:jj}
\end{figure}

It has been shown\footnote{There is a straightforward derivation for Josephson equations based on microscopic BCS theory, See for example Ref.~\cite{tinkham2004introduction}.} that, effectively, a JJ can be though of as a dissipationless nonlinear inductor which has the $I$-$V$ characteristics,
\begin{subequations}\label{eq:jjI}
\begin{eqnarray}
I&=&I_0 \sin(\delta)\\
V&=&\frac{\Phi_0}{2 \pi} \dot{\delta},
\end{eqnarray}
\end{subequations}
where $\delta=\theta_2-\theta_1$ and $I_0$ is a critical current above which the JJ becomes a normal dissipative junction. One can then infer the effective inductance of the Josephson junction is,
\begin{eqnarray}
V=L \frac{d I}{d t} \to \frac{\Phi_0}{2 \pi} \dot{\delta} = L I_0 \dot{\delta} \cos(\delta) \to L=\frac{\Phi_0}{2 \pi I_0  \cos(\delta) } \to L=\frac{L_{J0}}{\cos(\delta) },
\label{eq:jjL}
\end{eqnarray}
where $\Phi_0=\frac{h}{2 e}$ is the flux quantum, and we define $L_{J0}=\frac{\Phi_0}{2 \pi I_0}$ as the Josephson inductance at zero current. It is apparent that the Josephson inductance is a function of current $L=L(I)$. This dependence can be explicitly shown by using Equation~\eqref{eq:jjI}a in ~\eqref{eq:jjL},
\begin{eqnarray}
L=\frac{L_{J0}}{\sqrt{1-(\frac{I}{I_0})^2} } .
\label{eq:jjLI}
\end{eqnarray}
Moreover, one can use two JJs (assuming identical JJs) in a loop to effectively have a tunable JJ where the critical current can be tuned by passing an external flux $\Phi_{ext}$ through the loop,
\begin{eqnarray}\label{eq:JJ_squid}
I_0^{\mathrm{SQUID}}= 2I_0 | \cos(\frac{\pi \Phi_{ext}}{\Phi_0})|
\end{eqnarray}
where $I_0$ is the critical current of an individual junction. \\

\noindent\fbox{\parbox{\textwidth}{
\textbf{Exercise~6:} Derive Equation~\eqref{eq:JJ_squid}. What is the effective critical current for an asymmetric SQUID where two non-identical JJs are placed in a loop?
}} \vspace{0.25cm}

The total energy stored\footnote{Naturally, a JJ also has some small capacitance, but for our purposes and simplicity we ignore this since we are eventually going to shunt the JJ to a much larger capacitor to make a transmon qubit.} in a JJ can be calculated by adding up the energy changes $dU/dt= VI$ (assuming there was no current ($\delta=0$) at $t=-\infty$) and obtain\footnote{For a normal inductor the energy is simply $U=\int VI dt= \int L I dI = LI^2/2$. But for JJ the inductance $L$ is a function current $I$.},
\begin{eqnarray}
U &=& \int_{-\infty}^t I(t')V dt' = \frac{I_0 \Phi_0}{2 \pi} \int_{-\infty}^t  \sin(\delta) \dot{\delta} dt' \nonumber \\
&=&  \frac{I_0 \Phi_0}{2 \pi} \int_{-\infty}^t  \sin(\delta) d\delta = E_J[1-\cos(\delta)],
\label{eq:jjE}
\end{eqnarray}
where we define the Josephson energy $E_J=\Phi_0 I_0/2 \pi= \hbar I_0/2e$. In the next subsection, we will shunt a JJ by a capacitor and quantize the LC circuit (or JJ-C circuit). We will see that the parameter $\delta$ is the canonical position for that anharmonic oscillator. In the next chapter, we provide details from the experimental perspective, e.g. fabrication and characterization of a JJ.

\subsection{Transmon qubit}
The fact that the inductance of the JJ is a function of current passing through the JJ, makes it an interesting nonlinear element which can be leveraged for a qubit architecture. In particular, one can imagine shunting the JJ by a capacitor to have the anharmonic oscillator depicted in Figure~\ref{fig:transmon_simple}. 
\begin{figure}[ht]
\centering
\includegraphics[width = 0.3\textwidth]{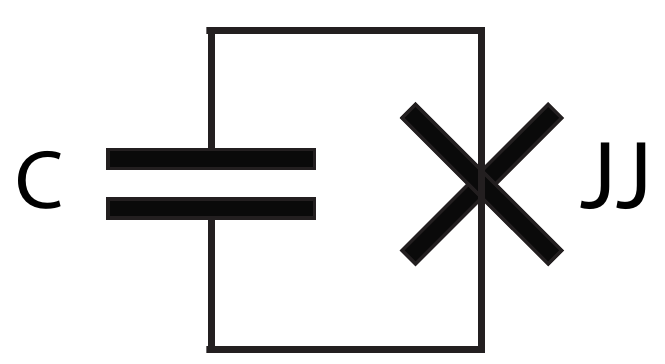}
\caption[Transmon circuit]{ {\footnotesize \textbf{Transmon circuit:} The transmon circuit consists of a JJ shunted by a relatively large capacitor so that $E_J\gg E_C$.}} 
\label{fig:transmon_simple}
\end{figure}
The total energy of the circuit is,
\begin{eqnarray}
H_{\mathrm{trans}}= \frac{Q^2}{2C} + E_J[1-\cos(\delta)],
\end{eqnarray}
where $Q$ is the total charge in the capacitor $C$ and we use Equation~\eqref{eq:jjE} for JJ energy. It is convenient to represent the total charge in capacitor in terms of number of Cooper pairs\footnote{When $I<I_0$, only pairs of electrons tunnel through the JJ insulating barrier, called Cooper pairs. Thus in this case it makes sense to represent charge in terms of the number of Cooper pairs, $m$.}, $Q= 2e m$. Therefore the total energy can be written in this form,
\begin{eqnarray}
H_{\mathrm{trans}}= 4 E_C m^2 + E_J[1-\cos(\delta)],
\end{eqnarray}
where we define the charging energy $E_C=e^2/2C$. The first terms is the kinetic energy stored in capacitor and last term is the potential energy stored in JJ (inductor). Similar to quantization of harmonic oscillator, here $m$ and $\delta$ are canonical momentum and position for the transmon circuit. Therefore, we may transition to the quantum regime by promoting them to be operators and then we arrive at the quantum Hamiltonian,
\begin{eqnarray}
\hat{H}_{\mathrm{trans}}= 4 E_C \hat{m}^2 + E_J (1-\cos \hat{\delta}).
\label{eq:trans_H0}
\end{eqnarray}
Now, we have a Hamiltonian for the transmon circuit. In order to find the energy transitions of the transmon, we need to find the eigenvalues and eigenstates for this Hamiltonian. This Hamiltonian has an analytic solution in the $\hat{\delta}$-basis\footnote{In $\delta$-basis you have $\hat{m}=i\hbar \frac{\partial}{\partial \hat{\delta}}$ then you obtain a solvable 2nd-order differential equation.} in terms of Matthieu functions (see for example Ref.~\cite{devoret2004shortreviwe}). More conveniently, one can truncate the Hilbert space and perform numerical diagonalization\footnote{Numerical calculation in number basis is more convenient  because the first term is diagonal and 2nd term is tri-diagonal, $ \langle m \pm 1 | \cos(\delta) | m \rangle =1/2$. Note, $e^{i\delta} |m\rangle = |m-1\rangle$.} in $\hat{m}$-basis.

In the limit of $E_J/E_C \gg 1$ which implies $\delta\ll1$, one may expand the last term up to the 4th-order of $\delta$ and obtain the harmonic oscillator Hamiltonian plus a nonlinear term,
\begin{eqnarray}
\hat{H}_{\mathrm{trans}}= 4 E_C \hat{m}^2 + E_J \delta^2/2 - E_J \delta^4/24 + \cdots
\label{eq:trans_H1}
\end{eqnarray}
This is convenient approximation because we can follow same procedure for harmonic oscillator quantization and use creation and annihilation operators. Looking at the first two terms in the Hamiltonian~\eqref{eq:trans_H1} in analogy to a harmonic LC circuit\footnote{For this analogy, consider a LC circuit energy as $E=\frac{m^2}{2C} + \frac{\delta^2}{2 L}$ where $m$ and $\delta$ are charge and flux respectively.} we have, 
\begin{subequations}
\begin{eqnarray}
4E_c &\leftrightarrow& \frac{1}{2C}\\
\frac{E_J}{2} &\leftrightarrow& \frac{1}{2 L}\\
\omega_J= \sqrt{8E_J E_c} &\leftrightarrow& \omega_{LC}=\frac{1}{\sqrt{LC}}
\end{eqnarray}
\end{subequations}
Similarly one can define creation and annihilation operators\footnote{ Here we have, $\hat{\delta}=  \sqrt{\frac{\hbar Z_R}{2}} (\hat{b} + \hat{b}^\dagger), \hat{m}= i  \sqrt{\frac{\hbar}{2 Z_R}} (\hat{b} - \hat{b}^\dagger)$ , where $Z_R=\sqrt{ \frac{8 E_C}{E_J}}$.} and write down the Hamiltonian~\eqref{eq:trans_H1} in terms of $\hat{b}$ and $\hat{b}^\dagger$,
\begin{eqnarray}
\hat{H}_{\mathrm{trans}}&=& \omega_J \hat{b}^\dagger \hat{b}  - \frac{E_C}{12} ( \hat{b} + \hat{b}^\dagger )^4 + ... \nonumber  \\
&=& \omega_J \hat{b}^\dagger \hat{b}  - \frac{E_C}{2} ( \hat{b}^\dagger \hat{b}^\dagger \hat{b}  \hat{b} + 2 \hat{b}^\dagger \hat{b} ) + ...,
\label{eq:trans_H_bb}
\end{eqnarray}
where the first term comes from first two terms in Equation~\eqref{eq:trans_H1} and the last terms comes from the third term in Equation~\eqref{eq:trans_H1}.\\

\noindent\fbox{\parbox{\textwidth}{
\textbf{Exercise~7:} Derive Equation~\eqref{eq:trans_H_bb}. Note that you will need normal ordering and the rotating wave approximation to ignore the terms that do not conserve the energy.
}} \vspace{0.25cm}

\begin{figure}[ht]
\centering
\includegraphics[width = 0.9\textwidth]{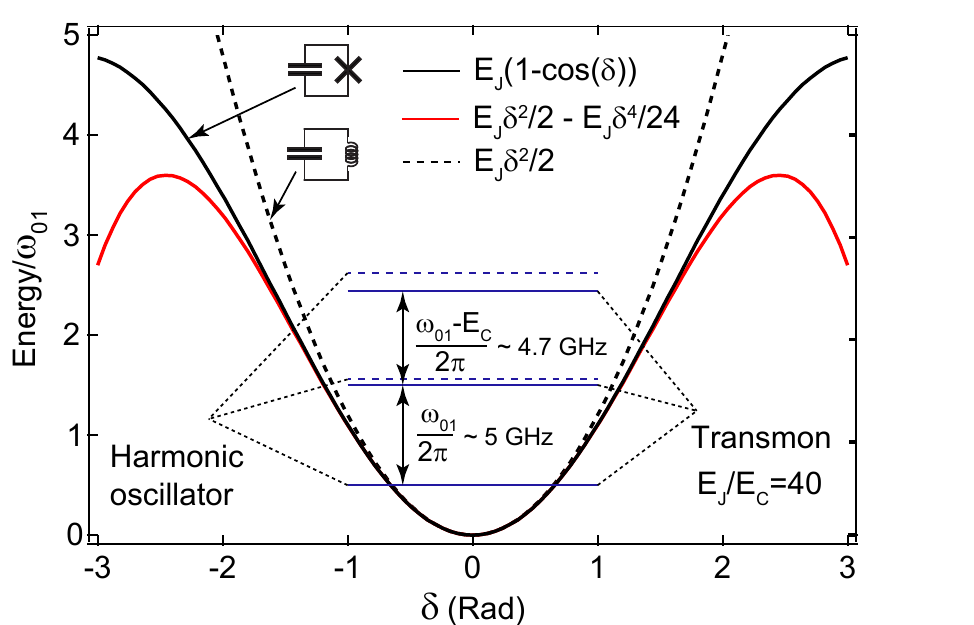}
\caption[Transmon energy levels]{ {\footnotesize \textbf{Transmon energy levels:} A typical transmon ($E_J/E_C=40$) potential (Eq~\eqref{eq:trans_H0} in solid black curve) in comparison with a nonlinear oscillator (Eq.~\eqref{eq:trans_H1} in the solid red curve) and a harmonic oscillator (dashed parabola). The first three energy levels are also depicted for the transmon (nonlinear oscillator) in comparison to the harmonic oscillator. The typical values for transition energies/frequencies are shown. Note $E_{01}=\hbar \omega_{01} $ and we set $\hbar=1$.}} 
\label{fig:anharm_energylevel}
\end{figure}

One can rearrange Equation~\eqref{eq:trans_H_bb} in this form,
\begin{subequations}
\begin{eqnarray}
\hat{H}_{\mathrm{trans}}&=& ( \omega_J - E_C) \hat{b}^\dagger \hat{b}  - \frac{E_C}{2} \hat{b}^\dagger \hat{b}^\dagger \hat{b}  \hat{b} \\
&=& \omega_{01} \hat{b}^\dagger \hat{b}  + \frac{\alpha}{2} \hat{b}^\dagger \hat{b}^\dagger \hat{b}  \hat{b},
\label{eq:trans_H_bb2}
\end{eqnarray}
\end{subequations}
where we arrive at a Hamiltonian for an anharmonic oscillator with a lower energy transition $\omega_{01}= \sqrt{8E_J E_C} -E_C$ and an anharmonicity $\alpha/2\pi=-E_C$ as in shown in Figure~\ref{fig:anharm_energylevel}.\\

\noindent\fbox{\parbox{\textwidth}{
\textbf{Exercise~8:} Find the first three eigenvalues for the anharmonic oscillator Hamiltonian \eqref{eq:trans_H_bb2}. You may use perturbation theory. In case you prefer to do this numerically is makes sense to do it for the original Hamiltonian~\eqref{eq:trans_H0}. 
}} \vspace{0.25cm}

With reasonable anharmonicity $\alpha/2\pi=E_{12}-E_{01}$ (typically $\alpha/2\pi\sim -300$ MHz) we can individually address the lower states and leave higher levels intact\footnote{This is true as long as the Rabi oscillation we induced in the lower transition is much less that anharmonicity, $\Omega_R \ll \alpha$.}. Therefore we consider a transmon circuit as a two level system which can be described as a pseudo-spin with the Pauli operator,
\begin{eqnarray}
\hat{H}_q=-\frac{ \omega_q}{2} \sigma_z,
\label{eq:Hq}
\end{eqnarray} 
where $\omega_q = \omega_{01}$ the lowest transition in the transmon circuit\footnote{The minus sign is because we use the NMR convention in which $\langle \sigma_z \rangle =1$ for the ground state.}.

\section{Qubit-cavity interaction} 
In previous sections, we quantized a single mode of the electromagnetic field for a cavity and showed that it results in a harmonic oscillator Hamiltonian (Eq.~\ref{eq:Haadag}). In this section, we consider only the lowest mode of the cavity ($m=1$) which has the minimum frequency. This mode has maximum electromagnetic field amplitude at the center of the cavity ($z=L/2$). Here, we study the interaction between this mode of the cavity (as a quantum harmonic oscillator)  and a two-level quantum system (qubit) which is represented by Hamiltonian~\eqref{eq:Hq}. Assume that we place the qubit right at the center of the cavity. The dimension of the qubit is much smaller than the dimension of the cavity therefore with a good approximation, the qubit only interacts with the electromagnetic field at $z=L/2$ as depicted\footnote{The assumption that the qubit interacts only with the electromagnetic field at the center of the cavity is a classical interpretation. In quantum picture, each photon is a packet of energy extended to the entire cavity. But this classical picture is very clear to convey the fact that by placing the qubit at the center of the cavity, statistically, the qubit experiences a stronger electromagnetic field.} in Figure~\ref{fig:qubit_cavity}.
\begin{figure}[ht]
\centering
\includegraphics[width = 0.68\textwidth]{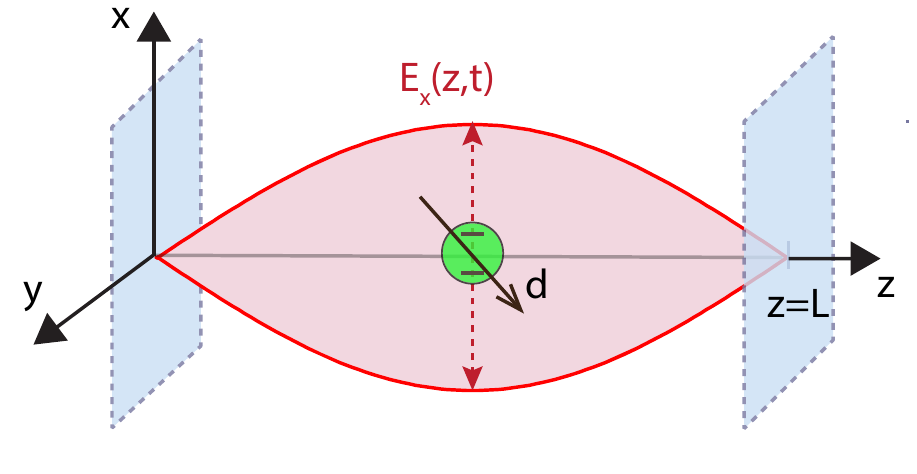}
\caption[The qubit-cavity interaction]{ {\footnotesize \textbf{The qubit-cavity interaction:} The qubit is placed at the center of the cavity where the electromagnetic field is maximum for the first mode of the cavity. The qubit interacts with the electric field via its electric dipole $d$.}} 
\label{fig:qubit_cavity}
\end{figure}

The qubit interacts via its electric dipole moment to the electric field of the cavity via the interaction Hamiltonian,
\begin{eqnarray}
H_{int}= - \hat{d} \ \cdot \ \hat{E}_x(\frac{L}{2},t) \hspace{0.1cm} , \mbox{where} \hspace{0.3cm}  \hat{d}=
  \left( {\begin{array}{cc}
   0 & d \\
   d^{*} & 0 \\
  \end{array} } \right).
\end{eqnarray}
The parameters $d$ is the magnitude of the dipole of the qubit which can be in any direction. Let's define $d_x$ as the magnitude of the qubit dipole aligned with electric field of the cavity. Then the effective dipole operator can be represented as $\hat{d}=d_x \sigma_x=d_x ( \sigma_+ + \sigma_-)$  where $\sigma_+$ ($\sigma_-$) are the raising (lowering) operators for the qubit. Without loss of generality, we can assume $d_x$ is real\footnote{Note, the complex $d_x$ means that the electric dipole has non-zero moment along $\sigma_y$.}. Then the interaction Hamiltonian reads,
\begin{eqnarray}
H_{int}= -  g (\hat{a}  + \hat{a}^{\dagger} )(\sigma_+  + \sigma_- ),
\label{eq:Hint1}
\end{eqnarray}
where we use Equation~\eqref{eq:Eaadag} and define $g=d_x \mathcal{E}_0$ to quantify the interaction strength or qubit-cavity coupling energy\footnote{If we place the qubit off-center the coupling  $g$ would be smaller. In fact, the placement of the qubit inside the cavity is, to some extent, a knob to adjust the qubit-cavity coupling.}.

\subsection{Jaynes-Cummings model} 
Now we have all the pieces to describe the combined qubit-cavity system. Note that the qubit Hamiltonian (Eq.~\ref{eq:Hq}) by itself has two eigenstates $\{ |g\rangle, |e\rangle\}$ corresponding to two eigenvalues (energies)  $\{\mp  \omega_q/2\}$. Similarly, a single cavity mode Hamiltonian (Eq.~\ref{eq:Haadag}) by itself has an infinite number of eigenstates $\{|n\rangle \}$ with eigenvalues $\{  \omega_c ( n+ 1/2) \}$ corresponding to $n$ photons in that mode. Here we are interested to know what are the eigenstates and eigenvalues of the hybrid system of the cavity and qubit combined via the interaction Hamiltonian (Eq.~\ref{eq:Hint1}). The total Hamiltonian\footnote{Here we refer to it as the Rabi Hamiltonian ---the JC Hamiltonian comes from the Rabi Hamiltonian once taking the RWA.} has three parts,
\begin{eqnarray}
H_{\mathrm{Rabi}}=  \omega_c (\hat{a}^{\dagger}\hat{a}  + \frac{1}{2} ) -\frac{1}{2}  \omega_q \sigma_z  -  g (\hat{a}  + \hat{a}^{\dagger} )(\sigma_-  + \sigma_+ ).
\label{eq:RabiH}
\end{eqnarray}
In the case of no interaction between qubit and cavity ($g=0$) the eigenstates of the qubit-cavity system are simply the tensor product of the cavity and qubit eigenstates $\{ |g\rangle|n\rangle , |e\rangle|n\rangle \}$ which are called bare states or the bare basis and, obviously, with eigenvalues that are simply the sum of eigenvalues for each qubit and cavity eigenstates, $\{ \pm  \omega_q/2 +  \omega_c (n+1/2) \}$.
\begin{eqnarray}
|g\rangle|0\rangle &\rightarrow& \mbox{{\small qubit in ground state, no photons in the cavity}}\\
|g\rangle|n+1\rangle &\rightarrow& \mbox{{\small qubit in ground state, $n+1$ photons in the cavity}}\\
|e\rangle|n\rangle &\rightarrow& \mbox{{\small qubit in excited state, $n$ photons in the cavity}}
\end{eqnarray}
However bare states no longer are the energy eigenstates for the system when the qubit and cavity interact ($g\neq0$). Yet, we can represent the total Hamiltonian in the bare basis and attempt to diagonalize it to find its eigenstates and eigenvalues. Before we do this, we simplify the interaction Hamiltonian by the rotating wave approximation (RWA). This approximation is valid in most practical situations where the coupling strength is much less than both the qubit and cavity frequency, $g\ll \omega_q, \omega_c$, and also $|\omega_c - \omega_q| \ll | \omega_c+\omega_q|$. Having this situation in mind, let's revisit the interaction Hamiltonian where we have four terms, 
\begin{eqnarray}
H_{int} \Rightarrow \hat{a}^{\dagger} \sigma_-   +   \hat{a} \sigma_+  + \hat{a}^{\dagger} \sigma_+  + \hat{a} \sigma_-
\end{eqnarray}
The first term describes `the decay of the qubit and creation of a photon for the cavity' and second term accounts for `an excitation of the qubit and annihilation of a photon in the cavity'. These processes somehow ``conserve" the total energy in the system since the energy change would be $\pm (\omega_c - \omega_q)$, which is much less that the total energy in the system even in the few photon regime where $E_{tot} \sim  \omega_c + \omega_q$. However, the last two terms correspond to `the excitation (decay) of the qubit and creation (annihilation) of a photon for cavity' which requires a relatively substantial energy change $\pm (\omega_c + \omega_q)$ in the system, especially when we have only a few photons in the system. This means that the last two processes are much less likely to occur compared to the first two processes so we can simply ignore those terms\footnote{One would expect RWA breaks in the regime of many photons. See for example \cite{khezri26beyand,sank16beyond} for beyond RWA.}. This also can be understood from energy-time uncertainty principle which implies that the last two processes happen on much faster time-scales and normally are averaged out compared to the first two processes\footnote{For example, see chapter~4 of the Ref.~\cite{gerr05} for more detailed discussion of RWA }. Therefore with this rotating wave approximation (RWA) we obtain the Jaynes-Cummings Hamiltonian,
\begin{eqnarray}
H_{\mathrm{JC}}=  \omega_c (\hat{a}^{\dagger}\hat{a}  + \frac{1}{2} ) -\frac{1}{2}  \omega_q \sigma_z  -  g ( \hat{a}^{\dagger} \sigma_-  + \hat{a} \sigma_+ ).
\label{eq:JCH}
\end{eqnarray}
Although the RWA simplifies the Hamiltonian, still we have to deal with an infinite dimensional Hilbert space (since the number of photons $n$ ranges from $0 \rightarrow \infty $) which means the Hamiltonian is a semi-infinite matrix which makes it tricky to diagonalize. Normally in such situation we truncate the Hilbert space at some point, but fortunately in this case we can go around this problem and diagonalize the Hamiltonian in the infinite dimension Hilbert space. If we use the bare basis to represent the $H_{JC}$ in the form of matrix we find,
\begin{eqnarray}
H_{JC} =  \left(
\begingroup\makeatletter\def\f@size{8}\check@mathfonts
\def\maketag@@@#1{\hbox{\m@th\large\normalfont#1}}%
 {\begin{array}{cccccc}
  \frac{1}{2} \omega_c- \frac{\omega_q}{2} & 0 & 0 & 0& 0 & 0\\
   0 & \frac{3}{2} \omega_c - \frac{\omega_q}{2}  & g & 0 & 0 & 0\\
   0 & g  & \frac{3}{2} \omega_c + \frac{\omega_q}{2} & 0 & 0 & 0\\
     &    &   & ... &   &  \\
   0 & 0 & 0 & 0  & (n +\frac{1}{2}) \omega_c - \frac{\omega_q}{2}& \sqrt{n+1}g \\
   0 & 0 & 0&0 & \sqrt{n+1}g &  (n +\frac{1}{2}) \omega_c + \frac{\omega_q}{2} \\
  \end{array} } \endgroup
  \right),
\end{eqnarray}
which shows the Hamiltonian is block-diagonal and all blocks follow a general form (except the first block which has only one element $\frac{1}{2} \omega_c- \frac{\omega_q}{2}$ corresponding to the absolute ground state of the system). Having a block-diagonal Hamiltonian makes it easy to find its eigenvalues. We only need to diagonalized individual blocks and the resulting eigenvalues of each block indeed are the eigenvalues of the entire Hamiltonian. For each block $M_n$ we have,
\begin{eqnarray}
M_n =   \left( {\begin{array}{cc}
 (n +\frac{1}{2}) \omega_c - \frac{\omega_q}{2} & \sqrt{n+1}g \\
 \sqrt{n+1}g &  (n +\frac{1}{2}) \omega_c + \frac{\omega_q}{2} \\
  \end{array} } \right),
\end{eqnarray}
where $n=1,2,\ldots$. The eigenstates of $M_n$ and $|g\rangle |0\rangle$ corresponding to $n=0$, form a complete set of eigenstates for the entire qubit-cavity system. For the eigenvalues we have,
\begin{eqnarray}
E_g &=& -\frac{\Delta}{2} \\
E\mp &=& (n+1)\omega_c  \mp \frac{1}{2} \sqrt{4g^2 (n+ 1) + \Delta^2} .\label{eq:dressEpm}
\end{eqnarray}
where $\Delta=\omega_q-\omega_c$. The eigenstate associated with each of these eigenvalues are called the \textit{dressed states} of the qubit and cavity, 
\begin{eqnarray}
|0,-\rangle &=& |g\rangle |0\rangle \\
|n,-\rangle &=& \cos(\theta_n) |g\rangle |n +1\rangle -\sin(\theta_n) |e\rangle |n\rangle \label{eq:dressm}\\
|n,+\rangle &=& \sin(\theta_n) |g\rangle |n +1\rangle + \cos(\theta_n) |e\rangle |n\rangle \label{eq:dressp}
\end{eqnarray}
where $\theta_n= \frac{1}{2} \tan^{-1} (2g \sqrt{n+1} / \Delta)$ which quantifies the ``level of hybridization". In the limit of $\Delta \rightarrow 0$ where qubit and cavity have a the same energy we have $\theta_n=\pi/4$ and the dressed states are in maximum hybridization, 
\begin{eqnarray}
|n, \mp \rangle &=& \frac{1}{\sqrt{2}} \left(  |g\rangle |n +1\rangle \mp  |e\rangle |n\rangle  \right),
\label{eq:polaritonmp}
\end{eqnarray}
which means each of the dressed states has a 50 \%-50 \% characteristic of the cavity photon and qubit excitations. These states are called \emph{polaritons}. The energy difference between the first two polariton states is $2 g$.

A nice way to look at dressed state energy levels is by comparing them to the corresponding uncoupled system energy levels, the bare states. For that, consider Figure~\ref{fig:avoidcross2} where we display the energy levels of an uncoupled qubit-cavity system compared to the dressed state energy levels for different values of the qubit-cavity detuning.
\begin{figure}[ht]
\centering
\includegraphics[width = 0.98\textwidth]{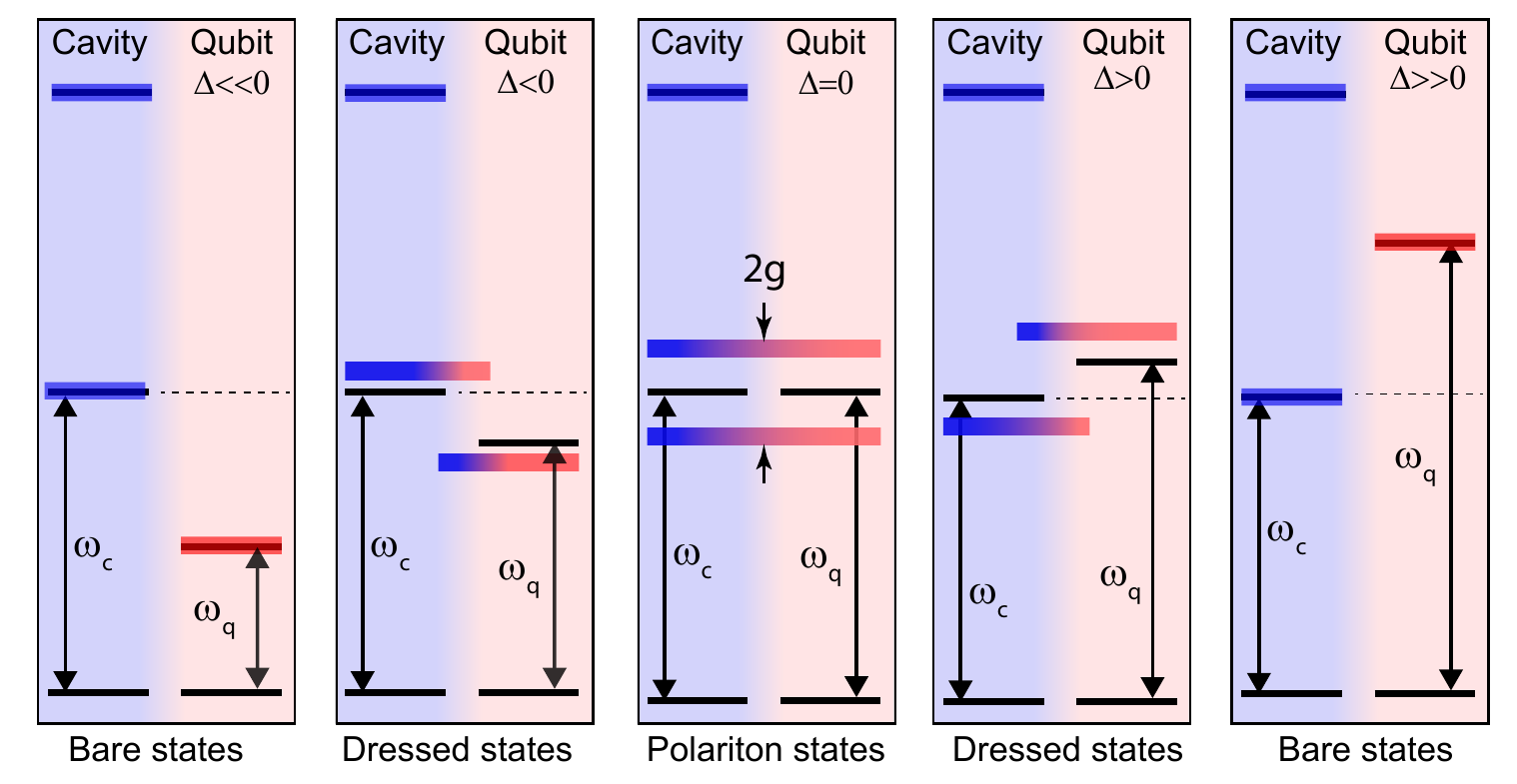}
\caption[Dressed states vs bare states]{ {\footnotesize \textbf{Dressed states vs bare states:} The panels illustrate the dressed states of the qubit-cavity system for different qubit-cavity detunings in comparison with the bare states (refer to the main text for a more detailed description). Note that this illustration is not accurate and lacks some details but we rather to avoid them here.} }
\label{fig:avoidcross2}
\end{figure}
The bare state energy levels are depicted by solid black lines. The dressed states are depicted by bars that are color-coded by blue (red) for cavity- (qubit-) like states. In the first panel, the qubit and cavity are far detuned ($\Delta \ll 0$) which means $\theta_n \simeq 0$ and the effective coupling is negligible. Therefore the dressed states energy levels almost overlap with the uncoupled cavity-quit state, the bare states (as depicted in panel 1). In the second panel, we change the energy level for the qubit. The detuning $\Delta$ is still negative but it is getting smaller and smaller in terms of magnitude. The dressed states start pushing away each other and deviate from the corresponding bare states. In this situation, $\theta_n \in(0, \pi/4)$ and the upper dressed state acquires some qubit character, and similarly, the lower dressed state acquire some photon character. In panel three $\Delta=0$ and the hybridization is its maximum level, $\theta_n=\pi/4$ and the dressed states (which we now call polaritons) push each other away and deviate maximally from the bare states. The separation between two polaritons is $2 g$. Now both polaritons have acquired equal photon and qubit character as depicted by color-coded bars in panel 3. If we further increase the energy level of the qubit (see panel 4) then again we get dressed states. Note that in panel 4, unlike in panel 3, the lower (upper) polariton has more photon (qubit) character. By increasing the detuning further, as in panel 5, we effectively decouple the qubit and cavity and the dressed states again approach the bare states. If we keep increasing the qubit frequency even further then the qubit energy will approach the higher level of the cavity and we would see another avoided crossing corresponding to $n=1$. Every time qubit level crosses one of the cavity levels, we may expect an avoided crossing and hybridization\footnote{Considering the higher energy levels of the cavity one might think that it is also possible that qubit level couples to two or multiple cavity energy levels at the same time. This is true, but usually, this effect is only significant when the qubit-cavity coupling is so strong ($g\sim \omega_c,\omega_q$) that qubit and cavity energy levels push each other even when they are far detuned. This regime is known as ultra strong coupling~\cite{niemczyk2010circuit,bosman2017multi}. But normally the coupling rate $g\ll \omega_c, \omega_q$. Therefore, in order to have hybridization the qubit energy has to be very close to the cavity energy ($\Delta\ll \omega_c,\omega_q$). In our case, we can safely assume that qubit effectively couples only to one cavity energy level at a time. However, I should warn you that in our description of the avoided crossing which is represented in Figure~\ref{fig:avoidcross2}, we have ignored the higher transmon energy levels which would make the situation much more complicated. Considering the transmon as a two-level system is good for intuition, but to be accurate one must include more transmon levels.}.

It is convenient to plot transition energy versus detuning since (as we will see in Chapter~3) we normally characterize the system by measuring the transition frequencies by doing spectroscopy. For example, when $n=0$ we have, 
\begin{eqnarray}
E_\mp - E_g &=& \omega_c  \mp \frac{1}{2} \sqrt{4g^2 + \Delta^2} +\Delta/2 .\label{eq:dresstrans}
\end{eqnarray}
In Figure~\ref{fig:avoidcross1}, we plot the energy $E_{\pm} - E_g $ versus detuning which clearly shows the avoided crossing. The transition energy levels are color coded so that again red (blue) is the qubit- (photon-) like transition.
\begin{figure}[ht]
\centering
\includegraphics[width = 0.48\textwidth]{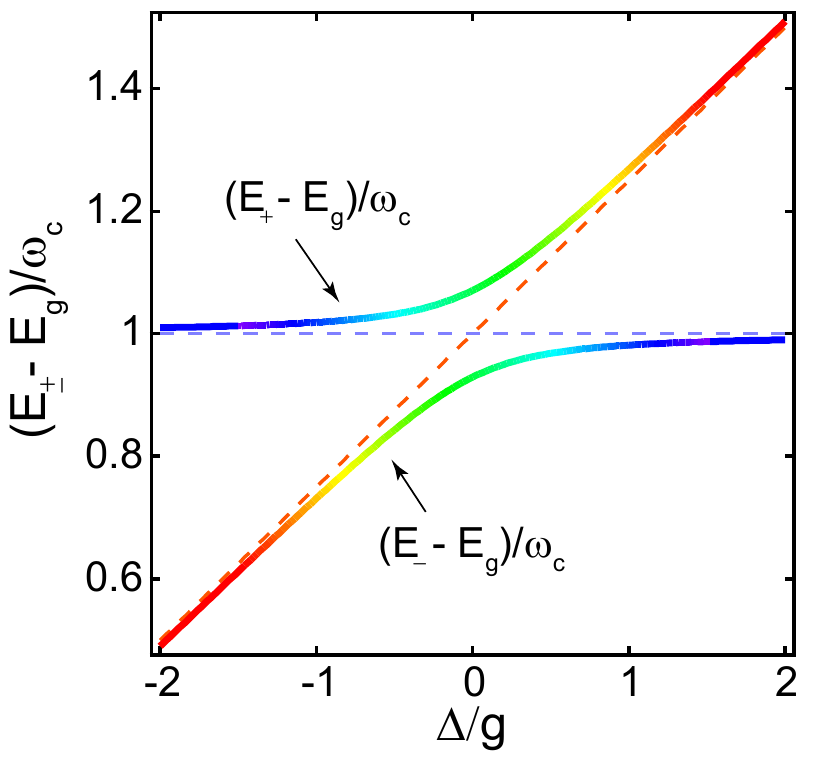}
\caption[Avoided crossing:]{ {\footnotesize \textbf{Avoided crossing:} The transition energy from higher and lower dressed states to the ground state versus the detuning $\Delta$. The transition energy is scaled by the energy of the cavity $\omega_c$ and the detuning is scaled by the coupling rate $g$}. The dashed lines indicate the bare states' transition. Note that you can somehow see a similar avoided-crossing curve in Figure~\ref{fig:avoidcross2} by connecting the upper (lower) dressed states in different detunings together.} 
\label{fig:avoidcross1}
\end{figure}

In this section, we learned that if we put a qubit inside a cavity, the energy levels hybridize and we have dressed states. Yet, just as we considered transmon as a two-level system (TLS) by addressing only lower transition, here also we consider the ground state and the lower dressed state as our new qubit.
\subsection{Dispersive approximation}
In this section, we perform another approximation to the interaction Hamiltonian. This approximation is valid in the regime that cavity and qubit are far detuned $\Delta\gg g$. In such situations, the interaction is relatively weak. In principle, in this regime, the cavity and qubit do not directly exchange energy unlike what we explicitly have in the interaction term\footnote{Note that this doesn't mean that in this limit the JC interaction term is not valid. It means that the effect of the coupling is so weak such that we can approximately represent the Hamiltonian in a simpler form.} in the JC Hamiltonian \eqref{eq:JCH}.
For that, consider the unitary transformation $$\hat{T}=e^{\lambda(\sigma_- a^{\dagger} - \sigma_+ a)},$$ where $\lambda=\frac{g}{\Delta}$. If we apply this transformation\footnote{Applying a unitary transformation is somehow a change of frame. So we do not add/remove any physics.} to the JC Hamiltonian~\eqref{eq:JCH} and use the Baker-Campbell-Hausdorff relation to evaluate all terms up to order $\lambda^2$ we have,
\begin{eqnarray}
\hat{T} \ \hat{H}_{JC} \ \hat{T}^{\dagger} =  \omega_c (\hat{a}^{\dagger}\hat{a}  + \frac{1}{2} ) -\frac{1}{2}  \omega_q \sigma_z -  \frac{g^2}{\Delta} \hat{a}^{\dagger} \hat{a}\sigma_z +  \frac{g^2}{2 \Delta} \sigma_z.
\label{eq:dispersive_trans}
\end{eqnarray}
We may ignore constant terms\footnote{The term $  \frac{g^2}{2 \Delta} \sigma_z$ (Lamb shift) is also a constant shift in qubit frequency that we can absorb it into $\omega_q$.} since these do not affect dynamics, and obtain the JC Hamiltonian in the dispersive limit,
\begin{eqnarray}
\hat{H}_{\mathrm{dis}}  =  \omega_c \hat{a}^{\dagger}\hat{a} - \frac{1}{2}  \omega_q \sigma_z -  \frac{g^2}{\Delta} \hat{a}^{\dagger} \hat{a}\sigma_z.
\label{eq:H_dis}
\end{eqnarray}
\noindent\fbox{\parbox{\textwidth}{
\textbf{Exercise~9:} Show that Equation~\eqref{eq:dispersive_trans} is true by using Baker-Campbell-Hausdorff relation, $$e^{\lambda \hat{B}} \hat{A} e^{-\lambda \hat{B}} = \hat{A} + \lambda [\hat{B},\hat{A}] + \frac{\lambda^2}{2!}  [\hat{B},  [\hat{B},\hat{A}]] + \mathcal{O}[\lambda^3],$$
and keeping the terms up to the order $\lambda^2$.
}} \vspace{0.25cm}

The dispersive Hamiltonian~\eqref{eq:H_dis} describes the situation were the cavity and qubit are far detuned and coupling is weak and dressed states are almost overlapping with the bare states (see Figure~\ref{fig:avoidcross2} panel 1). Yet, there is a very small interaction as described by the last term in Equation~\eqref{eq:H_dis}. In order to make better sense of this interaction, we re-arrange the terms in Equation~\eqref{eq:H_dis} as follows,
\begin{eqnarray}
\hat{H}_{\mathrm{dis}}  = ( \omega_c - \chi \sigma_z ) \hat{a}^{\dagger}\hat{a} - \frac{1}{2}  \omega_q \sigma_z,
\label{eq:H_dis_rearange}
\end{eqnarray}
where $\chi=\frac{g^2}{\Delta}$ is the dispersive shift or dispersive coupling rate\footnote{Note that we define $\Delta=\omega_q-\omega_c$ and usually we prefer to have $\omega_q<\omega_c$ because of the transmon higher levels and also to avoid coupling to higher frequency cavity modes~\cite{koch2007charge}. Therefore the dispersive coupling is often negative.}. We see that the dispersive interaction is manifested as a qubit-state-dependent frequency shift for the cavity. If the qubit is in the ground (excited) state $|g\rangle$ ($|e\rangle$) then $\langle \sigma_z \rangle =1$ ($\langle \sigma_z =-1 \rangle$) which means that the cavity frequency shifts by $+\chi$ ($- \chi$). Therefore one can detect this frequency shift for the cavity to determine the state of the qubit. 

Alternatively, one can rearrange the terms in (\eqref{eq:H_dis}) as,
\begin{eqnarray}
\hat{H}_{\mathrm{dis}}  = \omega_c  \hat{a}^{\dagger}\hat{a} - \frac{1}{2}  ( \omega_q +2  \chi \bar{n})  \sigma_z,
\label{eq:H_dis_rearange2}
\end{eqnarray}
and interpret the interaction as a shift in qubit frequency due to photon occupation ($\bar{n} =\hat{a}^{\dagger}\hat{a}$) in the cavity\footnote{In chapter 4 we will use this interpretation to calibrate dispersive shift and average photon number in the system.}.

\section{Dynamics of a driven qubit}
In this section we discuss some of the most basic and important dynamics of the qubit. Essentially, we want to know what happens to the qubit if we continuously drive it with a coherent signal. We may take two approaches to solve this problem. One approach is semi-classical, where we treat the coherent drive as a classical signal. The other approach is fully quantum, where we treat the drive as a coherent state of light. For most purposes, the semi-classical approach works perfectly fine and captures almost all the physics we are interested in. Therefore, we discuss the semi-classical approach (for fully quantum mechanical approach see Ref.~\cite{gerr05}).
\subsection{Rabi oscillations: The semi-classical approach}
We are interested in qubit dynamics and we ignore the cavity for now\footnote{We have qubit inside the cavity and the qubit and cavity are already hybridized and we consider lowest two levels of system (ground state and the lowest dressed-state, or polariton state) as our new qubit. Moreover we assume that the qubit drive is off-resonant with the cavity transition. Therefore in this situation we effectively have just a qubit. Although experimentally the cavity still plays a crucial rule in terms of noise protection and will be essential for qubit readout, this is not our focus in this section.}. With that, assume we have a qubit with Hamiltonian $\hat{H}_q = -  \omega_q \sigma_z/2$ and electric dipole moment $\hat{d}= \vec{d} \sigma_x$. The qubit interacts with the electric field of the coherent light (a classical signal) $E(t) = E \cos(\omega_d t)$ by the interaction Hamiltonian $H_{int}= -\vec{E} \cdot \hat{d}$. Therefore, for the total Hamiltonian we have,
\begin{eqnarray}
H_{\mathrm{semi-classic}}= -\frac{1}{2}  \omega_q \sigma_z  - E(t) \cdot \hat{d}.
\label{eq:RabiH2}
\end{eqnarray}
For simplicity we assume that the dipole moment of the qubit is aligned with the electric field. Therefore we obtain,
\begin{eqnarray}
H_{\mathrm{semi-classic}}= -\frac{1}{2}  \omega_q \sigma_z  - A \cos(\omega_d t) \sigma_x,
\label{eq:H_SC}
\end{eqnarray}
where $A=Ed$ quantifies how strong the interaction is. Now we want to know how the qubit evolves under this Hamiltonian. There are couple of ways we may solve this Hamiltonian. The first way is to solve the Schr\"odinger equation for this time-dependent Hamiltonian. We start with an \emph{ansatz} instead of starting from scratch. The idea is that if we have no electric field or turn off the interaction, then we know the solution for Hamiltonian~\eqref{eq:H_SC} would be $|\psi \rangle = C_g |g\rangle + C_e |e\rangle$ and its time evolution would be $|\psi (t) \rangle = C_g e^{+i\frac{\omega_q}{2}}|g\rangle + C_e e^{-i\frac{\omega_q}{2}} |e\rangle$ . Now, we hope to find the solutions for \eqref{eq:H_SC} in the form of,
\begin{eqnarray}
|\psi (t) \rangle = C_g(t) e^{+i\frac{\omega_q}{2}}|g\rangle + C_e(t) e^{-i\frac{\omega_q}{2}} |e\rangle,
\label{eq:ansatz}
\end{eqnarray}
where we just let the coefficients also be time-dependent. Now we plug this ansatz into the Schr\"odinger equation,
\begin{eqnarray}
i  \frac{\partial |\psi (t) \rangle }{\partial t}  = \left( -\frac{1}{2}  \omega_q \sigma_z  - A \cos(\omega_d t) \sigma_x \right)  |\psi (t) \rangle
\label{eq:H_ansatz}
\end{eqnarray}
By substituting Equation~\eqref{eq:ansatz} into~\eqref{eq:H_ansatz}, one can obtain two coupled ordinary differential equations (ODEs) for $C_g(t)$ and $C_e(t)$,
\begin{subequations}\label{eq:ODE1_rabi_classic_both}
\begin{eqnarray}
\dot{C}_g&=& i  A \cos(\omega_d t) e^{-i \omega_q t}  C_e, \label{eq:ODE1_rabi_classic}\\
\dot{C}_e&=&i  A \cos(\omega_d t) e^{+i \omega_q t}  C_g.
\label{eq:ODE2_rabi_classic}
\end{eqnarray}
\end{subequations}
In order to solve this analytically, we do a simplification which is nothing but the RWA. First, we expand $\cos(\omega_d t)e^{\pm i \omega_q t}= e^{i (\omega_d \pm \omega_q) t} + e^{-i (\omega_d \mp \omega_q) t}$, then argue that we are not interested in very short timescales in the dynamics. In fact, in practically, we normally are not sensitive to short timescales\footnote{Assuming that the qubit frequency and drive are both in range of $5$ GHz, then the fast oscillatory terms $e^{ \pm i (\omega_d + \omega_q) t}$ oscillate at the 100 picosecond timescale. We are normally interested in qubit dynamics at microsecond timescale. Even for fast 5~ns rotation pulses, many of these fast oscillations are averaged out.}. Therefore we ignore fast rotating terms $e^{ \pm i (\omega_d + \omega_q) t}$ that are comparatively slow to the rotating terms $e^{ \pm i (\omega_d - \omega_q) t}$. Then we have
\begin{subequations}\label{eq:ODE1and2_rabi_c_RWA}
\begin{eqnarray}
\dot{C}_g&=&i  A \  e^{-i (\omega_q -\omega_d) t}  C_e, \label{eq:ODE1_rabi_c_RWA}\\
\dot{C}_e&=&i  A \   e^{+i ( \omega_q - \omega_d) t}  C_g .
\label{eq:ODE2_rabi_c_RWA}
\end{eqnarray}
\end{subequations}
For the qubit initially in the ground state (initial conditions $C_g(0)=1$, $C_e(0)=0$) one can show that the solutions are,
\begin{subequations}\label{eq:ODE1_rabi_c_RWA2_both}
\begin{eqnarray}
C_g(t)&=& \frac{e^{-i \frac{\Delta_d}{2}t}}{\Omega_R} \left( \Omega_R \cos(\frac{\Omega_R}{2}t ) + i \Delta_d \sin(\frac{\Omega_R}{2}t) \right) \label{eq:ODE1_rabi_c_RWA2}\\
C_e(t)&=&i \frac{ A \  e^{+i \frac{\Delta_d}{2}t}}{\Omega_R}  \sin(\frac{\Omega_R}{2}t),
\label{eq:ODE2_rabi_c_RWA2}
\end{eqnarray}
\end{subequations}
where $\Delta_d=\omega_q-\omega_d$ and $\Omega_R=\sqrt{A^2 + \Delta_d^2}$. Having the solution for $|\psi(t)\rangle$, we can obtain the the evolution for any relevant observables. In order to see what the dynamics look like, we may look at the population of the qubit excited state,
\begin{eqnarray}
P_e(t)= |C_e(t)|^2&=& \frac{ A^2 \ }{\Omega_R^2}  \sin^2(\frac{\Omega_R}{2}t).
\label{eq:classic_pe}
\end{eqnarray}
As depicted in Figure~\ref{fig:pe_rabi_classic}, the qubit doesn't respond that much to a far detuned drive, but as the detuning gets smaller the oscillations grow. For an on-resonant drive ($\Delta_d=0$), we have slowest but highest contrast oscillations of the qubit populations.
\begin{figure}[ht]
\centering
\includegraphics[width = 0.9\textwidth]{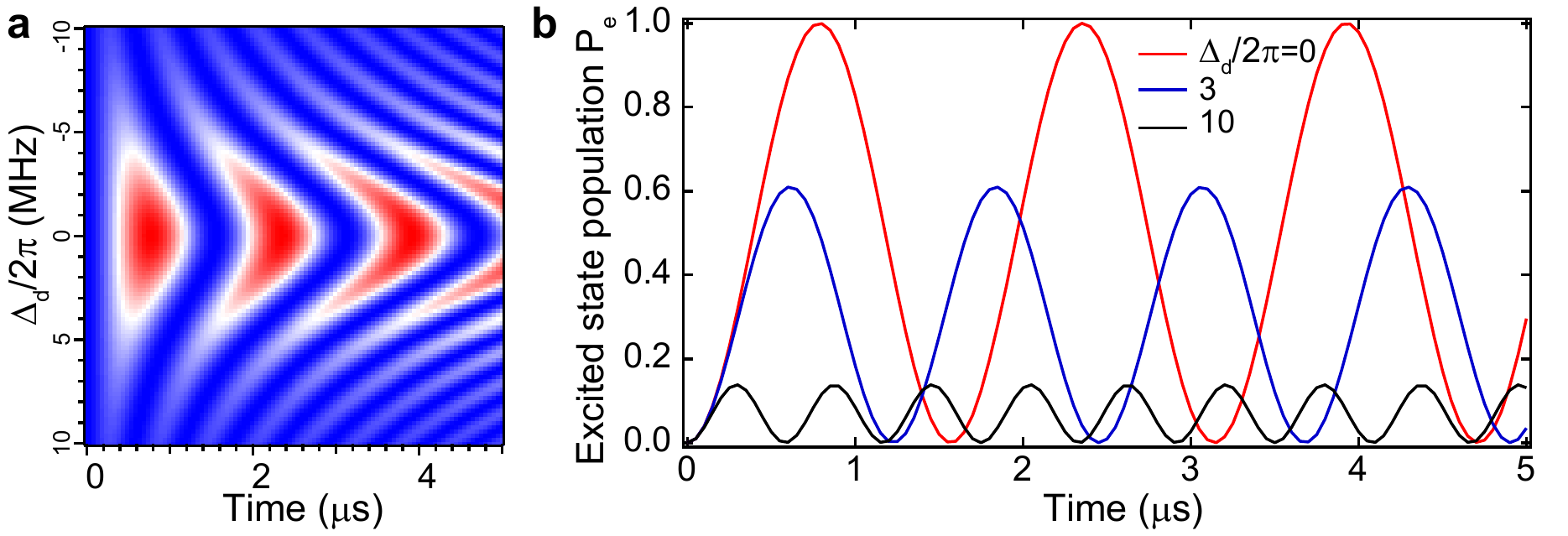}
\caption[Rabi oscillations]{ {\footnotesize \textbf{Rabi oscillations:} \textbf{a}, The chevron plot. The excited state population $P_e$ versus time for different detunings $\Delta_q$.  \textbf{b}, Three cuts from the chevron plot at different detuning values. The on-resonant drive gives the maximum contrast for the oscillations.}} 
\label{fig:pe_rabi_classic}
\end{figure}\\
The fact that we can fully rotate the qubit from the ground to excited by an on-resonant drive is very practical. All we need is to know how strong and how long to drive the qubit with light to put the qubit in the excited state\footnote{$\pi$ pulse calibration! We will see in next chapter how this is done in experiment.}.

\textbf{\emph{Rotating frame--}} There is a rather easy way to solve the Hamiltonian~\eqref{eq:H_SC} by going to a rotating frame of drive. That makes the Hamiltonian time-independent\footnote{This example would be useful to see how rotating frame works. Moreover, this solution will give better picture of detuned Rabi oscillations in the Bloch sphere.}. For this, we transform the Hamiltonian by a unitary operator $U(t)$. The Hamiltonian in the new frame can be evaluated by the following relation,
\begin{eqnarray}
\hat{\mathcal{H} }=&{U}(t)\hat{H}{U}^{\dagger}(t) - i {U}\dot{U}^{\dagger}.
\label{eq:RotatingFrame_Ut}
\end{eqnarray}
Now consider $U(t)=e^{ - i  \frac{\omega_d}{2} \sigma_z t }$, which basically transforms the Hamiltonian to a frame that rotates with the frequency of drive, $\omega_d$. One can show that the Hamiltonian $\eqref{eq:H_SC}$ in the rotating frame of the drive would be
\begin{eqnarray}
\hat{\mathcal{H} }=&- \frac{1}{2} \Delta_d  \sigma_z -  \frac{E d}{2}  \sigma_x,
\label{eq:RotatingFrame_H_sc}
\end{eqnarray}
which no longer has time dependence. Now we may diagonalize the Hamiltonian in qubit energy basis,
\begin{eqnarray}
\hat{\mathcal{H} }= \frac{1}{2} \left( {\begin{array}{cc}
 -\Delta_d & -A  \\
-A & +\Delta_d \\
  \end{array} } \right).
\label{eq:H_sc_Matrix}
\end{eqnarray}
to obtain its eigenvalues, $E_{\pm} = \pm\frac{1}{2} \sqrt{A^2 + \Delta_d^2}$ and eigenstates,
\begin{eqnarray}
|V_-\rangle &=&  \cos(\theta)  |e\rangle - \sin(\theta) |g\rangle,\\
|V_+\rangle &=&  \sin(\theta)  ) |e\rangle + \cos(\theta) |g\rangle,
\label{eq:eigen_Vpm}
\end{eqnarray}
where $\theta=\tan^{-1}(\frac{A}{\sqrt{A^2 + \Delta_d^2}-\Delta_d})$. Figure~\ref{fig:Vpm_bloch} demonstrates the eigenstate $|V_\pm\rangle$ in the Bloch sphere picture.
\begin{figure}
\centering
\includegraphics[width = 0.9\textwidth]{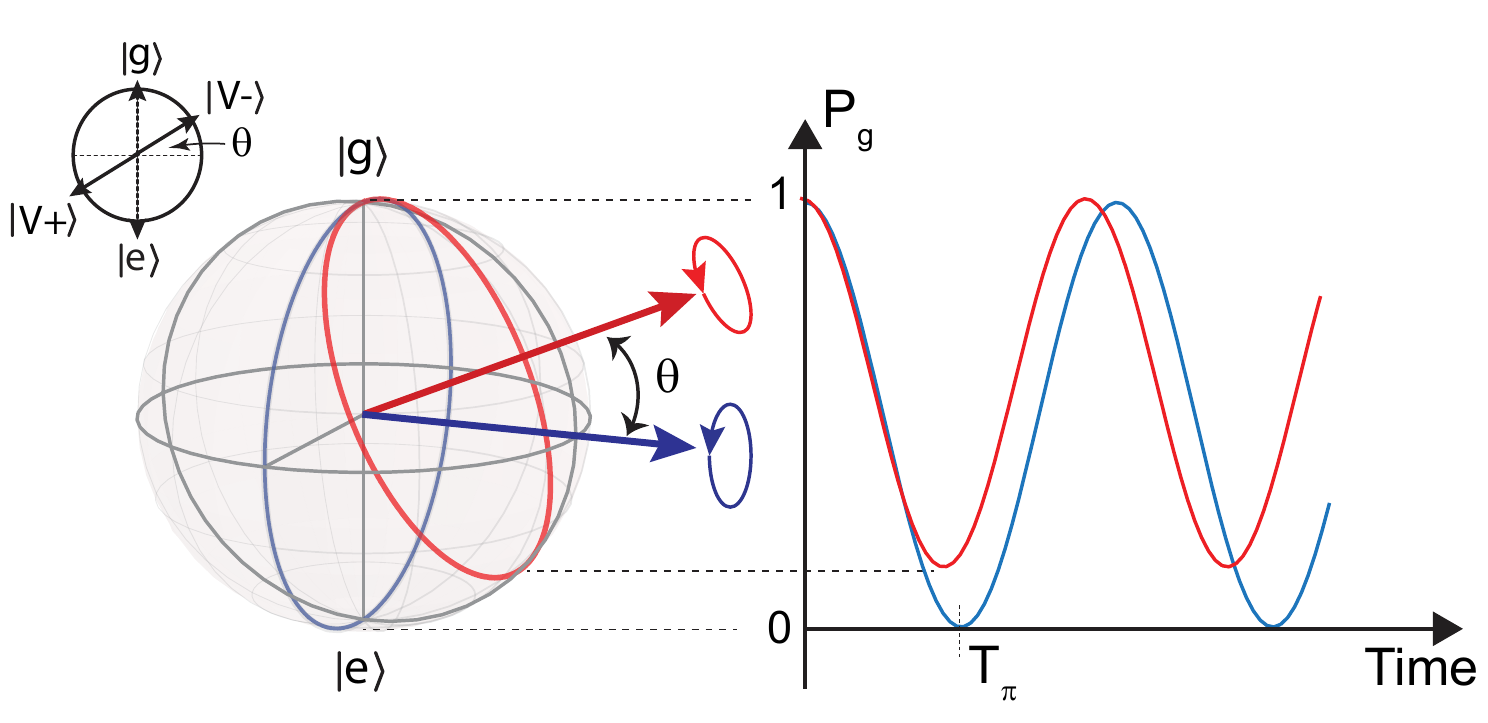}
\caption[Eigenstates on the Bloch sphere for a driven qubit]{ {\footnotesize  \textbf{Eigenstates on the Bloch sphere for a driven qubit:} The eigenstates $|V_\pm \rangle$ for a detuned driven qubit make angle $\theta$ respect to the equator in the Bloch sphere. This picture gives a better understanding of  why the population doesn't reach to the maximum value for a detuned drive.}} 
\label{fig:Vpm_bloch}
\end{figure}
The evolution of the system can be described by
\begin{eqnarray}
|\psi(t)\rangle= C_+ e^{-i E_+t}|V_+\rangle + C_- e^{-i E_-t}|V_-\rangle 
\label{eq:psi_t_vpm}
\end{eqnarray}
where $C_\pm$ are constants determined by the initial condition. We may rewrite Equation~\eqref{eq:psi_t_vpm} in terms of $|g\rangle$ and $|e\rangle$ using~\eqref{eq:eigen_Vpm},
\begin{eqnarray}
|\psi(t)\rangle &=& \left[ C_+ e^{-i E_+t} \sin(\theta)  + C_- e^{-i E_-t} \cos(\theta)  \right] |e\rangle  \nonumber \\ 
 &+& \left[ C_+ e^{-i E_+t} \cos(\theta)  - C_- e^{-i E_-t} \sin(\theta)  \right] |g\rangle
\label{eq:psi_t_ge}
\end{eqnarray}
For example for a qubit starting in the ground state, $|\psi(0)\rangle = |g\rangle$, we have $C_+=\cos(\theta), C_-=-\sin(\theta)$, then the time evolution would be,
\begin{eqnarray}
|\psi(t)\rangle= -i \sin(E_+ t) \sin(2 \theta) |e\rangle + \cos(E_+ t) \sin(2\theta) |g\rangle,
\label{eq:psi_t_ge2}
\end{eqnarray}
where we used the fact that $E_+=-E_-$. Once again we can calculate the expectation value for any observable. For example, the probability of being in the excited state would be\footnote{You may need convince yourself that $\sin^2[ 2 \tan^{-1} ( \frac{A}{\sqrt{A^2 + \Delta_d^2}-\Delta_d})] = \frac{A^2}{\sqrt{A^2 + \Delta_d^2}}$.}, 
\begin{eqnarray}
P_e=\sin^2(2\theta) \sin^2(E_+ t) = \frac{A^2}{A^2 + \Delta_d^2} \sin^2\bigg(\frac{\sqrt{A^2 + \Delta_d^2}}{2} t\bigg),
\label{eq:psi_t_ge3}
\end{eqnarray}
which is consistent with the result we had in lab frame where the Hamiltonian was time-dependent (see Eq.~\ref{eq:classic_pe}). But, this picture gives a visualization of why the population doesn't reach the maximum value for a detuned drive as illustrated in Figure~\ref{fig:Vpm_bloch}.

Generally, in the experiment we use an on-resonance drive to prepare the qubit states. 
Figure~\ref{fig:sum_up_rabi} summarizes our discussion of a driven qubit in the lab and rotating frame by showing Rabi oscillations of a qubit initialized in ground state for both on-resonant and detuned drives.
\begin{figure}[ht]
\centering
\includegraphics[width = 0.69\textwidth]{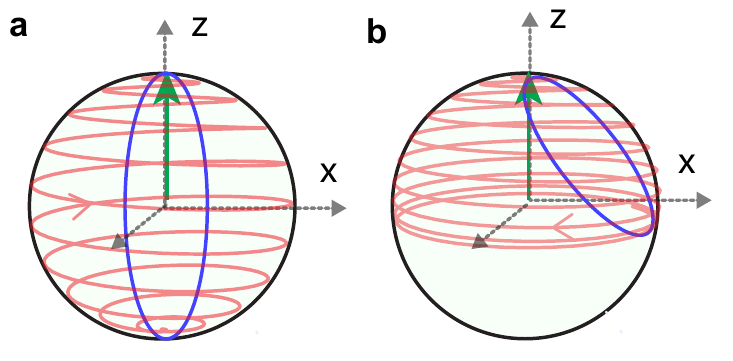}
\caption[Driven qubit evolution in the Bloch sphere]{ {\footnotesize \textbf{Driven qubit state evolution in the Bloch sphere:} The red (blue) line shows the evolution of a driven qubit in the lab frame (rotating frame of the drive) \textbf{a}, for an on-resonant drive and \textbf{b}, for a detuned drive.}} 
\label{fig:sum_up_rabi}
\end{figure}



\subsection{Dynamics in the presence of dissipation}
So far we assume that the qubit is an ideal closed system that undergoes unitary evolution given by the Schr\"odinger equation. However, in reality all systems either classical or quantum are open systems, meaning they are interacting with their environment to some extent. For quantum systems, this interaction degrades the peculiar quantum property of the system (e.g. superposition and entanglement) resulting in energy dissipation and decoherence. Dissipation is a curse in many applications of quantum information. However, dissipation is believed to be the essential piece for allowing the classical laws to emerge out of the underlying quantum laws.

For our purposes, there are two main mechanisms which we need to consider to have a more realistic picture of qubit dynamics: \emph{relaxation} and \emph{dephasing}.

\emph{\textbf{Relaxation}}-- Placing the qubit inside a cavity protects the qubit from environmental noise and limits the available modes the qubit can interact with. Still, the qubit finds some ways to relax its energy and decay to the ground state\footnote{Also, sometimes we intentionally provide the qubit with a decay channel to relax its energy.}. For example, when you prepare the qubit in the excited state, the qubit eventually decays to the ground state and relaxes its energy in the form of a photon to one of the unknown/known decay channels. This process of jumping\footnote{For now, we assume this process happens instantaneously which is a very reasonable and valid assumption. Yet, we will see later that the decay of an atom is not always jumpy.} from $|e\rangle \to |g\rangle$ happens in a random time. This process is not included in the Hamiltonian, therefore we need to account for that somewhat phenomenologically\footnote{In principle one can build these processes into the Hamiltonian if write down the Hamiltonian for the entire universe: qubit+cavity+environment.}.

Let's say you prepare the qubit in the excited state or in some superposition state $|\psi\rangle$. Having the qubit relax after a time, you are not sure if the qubit is still in state $|\psi\rangle$ or has relaxed into the ground state. You might have a mixed feeling about the state of the qubit and, in quantum mechanics, this is an absolutely legitimate feeling because the qubit is indeed in a \emph{mixed state} which can be described by a density matrix $\rho=a |\psi\rangle \langle \psi | + b |g\rangle \langle g | $. Where $a$ is the probability that qubit is still in state $|\psi\rangle$ and $b$ is the probability that qubit has decayed into ground state $|g\rangle$\footnote{Here comes a lesson that I learned very late: In quantum mechanics the state of the system is nothing but your state of knowledge about that system. The reality happens in your mind!?}. If you wait longer, the probability $a$ becomes smaller and smaller while the probability $b$ approaches to 1 meaning that you are getting certain that the qubit is in the ground state. The time scale that qubit spontaneously relaxes its energy is called the \emph{relaxation time} and is indicated by $T_1$. Experimentally, identifying the $T_1$ is a basic step of any qubit characterization process. As depicted in Figure~\ref{fig:T2_time}a, the $T_1$ time is the time by which the population of excited state decays to $1/e$ of its initial value, $P_e(t)=P_e(0)e^{-t/T_1}$. We will see soon how we systematically account for this process in the qubit dynamics. 

\begin{figure}[ht]
\centering
\includegraphics[width = 0.8\textwidth]{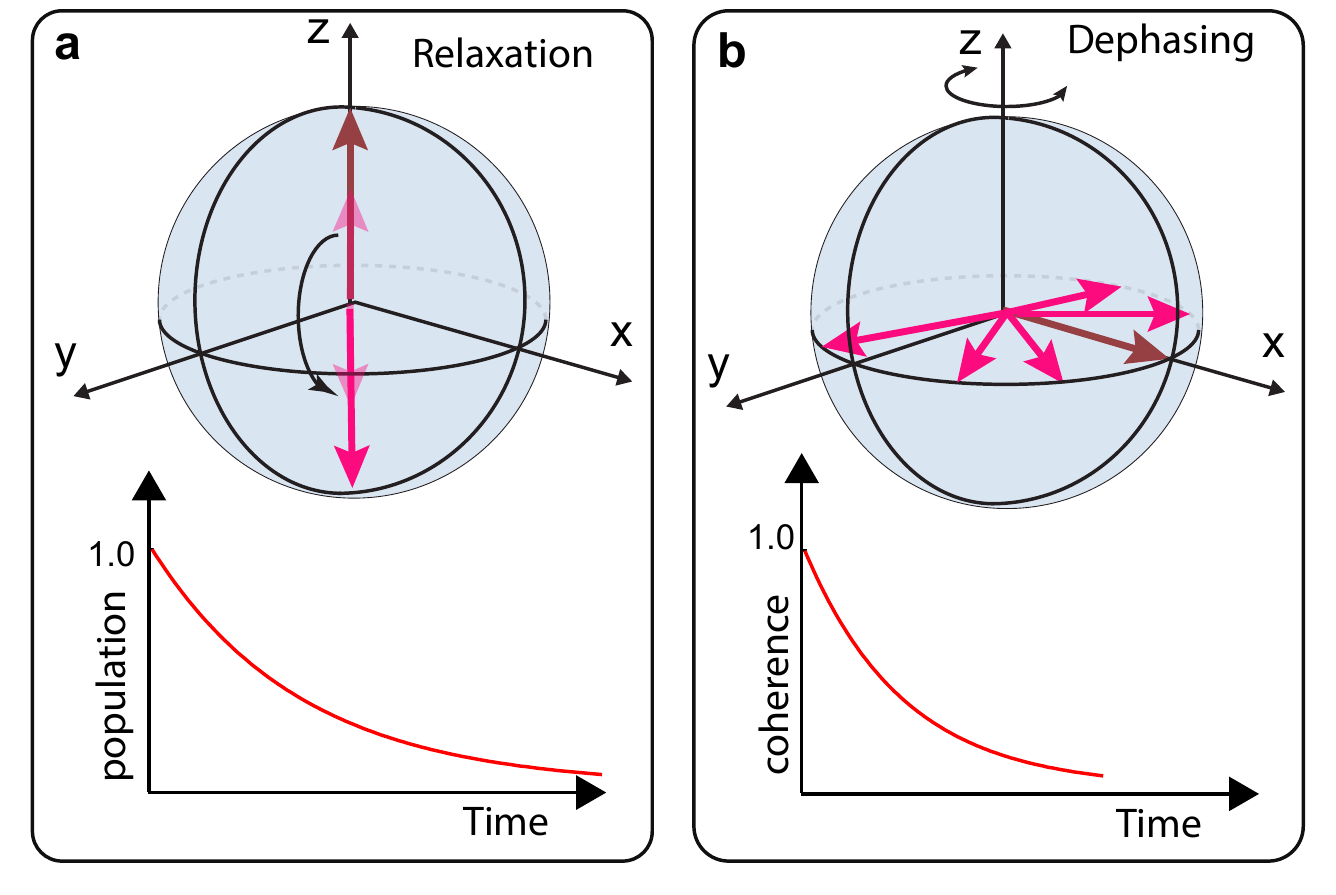}
\caption{ {\footnotesize  Relaxation and dephasing of a qubit.}} 
\label{fig:T2_time}
\end{figure}

\emph{\textbf{Dephasing}}-- There is another imperfection that affects the dynamics of a qubit. In reality, due to the various noise sources in the system, the frequency of the qubit shifts around stochastically. This imperfection doesn't cause the qubit to relax its energy but instead we lose track of the qubit resonance frequency and thus loose track of the phase of the qubit wavefunction. Considering the evolution of a superposition state in the lab frame, the qubit state rotates around the equator of the Bloch sphere. After time $t$ the phase of the qubit would be $\phi=\omega_q t$, however, if the qubit frequency stochastically jitters around $\omega_q$, then the final phase at time $t$  would be $\phi=\omega t + \zeta(t)$, where $\zeta$ is our uncertainty about the phase of the qubit which is growing in time\footnote{$\zeta(t)$ can be considered as a 1D-random-walk distribution.}. Therefore, after a time, again we have a mixed feeling about the state qubit and we lose the quantum coherences as depicted in Figure~\ref{fig:T2_time}b.
The time scale that the qubit loses its coherence is usually called the dephasing time or $T_2^*$ and it is characterized by a Ramsey measurement in experiment.
In the next section, we discuss how to systematically account for relaxation and dephasing in qubit dynamics.

\subsection*{Lindblad master equation }
In order to account for non-unitary and dissipative processes (e.g. dephasing and relaxation) in qubit dynamics, we consider the Heisenberg picture where the unitary evolution of density matrix $\rho$ is described by,
\begin{eqnarray}
\dot{\rho}= -i [\hat{\mathcal{H} }, \rho],
\label{eq:Heisen_lindblad}
\end{eqnarray}
where $\hat{\mathcal{H}}$ is the driven qubit Hamiltonian in the rotating frame---see Eq~\eqref{eq:RotatingFrame_H_sc}. Equation~\eqref{eq:Heisen_lindblad} is equivalent to the Schr\"odinger equation except with the density matrix approach we can also describe unitary evolution for a mixed state. Moreover, the Heisenberg picture allows us  to add more terms for $\dot{\rho}$ to describe other non-unitary processes for the system like dephasing and relaxation\footnote{Normally these non-unitary processes are positive and trace preserving.}. In general, we have
\begin{eqnarray}
\dot{\rho}= -i [\hat{\mathcal{H} }, \rho] + \sum_i \left( L_i^{\dagger} \rho L_i -\frac{1}{2} \{ L_i^{\dagger} L_i , \rho \} \right),
\label{eq:Lindblad_gen}
\end{eqnarray}
where each $L_i$ is an ``Lindbladian" operator describing a specific non-unitary process. For example, the Lindbladian operator for the relaxation process of a qubit is $L_{\mathrm{relaxation}}=\sqrt{\gamma} \sigma_-$ where $\gamma=1/T_1$ accounts for the rate in which the qubit decays. The dephasing Lindbladian operator $L_{\mathrm{dephasing}}=\sqrt{\gamma_\phi}\sigma_z$ where $\gamma_\phi=1/T_2^*$ quantifies in which rate the qubit dephases and we loose coherences.\\

\noindent\fbox{\parbox{\textwidth}{
\textbf{Exercise~10:} Explicitly represent the Linblad Master equation \eqref{eq:Lindblad_gen} in terms of Bloch components $x=\mathrm{Tr}(\rho \sigma_x), y=\mathrm{Tr}(\rho \sigma_y),,z=\mathrm{Tr}(\rho \sigma_z)$ for $\hat{\mathcal{H}}= -  \Omega_R \sigma_y/2$ in presence of relaxation and dephasing. Now you may solve these equations to obtain the evolution for the qubit.
}} \vspace{0.25cm}

In Figure~\ref{fig:sum_up_depha_relax} we plot the evolution of a driven qubit in the presence dephasing and relaxation. We will return to the Lindbladian evolution and non-unitary evolution in Chapter~4 when we discuss continuous measurements.
\begin{figure}[ht]
\centering
\includegraphics[width = 0.6\textwidth]{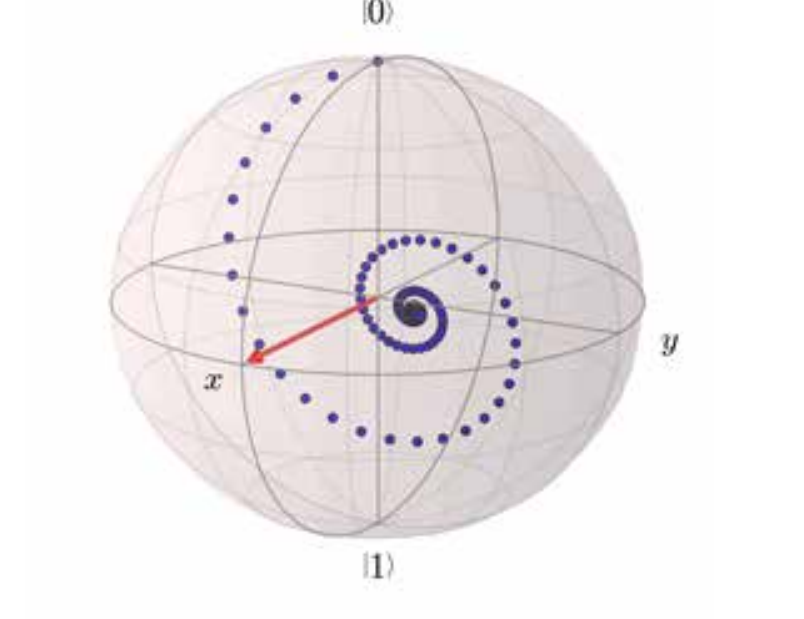}
\caption[Dephasing and relaxation for the qubit]{ {\footnotesize \textbf{Dephasing and relaxation for the qubit:} The solution of the Lindblad equation for a driven qubit in presence of dephasing and relaxation.}} 
\label{fig:sum_up_depha_relax}
\end{figure}

\chapter{Superconducting Quantum Circuits\label{ch3}} 

The aim of this chapter is to make a clear connection between the theoretical concepts introduced in the previous chapter and their experimental realization. I will discuss {measurements with} superconducting circuits including, transmon qubits, 3D cavities, and parametric amplifiers from the experimental point of view. I will try to give a clear explanation of the basic procedures of fabrication and characterization.

\section{Cavity \label{section:cavity}}
In the previous chapter, we discussed a 1D cavity by considering two perfectly conducting walls separated by a distance $L$. Aluminum is a good choice for these walls since it becomes superconducting below $1.2$ K and can be used to realize a high quality factor cavity. Copper can also be used when we need to thread external magnetic flux {through the cavity to tune the qubit resonance.}

\begin{figure}[ht]
\centering
\includegraphics[width = 0.98\textwidth]{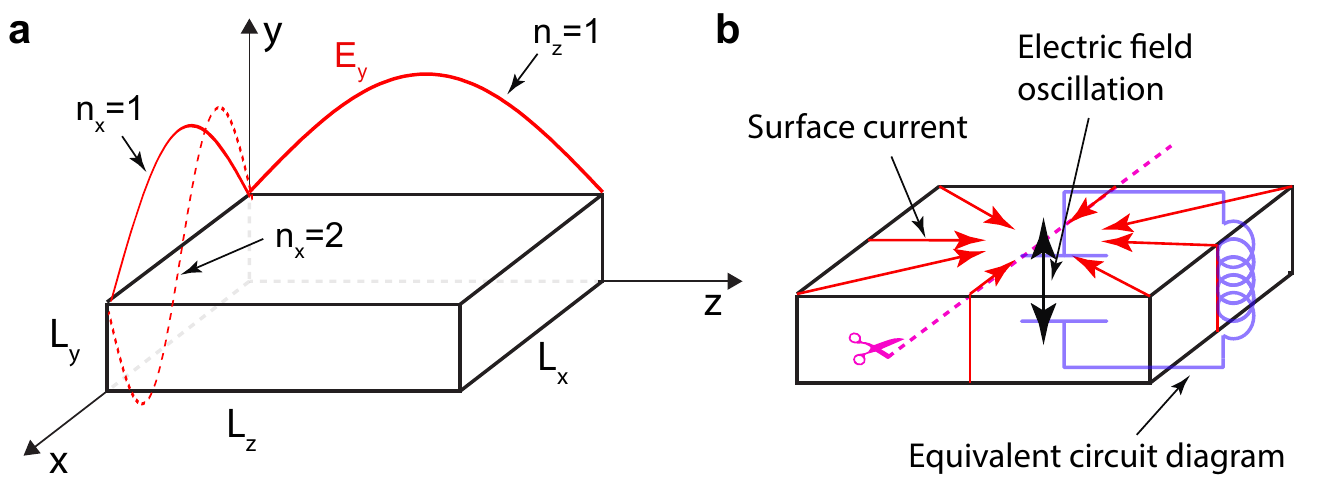}
\caption[TE$_{101}$ mode in rectangular 3D cavity]{ {\footnotesize \textbf{TE$_{101}$ mode in rectangular 3D cavity:} \textbf{a}, The typical cavity geometry is shown. Red lines shows the electric field profile along $z$,$x$ spatial directions for the TE$_{101}$ mode. The electric field oscillations are maximum at the center if the cavity. \textbf{b}, The surface current oscillations (red arrows) have been depicted for the TE$_{101}$ mode. The induced charges are maximum at the center (opposite charges at the top and bottom of the cavity). The equivalent circuit diagram is shown for a section of the cavity. The cut-line (magenta dashed line) indicates one of the planes where the surface current is tangential.}} 
\label{fig:rect_cav}
\end{figure}

Although we induced the cavity in 1D, it is trivial to extend the result to 3D. One can show that for a 3D cavity described by $L_x,L_y,L_z$ dimensions and depicted in Figure~\ref{fig:rect_cav}, the Equation~\ref{eq:EandB} simply generalized to,
\begin{subequations}\label{eq:EB3d}
\begin{eqnarray}
E(\vec{r},t) &=& \mathcal{E} \ q(t) \sin(\vec{k}\cdot \vec{r}) \label{eq:Ex3d}\\
B(\vec{r},t) &=& \mathcal{E} \  \frac{\mu_0 \varepsilon_0}{k}  \dot{q}(t) \cos(\vec{k}\cdot \vec{r}), \label{eq:By3d}
\end{eqnarray}
\end{subequations}
where $\vec{k}=(n_x \pi /L_x , n_z \pi /L_z,n_z \pi /L_z)$ and $\vec{r}=(x,y,z)$ and the corresponding resonance frequency of modes are,
\begin{eqnarray}
f= \omega_c/2\pi =  \frac{c}{2} \sqrt{ (\frac{n_x }{L_x})^2 + (\frac{n_y}{L_y})^2  + (\frac{n_z \pi}{L_z})^2   } ,\label{eq:res_freq}
\end{eqnarray}
where $c$ is the speed of light inside the cavity. Each mode is described by a set of integers ${n_x,n_y,n_z}$, for example, TE$_{101}$ corresponds to a mode with an the electric field profile that has one anti-node in $x$ and $z$ directions (depicted in Figure~\ref{fig:rect_cav}a~\cite{pozar}). 

Thus we have spatially distributed electromagnetic modes inside a cavity, and apart from that, all the quantum mechanical descriptions are essentially identical. Furthermore, we are still interested only in one of these modes. Therefore we consider only the lowest mode of the cavity and choose the dimensions so that the other higher modes' frequencies are far away from the lowest frequency. The dimensions that we normally use are {$L_z \sim L_x \sim 3.0$~cm} and $L_y \sim 0.8$~cm which gives a cavity frequency $\omega_c/2\pi \sim$7 GHz for the TE$_{101}$ mode. Figure~\ref{fig:rect_cav}b shows the surface current and the equivalent circuit diagram for the TE$_{101}$ mode\footnote{For a better circuit diagram visualization, you can replace the inductor with a wire and imagine that the loop has some effective inductance. In this way, the loop magnetic field gives the correct direction for the cavity magnetic field at that cross-section.}.

Although it is easy to analytically calculate the resonance frequency for the simple geometries like a rectangular or cylinder cavity, one can use {numerical simulations} to get more realistic predictions {since they can account} for geometry imperfections, input-output connectors, and qubit chip\footnote{Moreover, simulation gives you to have access to more detailed information about the electromagnetic field distribution inside the cavity.}. Figure~\ref{fig:cavity_3d} shows a simulation result for cavity transmission by considering other components using {Ansys} HFSS.
\begin{figure}[ht]
\centering
\includegraphics[width = 0.98\textwidth]{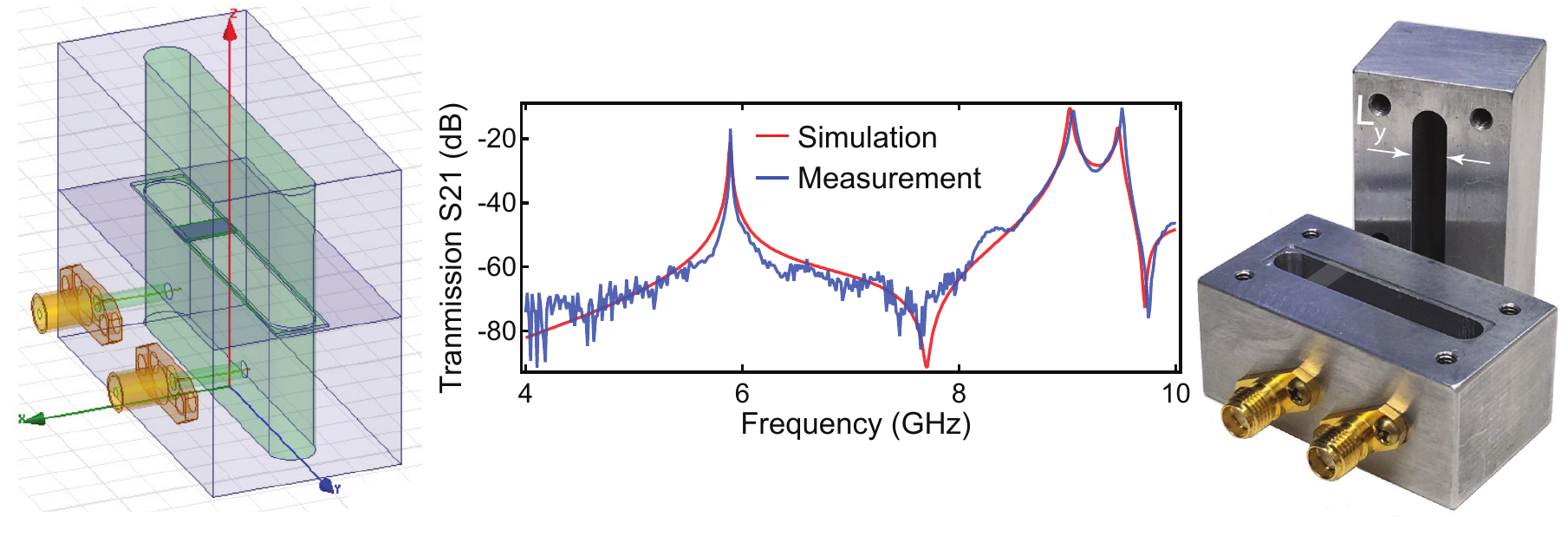}
\caption[HFSS simulation for cavity transmission]{ {\footnotesize \textbf{HFSS simulation for cavity transmission:} The cavity transmission is simulated by HFSS (red curve) which is in agreement with actual measurement (blue curve).}} 
\label{fig:cavity_3d}
\end{figure}
The simulation is in a good agreement with actual transmission measurement of a similar cavity\footnote{Although, the details of the input/output connectors and pins (e.g. length, shape, soldering parts, ...) are may not have significant contribution to the cavity frequency, they significantly affect the cavity quality factor because they can dramatically alter the characteristic impedance of the ports.}.

In order to fabricate a cavity, we literally machine a cavity in two chunks of aluminum as depicted in Figure~\ref{fig:cavity_3d}c\footnote{Symmetrical pieces are not only convenient in terms of the fabrication, but also in this geometry, symmetrical pieces minimize the adverse effect of imperfections where two pieces are connected since the surface current doesn't need to pass between pieces at all. Note the cut-line (magenta dashed line) in Figure~\ref{fig:rect_cav}b.}.

\emph{Cavity linewidth-} The cavity linewidth $\kappa$, can be determined by measuring the transmission through the cavity. As depicted in Figure~\ref{fig:cavity_kappa}a, we use a vector network analyzer (VNA) and record S12 (or S21) transmitted power versus the frequency. The parameter $\kappa$ is roughly the frequency bandwidth by which the transmitted power drops by 3 dB a depicted in \ref{fig:cavity_kappa}b. More rigorously, one can fit the transmitted power (in linear scale) to a Lorentzian function $F(x)=\frac{A}{(x-f_c)^2 + \kappa^2/4}$ and extract the cavity frequency $f_c=\omega_c/2\pi$ and cavity linewidth $\kappa$ which is the FWHM of the Lorentzian fit. We will see in Chapter 4, the parameter $\kappa$ quantifies how much signal we get from the cavity and this value needed for calibration of the quantum efficiency during measurement.
\begin{figure}[ht]
\centering
\includegraphics[width = 0.98\textwidth]{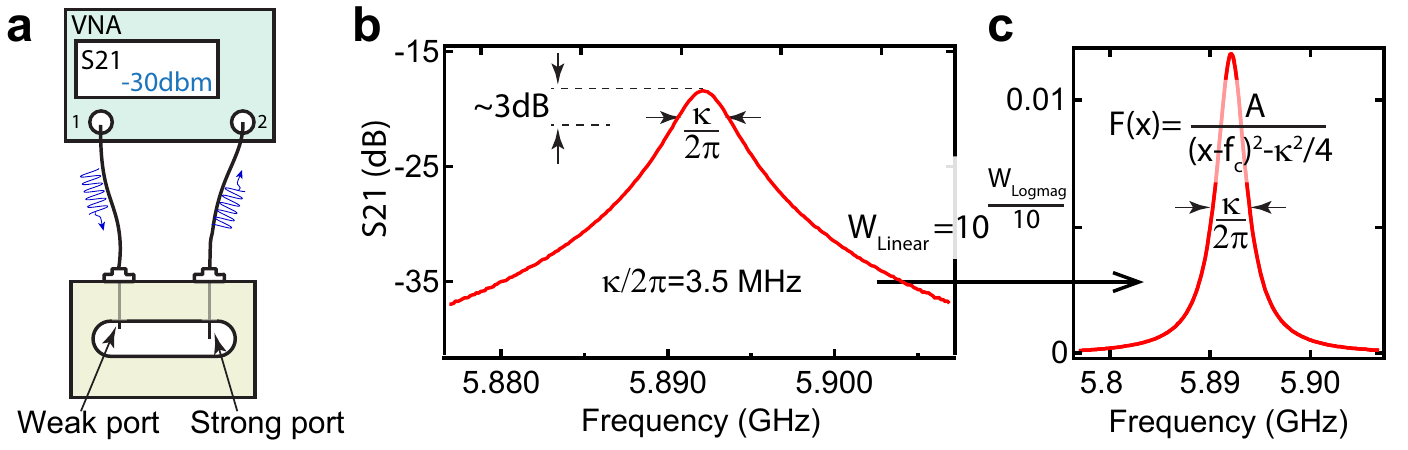}
\caption[The cavity linewidth characterization]{ {\footnotesize \textbf{The cavity linewidth characterization:} The cavity linewidth $\kappa$ can be quantified by measuring the scattering parameters of the cavity. \textbf{a},\textbf{b} In the transmission measurement S21, the cavity linewidth can be estimated by the bandwidth that the transmission signal drops by 3 dB. \textbf{c}, More carefully, one can scale the transmission in the linear form and fit to the Lorentzian function. The FWHM of the Lorentzian function would be the cavity linewidth.}} 
\label{fig:cavity_kappa}
\end{figure}

\emph{Cavity phase shift-} It is worth discussing the cavity phase shift across resonance. As depicted in Figure~\ref{fig:cavity_chi_shift}, the phase of the transmitted signal shifts by $\pi$ across the resonance of the cavity\footnote{The reflected signal acquires a $2\pi$ phase shift across the cavity resonance. Does this mean it is better to use reflection to detect the phase shift?} which can be represented by,
\begin{eqnarray}
\theta=\arctan\left[\frac{2}{\kappa} (\omega_c-\omega)\right]  \xrightarrow{\omega \simeq \omega_c} \theta = \frac{2}{\kappa} (\omega_c - \omega),
\label{eq:ref_phase_cavity}
\end{eqnarray}
which varies almost linearly around the cavity resonance frequency with slope $2/\kappa$ which quantifies the sensitivity to the frequency by measuring the phase of the transmitted (or reflected) signal\footnote{We will see in Chapter~4 that the cavity phase shift and cavity linewidth come into play for describing {continuous measurement in terms of POVMs.}}.
\begin{figure}[ht]
\centering
\includegraphics[width = 0.48\textwidth]{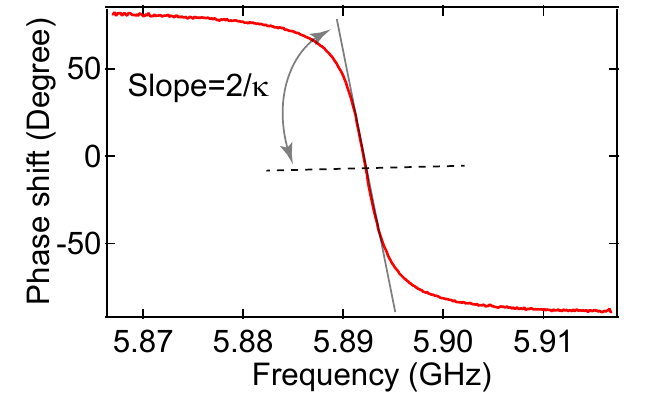}
\caption[The cavity phase shift across the resonance]{ {\footnotesize \textbf{The cavity phase shift across the resonance:} The transmitted signal through the cavity experiences a $\pi$ phase shift across the resonance of the cavity. Near resonance, the phase shift is approximately linear with a slope of $2/\kappa$.}} 
\label{fig:cavity_chi_shift}
\end{figure}

\emph{Cavity internal and external quality factors-} The external quality factor $Q_\mathrm{ext}$ can be adjusted by the length of {the input and output port pin antennas}. Normally, we have two pins corresponding to the weak port and strong ports. The weak port is used as an input (qubit manipulation and readout) and usually has $\sim$ 100 times weaker coupling to the cavity compared to the strong port. The lengths of the port antennas determine the coupling to a $50\,\Omega$ transmission line, determining the external quality factor $Q_\mathrm{ext}$.
The internal quality factor (often called unloaded quality factor) has to do with the losses due to the cavity itself, e.g. absorption of photons by the cavity. The total cavity quality factor is then,
\begin{eqnarray}
\frac{1}{Q_{\mathrm{tot}}} = \frac{1}{Q_{\mathrm{int}}}  + \frac{1}{Q_{\mathrm{ext}}}. 
\label{eq:Q_total}
\end{eqnarray}
Often, the deliberate coupling to the outside is dominant $Q_{\mathrm{int}} \gg Q_{\mathrm{ext}}$. Therefore the total quality factor is almost equal to the external quality factor. Note, that $Q_{\mathrm{tot}}=\omega_c/\kappa$.

For a careful characterization of the internal quality factor and the input and output coupling strengths, one can perform reflection measurements on each port (while the other port is terminated by 50 $\Omega$). By analyzing the amplitude and phase of the reflected signals, one can obtain both the internal and external quality factors for the cavity and characterize the coupling strength for each port (see Refs.~\cite{megrant2012planar,khalil2012analysis}).

\section{Qubit}
In Chapter 2 we studied the Josephson junction and the transmon qubit from a theoretical perspective. Here we discuss the fabrication and characterization of Josephson junctions and transmon circuits.

\subsection{Transmon fabrication:} 
Josephson junctions can be fabricated by evaporation of aluminum on a silicon wafer using an electron beam evaporator which allows for directional evaporation. The common technique for JJ fabrication is the double-angle evaporation technique which utilizes the evaporation directionality for fabrication.  A typical procedure for the JJ fabrication includes; spin-coating e-beam resists on a silicon wafer, e-beam lithography, development, pre-cleaning, double-angle evaporation, lift-off, and post-cleaning.

\textbf{\emph{e-beam resist-}} We use a stack of two resists for junction fabrication. The bottom layer normally is a relatively thick ($\sim 1\mu$m) and soft resist (MMA). In contrast, the top layer is a relatively thin ($\sim 300$ nm) and hard (e.g. ZEP) resist. The reason for this choice of resist staking is to achieve a wide undercut which is convenient for clean lift-off as depicted in Figure~\ref{fig:liftoff}. It also enables for a suspended bridge needed for junction fabrication (Fig.~\ref{fig:liftoff}c).
\begin{figure}[ht]
\centering
\includegraphics[width = 0.98\textwidth]{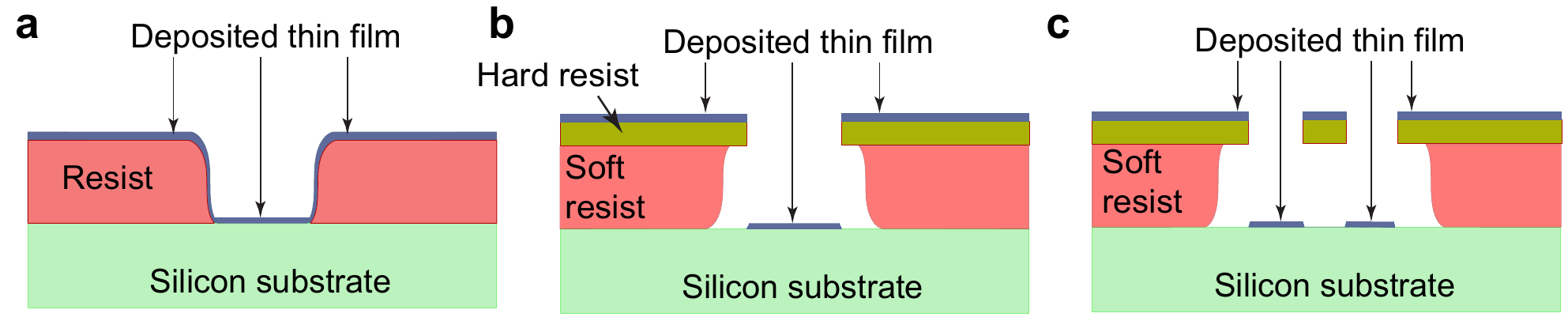}
\caption[Double stack e-beam resist]{ {\footnotesize \textbf{Double stack e-beam resist:} \textbf{a}, With a single layer of resist, it is often difficult to get small and clean patterns. Mainly because the wall of the resist may also get deposited which connects top layers to the bottom layers which make it difficult to properly lift-off the resist without peeling off the actual pattern. \textbf{b}, This issue can be avoided by using two layers of the resist. The top layer provides sharp edges as a mask and the bottom layer act as a spacer. The proper amount of undercut aids the liftoff process. \textbf{c}, Moreover, the undercut of the lower resist can be used to created suspended resist (free-standing bridge) which is used for the JJ fabrication in double angle evaporation.}} 
\label{fig:liftoff}
\end{figure}
The resist layers can be coated on the substrate by spin-coating. The thickness of the layers is controlled by {the spinning velocity, the total time of spinning, and the viscosity of the resist}. A typical spin-coating recipe for MMA/ZEP double stack resist is displayed in Table~\ref{table:resist}.

\begin{table}[ht]
\centering
\begin{tabular}{|c|c|}
\hline
Step~1& MMA spin-coat, 3000 rpm, 60 seconds\\
Step~2&Soft bake for 5 minutes,200$^\circ$C\\
Step~3& ZEP spin-coat, 3000 rpm, 60 seconds\\
Step~4&Soft bake for 3 minutes,180$^\circ$C\\
\hline
\end{tabular}
\caption[Double-stack e-beam MMA/ZEP resist spin coating recipe
]{{\footnotesize \textbf{Double-stack e-beam MMA/ZEP resist spin coating recipe}}}\label{table:resist}
\end{table}
\emph{\textbf{Electron-beam lithography-}} We use a 30 keV focused beam of electrons in a scanning electron microscope (SEM) to pattern the resist. The SEM is controlled by Nanometer Pattern Generation Software (NPGS). For fine features, we need to have a good focus of the electron beam. {To achieve a good focus,} we use gold particles which are easily detectable in the SEM for in-situ focus calibration\footnote{We drop gold particles close to the edge of the sample and try to get the best focus at each point. NPGS uses the focus point data and extrapolates the focus settings for the entire chip. One can also use single point focus and move the beam by $\sim$1000 $\mu$m and write the pattern with the same focus settings. Ideally, for the transmon junction, one should be able to distinguish $\sim$5 nm gold particles at each focus point.}.
The transmon pattern is designed in `Design CAD' software using polygons in different layers\footnote{NPGS allows for different expose/focus setting for each layer. Therefore, with a multi-layer pattern, we can optimize the {exposure} time.}. A simple transmon pattern design in Design CAD software has been shown in Figure~\ref{fig:designcad}. Each layer represented in different color. 
\begin{figure}[ht]
\centering
\includegraphics[width = 0.98\textwidth]{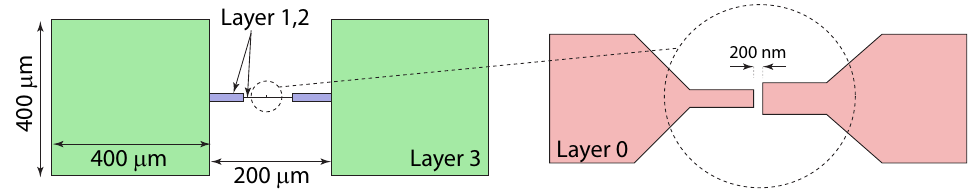}
\caption[A simple design for transmon qubit]{ {\footnotesize \textbf{A simple design for transmon qubit:} A qubit design consists of few polygons in different layers in DesignCAD software. The free-standing bridge design for a JJ and few micrometer extremities are shown in red (Layer~0). The connector lines in the two steps (Layer~1,2) shown in blue and capacitor pads (Layer~3) in green. The corresponding e-beam dosage for each layer is displayed in Table~\ref{table:SEMdose}.}} 
\label{fig:designcad}
\end{figure}

We use {a higher magnification} and lower dosage for finer features. Table~\ref{table:SEMdose} displays typical focus and dosage for each layers.
\begin{table}[ht]
\centering
\begin{tabular}{|c|c|c|c|}
\hline
Layer~\# & smallest feature size & SEM {magnification} & e-beam current  \\
\hline
Layer~0 & $<$ 200 nm & 1300X & 30 pA  \\
Layer~1 & $\sim$ 1 $\mu$m & 600X & 220 pA  \\
Layer~2 & $\sim$ 10 $\mu$m & 200X &600 pA  \\
Layer~3 & $>$100 $\mu$m  & 50X & 10000 PA  \\
\hline
\end{tabular}
\caption[The NPGS setting for a 30 KV electron beam]{{\footnotesize \textbf{The NPGS settings for s 30 keV electron beam}.}}\label{table:SEMdose}
\end{table}

\emph{\textbf{Resist development and pre-cleaning-}} The development recipe has three steps. We use an ice bath to bring the developer's temperature down to {$\sim 0^\circ$C} to slow down the development process. Figure~\ref{fig:icebath} demonstrates the development recipe.
\begin{figure}[ht]
\centering
\includegraphics[width = 0.8\textwidth]{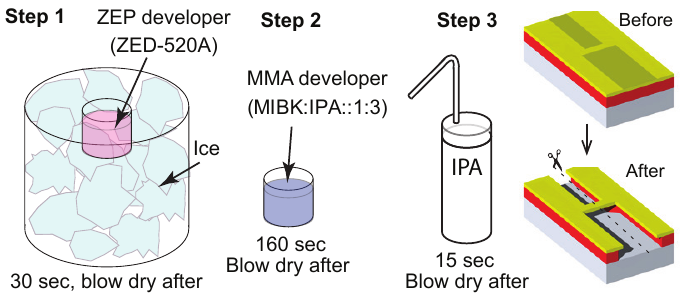}
\caption[e-beam resist development recipe]{ {\footnotesize \textbf{e-beam resist development recipe:} \textbf{Step~1}, ZEP developer in ice bath, wait for developer to cool down to T$\sim 1^\circ$C. Plunge the sample into the beaker for 30 seconds, then blow dry immediately. \textbf{Step~2}, MMA developer for 160 seconds and blow dry afterward. \textbf{Step~3}, Rinse with IPA for 15 seconds. The left panel shows the JJ area in the simple transmon design (Fig.~\ref{fig:designcad}) before and after development. Note the undercut and the suspended bridge in the middle.}} 
\label{fig:icebath}
\end{figure}
After the development we may use `oxygen plasma cleaning' to further remove resist residue from the substrate surface.

\emph{\textbf{Electron-beam evaporation-}} We use a double angle evaporation method to fabricate JJs as depicted in Figure~\ref{fig:qubit_fab_2angle}. The {transmon capacitor pads are} also fabricated {during this} process. The thickness of the aluminum film is normally 30 nm for the lower layer and 60 nm for the top layer and there is $\sim1$ nm thick of aluminum oxide layer grown between two layers.
\begin{figure}[ht]
\centering
\includegraphics[width = 0.98\textwidth]{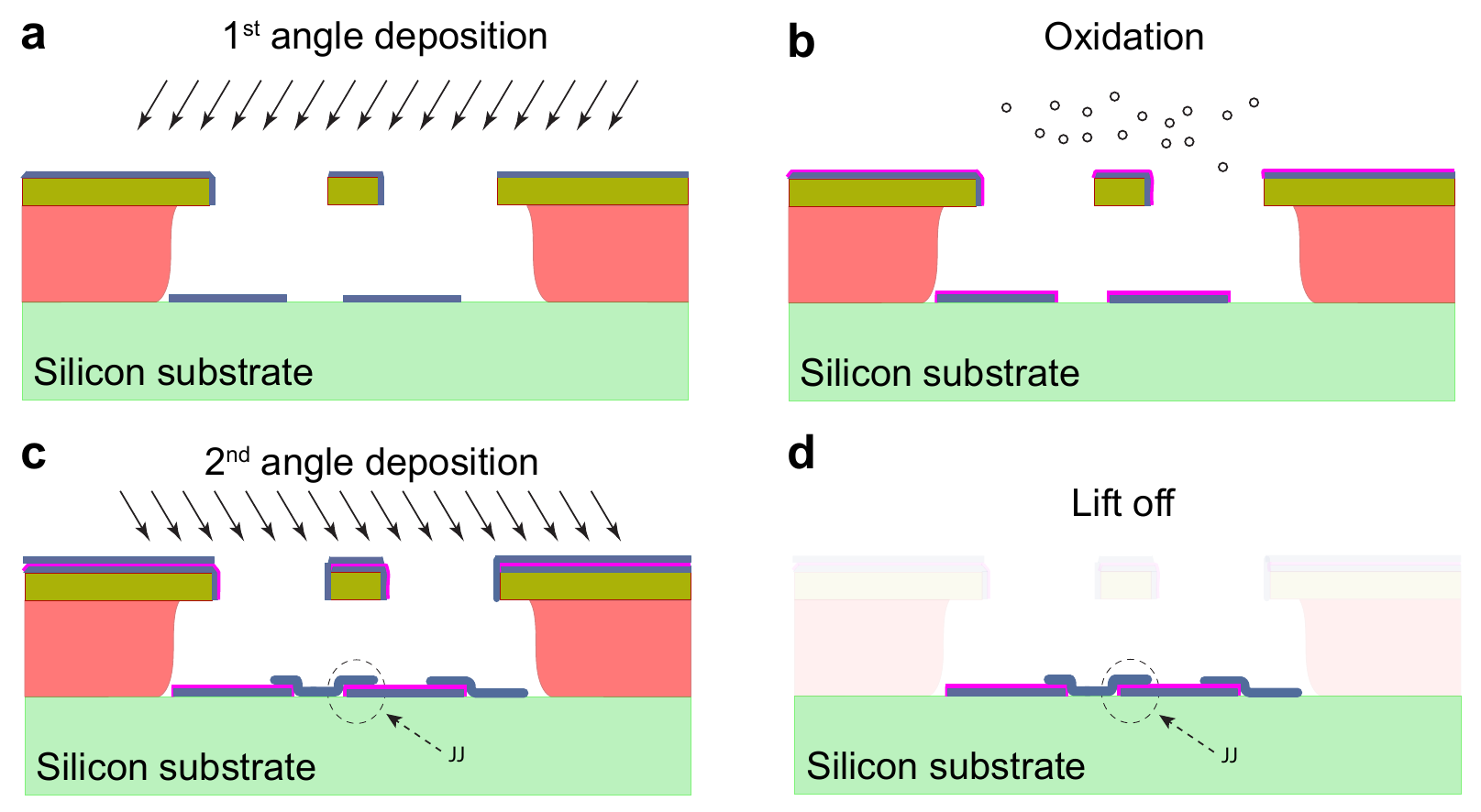}
\caption[Double-angle evaporation and Josephson junction fabrication]{ {\footnotesize \textbf{Double-angle evaporation and Josephson junction fabrication:} Considering the indicated cross-section of the freestanding bridge in Figure~\ref{fig:liftoff} we use double angle evaporation to fabricate the JJ. \textbf{a}, The first layer of aluminum evaporation is about 30 nm. \textbf{b}, Introducing the oxygen mixture to form a thin layer of aluminum oxide $\sim 1$ nm as the insulator. \textbf{c}, The second layer of aluminum evaporation is about 60 nm at the opposite angle {normal to the substrate surface}. \textbf{d}, Removing the resists and the {deposited aluminum} in a lift-off process.}} 
\label{fig:qubit_fab_2angle}
\end{figure}

\emph{\textbf{Lift-off and post-cleaning-}} We use acetone in temperature $T\sim 60^\circ$C for 40 minutes to dissolve the resist which leaves behind the transmon circuit on the substrate. Figure~\ref{fig:qubit_fab} shows the SEM image of the final transmon circuit and the JJ.
\begin{figure}[ht]
\centering
\includegraphics[width = 0.9\textwidth]{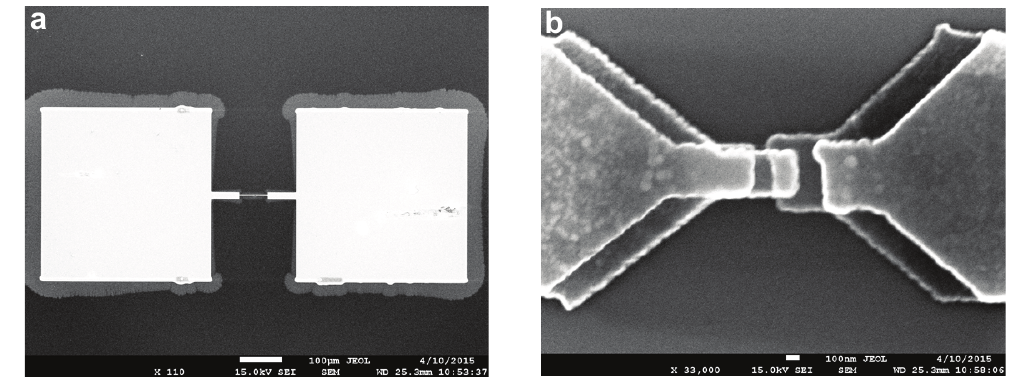}
\caption[Qubit pattern SEM]{ {\footnotesize \textbf{Qubit pattern SEM}.}} 
\label{fig:qubit_fab}
\end{figure}

\subsection{JJ characterization}
For the qubit design, we have a couple of considerations. First, the qubit frequency and its anharmonicity need to be in the proper range. We would like to have anharmonicity somewhere in the range $200-300$ MHz. According to Equation~\ref{eq:trans_H_bb2}, the anharmonicity is determined by the energy associated to the shunted capacitor, $E_C = \frac{e^2}{2C} $. This capacitance mostly comes from the transmon pads. Therefore $E_C$ can be set based on the design of the transmon pads (the size and separation of pads).\\

\noindent\fbox{\parbox{\textwidth}{
\textbf{Exercise~1:} What is the capacitance between two sheets of perfect conductors separated in the horizontal orientation by $2a$ in a homogeneous medium. What if the medium has two different dielectric constants on each side as depicted?

\centering \includegraphics[width = 0.6\textwidth]{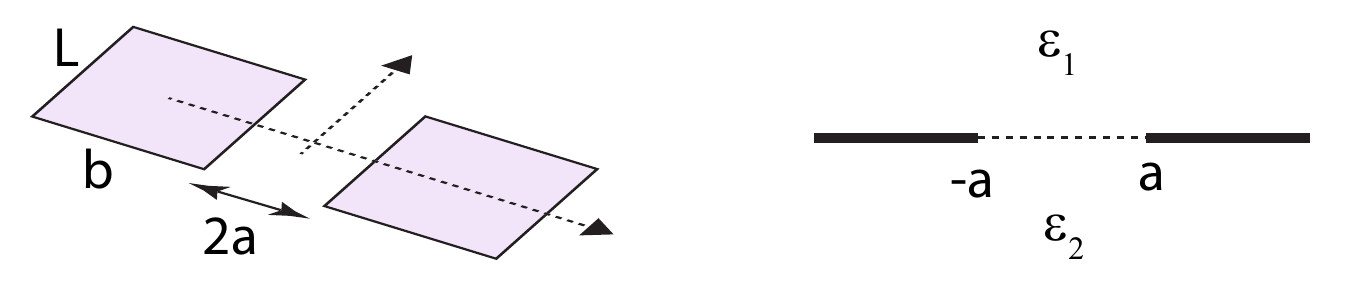}
}} \vspace{0.25cm}

Using HFSS simulation, the capacitance of our normal design ( see Fig.~\ref{fig:qubit_fab}a, pad size $400\times400 \ \mu$m separated by $200\ \mu$m, with connection arms) is about $C=0.057$ pF. The contribution of $C_J$ in negligible (estimated to be about $0.35$ fF  for $200 \times 100$ nm JJ area, assuming oxide layer thickness $\sim 1$ nm).

\begin{figure}[ht]
\centering
\includegraphics[width=0.5\textwidth]{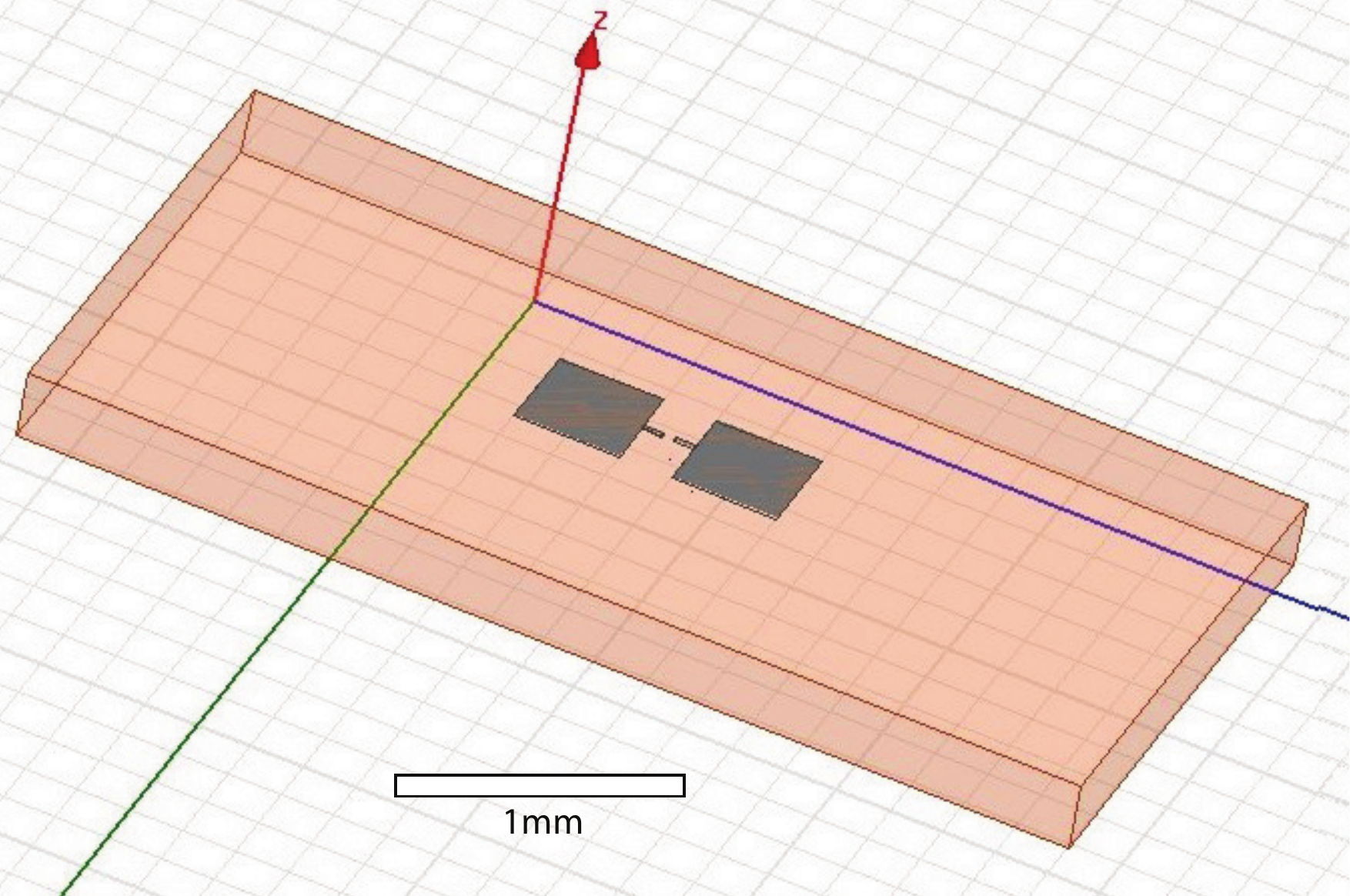}
\caption[The HFSS simulation for the transmon shunting capacitor]{{\footnotesize \textbf{The HFSS simulation for the transmon shunting capacitor:} Using HFSS, the shunting capacitance is calculated to be $C=0.057$ pF (for the simple design shown in Figure~\ref{fig:designcad}) which results in $E_J \sim 340$ MHz.}}
\label{fig:td1}
\end{figure}

For a certain qubit design $E_C$ is fixed and for a certain qubit frequency $\omega_{q}=\sqrt{8E_J E_c} - E_c$, the only knob is the Josephson energy $E_J=\frac{\hbar}{2e}I_c $, where the critical current is a function of the junction area and the thickness of the oxide layer, $I_c \propto \mathrm{area}/ \mathrm{oxide \  thickness}$. Therefore, having the right critical current is critical. Fortunately, there is a very useful relationship between the resistance of a JJ at room temperature $R_n$ and the JJ critical current,
\begin{eqnarray}
I_c=\frac{\pi\Delta(0)}{2 e R_n},
\end{eqnarray}
where $\Delta(0) \sim170 \  \mu$eV is the aluminum superconducting energy gap at zero temperature\footnote{Basically the gap energy depends on the temperature, $ \Delta(T)= \Delta(0) \tanh[\frac{\Delta(T)}{2k_B T}]$. However since the qubit will be operated at temperature close to zero, $T \ll T_c$, therefore with good approximation $\Delta(T) \simeq \Delta(0)$.}. The normal resistance of the junction $R_n$ can be measured by sending a probe current  $I_\mathrm{prob}$ through the junction and reading the voltage $V_\mathrm{probe}$ across the junction.
With this room temperature resistance measurement, and with our prior knowledge about the $E_C$ (either from previous transmon measurements or simulation) we can estimate the frequency of the qubit before the cool-down. The estimation for transmon energy transition would be,
\begin{eqnarray}
E_{01}&=&  \sqrt{\frac{h \Delta(0) }{C R_n}}-\frac{e^2}{2C} \nonumber\\
f=\omega_{01}/(2\pi) &=&  \sqrt{\frac{\Delta(0) }{h C R_n}}-\frac{e^2}{2 h C} ,
\label{eq:jj_brob}
\end{eqnarray}
where $h$ is Planck's constant\footnote{Here, we would rather $\hbar$ and treat it carefully, since $E_C$ doesn't explicitly depend on $\hbar$.}. For example, in order to have qubit frequency around $\omega_{01}/(2\pi) \sim 6$ GHz with our normal transmon geometry ($C \sim 0.057$ pF see Fig.~\ref{fig:td1}), the critical current should be $I_c \sim 0.015 \  \mu$A ($R_n=18$ k$\Omega$).

\section{Qubit-cavity system characterization}
In this section, we discuss a typical qubit characterization procedure. Here are the typical steps before the cool-down:
\begin{itemize}
\item Josephson junction room temperature resistance probe,
\item Choosing a proper cavity and weak/strong pin length adjustment.
\item The qubit placement inside the cavity.
\item Characterizing the cavity transmission (qubit chip included).
\end{itemize}
Then we put the cavity-qubit system inside the fridge and cool them down. The minimum circuitry inside the fridge is depicted in Figure~\ref{fig:exp_setup_char}.
The main qubit characterization includes five basic experiments. 
\begin{itemize}
\item One-tone spectroscopy, or ``punch-out".
\item Two-tone spectroscopy.
\item Rabi measurement.
\item {$T_1$} measurement.
\item Ramsey measurement ($T_2^*$).
\end{itemize}
The first two experiments are in the frequency domain which means we only look at scattering parameters of the system for characterization. However, the last three experiments {are measured} in the time domain and involve preparation and readout of the qubit state. In the following sections, we discuss how these are performed in the lab and what we learn from each experiment in more detail.

\begin{figure}[ht]
\centering
\includegraphics[width = 0.7\textwidth]{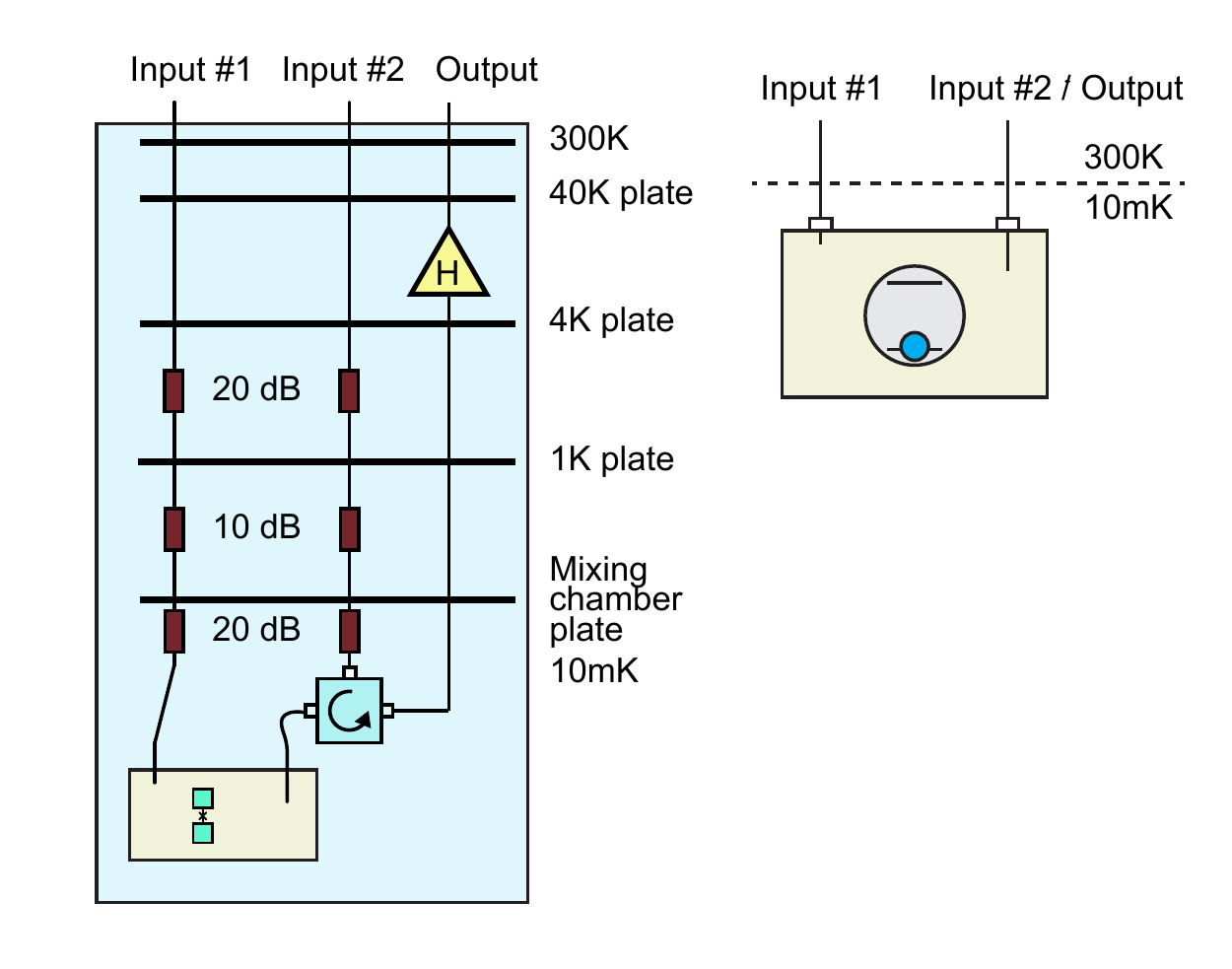}
\caption[The minimum experimental setup for basic qubit characterization]{ {\footnotesize \textbf{The minimum experimental setup for basic qubit characterization:} The input lines can be used for qubit manipulation signals and cavity probe signals. Note that we don't get any signal back from input lines (because of $\sim 2\times 50$ dB of attenuation). However, because we have a circulator connected to the strong port, the reflection off of the strong port can be measured by sending the signal from the input~\#2 and receiving it back from the output. We will refer to the fridge circuitry (depicted in left) by the short version (depicted in right) throughout this document.}} 
\label{fig:exp_setup_char}
\end{figure}

\subsection{One-tone spectroscopy: ``punch-out"}
The first step is to check if the qubit is ``alive" or not. For that, we need to check whether the cavity frequency shifts based on the state of the qubit. Of course, at this point, we don't know the qubit frequency to carefully manipulate it. Fortunately, we don't need it to know what is the qubit frequency to check if it is there.  One way to think about that is if the qubit is coupled to the cavity, {the cavity becomes hybridized with the qubit and} we should be able to detect a little bit of nonlinearity in the cavity.

All we need to {do is} compare the transmission (or reflection) of the cavity in low power versus high-power and see if the frequency of the cavity shifts.
When we probe the cavity with very low power we are pretty sure that qubit is in its ground state\footnote{If you have not convinced, consider that we only sweep the VNA frequency across the cavity resonance frequency so that VNA span is $\sim \kappa$. Considering the avoided crossing picture and the fact that $g\gg\kappa$, therefore, the qubit couldn't be in this region. So we are pretty much sure that we are not driving the qubit in this situation.}. Therefore we measure the resonance frequency of cavity when qubit is in the ground state.
Next, we turn up the power of the VNA to a very high power. In this case, we send a huge amount of photons into the cavity which essentially overwhelms the qubit. Basically, the driving amplitude is so high that the induced current exceeds the critical current of the junction. Practically, in such a high power regime, we measure the bare cavity frequency. Now if there is a working qubit inside the cavity we can see the cavity frequency shift and we say that the cavity ``punched out". 
\begin{figure}[ht]
\centering
\includegraphics[width = 0.98\textwidth]{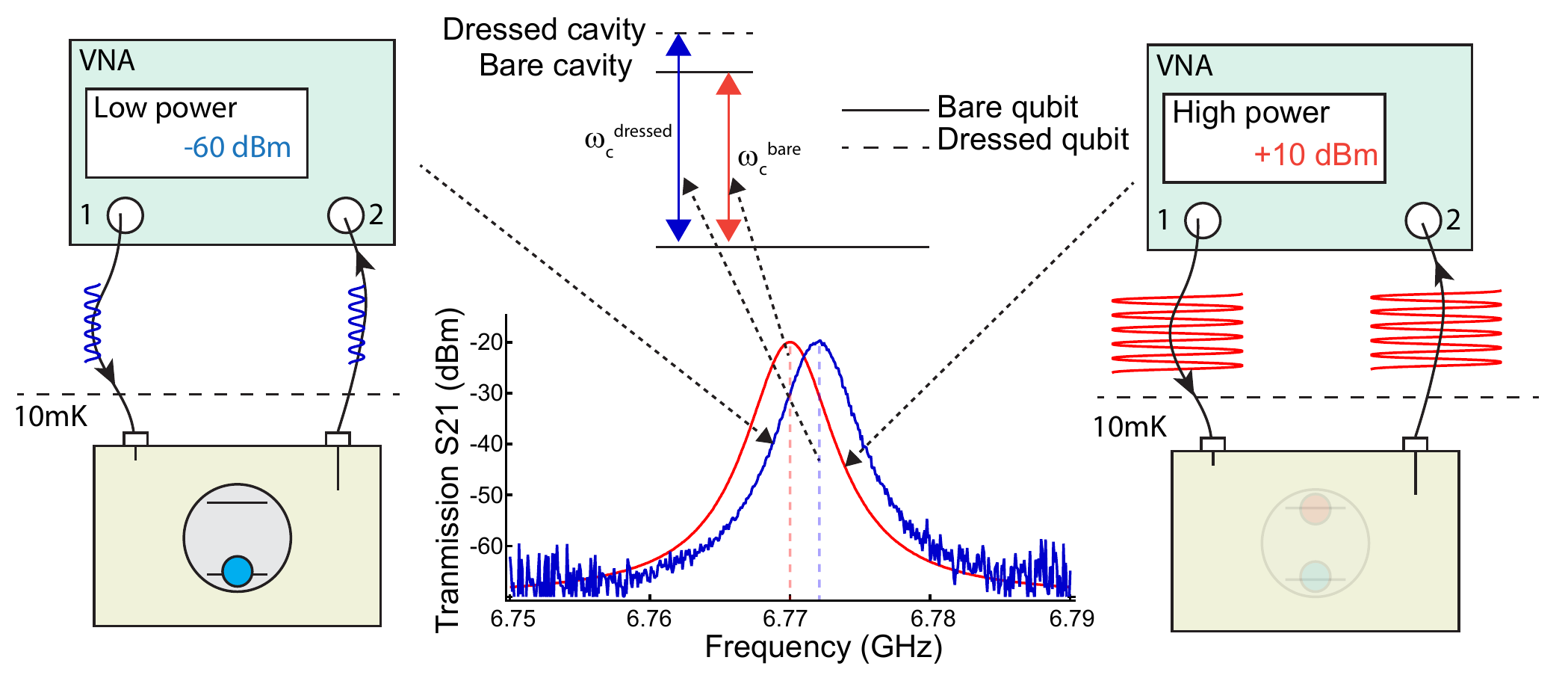}
\caption[The ``punch-out" measurement] {\footnotesize {\textbf{The ``punch-out" measurement:} The low (high) power transmission of the cavity indicated by the blue (red) trace. For low power probe signal, the qubit remains in the ground state and we essentially measure the dressed cavity transition indicated by the blue double-sided arrow. In high power case, the is qubit ``washed out" and we essentially measure the bare cavity transition as depicted by the red double-sided arrow.}} 
\label{fig:punch_out}
\end{figure}
There is one more piece of information we can get from the punch-out measurement. If the high-power frequency shifts to a lower frequency, we infer that qubit frequency is below the cavity and vise versa. A bigger shift means that the cavity and qubit are {more strongly coupled}\footnote{Note, the placement of the qubit inside the cavity also affects the coupling and consequently the punch-out shift.}. One can consider the phase shift {of the cavity resonance} as a rough estimation of $\chi=g^2/\Delta$ but a careful characterization of $\chi$ can be done with time domain measurements.

If the qubit bare frequency happens to be very close to the cavity bare frequency $\Delta<g$, then the qubit and cavity may enter the polariton regime and you may clearly see two peaks (separated by $\sim 2g$) in the cavity at low power transmission\footnote{In this case you are directly resolving polariton qubit.}. If the high power peak (bare cavity) is exactly on the middle of low-power peaks, then qubit and cavity are exactly on-resonance and the separation is exactly\footnote{Therefore, one can use a tunable qubit to directly measure the effective coupling strength $g$.} $2g$. 

\subsection{Two-tone spectroscopy}
Knowing that the qubit is working, the next step is to find the qubit frequency. The idea is to continuously send a weak microwave signal to the cavity {at the low power cavity resonance}---the cavity frequency when the qubit is in the ground state---and probe the cavity transmission. Therefore, we constantly receive a high transmission signal because we probe at the resonance of the cavity. While this first tone is on, we start sending another microwave signal (labeled as BNC\footnote{`BNC' is simply the name of the generator we normally use in the lab.}) into the cavity. {We sweep the frequency of the probe tone BNC} and monitor the cavity transmission as depicted in Figure~\ref{fig:spectroscopy}. During the sweep, once the BNC frequency hits the qubit transition frequency (BNC$=\omega_q$), it excites the qubit\footnote{Actually the BNC signal drives Rabi oscillations on the qubit. We saw in previous chapter (Eq.~\ref{eq:psi_t_ge3}) that the qubit reaches maximum excitation ($P_e=1$) only if the drive is on-resonance with the qubit.} therefore the state of the qubit is no longer in ground state (on average) and that causes a shift in the cavity frequency\footnote{Remember interaction Hamiltonian in the dispersive regime~(Eq.~\ref{eq:H_dis_rearange}) which results in  the qubit-state-dependent frequency for the cavity.}. Now, because the VNA frequency (which is fixed) is no longer {resonant with the} cavity, the transmitted power drops\footnote{If BNC hits the cavity frequency (which is not shown here) we may also see a dip in transmission. However, we never mistake that with the qubit dip because that happens exactly at the VNA frequency. This dip could be due to some nonlinearity induced into the cavity but it is more likely to be the amplification chain saturation. Which means, when we add the relatively high power BNC signal on top of VNA and both highly transmitted to amplifiers, the amplifiers may be saturated and this effect may show up as a dip in the VNA trace.} as depicted in Figure~\ref{fig:spectroscopy}b.
\begin{figure}[ht]
\centering
\includegraphics[width = 0.98\textwidth]{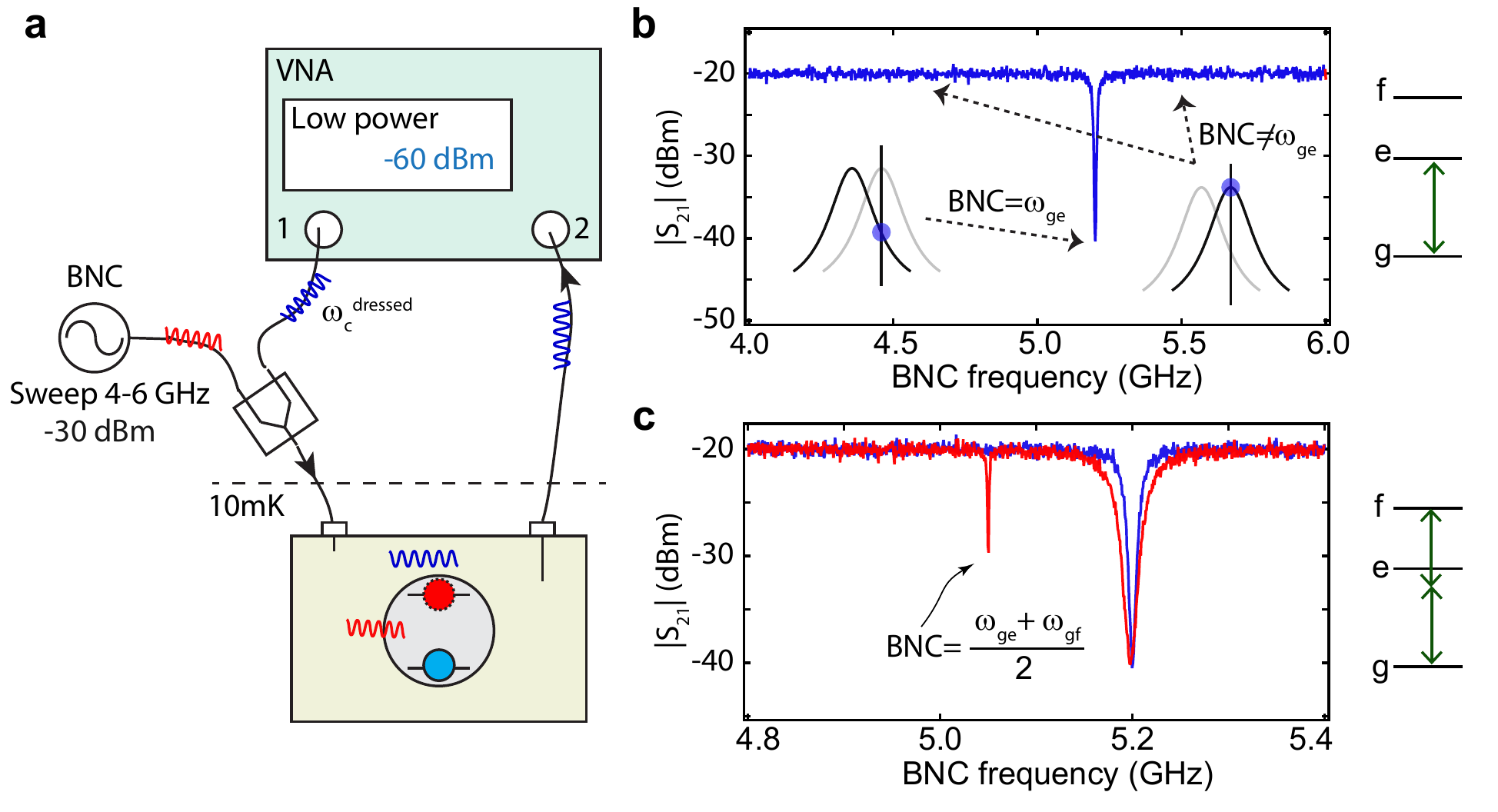}
\caption[Two-tone spectroscopy]{ {\footnotesize \textbf{Two-tone spectroscopy:} \textbf{a}, The schematics for experimental setup for two-tone spectroscopy. A constant fixed frequency signal from the VNA at the cavity low power probes the cavity while a signal from the BNC sweeps across different frequencies from 4 to 6 GHz. \textbf{b}, The cavity transmission versus BNC frequency shows a dip at qubit frequency. \textbf{c}, With higher power for BNC the two-photon transition is also detectable (red trace). Note that the qubit transition is power broadened.}} 
\label{fig:spectroscopy}
\end{figure}
If we increase the BNC signal amplitude (by $\sim 10$ {dBm}), we can also see a second dip at a slightly lower frequency. This dip corresponds to the process by which two photons of drive excited the qubit from the ground state to second excited state\footnote{Transition from $|g\rangle \to |f\rangle$ is a two-photon process which is less probable compared to a one-photon process. Therefore a higher power is needed to drive that transition.}, $|g\rangle \to |f\rangle$ as depicted in Figure~\ref{fig:spectroscopy}c. The second dip gives a useful piece of information which allows us to simply calculate the transmon anharmonicity, {$\omega_{eg} - \omega_{fe}= 2 (\omega_1- \omega_2)\simeq2(5.2-5.05)\,\text{GHz}=300$ MHz in this case}.

We just discussed spectroscopy in transmission mode. Equivalently, we may {use reflection off the cavity} for spectroscopy\footnote{The are a couple of reasons we may want to use reflection {for} spectroscopy. First, we might have a limited number of input lines so might not have a weak port for the cavity. Or, sometimes we have low signal-to-noise in transmission and we might have a better chance looking at the reflected phase. Also, we sometime may not have enough power from the weak port and we can use reflection port which has much stronger coupling to the cavity.}. In reflection, most of the signal is reflected from the cavity, therefore there is not much information in the magnitude of the signal. But, we can look at the phase of the reflected signal which is sensitive to the cavity resonance frequency\footnote{In spectroscopy by reflection you may get a dip or peak depend on the delay offset of the VNA.}.

\subsection{Time domain measurement: basics}
In this section, we will discuss the time domain measurement of the qubit. Time domain measurements require initialization, preparation and manipulation of the qubit state, and readout. In what follows, we briefly discuss these three steps.

\emph{\textbf{Initialization--}} In our case the initialization is quite simple. The lifetime of the system is on the order of tens of microseconds, therefore all we need to do is leave the qubit for some amount of time ($\sim$100 microseconds) to make sure it is in ground state with fairly high fidelity\footnote{In our case this fidelity is about 97\% which depends on the effective temperature of the system. One may calculate the probability of thermal excitations for the qubit $P_e^{\mathrm{thermal}} = \exp(-\hbar \omega/ K_B T)$ given the temperature of the fridge $T$ and energy of the qubit $\omega_q$. However, the effective temperature for our system is slightly higher than the physical temperature of the fridge as we normally measure $P_e^{\mathrm{thermal}} =0.03$ at 10 mK.}.

\emph{\textbf{Preparation/Manipulation--}} Unlike the spectroscopy measurements where we constantly send signals and measure the scattering parameters, in time domain measurements we need to carefully send signals to the system with accurate timing and proper duration. This means we need to be able to switch the signals on and off with reasonable accuracy. We use analog RF $I/Q$ \textit{mixers} to perform switching. 
Mixers can be used for modulating and demodulating. For switching purposes, we use mixers as a modulator\footnote{We will see that for readout purposes we use mixers as a demodulator.}. As depicted in Figure~\ref{fig:mixer}a, a typical $I/Q$ mixer, ideally, multiplies the LO signal by signals in port $I$ and $Q$ with 90-degree phase difference\footnote{We will see later that this 90-degree phase difference has a very important role in the qubit preparation and tomography.}. 
\begin{figure}[ht]
\centering
\includegraphics[width = 0.7\textwidth]{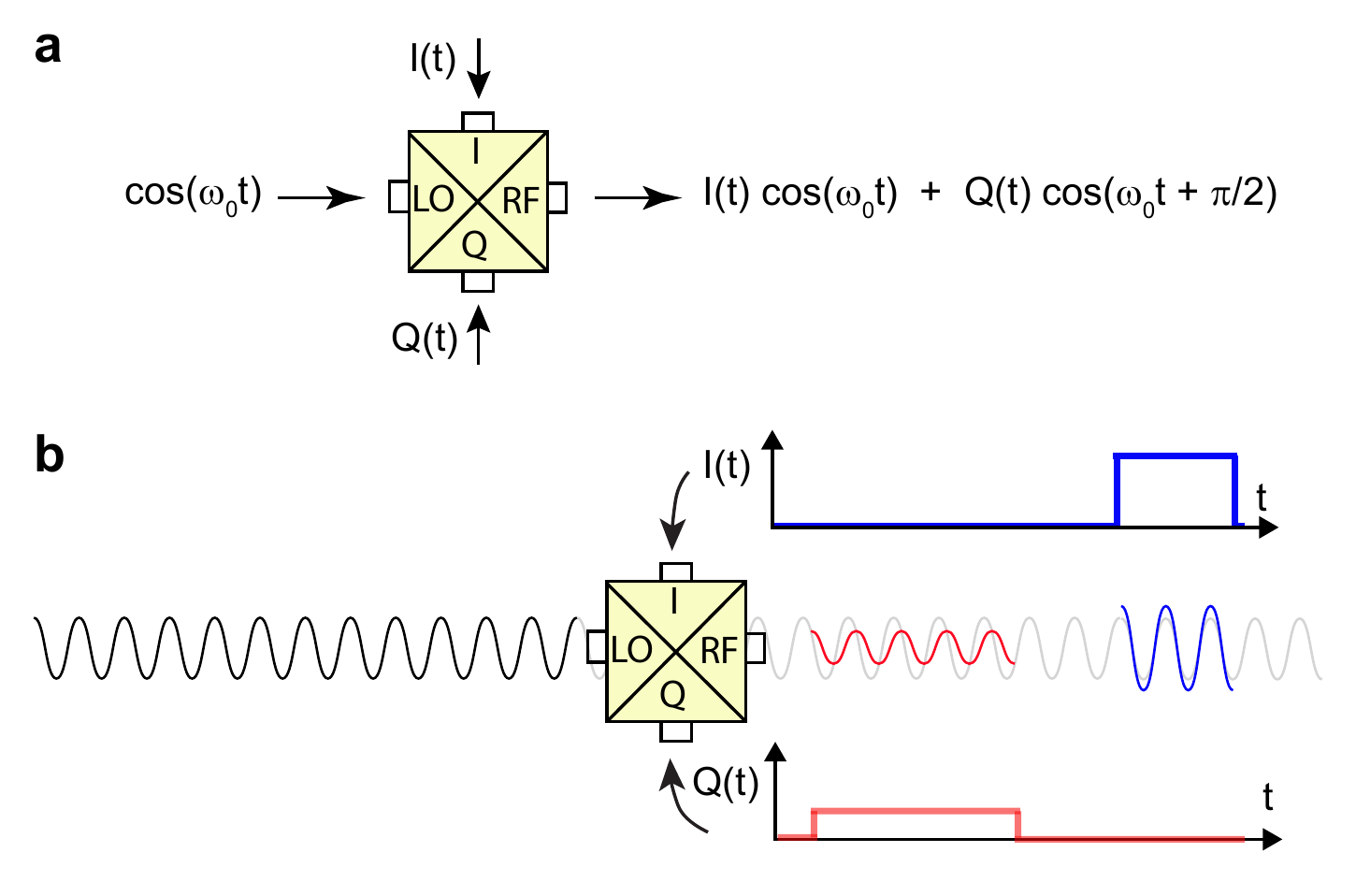}
\caption[$I/Q$ mixer]{ {\footnotesize \textbf{$I/Q$ mixer:} \textbf{a}, An $I/Q$ mixer can be used as a modulator. Note that the outputs correspond to $I$ and $Q$ pulses are out of phase by 90 degrees. \textbf{b}, The  DC pulse in $I/Q$ ports can be used to control (switch) a continuous signal. The blue (red) DC pulse is the input for port $I$ (port $Q$) and its corresponding output is in-phase (90 out-of-phase) with respect to the local oscillator.}} 
\label{fig:mixer}
\end{figure}
Therefore one can use an $I/Q$ mixer in modulation mode to switch a continuous signal. For that, we send the continuous signal to LO port and we switch it by a DC pulse on port $I$ or $Q$. The RF port output is only a segment of the continuous signal LO, as gated by the DC pulses depicted in Figure~\ref{fig:mixer}b. This technique gives us enough control to prepare and manipulate the qubit via Rabi oscillations of the qubit. For example, If we choose the LO frequency to be the qubit frequency then by applying a DC pulse to the port $I$, the pulse rotates the qubit. Thus by choosing a proper duration and amplitude of the DC pulse, we can prepare the qubit in the excited state or a superposition state. Figure~\ref{fig:rabi_along_x_y} demonstrates qubit preparations for the excited state and superposition states where we define pulses in port $I$ (port $Q$) to rotate the qubit along $x$-axis ($y$-axis)\footnote{Of course there is no preferred direction for the qubit as $x$ and $y$. However, the first pulse (first rotation) in each run of the experiment sets a clock reference, determining the rotating frame of the coherent drive. Subsequent signals will rotate the qubit along the same axis if it is in-phase with the first pulse and will rotate in a different axis if it is out-of-phase with respect to the first rotation pulse.}.
\begin{figure}[ht]
\centering
\includegraphics[width = 0.9\textwidth]{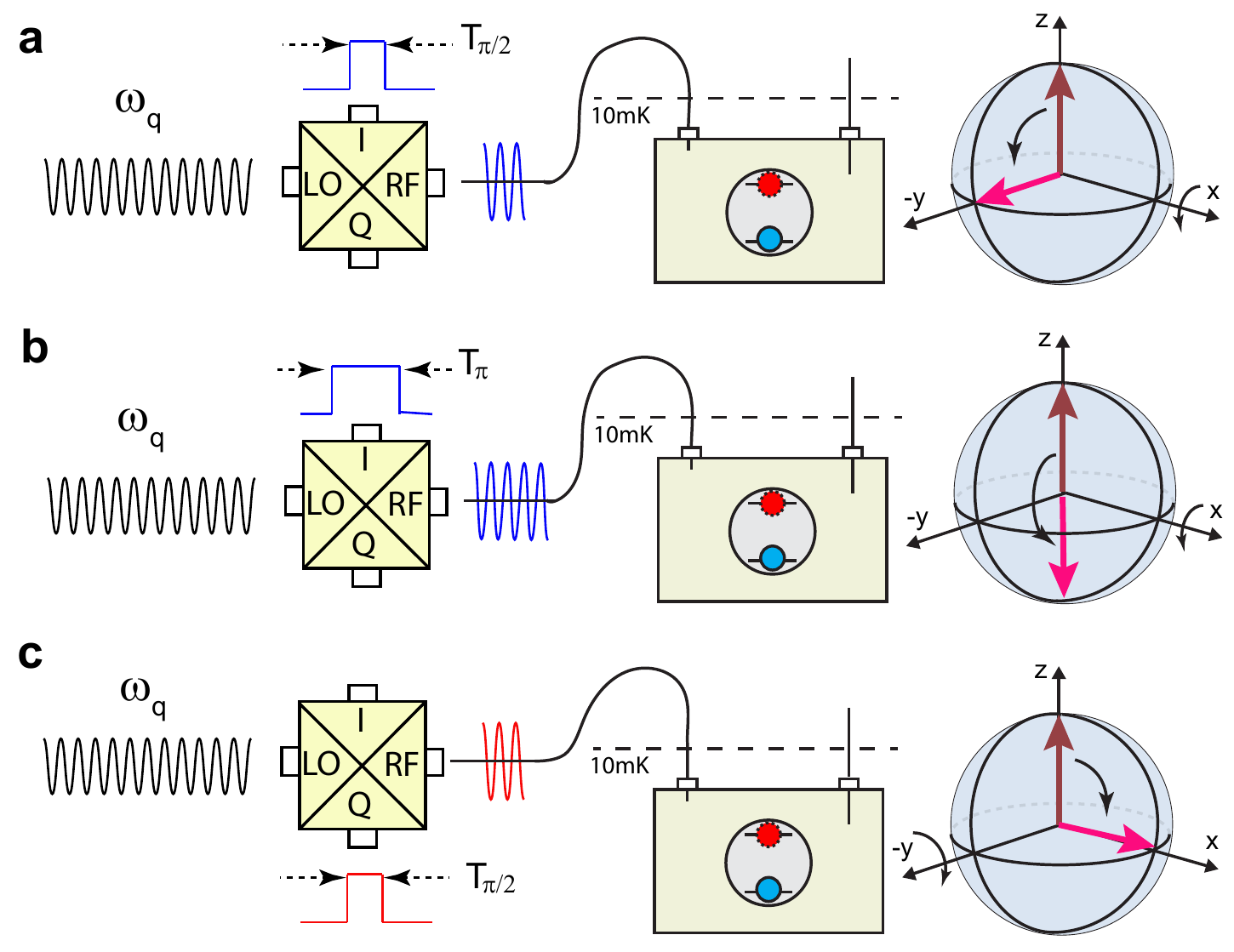}
\caption[Qubit rotation pulses]{ {\footnotesize \textbf{Qubit rotation pulses:} Pulses in the $I,Q$ ports results in rotation in $x,y$ directions in the Bloch sphere. \textbf{a}, A $\pi/2$ pulse in port $I$ rotates the qubit along the $x$ direction and prepares a superposition state along $y$ axis. \textbf{b}, $\pi$ pulse prepares the qubit in the excited state. \textbf{c}, A $\pi/2$ pulse on $Q$ rotates the qubit around the $y$ axis and prepares the qubit in a superposition along $x$ axis.}} 
\label{fig:rabi_along_x_y}
\end{figure}

\emph{Single sideband modulation--} In practice, using DC pulses in $I/Q$ ports to manipulate the qubit has two drawbacks due to the mixer nonlinearity. First, the mixer may not exactly provide 90-degree phase difference between $I/Q$ ports which is not convenient and requires careful corrections for tomography results. The second drawback is that even when we do not apply a pulse to the $I/Q$ ports there may be some {signal} leakage from the LO port to the RF port. This leakage can be minimized by adding DC offsets to the $I$ and $Q$ inputs. But even very small leakage constantly drives the qubit and causes imperfections in the experiment. One way around this issue is to employ a single sideband modulation (SSB) technique for the qubit pulses. The idea is to set LO frequency to be $\omega_q \pm \Omega_{\mathrm{SSB}}$ where $ \Omega_{\mathrm{SSB}} \sim 100$-$500$ MHz is the sideband frequency. Then, for $I/Q$ ports we apply signals with the frequency of $\Omega_{\mathrm{SSB}}$ with $\pm$90 degrees out-of-phase to up-convert (down-convert) the LO to qubit frequency as depicted in Figure~\ref{fig:SSB} for the up-converting case. 
\begin{figure}[ht]
\centering
\includegraphics[width = 0.8\textwidth]{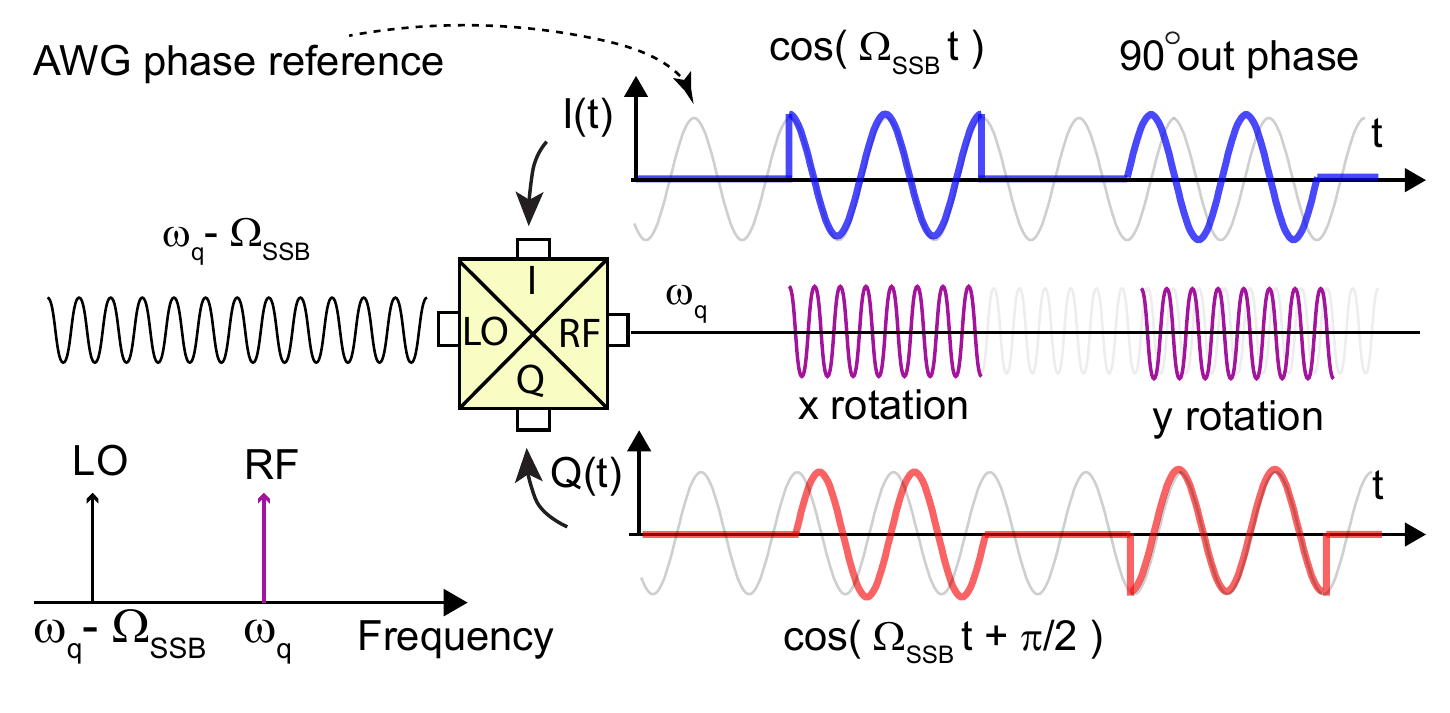}
\caption[Single sideband modulation (SSB)]{{\footnotesize \textbf{Single Sideband Modulation (SSB):} By up-converting the LO signal by $\Omega_{\mathrm{SSB}}$ the mixer outputs at qubit frequency. The relative phase of SSB pulses determines the phase of the output signal thus the direction of the rotation for the qubit.}} 
\label{fig:SSB}
\end{figure}
SSB solves both drawbacks we had with DC pulsing technique. In this case, we don't need to worry about ``mixer non-orthogonality" because the phase of the output pulse is set by the phase of the $I/Q$ signal\footnote{The mixer non-orthogonality, in this case, may cause some issues with carrier leakage or leakages in opposite sideband and higher harmonics. But that can be compensated by adjusting the phase in AWG pulses.}. Therefore, the phase stability of the arbitrary waveform generator (AWG) sets our rotation axis accuracy, which is normally good enough for sideband frequencies of a few hundred MHz. As it is shown in Figure~\ref{fig:SSB} the first pulses in $I/Q$ port defines the preferred axis (we defined this to be the $x$ axis). The phase of subsequent pulses referenced by AWG determines the rotation axis as it is shown for 90 out-of-phase pulse which results in a qubit rotation along $y$ axis.

Moreover, we don't need to worry about small leakage of LO to RF because the LO frequency is off-resonant by $\Omega_\mathrm{SSB}$ and won't disturb the qubit.

\emph{\textbf{Readout--}} We discussed how to manipulate the qubit state by sending signals with frequency $\omega_q$ into the system to rotate the qubit. Here we discuss how to actually measure the qubit state by sending a signal at the cavity frequency $\omega_c$.
As we discussed in the previous chapter, the frequency of the cavity depends on the state of the qubit. Therefore, the natural way to detect the state of the qubit is to measure the phase shift across the cavity by using homodyne measurement. We take two copies of a coherent signal {that has the frequency of the cavity}\footnote{This frequency is the cavity frequency at low power meaning the dressed-cavity frequency. When we use the cavity dressed state frequency for readout we usually need to send it at low-power and this readout is called ``low-power readout''. Often the fidelity of low-power readout is not very good unless we use a parametric amplifier. However, one can also use cavity bare-frequency and again see the phase shift and detect the state of the qubit. This method essentially uses the nonlinearity induced by the qubit-cavity interaction and requires more power. This method is called high-power readout~\cite{reed2010} and does not require a parametric amplifier.}) and whenever we decide to readout the state of the qubit, we send one of the copies to the cavity and take the transmitted (reflected) signal back and demodulate it with the other copy\footnote{Basically, the idea is to make a ``microwave interferometer" except here instead of adding signals together we multiply them together.}. The demodulation results in a DC signal corresponding to the phase difference between two copies. By reading the DC signal we can infer whether the cavity frequency shifted up or down. Figure~\ref{fig:readout} demonstrates the readout process.
\begin{figure}[ht]
\centering
\includegraphics[width = 0.98\textwidth]{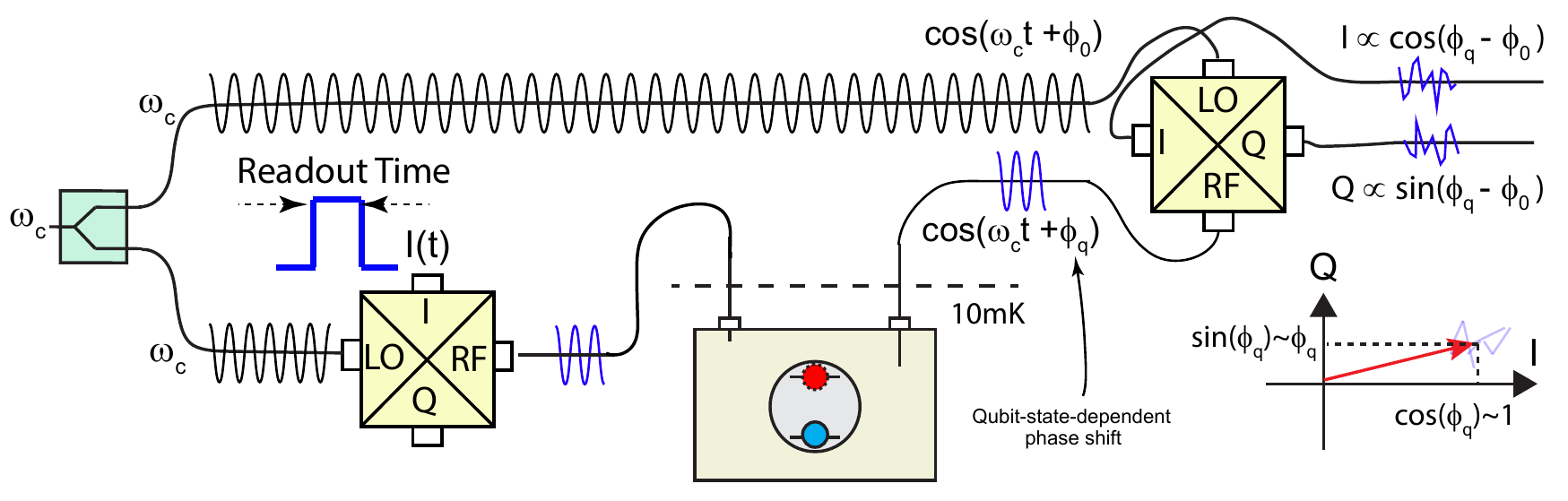}
\caption[Qubit state readout, homophone detection]{ {\footnotesize \textbf{Qubit state readout, homophone detection:} The qubit-state-dependent phase shift is detected by comparing the signal which passes the cavity with the reference signal.}} 
\label{fig:readout}
\end{figure}
The phase $\phi_0$ accounts for the fixed phase difference between two signals due to the different path length and can be set to zero by adding a phase shifter ($\phi_0=0$). Therefore the demodulation gives $I=\cos(\phi_q)$ and $Q=\sin(\phi_q)$. For most practical situations the phase shift is small $\cos(\phi_q) \simeq 1$, and therefore the phase shift (information about qubit state) is encoded in only in one quadrature of the reflected signal $Q=\sin(\phi_q) \simeq \phi_q$. The entire readout process can be simply represented in a phasor diagram (see Figure~\ref{fig:phasor_readout}).
\begin{figure}[ht]
\centering
\includegraphics[width = 0.98\textwidth]{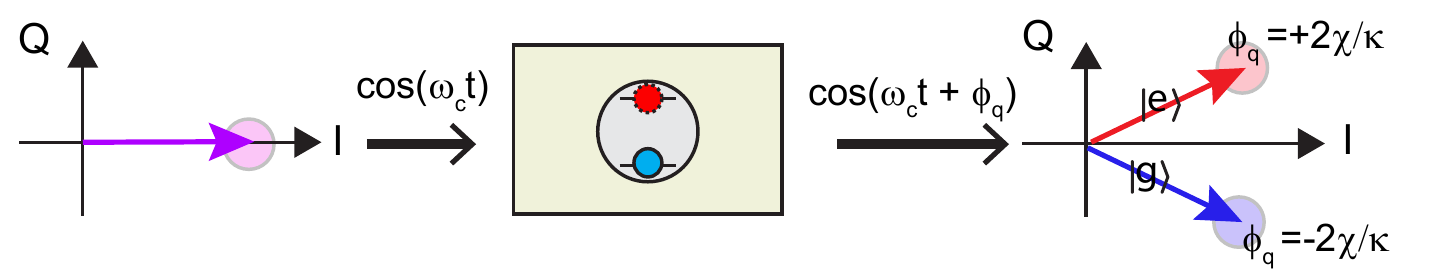}
\caption[Readout in phase space representation]{ {\footnotesize \textbf{Readout in phase space representation:} A coherent signal with a minimum uncertainty in each quadrature probes the cavity whose frequency shifts by $\pm \chi$ depends on the state of the qubit. The transmitted (reflected) signal acquires a state-dependent-phase shift $\pm 2\chi/\kappa$ as discussed in Section~\ref{section:cavity}.}} 
\label{fig:phasor_readout}
\end{figure}
In order to measure the qubit state, we need to repeat the experiment $N$ times ($N>100$) and each time we detect the phase shift by a value in the $Q$ quadrature. For positive (negative) values we assign $-1$ $(+1)$, indicating that we have found the qubit in the excited (ground) state. After repeating the experiment $N$ times, we gather these statistics and report the population for ground and excited state as 
\begin{eqnarray}
P_g=\frac{N_+ }{N_+ + N_-} \pm  \sqrt{\frac{N_+ N_-}{N^3}}  \hspace{0.5cm} , \hspace{0.5cm} P_e=1-P_g,
\end{eqnarray}
where $N_\pm$ is the number of the experiments where we found the qubit in ground (excited) state inferred by positive (negative) values for $Q$. The error $\sqrt{(N_+ N_-)/N^3}$ is calculated based on binomial error.
\subsection{Rabi measurements}
Now we are ready to discuss Rabi measurements. Once we know the qubit frequency from spectroscopy, then the first experiment in the time domain is to see Rabi oscillation, since this experiment doesn't require any pulse calibration. The idea is to send a qubit signal (with duration $t$) to the system and right after that send the cavity readout pulse and readout the state of the qubit. Normally, we repeat this experiment for different durations\footnote{Normally we sweep $t$ from 0 to 100 ns in 100 steps. In each step the experiment is repeated $N \sim$ 200 times to have a reasonably small binomial error so we can see clear Rabi oscillations.} $t$ and for each timestep, we repeat the experiment $N$ times.
\begin{figure}[ht]
\centering
\includegraphics[width = 0.98\textwidth]{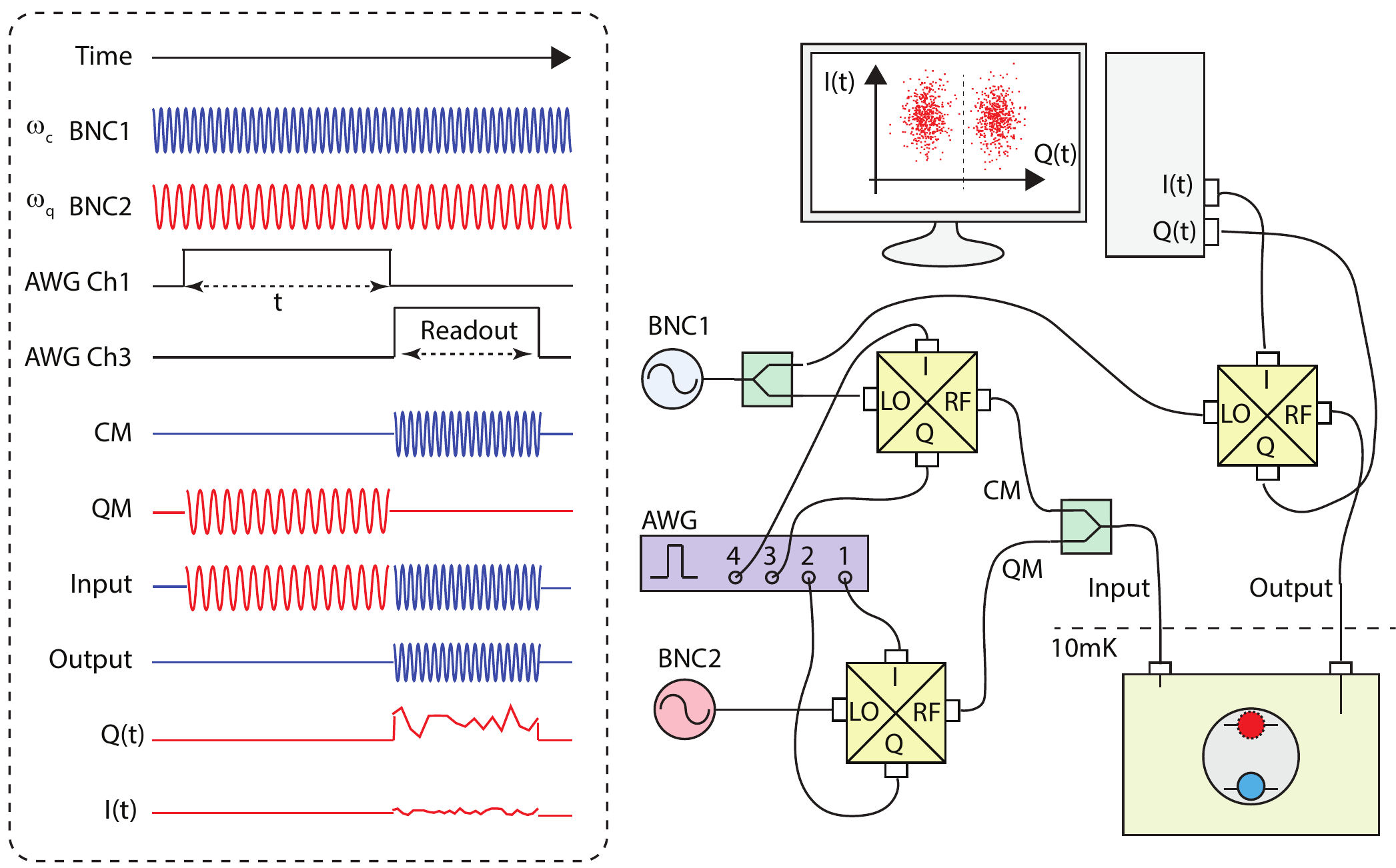}
\caption[Rabi measurement]{ {\footnotesize \textbf{Rabi measurement:} The sequence for the measurement of Rabi oscillations and the typical room temperature circuitry are shown.}} 
\label{fig:rabi}
\end{figure}
Figure~\ref{fig:rabi} summarizes the Rabi experiment setup and procedure. As we discussed in the previous section, in order to control the qubit signal, we use a mixer as a switch which is controlled by DC pulses from an arbitrary waveform generator (AWG). Another mixer is used to control the cavity with the same technique to perform homodyne measurement on the cavity and detect the phase shift to readout the state of the qubit. Figure~\ref{fig:rabi100ns} shows a typical Rabi oscillation.
\begin{figure}[ht]
\centering
\includegraphics[width = 0.98\textwidth]{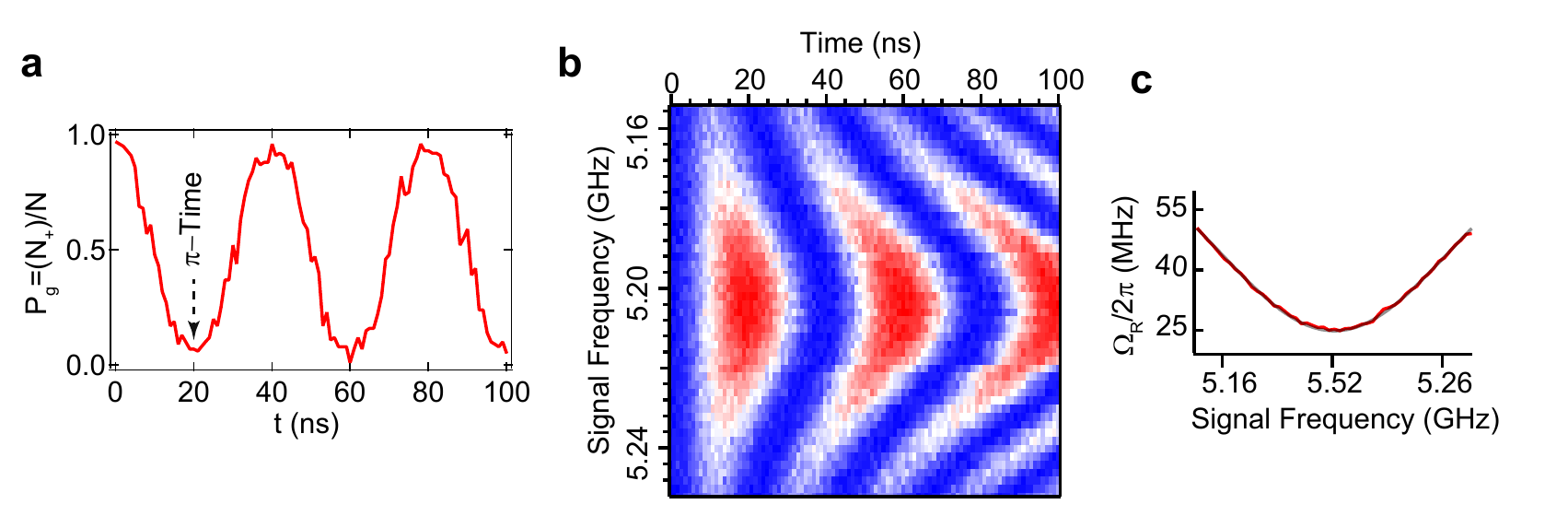}
\caption[Chevron plot]{ {\footnotesize \textbf{Chevron plot:} \textbf{a}, A typical result of Rabi oscillation measurements. \textbf{b}, By repeating the Rabi measurement for different qubit frequencies we obtain the ``chevron plot" which can be used to calibrate the qubit frequency. \textbf{c}, The Rabi oscillation frequency versus qubit pulse frequency. As we discussed in Chapter~2, the minimum oscillation rate corresponds to the maximum contrast for Rabi oscillations which happens when the qubit pulse is on-resonance with qubit frequency.}} 
\label{fig:rabi100ns}
\end{figure}
The reason we use a rather short Rabi time (100-200 ns) for Rabi measurement is because we will use this Rabi sequence to calibrate preparation pulses ($\pi$-pulse and $\pi/2$-pulse) which are normally short pulses\footnote{Moreover it is wise to start off by a short sequence for the experiment because we might have a qubit with short decoherence times or there might be some calibration issue in the system which could make it hard to see the qubit evolution at longer qubit evolution times.}. In this step, we may also tweak the readout power, frequency, and phase to maximize the oscillation contrast as depicted in Figure~\ref{fig:rabi100ns}a.

In order to calibrate $\pi$-pulses, we first need to make sure that the qubit frequency is accurate and oscillations are on-resonance with the qubit. One way to check this is to sweep the qubit frequency while performing Rabi oscillation measurements. The resulting 2D color plot is called ``chevron plot". By fitting a sine-wave to the oscillations, one can find the best estimate of the minimum oscillation frequency which is the qubit frequency\footnote{Rabi spectroscopy is not super sensitive to the detuning in the regime of fast Rabi oscillation. Moreover, one might think that with a stronger drive we might also stark shift the qubit. So, one would think in order to find qubit frequency it is better to have longer Rabi sequence and slower Rabi oscillation to improve precision. But since our main concern is $\pi$-pulse calibration, it makes sense to do Rabi spectroscopy (chevron plot) with actual power that we are going to use for the $\pi$-pulses.}. Later we will see that with Ramsey measurement we can have a better estimation of the qubit frequency. After doing this calibration, we know the power, frequency and the duration for $\pi$-pulse and $\pi/2$-pulse.

\subsection{$T_1$ Measurement}
One of the main characteristics of a qubit is its relaxation time. In order to measure the lifetime of the qubit we do the following sequence: 1) We prepare the qubit in the excited state by sending a $\pi$-pulse to the qubit. 2) We wait for time $t$, then 3) we measure the qubit state by sending readout pulse to the cavity. Therefore we use the same setup that we had in for Rabi oscillation measurements (see Figure~\ref{fig:rabi} for schematics) but we use a slightly different sequence from the AWG. Figure~\ref{fig:T1_measurement} summarizes the sequence and a result we get from a $T_1$ measurement.
\begin{figure}[ht]
\centering
\includegraphics[width = 0.98\textwidth]{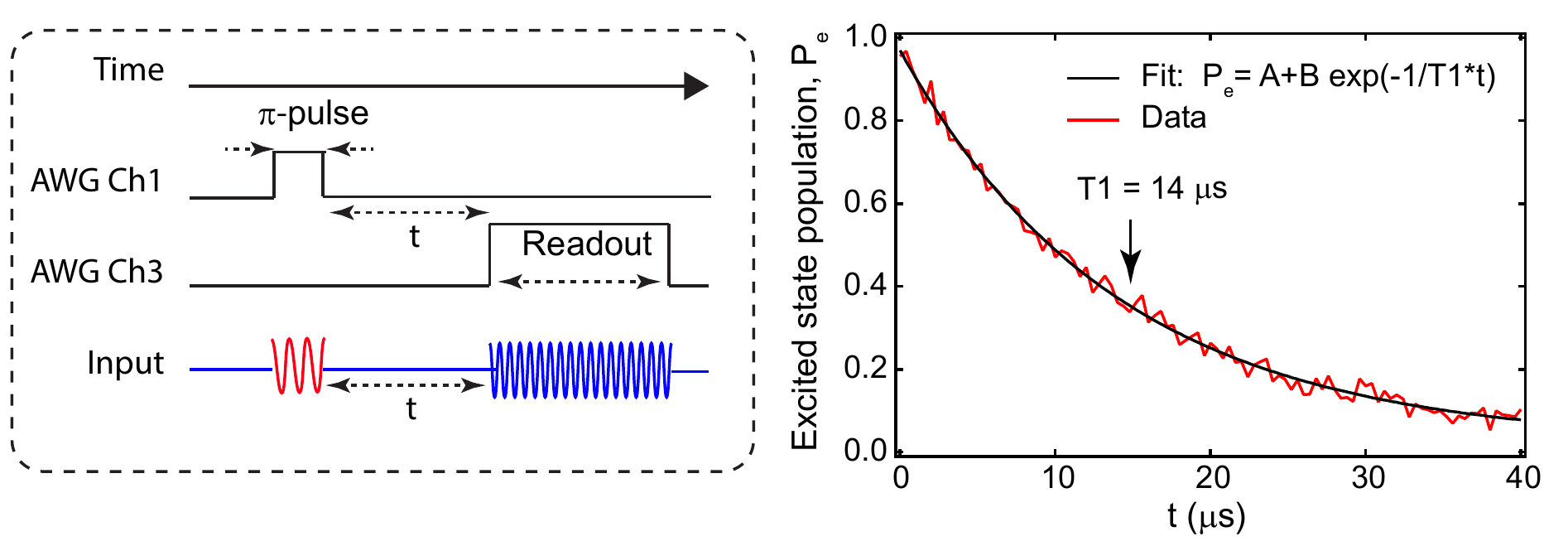}
\caption[$T_1$ measurement]{ {\footnotesize \textbf{$T_1$ Measurement:} The $T_1$ measurement sequence (consider the experimental setup depicted in Figure~\ref{fig:rabi}) and a typical result. The qubit lifetime (relaxation time $T_1$) can be measured by fitting the result  to an exponential decay function.}} 
\label{fig:T1_measurement}
\end{figure}

\subsection{Ramsey Measurement ($T_2^*$)}
With Ramsey measurement, we characterize the dephasing time for the qubit, $T_2^*$. For that, again we use the same setup as we had for measurements of Rabi oscillations (see Figure~\ref{fig:rabi}). The Ramsey sequence follows as: 1) prepare the qubit in superposition state $1/\sqrt{2}(|g\rangle + i | e\rangle)$ by applying a $\pi/2$-pulse to the qubit. 2) Wait for a time, then $t$ 3) apply another $\pi/2$-pulse to bring back the qubit to ground state\footnote{If the last $\pi/2$-pulse has the same phase as the first $\pi/2$-pulse we will put the qubit in the excited state. If we do negative pulse (opposite rotation) we bring the qubit back to ground state. Either way is fine, all we need is to bring the qubit to an eigenstate.} and immediately 4) perform readout. The sequence for Ramsey measurement and the result has been shown in fig~\ref{fig:ramsey}.
\begin{figure}[ht]
\centering
\includegraphics[width = 0.98\textwidth]{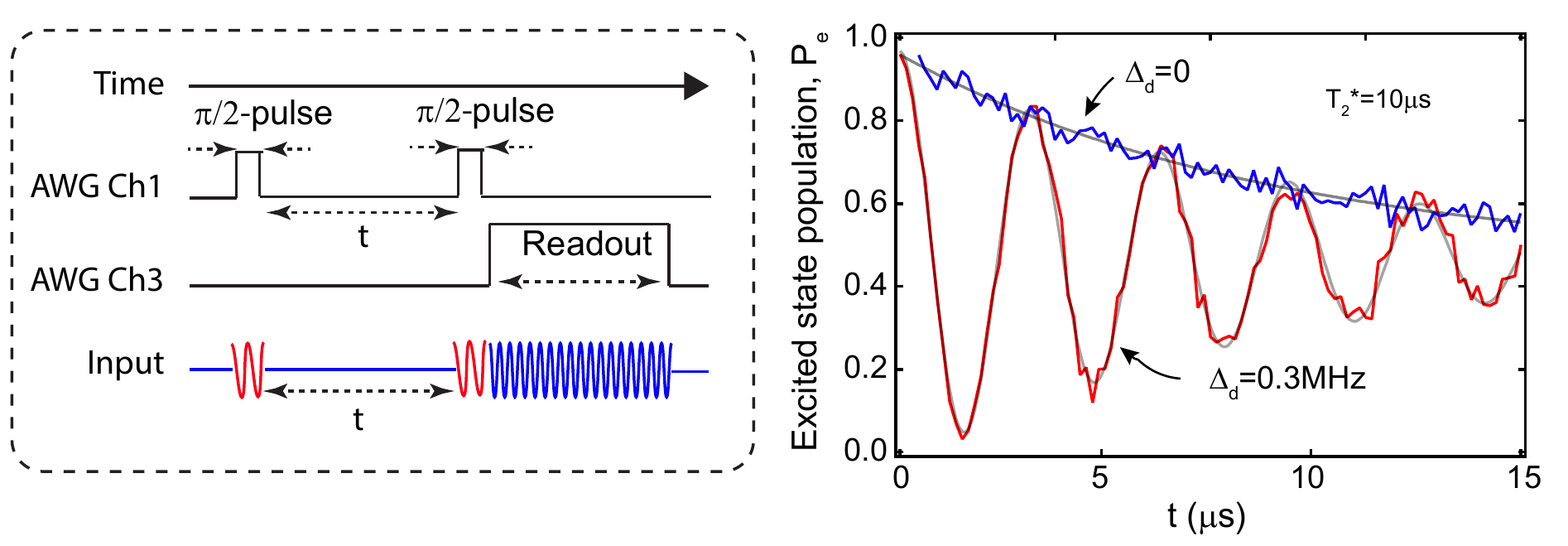}
\caption[Ramsey measurement]{ {\footnotesize \textbf{Ramsey measurement:} The sequence for the Ramsey measurement (consider the experimental setup depicted in Figure~\ref{fig:rabi}) and the typical Ramsey result. The blue (red) curve is when the pulses are on-resonance (0.3 MHz off-resonance) with the qubit frequency. The dephasing time $T_2^*$ can be determined by fitting the data to exponentially decaying sine function $F(t)=A+B\sin(2 \pi \Delta_d t+\phi) \exp(t/T_2^*)$.}} 
\label{fig:ramsey}
\end{figure}

\subsection{Full state tomography}
The qubit readout projects the state of the qubit along $z$ axis. Therefore one can determine the expectation value for $\sigma_z$ operator as,
\begin{equation}
z=\langle \sigma_z \rangle = P_g-P_e = \frac{N_+ - N_-}{N_+ + N_-} \pm 2 \sqrt{ \frac{N_+ N_-}{N^3}}.
\end{equation}
Similarly, one can determine the expectation value for $\sigma_x$ and $\sigma_y$ as well by applying a 90 degree rotation pulse along $y$ and $x$ respectively right before the readout\footnote{This is exactly what we do in Ramsey where we prepare the qubit in $x$ and measure $\langle x\rangle$.}. Therefore, a full tomography sequence has 3 copy of each sequence with no rotation ($z$), $\pi/2$-rotation in phase ($x$), and $\pi/2$-rotation 90 degree out phase ($y$) right before the readout as depicted in figure~\ref{fig:state_tomog}.
\begin{figure}[ht]
\centering
\includegraphics[width = 0.9\textwidth]{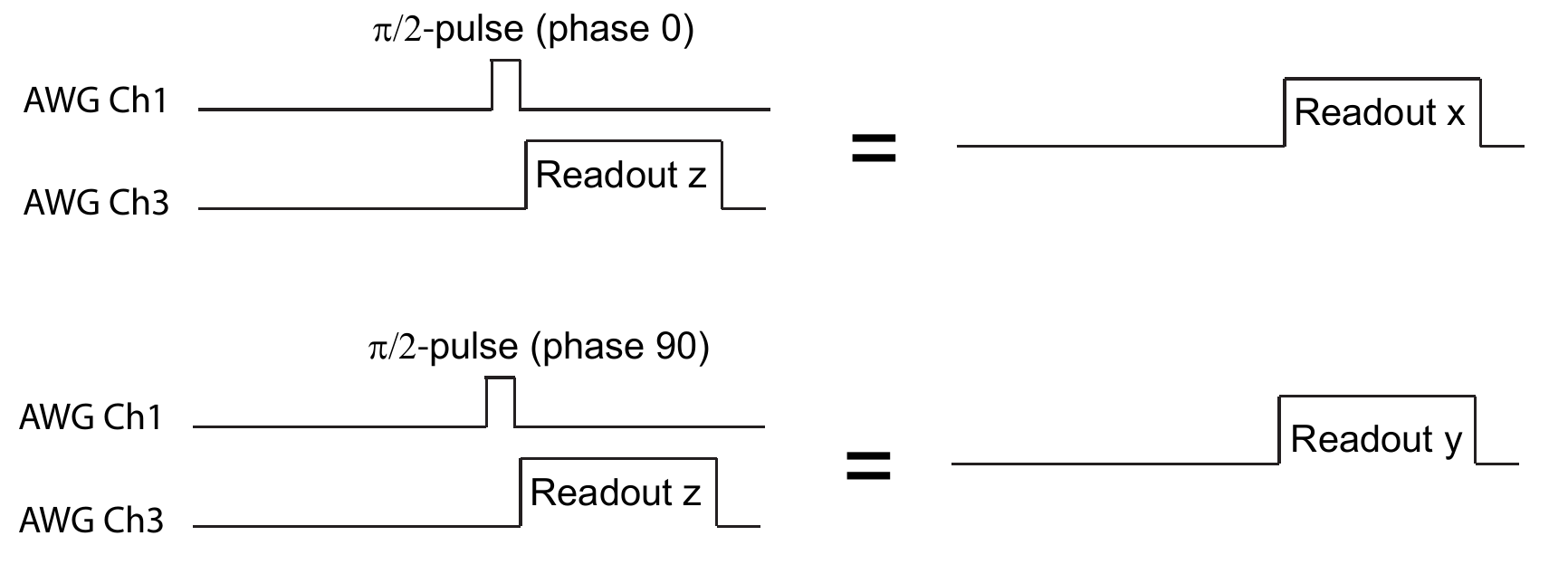}
\caption[Full state tomography readout pulses]{ {\footnotesize \textbf{Full state tomography readout pulses:} A readout pulse without any tomographic pulse gives the expectation value for $\sigma_z$ since it projects the qubit in ground or excited state. A readout pulse augmented by tomographic pulse can be used to measure the expectation values for $\sigma_x, \sigma_y$ or any arbitrary basis in the Bloch sphere.}} 
\label{fig:state_tomog}
\end{figure}
The expectation values for $\sigma_x$ and $\sigma_y$, can be determined in a same way,
\begin{eqnarray}
x=\langle \sigma_x \rangle = P_g^{(x)}-P_e^{(x)} = \frac{N_+ - N_-}{N_+ + N_-}\\
y=\langle \sigma_y \rangle = P_g^{(y)}-P_e^{(y)} = \frac{N_+ - N_-}{N_+ + N_-},
\end{eqnarray}
where superscripts$\ ^{(x)}$ and $\ ^{(y)}$ indicate that the readout has in-phase and out-of-phase rotation pulse respectively.

\section{Josephson Parametric Amplifier}
The signals that carry quantum information at cryogenic temperature are feeble microwave signals. Especially for weak measurement, these signals often contain $\sim 1$ photon per microsecond or less on average. Therefore measurement signals need to be amplified before processing at room temperature where they would otherwise be contaminated by thermal noise. Amplification is essential and often multiple steps of amplification are needed.
However, amplifiers add some noise to the signal at each step of amplification.  The added noise is not just a technical subtlety but it is rather a fundamental property of quantum mechanics~\cite{slichterthesis}.

In this section I follow the discussion in Ref.~\cite{slichterthesis} and briefly discuss the Josephson parametric amplifier and phase sensitive amplification. I will try to connect the discussion to the previous chapters and add some points from the experimental perspective. For a detailed study of noise and amplification see the nice discussions in Ref. \cite{slichterthesis}.

\subsection{Classical nonlinear oscillators\label{Subsection:JPAtheory}}

Similar to the transmon circuit discussed in Chapter~2, the Josephson parametric amplifier (JPA) is a nonlinear oscillator except that the critical current is much higher for a JPA\footnote{Typically $I_0^{\mathrm{(JPA)}} \sim10 I_0^{\mathrm{(Transmon)}}, (E_J/E_c)^{\mathrm{(JPA)}}=100 (E_J/E_c)^{\mathrm{(Transmon)}}$.} compared to the transmon. This means that for JPA there are many more energy levels bound in the potential. Moreover, a higher critical current means a weaker nonlinearity. Therefore a JPA can be treated classically as an oscillator which has a weak nonlinearity. Figure~\ref{fig:JPA_simple} shows the schematic for a JPA where we include the current source and corresponding impedance $Z_0$ to drive (pump) the JPA\footnote{Remember, we drive the transmon through the coupling between transmon electric dipole and the electric field. Here we directly drive the JPA by connecting it to a current source via an effective impedance $Z_0$.}.
\begin{figure}[ht]
\centering
\includegraphics[width = 0.69\textwidth]{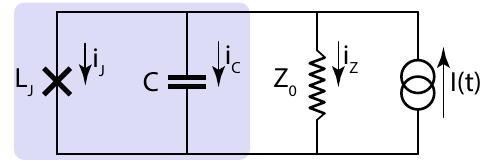}
\caption[JPA Schematic]{ {\footnotesize \textbf{JPA Schematic:}  A Josephson parametric amplifier can be considered as a nonlinear LC oscillator (shaded region) connected to a current source via impedance $Z_0$ .}} 
\label{fig:JPA_simple}
\end{figure}
For the currents flowing in the circuit we have,
\begin{subequations}\label{eq:LPA_ODE1}
\begin{eqnarray}
i_J + i_C + i_Z= I(t)\\
I_0 \sin(\delta) + C\dot{V} + \frac{V}{Z_0}= I(t)\\
\xrightarrow{V=V_J=\frac{\Phi}{2 \pi} \dot{\delta}} I_0 \sin(\delta) + C \frac{\Phi_0}{2\pi} \ddot{\delta} + \frac{\Phi_0}{2 \pi Z_0} \dot{\delta}= I(t),
\end{eqnarray}
\end{subequations}
where we use Josephson relations for $i_J$ and $V$ which is the voltage across the components\footnote{The voltage across the parallel components are equal, we substitute its value by the voltage across the JJ.}. We drive the JPA with a coherent classical signal $I(t)=I_{p} \cos(\omega_{p} +\phi_{p})$ and may assume $I_p<I_0$ which insures that $i_J<I_0$. However, we are interested to the regime where the $i_J\ll I_0$ which means the $\delta$ is small enough so that we can expand the $\sin(\delta)$ to up to order $\delta^3$,
\begin{subequations}\label{eq:Duffing_JPA1}
\begin{eqnarray}
I_0 (\delta - \delta^3/6) + C \frac{\Phi_0}{2\pi} \ddot{\delta} + \frac{\Phi_0}{2 \pi Z_0} \dot{\delta}&=& I_p \cos(\omega_p + \phi_p)\\
\to  \ddot{\delta} + 2\Gamma \dot{\delta} +  \omega_0^2 (\delta - \delta^3/6) &=& \omega_0^2 \frac{I_p}{I_0} \cos(\omega_p + \phi_p),
\end{eqnarray}
\end{subequations}
where $\omega_0=\sqrt{\frac{2 \pi I_0}{C \Phi_0}}$ is the natural frequency\footnote{We may call this the ``unloaded" frequency of the JPA when it is disconnected from load $Z_0$, or $Z_0 \to \infty$.} of the JPA resonator (shaded region in Figure~\ref{fig:JPA_simple}). The parameter $\Gamma=1/(2CZ_0)$ accounts for damping of the resonator due to the coupling to the $Z_0$. 

The Equation~\eqref{eq:Duffing_JPA1}b is the well-known \emph{Duffing Equation} which appears in many nonlinear situations ranging from the pendulum to harmonic frequency generation in nonlinear optics. There are variety of methods for solving the Equation~\eqref{eq:Duffing_JPA1}b. Here I follow the method in Ref~\cite{slichterthesis}. 

We set the phase of the pump as a reference $\phi_p=0$ and consider the solution for $\delta$ to have both components; in-phase and out-of-phase with the pump. Therefore we use the following ansatz,
\begin{eqnarray}\label{eq:ansatz_JPA} 
\delta=\delta_0 \cos(\omega_p + \theta) = \delta_{\parallel} \cos(\omega_p t) + \delta_{\perp} \sin(\omega_p t),
\end{eqnarray}
where the $\delta_{\parallel} (\delta_{\perp})$ is the amplitude of in-phase (out-of-phase) ``oscillations" in the JPA circuit\footnote{Note that $\delta$ is `the difference between the phases of the superconducting order parameter on each side of the junction'. But it easily parameterizes  the current and voltage in the circuit as we shall see in Equation~\eqref{eq:LPA_ODE1}. So we may simply refer to it as the ``oscillation" in the circuit for now.} By plugging the ansatz~\eqref{eq:ansatz_JPA} into \eqref{eq:Duffing_JPA1}b, we have
\begin{eqnarray}\label{eq:Duffing_JPA2} 
(\omega_0^2-\omega_p^2) \left[ \delta_{\parallel} \cos(\omega_p t) + \delta_{\perp} \sin(\omega_p t) \right] \hspace{7cm} \nonumber  \\
+ 2\Gamma \left[  \omega_0 \delta_{\perp} \cos(\omega_p t)  -\omega_0 \delta_{\parallel} \sin(\omega_p t)   \right]  \hspace{6cm} \nonumber   \\
- \omega_0^2/6 \left[   \delta^3_{\parallel} \cos^3(\omega_p t) + \delta^3_{\perp} \sin^3(\omega_p t) \right]  \hspace{5cm} \nonumber  \\
 -  \omega_0^2/2 \left[ \delta^2_{\parallel} \delta_{\perp} \cos^2(\omega_p t)  \sin(\omega_p t)  +  \delta_{\parallel} \delta^2_{\perp} \cos(\omega_p t)  \sin^2(\omega_p t)  \right]  \hspace{0cm} \nonumber  \\
 = \omega_0^2 I_d/I_0 \cos(\omega_p t). \hspace{3cm}
\end{eqnarray}
Now we apply the RWA for fast oscillating terms\footnote{For example; $\cos^3(\omega_p t) \to 3/4 \cos(\omega_p t) \  \mathrm{and} \  \cos^2(\omega_p t)\sin(\omega_p t) \to 1/4 \sin(\omega_p t)$.} to obtain the following equations,
\begin{eqnarray}\label{eq:Duffing_JPA3} 
(\omega_0^2-\omega_p^2)  \delta_{\parallel} + 2\Gamma  \omega_0 \delta_{\perp} - \omega_0^2/8 \left[   \delta^3_{\parallel}  +  \delta_{\parallel} \delta^2_{\perp}) \right] &=& \omega_0^2 I_p/I_0 \\
(\omega_0^2-\omega_p^2)  \delta_{\perp} - 2\Gamma  \omega_0 \delta_{\parallel} - \omega_0^2/8 \left[   \delta^3_{\perp}  +  \delta_{\perp} \delta^2_{\parallel} \right] &=& 0.
\end{eqnarray}
One can rearrange the above equations in terms of the quality factor of the resonator $Q=\omega_0 Z_0 C$ and dimensionless detuning $\tilde{\Delta}= 2Q (1-\omega_p/\omega_0)$,
\begin{subequations}\label{eq:Duffing_JPA4}
\begin{eqnarray} 
\delta_{\perp} + \delta_{\parallel} \left[  \tilde{\Delta} - \frac{Q}{8}( \delta^2_{\parallel} +  \delta^2_{\perp}) \right] &=& Q I_p/I_0 \\
-\delta_{\parallel} + \delta_{\perp} \left[  \tilde{\Delta} - \frac{Q}{8}( \delta^2_{\parallel} +  \delta^2_{\perp} )\right] &=& 0.
\end{eqnarray}
\end{subequations}
Figure~\ref{fig:JPA_tilt2} shows the numerical solution to Equation~\ref{eq:Duffing_JPA4} for $\delta_0^2$ and $\theta$ versus dimensionless detuning $\tilde{\Delta}$. Note, $\delta_0^2=\delta_{\perp}^2 + \delta_{\parallel}^2,\ \theta=\mathrm{atan}[\delta_{\perp}/\delta_{\parallel}]$.
\begin{figure}[ht]
\centering
\includegraphics[width = 0.98\textwidth]{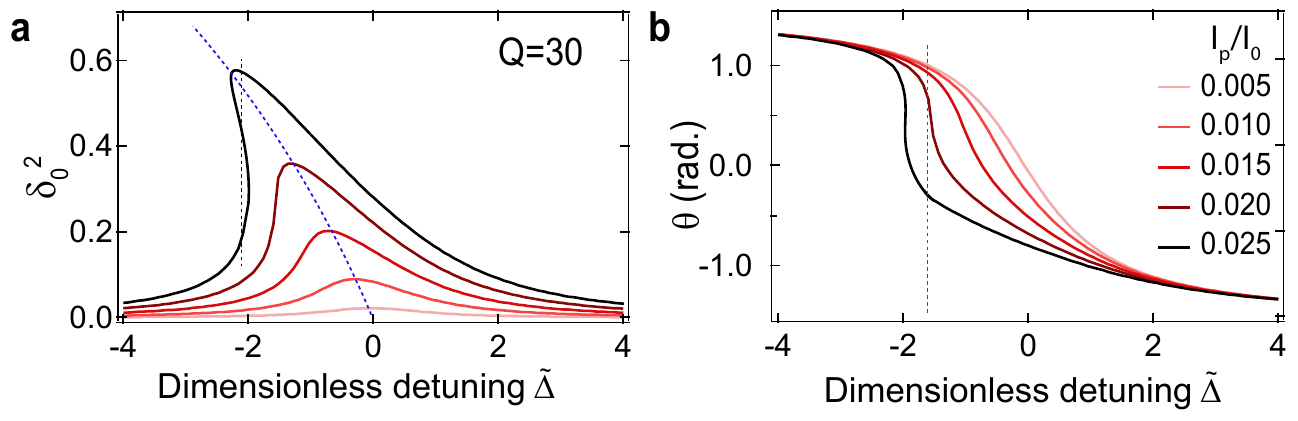}
\caption[Duffing resonator response]{{\footnotesize \textbf{Duffing resonator response:} \textbf{a}. Solutions to Equations~\ref{eq:Duffing_JPA4}. \textbf{a} The resonance frequency decreases by increasing the power and ultimately the system enters the bi-stable regime where there are more that one solution to the Equation~\ref{eq:Duffing_JPA4}. \textbf{b}, This bifurcation behavior also can be seen in the phase response of the oscillator. Right before the bifurcation, the system exhibits sharp response respect to the detuning and power. The sensitivity for power is more clearly shown in Figure~\ref{fig:JPA_pup_transfer}}}
\label{fig:JPA_tilt2}
\end{figure}
Unlike in a linear resonator (e.g. a bare cavity\footnote{For a cavity which hybridized with a qubit, you may see similar nonlinear behavior in the punch-out experiment during the transition between low to high power.}) where the frequency is independent of the power, for nonlinear resonator the frequency decreases for higher driving power (Fig.~\ref{fig:JPA_tilt2}).

After certain power, the JPA bifurcates signaling that the system has more than one steady state. For amplification purposes, the ``sweet spot'' is right before this bifurcation where the system exhibits maximum sensitivity as manifest by a sharp slope in phase. 

In order to understand how an amplifier works at the sweet spot, let's look at the phase $\theta$ versus deriving power $I_p/I_0$ as depicted in Figure~\ref{fig:JPA_pup_transfer}.
\begin{figure}[ht]
\centering
\includegraphics[width = 0.8\textwidth]{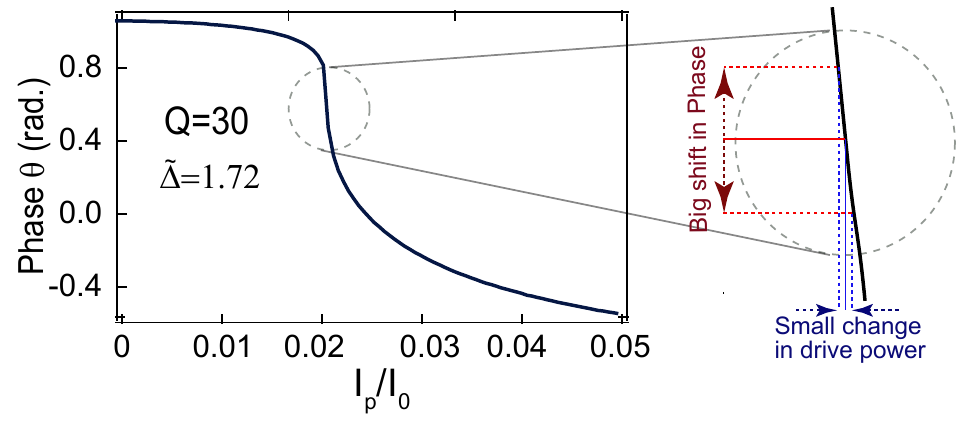}
\caption[JPA transfer function]{{\footnotesize \textbf{JPA transfer function:} The change in phase of the oscillations in JPA versus the power of the drive. This curve is a cross section from Figure~\ref{fig:JPA_tilt2}b at $\tilde{\Delta}=1.72$ (red dashed line), shwing a sharp response of the phase to the small changes in the power which is the essence of the JPA amplification.}} 
\label{fig:JPA_pup_transfer}
\end{figure}
Figure~\ref{fig:JPA_pup_transfer} is a cross section of Figure~\ref{fig:JPA_tilt2}b at $\tilde{\Delta}=1.72$ (red dashed line). The fact that the phase response is also sharp with respect to the small change in the power right before the bifurcation power is the essence of JPA amplification\footnote{Note that $\theta$ here is the phase of the oscillation inside the resonator. Using the input-output relation, one can show that the similar behavior also manifested at the phase of the reflected signal from the resonator (e.g see chapter~3 in Ref.~\cite{drummond2013quantum}).}.\\
 
\noindent\fbox{\parbox{\textwidth}{\textbf{Exercise~2:} Use the input-output theory and show that the similar sensitivity manifested in the reflected signal off of the JPA. Show that the phase of the reflected signal dramatically changes by a small change in the signal power.
}} \vspace{0.25cm}

\subsection{Paramp operation}

Now we turn to connect our understanding of the JPA and its transfer function to the actual situation that happens in the experiment. Figure~\ref{fig:paramp_added} displays the minimum circuitry inside the fridge when we add the paramp in the line. 

\begin{figure}[ht]
\centering
\includegraphics[width = 0.9\textwidth]{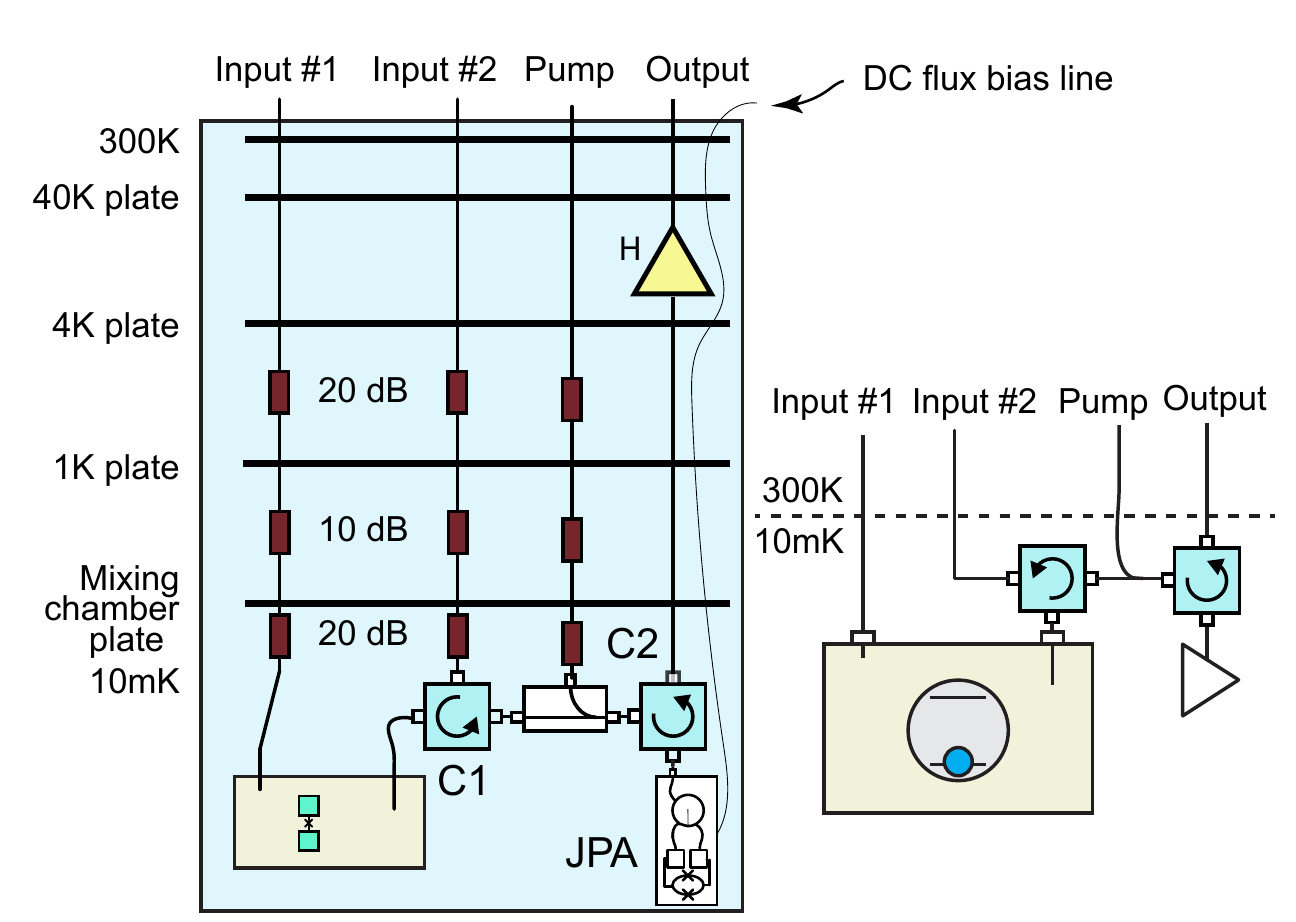}
\caption[The minimum experimental setup with paramp]{ {\footnotesize \textbf{The minimum experimental measurement setup with paramp:} The input lines can be used for qubit manipulation signals and cavity probe signals. An additional input line is dedicated to the paramp pump. The pump signal passes by a directional coupler and circulator to reach the paramp and the reflected signal (which have acquired a phase shift) goes to the output line. A DC line also is used to flux bias the paramp.}} 
\label{fig:paramp_added}
\end{figure}
We use an input line to send the pump signal to the paramp. Ideally, the pump (which is relatively strong coherent signal) should be isolated from the qubit system. A directional coupler and a circulator (the circulator C1 in Figure~\ref{fig:paramp_added}) prevent the pump signal from entering the cavity\footnote{Practically, sometimes multiple circulators are used for more isolation.}. The other circulator (C2 in Figure~\ref{fig:paramp_added}) directs the incoming pump signal to the paramp and sends the reflected signal to the output line.

\subsubsection{Paramp calibration: single pump}
The first step in the paramp setup is to find the paramp resonance frequency and tune it to the frequency where we wanted to operate the paramp. This can be done by looking at the phase of the reflection tone off of the paramp as depicted in Figure~\ref{fig:single_pump}a\footnote{The nonlinear response of the paramp helps to find its resonance. Similar to the punch out experiment, one would see a shift in the phase resonance by increasing the pump power.}. Normally the phase response of the paramp shifts down by increasing the probe tone power (which somewhat acts as a pump as well).

After we convince ourselves that we have a resonance frequency of the paramp at the right place, we start pumping that with a separate signal generator and use VNA as a weak signal probe as depicted in Figure~\ref{fig:single_pump}b. By increasing the power of the pump and adjusting the frequency of the pump and several back-and-forths we should be able to see the gain profile. Once we see some gain we may tweak the pump power and frequency and even the flux bias to optimize the gain profile. Normally a symmetrical Lorentzian shape gain profile (20 dB peak/ 100 MHz bandwidth) is desirable as depicted by the dashed line in Figure~\ref{fig:single_pump}b.  
\begin{figure}[ht]
\centering
\includegraphics[width = 0.98\textwidth]{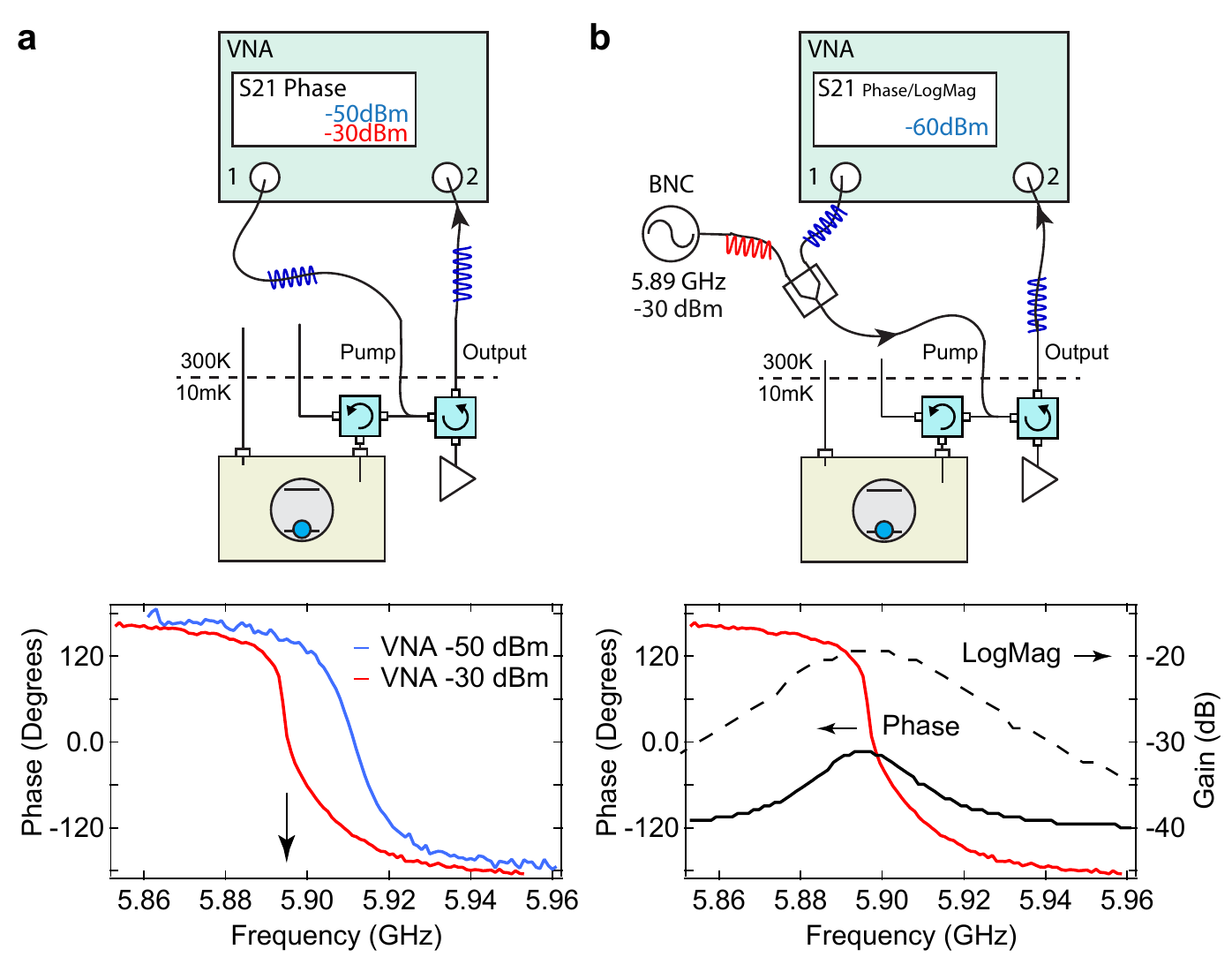}
\caption[The paramp single pump operation]{ {\footnotesize \textbf{The paramp single pump operation:} \textbf{a}, By looking at the phase response of the JPA, we obtain a rough estimation of proper power and flux bias to have sharp behavior at a desired frequency. \textbf{b}, Then we add a pump and set the power and frequency to the estimated values. We should be able to see a small amount of gain by adjusting the power. Then we fine tune the power, flux bias, pump frequency, and pump phase to obtain a reasonable amount of gain $\sim 20$ dB and bandwidth $\sim 100$ MHz.}} 
\label{fig:single_pump}
\end{figure}

\subsubsection{Paramp calibration: double pump}
The single pump paramp operation is not the best way to pump the paramp for practical reasons. Mainly because the isolation provided by the directional coupler and circulator is not perfect. Therefore the pump signal can leak into the cavity. This issue is important when the pump frequency is the same as the cavity frequency. This happens in the situation of weak z-measurement where the pump and cavity come from a same generator. In low-power measurement, any leakage of photons in cavity frequency dephases the qubit.

In order to get around this issue, we use a ``double-pump" technique. In this case instead of pumping the paramp with the frequency of $\omega_c$ we use two pumps at $\omega_c \pm \Omega_{SB}$ where $\Omega_{SB}$ is the sideband frequency which is typically between 100-500 MHz. With the double pump technique, the paramp effectively works at the cavity frequency but the pump signals are off-resonance with the cavity.

For the double pump paramp operation, we first operate the paramp in the single-pump mode and then simply modulate the pump tone by $\Omega_{SB}$. Normally we should see the gain profile by adjusting the pump power\footnote{Often double pump needs higher power but will give more stable performance for the paramp.}. Figure~\ref{fig:double_pump} demonstrates the schematics for the double pump operation.
\begin{figure}[ht]
\centering
\includegraphics[width = 0.98\textwidth]{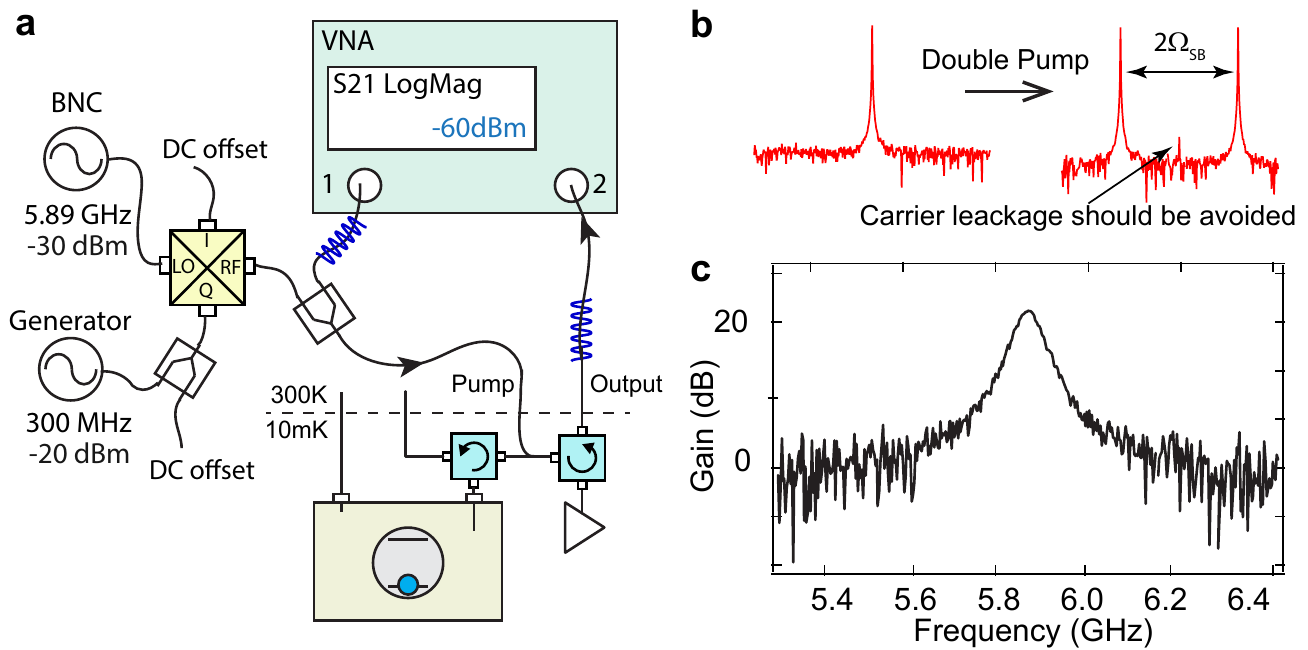}
\caption[The paramp double pump operation]{ {\footnotesize \textbf{The paramp double pump operation.} }} 
\label{fig:double_pump}
\end{figure}

\subsection{Phase-sensitive amplification: phase vs amplitude}
As we discussed in Subsection~\ref{Subsection:JPAtheory} The essence of JPA amplification is that right before the bifurcation the reflected phase is very sensitive to the pump power as depicted in Figure~\ref{fig:JPA_pup_transfer}. In the phase sensitive mode of amplification, where the signal and pump have the same frequency\footnote{Not only do the signal and pump have the same frequency, but their phases are fully correlated. Practically they come from a same generator.}, and we can amplify one of the quadratures and de-amplify the other quadrature. In Figure~\ref{fig:amp_vs_phase_JPA} we demonstrate the situation for amplification of each quadrature.
\begin{figure}[ht]
\centering
\includegraphics[width = 0.98\textwidth]{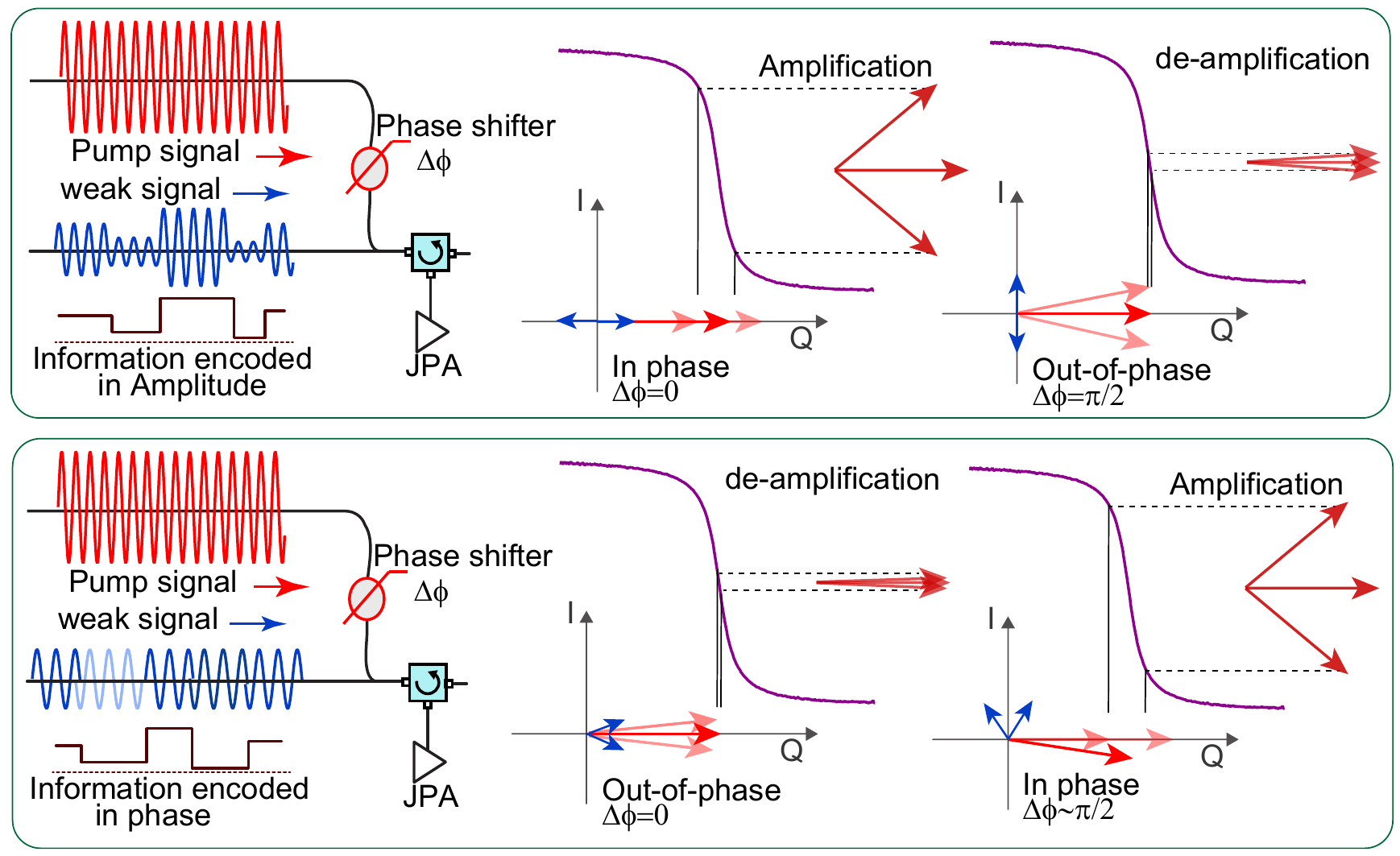}
\caption[Phase sensitive amplification]{ {\footnotesize \textbf{Phase sensitive amplification:} The upper (lower) panel demonstrates the paramp operation for amplification when the information encoded in the amplitude (phase) of the signal.}} 
\label{fig:amp_vs_phase_JPA}
\end{figure}
Note, that in case of qubit readout, or weak z-measurement, the information is encoded in the phase of the signal, but in the case of homodyne detection of qubit spontaneous emission, the information is encoded in the amplitude of the signal. We will discuss the qubit measurement more fully in Chapter~4. We will also discuss more practical situations and some consideration to improve the performance of the paramp (e.g. ``dumb-signal" cancellation).


\chapter{Quantum Measurement} \label{ch4}
The concept of measurement is very important in all disciplines of science and technology. However, measurement is a crucial concept in the science of quantum mechanics. This is not simply because quantum systems are small and delicate, but this is because measurement fundamentally disturbs the quantum system.\\
In this chapter, we will discuss the basics of quantum measurement in a pedagogical manner. This chapter includes the basic notion of projective measurement and more generalized types of measurement, including weak measurement. We will discuss continuous measurement and the stochastic master equation for the qubit-cavity system introduced in the previous chapters.
 
\section{Projective measurement}
Consider a quantum two-level system represented by the Hamiltonian $\hat{H} = -\omega_q \sigma_z/2$ with eigenstates $|\pm z\rangle$ and eigenvalues $\mp\frac{\omega_q}{2}$. The (pure) state of the system $|\psi \rangle$ is described by a normalized vector in Hilbert space which can be visualized as a vector pointing from the center of the Bloch sphere to a certain point on its surface\footnote{For a qubit system the Hilbert space is 2D. Note that the surface of the Bloch sphere is a 2D manifold.} with unit radius (Fig.~\ref{fig:bloch}).
\begin{figure}
\centering
\includegraphics[width = 0.35\textwidth]{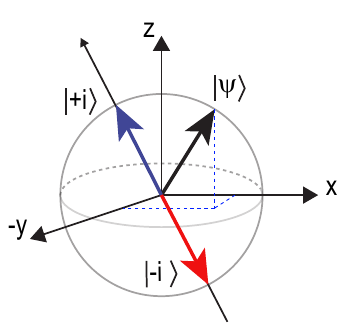}
\caption[Bloch sphere]{ {\footnotesize \textbf{Bloch sphere:} Projective measurement of the state $|\psi\rangle$ in the i-basis. The red and blue arrows indicates the backaction of the measurement.  }} 
\label{fig:bloch}
\end{figure}

In this visualization, a projective measurement can be thought to project the state $| \psi \rangle$ into a specific basis (direction). A projective measurement along the $i$-basis (where $i$ can be any direction but we mostly consider $i=x,y,z$) can be described by two projection operators $\hat{\Pi}^{\pm i} =\mid\pm i\rangle \langle \pm i |$. Every time we perform a projective measurement in $i$-basis, we collapse the state of the qubit and find it either in the $|+ i \rangle$  or $|- i \rangle$ (Fig.~\ref{fig:bloch}).

At this point, one may ask why we consider this destructive operation as a measurement. The point is that the probability of the state being collapsed into $|\pm i \rangle$ is related to the state $| \psi \rangle$. To understand this, it is convenient to represent the qubit state in the measurement basis $\{ |\pm i \rangle  \}$, 
\begin{eqnarray}
|\psi \rangle =   c_+ |+i \rangle + c_- |-i \rangle.
\end{eqnarray} 
According to \emph{Born's rule}, the probability of finding the qubit in the states $ |\pm i \rangle $ are $P_{\pm}= \langle \psi| \Pi^{\pm z}| \psi \rangle= |\langle \pm i | \psi \rangle|^2 = |c_\pm|^2$. Therefore, if we perform a projective measurement for $N\gg1$ copies of $|\psi \rangle$ (or repeat the same experiment $N$ times), we would get $N_\pm \simeq P_\pm N$ times result $ \pm 1$ indicating that we collapse the qubit state into $| \pm i \rangle$. Therefore we figure out the ratio $P_+/ P_-$ (note we also know that $P_+ + P_- =1$). Since the $c_\pm$ are complex numbers, we still need to figure out the relative phase between the eigenstates $|\pm i\rangle$. To find the relative phase we must perform another set of projective measurements along another basis $ j \neq i$. The best choice for the second basis is when  $|\langle \pm j | \pm i \rangle|^2=1/2 $.

We now proceed with an example: Consider a projective measurement in the $z$-basis, $\Pi^{\pm z}=|\pm z \rangle \langle \pm z|$ on a qubit state
\begin{eqnarray}
| \psi \rangle= \frac{1}{\sqrt{2}}( | +z \rangle + | -z \rangle).
\end{eqnarray}
Assume that the measurement apparatus outputs a signal $V= \pm 1$ when the state is projected to the state $|\pm z\rangle$\footnote{We discuss this in the previous chapter for qubit readout measurement. In the next section, we will discuss the mechanism by which the measurement outcome is actually is generated for general measurements.}. Since the qubit is initially prepared in an equal superposition of the state of the measurement basis, neither measurement outcome is more likely than the other. The qubit will be collapsed into $| \pm z \rangle$ with the probability of $P_\pm =\langle \psi| \Pi^{\pm z}| \psi \rangle=1/2$ and outputs $\pm 1$. But after the measurement, we know certainly that the qubit state is $|\pm z\rangle$. If we were to make another measurement, we would find the same result. This means we have gained information. However, we are now completely uncertain about the measurement result in the $x$-basis because $| \pm z \rangle= \frac{1}{\sqrt{2}}( | +x \rangle \pm | -x \rangle)$ . Therefore we have gained full information in $z$-basis but lost all information in $x$-basis. This is a consequence of the Heisenberg uncertainty principle; we can not be certain about two non-commuting observables at the same time\footnote{This doesn't mean we can not perform measurements on two non-commuting observables at the same time, see Ref.~\cite{hacohen2016quantum}}.

In a more general case, the qubit state can be a mixture of $|\psi_n \rangle$'s with the probabilities $P_n$,  which no longer can be written as a single state vector $| \psi\rangle$. However, this can be represented by a density matrix,
\begin{eqnarray}
\rho = \sum_n P_n |\psi_n \rangle \langle \psi_n |.
\end{eqnarray}
The density matrix $\rho$ fully represents our state of knowledge about the system by accounting for both the quantum superposition and classical uncertainty of the system. One can simply visualize the mixed state as a vector (norm$<$1) obtained by a weighted average over states with classical uncertainty. Similarly, projective measurement projects the mixed state along the measurement basis; 
\begin{eqnarray}
\sum_n P_n |\psi_n \rangle \langle \psi_n | \to  | \pm i  \rangle \langle \pm i |.
\end{eqnarray}

Here we specifically focus on a projective measurement along $z$-basis, which is the energy eigenbasis for the qubit. For example, we represent the density matrix $\rho$ in the measurement basis $\{ | \pm z \rangle \}$ we have,
\begin{eqnarray}
\rho &=&  P_{++} |+z\rangle \langle +z | + P_{--} |-z \rangle \langle -z | \nonumber \\
&+& P_{+-} |+z \rangle \langle -z | + P_{-+}|-z \rangle \langle +z |\\
&=&   \left( {\begin{array}{cc}
   P_{++}  & P_{+-} \\
   P_{-+} & P_{--} \\
  \end{array} } \right).
\end{eqnarray}
The diagonal element $P_{++}$  $(P_{--})$ is a real number that represents the probability of projecting the qubit into $|+z\rangle$ $(|-z\rangle)$ where $P_{++} + P_{--}=1$. The off-diagonal elements are complex numbers and represent quantum coherences and we have $P_{+-}^*=P_{-+}$. Therefore, a density matrix, in general, has three independent unknowns which means three sets of projective measurements (e.g. in the $x$, $y$, and $z$ bases) are needed to fully characterize the state of the qubit\footnote{Recall the full state tomography discussion in the previous chapter. We will discuss full state tomography in this chapter more fully.}.

Experimentally, we are usually able to project the qubit only along its energy eigenbasis. But we can add unitary rotations before projection along $z$ to realize an effective projective measurement along any arbitrary basis as discussed in the previous chapter. Projective measurements in the $z$-basis give us the ratio $P_{++}/P_{--}$, and since the probabilities must add to one, we obtain diagonal elements. Projective measurements in the $x$-basis ($y$-basis) give us $\mathcal{R}e[ P_{+-} ]$ ($\mathcal{I}m[ P_{+-} ]$).

\section{Generalized measurement in the $\sigma_z$ basis\label{section:generalized_m}}
In this section, we discuss quantum measurement in a more detailed approach allowing us to study generalized quantum measurement. Here we introduce the discussion in close relation to an actual lab experiment.
\subsection{Simple Model\label{subsection:simple_model}}
A quantum measurement is normally modeled by a system $\mathcal{S}$ with Hamiltonian $H_\mathcal{S}$ and a meter $\mathcal{M}$ with Hamiltonian $H_\mathcal{M}$. The measurement is performed by turning ``ON" the interaction between the meter and system, $H_{\mathrm{int}}$, for a certain time $t$ which entangles the state of the qubit with the state of the meter. By performing a subsequent measurement on the meter we collapse the entanglement and gain information.

Let's first study this by a simple model\footnote{I follow Andrew Jordan's discussion from the KITP conference, 2018.}. Consider Figure~\ref{fig:QM_simple_model} where the system is a qubit $H_\mathcal{S}=-\omega_q \sigma_z/2$ probed by a free particle\footnote{A free particle described by a Gaussian wave packet, which has minimum uncertainty in both position and momentum.} $H_\mathcal{M}=\hat{P}^2/2m$ passing by the qubit. Once the free particle is at a minimum distance from the qubit, they interact by $H_{\mathrm{int}}=-g \sigma_z \otimes \hat{P} \delta(t)$. Note that we assume the interaction is instantaneous and happens only at time $t=0$ when the particle has reached a minimum distance from the qubit. The parameter $g$ is measurement strength and in our model one can think of it as a measure of how close the particle passes by the qubit\footnote{smaller distance causes a stronger push and pull (larger $g$) which results in larger separation on the screen.}. 

Here we assume that we have minimum uncertainty in position and momentum of the particle which results in Gaussian distributions in the screen\footnote{More realistically one can assume that the interaction happens in a time scale T around $t=0$, in that case then effective coupling would be $g=\int_{-T/2}^{+T/2} g(t) dt$ and the separation is $\propto 2g$. Therefore the measurement strength depends on both interaction strength and interaction time. But here we assume that $g(t)=g \delta(t)$, meaning that the qubit and particle intact only at time $t=0$ when they have minimum distance.}.

\begin{figure}[ht]
\centering
\includegraphics[width = 0.8\textwidth]{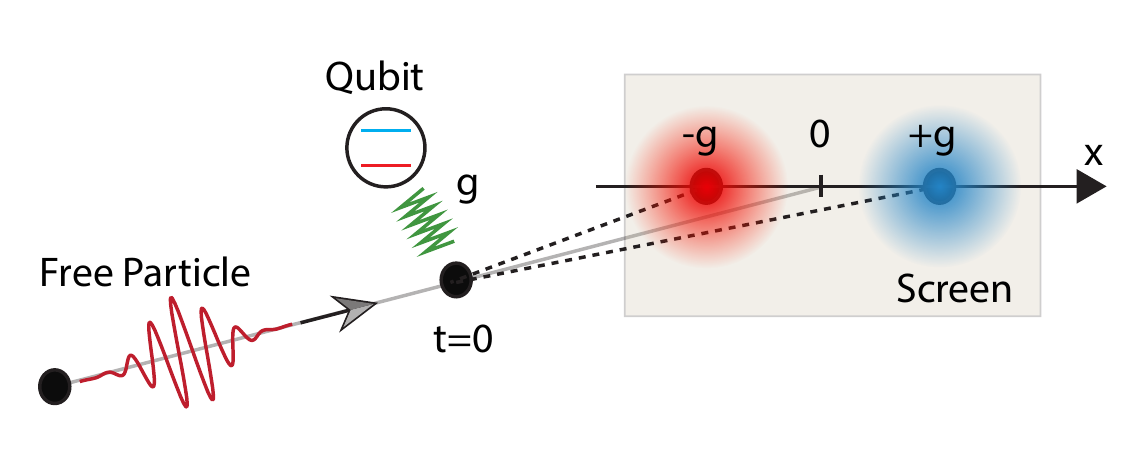}
\caption[Quantum measurement: simple model]{ {\footnotesize \textbf{Quantum measurement, simple model:} A free particle passes by and interacts with a qubit. The interaction is in the form of a push or pull depending on the state of the qubit. The position of the particle when it hits the screen tells us about the state of the qubit. The free particle has its natural Gaussian distribution. The separation between the two distributions is proportional to the interaction strength and the interaction time.}} 
\label{fig:QM_simple_model}
\end{figure}

Now, assume the qubit is initially in state $\psi=\alpha |0\rangle + \beta |1\rangle$ and the meter is in the state $\Phi$ which can be effectively represented as,
\begin{eqnarray}
\Phi= N e^{-\frac{x^2}{4 \sigma^2}},
\end{eqnarray}
where $\sigma$ quantifies the minimum fluctuation in position for the particle\footnote{Here we skipped irrelevant details of the free particle wave function and dynamics. Basically, the free particle is a wave packet moving along $z$ direction, $\Phi (r,t)= N \exp(k\cdot r - \omega_p t) \exp(r^2/4 \sigma^2)$ where $k\cdot r=k_z z$. Upon the interaction with the qubit at $t=0, z=0$, the particle gets pulled or pushed and obtains little bit of momentum along $x$ direction as well, $k\cdot r=k_z z + k_x x $. We only care about the particle position at the screen, so we describe the particle state by its position in the $x$-direction at $z=L$ where the screen is located.}. The qubit and particle interact at $t=0$ and the total system (qubit+meter) evolves under unitary evolution,
\begin{eqnarray}
\mathcal{U}_{\mathrm{tot}}= e^{+i g \sigma_z \otimes \hat{P}}
\end{eqnarray}
which entangles the qubit and particle state. Therefore, the state of total system would be,
\begin{eqnarray}
\Psi_{\mathrm{tot}} &=& \mathcal{U}_{\mathrm{tot}} \psi \otimes \Phi \nonumber \\
&=& \mathcal{N} \left[ \alpha |0\rangle \exp(- \frac{(x-g)^2}{4 \sigma^2}) + \beta |1\rangle \exp(- \frac{(x+g)^2}{4 \sigma^2}) \right].
\label{eq:qs_tildaO}
\end{eqnarray}
If we then measure the position that particle landed on the screen and found that it to be $x=\tilde{x}$ (the wave function of meter collapses) then our state of knowledge about the state of qubit would be,
\begin{eqnarray}
\psi &=& \tilde{N} \left[  \alpha \exp(+ \frac{g \tilde{x}}{2 \sigma^2})  |0\rangle  + \beta \exp(- \frac{g \tilde{x}}{2 \sigma^2}) |1\rangle  \right].
\label{eq:qs_tilda}
\end{eqnarray}
Where $\tilde{N}$ is a normalization constant. Therefore we learn about the state of the qubit via an indirect measurement. This type of measurement is more general than projective measurement.  As the measurement strength becomes stronger, we approach projective measurement, and if the measurement strength is weak, we are in the weak measurement limit. Now we interpret the result of Equation~\eqref{eq:qs_tilda} in these two limits.

\emph{Projective measurement limit}- Now consider the situation where $g\gg \sigma$ which means the measurement is strong such that two distributions are well separated with negligible overlap. That means we are pretty sure about which distribution $\tilde{x}$ belongs to, $\tilde{x} \sim +g $ or $\tilde{x} \sim -g $ as depicted in Figure~\ref{fig:QM_simple_model_s}.
\begin{figure}[ht]
\centering
\includegraphics[width = 0.8\textwidth]{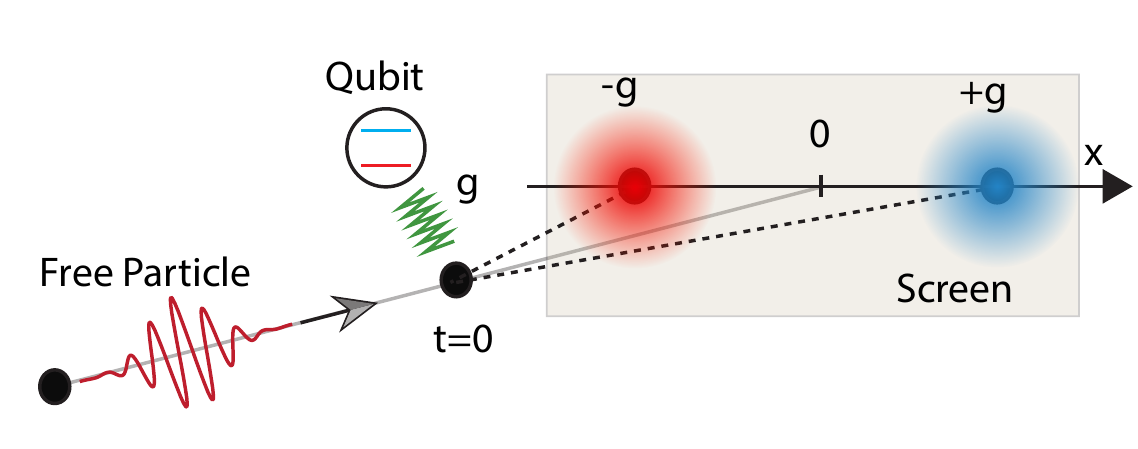}
\caption[Strong measurement]{ {\footnotesize \textbf{Strong measurement:} If the particle strongly interacts with the qubit, the separation between two distributions would be large enough so that we can readout the qubit state just by knowing in which distribution the particle has landed.}} 
\label{fig:QM_simple_model_s}
\end{figure}
Therefore one of the terms in \ref{eq:qs_tilda} is suppressed.
\bse \label{eq:qs_tilda_s}
\begin{eqnarray}
\tilde{x} \sim +g \xrightarrow{ g\gg \sigma} \psi =  |0\rangle, \\
\tilde{x} \sim -g \xrightarrow{ g\gg \sigma} \psi =  |1\rangle.
\end{eqnarray}
\ese
This means that the qubit wave function is also projected to one its eigenstates in this limit. In this case it is easy to define a threshold at $x_{\mathrm{th}}=0$ by which two histograms and completely separated. If $\tilde{x}>x_{\mathrm{th}}$ ($\tilde{x}<x_{\mathrm{th}}$) we realize that the qubit has been projected into the ground (excited) state.

\emph{Weak measurement limit}- Now consider the situation $g<\sigma$, meaning that the two distributions significantly overlap. Now if we obtain $\tilde{x}$, we are not sure which distribution $\tilde{x}$ belongs to. Yet, based on Equation~\eqref{eq:qs_tilda}, our knowledge about the qubit state is updated. If $\tilde{x}$ is positive (negative) the qubit state shifts more toward the ground (excited) state because one of the terms dominates over the other in Equation~\eqref{eq:qs_tilda}. Therefore by this measurement we slightly disturb the qubit but still we know what is the qubit state because we have measured that disturbance.
\begin{figure}[ht]
\centering
\includegraphics[width = 0.8\textwidth]{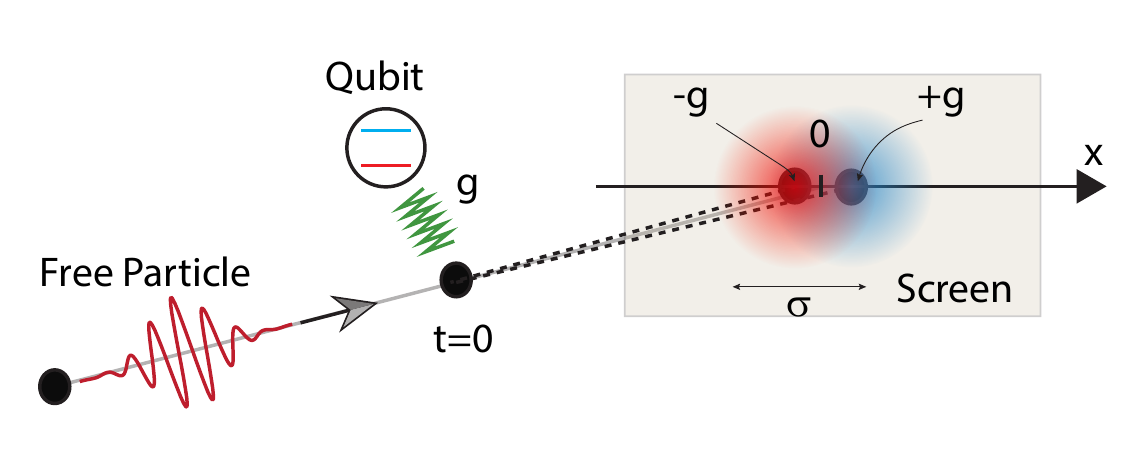}
\caption[Weak measurement]{ {\footnotesize \textbf{Weak measurement:} In the limit that the particle weakly interacts with the qubit, the separation between the two distributions would be smaller. This gives partial information about the state of the qubit.}} 
\label{fig:QM_simple_model_w}
\end{figure}
In the next section we will discuss this type of measurement more rigorously.

\subsection{POVM\label{subsection:POVMz}}
In the previous section, we studied a general type of measurement which is indirect and applies to a wide range of measurements. Formally, this type of measurement can be described in terms of POVMs\footnote{POVM stands for `positive operator-valued measured'.}. For that, let's revisit the result we had in the previous subsection in terms of the density matrix. For a projective measurement, the qubit state which can be described by $\rho=\sum_n |\psi_n\rangle \langle \psi_n|$ undergoes projection to one of the eigenstate
\bse\label{eq:density_projection}
\begin{eqnarray}
\sum_n |\psi_n\rangle \langle \psi_n| \xrightarrow{\tilde{x}>x_{\mathrm{th}}}  |0\rangle \langle 0 | \\
\sum_n |\psi_n\rangle \langle \psi_n|  \xrightarrow{\tilde{x}<x_{\mathrm{th}}} |1\rangle \langle 1 |,
\end{eqnarray}
\ese
which means the final density matrix is the result of acting with the projector $\Pi_n$ on the initial density matrix,
\bse
\begin{eqnarray}\label{eq:density_projectors}
\rho &\xrightarrow[\Pi_0 = |0 \rangle \langle 0|]{\tilde{x}>x_{\mathrm{th}}}&  \frac{\Pi_0 \rho \Pi_0 }{\mathrm{Tr}[\Pi_0 \rho \Pi_0]}, \mathrm{with \  probability} \  P_0=\mathrm{Tr}[\Pi_0 \rho \Pi_0]\\
\rho &\xrightarrow[\Pi_1 = |1 \rangle \langle 1|]{\tilde{x}<x_{\mathrm{th}}}&  \frac{\Pi_1 \rho \Pi_1 }{\mathrm{Tr}[\Pi_1 \rho \Pi_1]} , \mathrm{with \  probability} \  P_1=\mathrm{Tr}[\Pi_1 \rho \Pi_1],
\end{eqnarray}
\ese
where $\mathrm{Tr}[\Pi_n \rho \Pi_n]$ in denominator is the normalization factor. Note that we have $\sum_n \Pi_n=|0\rangle \langle0| + |1\rangle \langle 1|=\mathbb{1} $. 

In a more general manner,  one can describe partial measurements (including weak and strong measurements) by a set of operators $\Omega_n$ which obey the constraint $\sum_n \Omega_n^{\dagger} \Omega_n = \mathbb{1}$. In this case we have similar operations
\begin{eqnarray}
\rho &\xrightarrow[\Omega_n ]{n\mathrm{th\ outcome}}&  \frac{\Omega_n \rho \Omega_n^{\dagger} }{\mathrm{Tr}[\Omega_n \rho \Omega_n^{\dagger}]}, \mathrm{with \  probability} \  P_n=\mathrm{Tr}[\Omega_n \rho \Omega_n^{\dagger}],
\label{eq:density_partial}
\end{eqnarray}
except now $\Omega_n$ is not necessarily a projector. Actually $\Omega_n$ is called POVM and, in general, can be described by a weighted sum over all projection operators. For example the POVM corresponding to the general measurement discussed in our model (Eq.~\ref{eq:qs_tildaO}) can be described by
\begin{eqnarray}
\Omega_{\tilde{x}}= \mathcal{N}  \left[ \exp(-\frac{ (\tilde{x} -g)^2}{4 \sigma^2}) |0\rangle \langle 0| + \exp(- \frac{ (\tilde{x} + g)^2}{4 \sigma^2}) |1\rangle \langle 1| \right].
\label{eq:POVM0}
\end{eqnarray}
One can check that $\Omega_{\tilde{x}}$ acting on $\rho=|\psi \rangle \langle \psi |$, according to the Equation~\eqref{eq:density_partial}, results in Equation~\eqref{eq:qs_tilda}. Note, the measurement outcome $\tilde{x}$ is a continuous variable indicating the position the particle detected on the screen in our model, but can be a discrete value depending on the type of apparatus one uses for the meter\footnote{We will see later in this chapter that this value, in our experiment, is a ``semi-continuous" digitized homodyne voltage.}.

\subsection{POVM in terms of physical parameters }
Now we translate our model into the language of cavity QED and describe the actual weak measurement that we perform on the qubit in experiment. In our case the qubit is probed by a microwave coherent signal. So essentially we need to replace wave packet of the free particle with a coherent signal. There is, of course, a exact correspondence between a Gaussian wave packet and a coherent signal. As depicted in Figure~\ref{fig:povm_QED} the coherent signal is initially prepared along quadrature $I$. It has minimum uncertainty along each canonical position and momentum\footnote{Remember in our model, the wave packet also has minimum uncertainty for position and momentum.}. When the signal passes the cavity and interacts with the qubit, it acquires a phase shift which depends upon the state of the qubit. As we discussed in Chapter~2, the phase shift of the coherent signal can be translated to a displacement in $IQ$ plane.
\begin{figure}[ht]
\centering
\includegraphics[width = 0.98\textwidth]{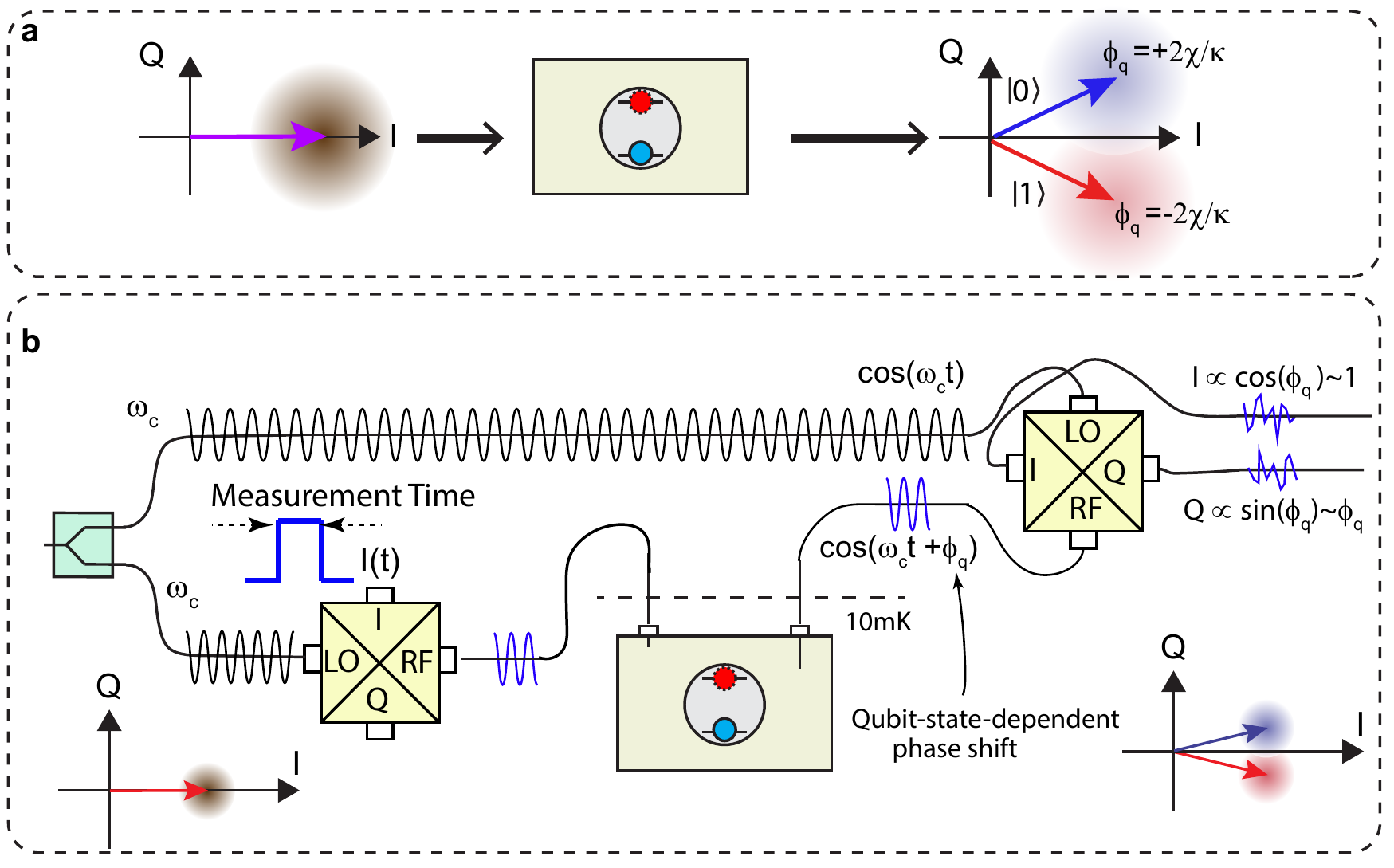}
\caption[Weak measurement cQED]{ {\footnotesize \textbf{Weak measurement cQED:} \textbf{a}, The qubit-state-dependent phase shift of the cavity probe signal. \textbf{b}, A more realistic schematic for the phase shift detection.}} 
\label{fig:povm_QED}
\end{figure}
Assuming the phase shift happens along the $Q$ quadrature\footnote{This can be done in the experiment by adjusting the phase of the probe signal.}, the measurement outcome is a signal $\tilde{V}$ that we obtain for $Q$ quadrature of the homodyne measurement. The corresponding POVM would be very similar to our model,
\begin{eqnarray}
\Omega_{\tilde{V}}= \mathcal{N}  \left[ \exp(-\frac{ (\tilde{V} -g)^2}{4 \sigma^2}) |0\rangle \langle 0| + \exp(- \frac{ (\tilde{V} + g)^2}{4 \sigma^2}) |1\rangle \langle 1| \right].
\label{eq:POVM}
\end{eqnarray}
However we need to figure out $g$ and $\sigma$ in terms of actual parameters in the measurement\cite{gambetta2008quantum}. As we discussed in Chapter~2,  the (dimension-less) variance of coherent state in each quadrature is $1/4$ which is the minimum fluctuation (see Exercise~7 of Chapter~2). However in the actual experiment we collect the signal for a certain amount of time $\Delta t$ with collection efficiency of $\eta$, which means the actual variance we have in the experiment is $\sigma^2=1/(4 \eta \kappa \Delta t)$ where $\kappa$ is the cavity linewidth\footnote{You may think it as a shot noise improvement in the variance. We get $ \eta \kappa \Delta t $ amount of signal from the cavity during the measurement which improves the uncertainty by a factor of $1/\sqrt{\eta \kappa \Delta t}$.}. The separation between the two Gaussian distributions results from the phase shift in the cavity frequency and also the number of photons inside the cavity, as depicted in Figure~\ref{fig:povm_QED}. We have a $2 \chi$ frequency shift of the cavity resonance frequency which, in the limit of $\chi\ll \kappa$, resulting in a phase shift of $4 \chi/\kappa$ of the cavity probe (see the discussion in Chapter~3 for the cavity phase shift). Therefore the separation $2g= 4 \chi \sqrt{\bar{n}}/\kappa$ where $\sqrt{\bar{n}}$ accounts for phasor vector length in $IQ$ plane as depicted in Figure~\ref{fig:povm_QED}. So we have,
\begin{equation}
\Omega_{\tilde{V}}= \mathcal{N}  \left[ e^{-\eta \kappa\Delta t (\tilde{V} -2 \chi \sqrt{\bar{n}}/\kappa)^2} |0\rangle \langle 0| + e^{-\eta \kappa\Delta t (\tilde{V} +2 \chi \sqrt{\bar{n}}/\kappa)^2} |1\rangle \langle 1| \right].
\label{eq:POVM2}
\end{equation}
We also define the signal-to-noise ratio (SNR) to be,
\begin{eqnarray}
S=\left(\frac{2g}{\sigma}\right)^2= \frac{64 \chi^2 \bar{n} \eta \Delta t}{\kappa }.
\label{eq:SNR}
\end{eqnarray}
Normally in this experiment, the separation between two Gaussian distributions is always scaled\footnote{In our model this can be adjusted by the position of the screen and in actual experiment we can simply amplify/attenuate the signal or just simply scale the signal after digitization. Note that scaling doesn't change the SNR.} to be $2 g=2$. This means, one can rewrite the POVM as
\begin{equation}
\Omega_{\tilde{V}}= \mathcal{N}  \left[ e^{- 4 \chi^2 \bar{n} \eta \Delta t / \kappa  (\tilde{V} -1)^2} |0\rangle \langle 0| + e^{- 4 \chi^2 \bar{n} \eta \Delta t / \kappa (\tilde{V} +1)^2} |1\rangle \langle 1| \right].
\label{eq:POVM3}
\end{equation}
where now $\tilde{V}$ is the scaled signal and the variance of the scaled signal is $\sigma^2=\kappa/(16 \chi^2 \bar{n} \eta \Delta t)$.

It is convenient to define $k=4 \chi^2 \bar{n}/ \kappa$ as the measurement strength\footnote{The definition for measurement strength $k$ seems to be different from literature by a factor of two. This wouldn't be an issue if we scaled the signal consistently. Here I define $k$ such that the the measurement operator in the Lindblad equation is exactly $\sqrt{k} \sigma_z$.} which quantifies how strong we are measuring the system regardless of the measurement time and efficiency.
\begin{equation}
\Omega_{\tilde{V}}= \mathcal{N}  \left[ e^{- k \eta \Delta t (\tilde{V} -1)^2} |0\rangle \langle 0| + e^{- k \eta \Delta t (\tilde{V} +1)^2} |1\rangle \langle 1| \right].
\label{eq:POVM4}
\end{equation}
We also define a characteristic measurement time $\tau=1/(4 k \eta)$ which quantifies how long we should collect the signal to achieve $\sigma^2=1$ (SNR=4).

To sum up this discussion, Equation~\eqref{eq:POVM4} describes weak measurement on the qubit state $|\psi\rangle \to \Omega_{\tilde{V}} |\psi \rangle $ or in a more general form,
\begin{equation}
\rho \to \frac{\Omega_{\tilde{V}} \rho \Omega_{\tilde{V}}^{\dagger}}{\mathrm{Tr}[\Omega_{\tilde{V}} \rho \Omega_{\tilde{V}}^{\dagger}]}
\label{eq:sum_up_POVM}
\end{equation}
and we obtain the signal $\tilde{V}$ with a probability
\begin{equation}
P(\tilde{V})=\mathrm{Tr}[\Omega_{\tilde{V}} \rho \Omega_{\tilde{V}}^{\dagger}]=\rho_{00} e^{-2 k \eta \Delta t (\tilde{V} -1)^2} + \rho_{11} e^{-2 k \eta \Delta t (\tilde{V} +1)^2},
\label{eq:sum_up_POVM2}
\end{equation}
where $\rho_{00} \  (\rho_{11})$ is the probability for the qubit to be in the ground (excited) state before the measurement\footnote{Note that there is a factor of 2 difference between exponents in Equation~\eqref{eq:POVM4} which is an operator and Equation~\eqref{eq:sum_up_POVM2} which is a probability distribution.}.

\section{Continuous measurement in $\sigma_z$ basis\label{section:CWM}}
In the previous section we studied how the state of the qubit changes under a generalized measurement for a time $\Delta t$. In this section we are going to study continuous monitoring of the qubit state in the limit of very weak measurement.

For that we start from the probability distribution of signal $\tilde{V}$ Equation~\eqref{eq:sum_up_POVM2}. In the limit of very weak measurement, $\Delta t \to 0$, the variance of the distributions $\sigma^2= 1/(4 k\eta\Delta t) \gg 1$ which means that the two distributions almost overlap as depicted in Figure~\ref{fig:overlap_PV}. In this limit one can show that,
\begin{equation}
P(\tilde{V}) \simeq e^{-2 k \eta \Delta t (\tilde{V} - \rho_{00} + \rho_{11})^2} = e^{- 2k \eta \Delta t (\tilde{V} - \langle \sigma_z \rangle )^2},
\label{eq:POVM_sigz}
\end{equation}
which means we can replace two distributions with one distribution which is centered at $\langle \sigma_z \rangle$ as depicted in Figure~\ref{fig:overlap_PV}.
\begin{figure}[ht]
\centering
\includegraphics[width = 0.9\textwidth]{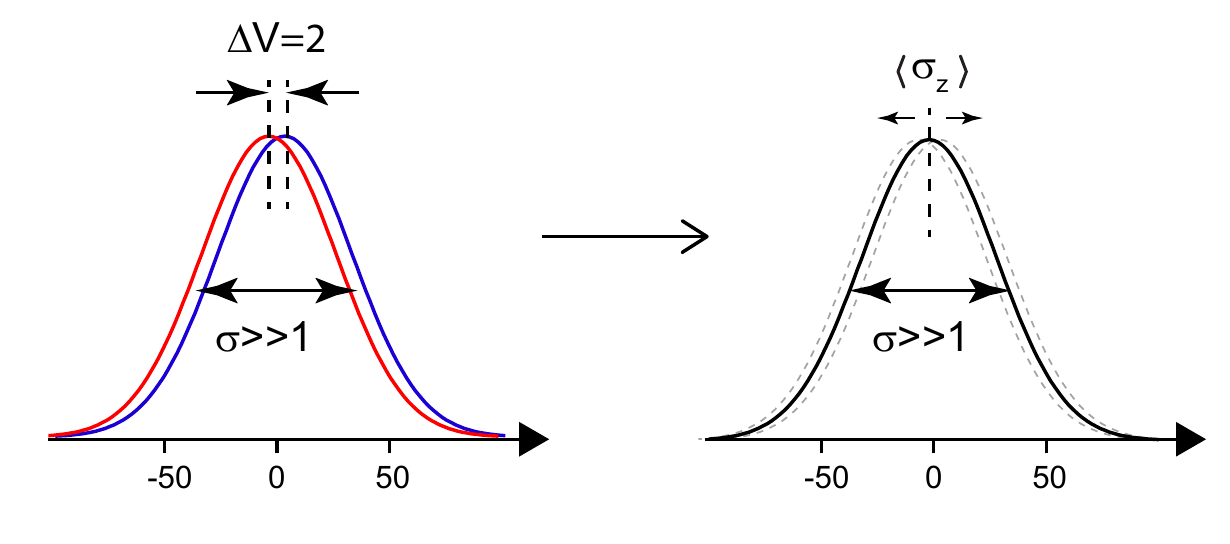}
\caption[Measurement signal distribution in the weak limit]{ {\footnotesize \textbf{Measurement signal distribution in the weak limit:} In the limit of weak measurement, the separation between the two distributions is much smaller than the variance of distributions. In such a case, we can approximate the two distributions by a distribution centered at $\langle \sigma_z \rangle$ as shown in Exercise~1.}} 
\label{fig:overlap_PV}
\end{figure}
Consequently the measurement operator $\Omega_{\tilde{V}}$ in Equation~\eqref{eq:POVM4} can be represented in a compact form up to a renormalization constant,
\begin{equation}
\Omega_{\tilde{V}} \simeq e^{-  k \eta \Delta t (\tilde{V} -\hat{\sigma}_z)^2}
\label{eq:POVM_sigzO}
\end{equation}

\noindent\fbox{\parbox{\textwidth}{
\textbf{Exercise~1:} Verify Equation~\eqref{eq:POVM_sigz} by expanding Equation~\eqref{eq:sum_up_POVM2}. For that you need to show (up to a normalization constant) that $$p e^{-(x-1)^2/a^2} + q e^{-(x+1)^2/a^2} \xrightarrow[p+q=1]{a\gg 1}  e^{-(x-p+q)^2/a^2}$$. 
}} \vspace{0.25cm}

The fact that the measurement signal has a Gaussian distribution centered on $\langle\sigma_z \rangle$ means one can think of the measurement signal as a noisy estimate of $\langle\sigma_z \rangle$ which can be represented as\footnote{In fact, this interpretation in Equation~\eqref{eq:nosy_estimate} comes in very handy for simulating quantum trajectories.},
\begin{equation}
\tilde{V} = \langle \sigma_z \rangle + \frac{d\mathcal{W}}{\sqrt{4 k \eta} \Delta t}
\label{eq:nosy_estimate}
\end{equation}
where $d\mathcal{W}$ is a \emph{Wiener} increment which is a zero-mean Gaussian random variable with variance of $\Delta t$.

\subsection{Stochastic Schr\"odinger equation\label{subsection:SSE}}
Now we study qubit evolution under the measurement operator $\Omega_{\tilde{V}}$. I follow the discussion in Ref.~\cite{jaco06}. For now we assume $\eta=1$ and consider a normalized qubit pure state $|\psi(t)\rangle$ at time $t$. The qubit state at a later time $t+\Delta t$ would be,
\bse
\begin{eqnarray}
|\psi(t+\Delta t)\rangle &=& \Omega_{\tilde{V}} |\psi(t)\rangle \\
&\propto&  e^{-  k  \Delta t (\tilde{V} -\hat{\sigma}_z)^2} |\psi(t)\rangle\\
&\propto&  e^{- k  \Delta t (\hat{\sigma}_z^2 -2 \tilde{V} \sigma_z)} |\psi(t)\rangle,
\label{eq:Schro_SME}
\end{eqnarray}
\ese
where we ignore the constant term proportional to the $\tilde{V}^2$) in the exponent since we are eventually going to renormalize $|\psi(t+\Delta t)\rangle$. We now substitute $\tilde{V}$ from Equation~\eqref{eq:nosy_estimate} (for now $\eta=1$), 
\begin{eqnarray}
|\psi(t+\Delta t)\rangle = \exp(-  k  \Delta t \hat{\sigma}_z^2 +2 k  \Delta t  \hat{\sigma}_z \langle \sigma_z \rangle  +\sqrt{ k}   \hat{\sigma}_z  d \mathcal{W} ) |\psi(t)\rangle.
\label{eq:Schro_SME2}
\end{eqnarray}
Now we replace $\Delta t \to dt $ implying the continuous limit and expand the exponential but only keeping terms up to first order in $dt$,
\begin{eqnarray}
\resizebox{.85\hsize}{!}{$|\psi(t+d t)\rangle = \left( 1 - k  d t \hat{\sigma}_z^2 +2k  d t  \hat{\sigma}_z \langle \sigma_z \rangle  +\sqrt{ k}   \hat{\sigma}_z  d \mathcal{W} + \frac{k}{2} \sigma_z^2  (d\mathcal{W})^2 \right) |\psi(t)\rangle,$}
\label{eq:Schro_SME25}
\end{eqnarray}
then we replace $(d \mathcal{W})^2 =dt$ according to stochastic calculus (It\^o rule)\footnote{The Wiener increment $d\mathcal{W}$ has dimension of $\sqrt{t}$, see Ref. \cite{jaco06} for more details.} and arrive at,
\begin{eqnarray}
|\psi(t+d t)\rangle = \left( 1 - \frac{k}{2}  \hat{\sigma}_z [ \hat{\sigma}_z - 4 \langle \sigma_z \rangle ] dt  +\sqrt{ k}   \hat{\sigma}_z  d \mathcal{W} \right)   |\psi(t)\rangle.
\label{eq:Schro_SME3}
\end{eqnarray}
Now we need to normalize the state $|\psi(t+d t)\rangle$ because, so far, we have ignored normalization constants. One can show that 
\begin{eqnarray}
\langle \psi(t+d t)|\psi(t+d t)\rangle = 1+4 k \langle \sigma_z \rangle^2 dt + \sqrt{4 k} \langle \sigma_z \rangle d \mathcal{W} + \mathcal{O}[t]^{3/2},
\label{eq:SSE_norm_cal}
\end{eqnarray}
where we just keep terms up to the first order in $dt$ and second order in $d\mathcal{W}$. By using a binomial expansion, one can show that
\begin{eqnarray}
\left[ \langle \psi(t+d t)|\psi(t+d t)\rangle \right]^{-\frac{1}{2}}= 1- \frac{k}{2} \langle \sigma_z \rangle^2 dt - \sqrt{ k} \langle \sigma_z \rangle d \mathcal{W} + \mathcal{O}[t]^{3/2}.
\label{eq:SSE_norm_cal2}
\end{eqnarray}
When we multiply the Equation~\eqref{eq:SSE_norm_cal2} by Equation~\eqref{eq:Schro_SME3} (and again keep terms up to $dt$ and $(d\mathcal{W})^2$), we obtain the normalized Stochastic Schr\"odinger Equation (SSE),
\begin{eqnarray}
d |\psi(t) \rangle = \left(  - \frac{k}{2} (\sigma_z - \langle \sigma_z \rangle )^2 dt + \sqrt{ k} (\sigma_z - \langle \sigma_z \rangle ) d \mathcal{W} \right)  |\psi(t) \rangle,
\label{eq:Schro_norm}
\end{eqnarray}
where we define $d |\psi(t) \rangle  =  |\psi(t + dt) \rangle  - |\psi(t) \rangle $. For a given measurement record $\{\tilde{V}\}$ one can infer $d\mathcal{W}$ (see Equation~\ref{eq:nosy_estimate}) and integrate this equation to obtain the qubit pure state evolution under measurement with perfect efficiency.

For example, the evolution of a qubit initialized in a pure state $\psi(0)=\alpha_0 |0\rangle + \beta_0 |1\rangle$ and subject to continuous measurement for time $T$ can be obtained by integrating the Equation~\eqref{eq:Schro_norm} as follows,
\begin{eqnarray}
\resizebox{.98\hsize}{!}{$
\left[ {\begin{array}{c}
   \alpha_{i+1} \\
   \beta_{i+1} \\
  \end{array} } \right] =  \left[ {\begin{array}{c}
   \alpha_{i} \\
   \beta_{i} \\
  \end{array} } \right]  - \frac{k}{2} dt \left[ {\begin{array}{cc}
   (1-z_i)^2 & 0\\
   0 & (1+z_i)^2\\
  \end{array} } \right] \left[ {\begin{array}{c}
   \alpha_{i} \\
   \beta_{i} \\
  \end{array} } \right]   + d\mathcal{W}_i \sqrt{ k}  \left[ {\begin{array}{cc}
   1-z_i & 0\\
   0 & -1-z_i\\
  \end{array} } \right] \left[ {\begin{array}{c}
   \alpha_{i} \\
   \beta_{i} \\
  \end{array} } \right], $} \nonumber\\
\   
\label{eq:SSE_update}
\end{eqnarray}
where $z_i=|\alpha_i|-|\beta_i|^2$ and $[i=1,2,...,N]$ where $N=T/dt$. Therefore, given the initial values $\alpha_0, \beta_0$, one can update the next values using the measurement record $d \mathcal{W}_i$ at each step\footnote{In the actual experiment we obtain $V_i= z_i + d\mathcal{W}/\sqrt{4 k}dt$ as the measurement signal (See, Eq.~\ref{eq:nosy_estimate}). So, one can rewrite \ref{eq:Schro_norm} in terms of $V_i$. But since efficient detection is not practical, we leave the \ref{eq:Schro_norm} in this form which is more convenient for simulating a quantum trajectory because one can simply use a Gaussian noise generator of variance $dt$ to generate $d\mathcal{W}$. Although for simulation purposes, the unnormalized version of Equation~\eqref{eq:Schro_SME3} works even better since one can manually normalize the state at each step.} and reconstruct the \emph{quantum trajectory} as depicted in Figure~\ref{fig:SSE_update} for $N=101, dt=0.01,k=1, \alpha_0=\beta_0=1/\sqrt{2}$.
\begin{figure}[ht]
\centering
\includegraphics[width = 0.9\textwidth]{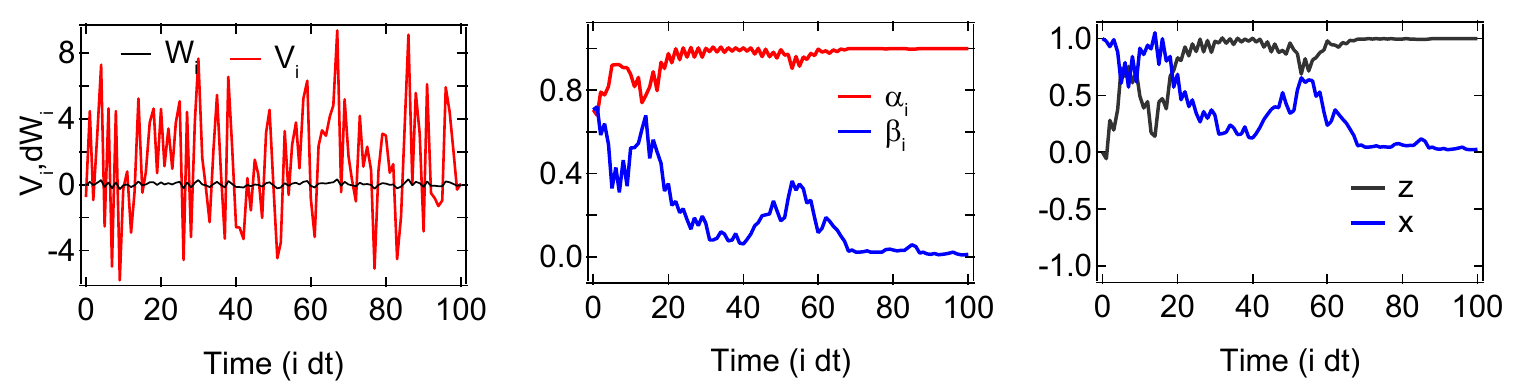}
\caption[SSE update trajectory]{ {\footnotesize \textbf{SSE update trajectory:} The measurement signal $V_i$ is used to infer $d\mathcal{W}_i$. The pure state evolution can then be reconstructed with the SSE. In the right panel, the evolution is represented in terms of Bloch coordinates.}} 
\label{fig:SSE_update}
\end{figure}
One can also add any unitary dynamics\footnote{For example, for adding Rabi oscillation to the dynamics one also needs to consider terms like $\dot{\alpha} = i \Omega_R \beta_i$ and $\dot{\beta} = i \Omega_R \alpha_i$ in the state update. Similar to Equation~\eqref{eq:ODE1_rabi_c_RWA} and Equation~\eqref{eq:ODE2_rabi_c_RWA}.}, the SSE is used to describe more general dynamics of the qubit state evolution.

\subsection{Stochastic master equation}
The SSE in the form of  (Eq.~\ref{eq:Schro_norm}) is only applicable for pure state evolution. However one can generalize this equation to describe mixed state evolution as well. An easy way to obtain the generalized form is to represent (Eq.~\ref{eq:Schro_norm}) in terms of density matrix\footnote{The trick here is that we use a pure state $|\psi\rangle$ to obtain the SME, and once we arrived at the equations for the SME in terms of density matrix, these equations can be applied to any density matrix, either pure or mixed. One can start from the beginning of the subsection \ref{subsection:SSE} and follow the steps in density matrix formalism and directly obtain the SME.} in this form,
\begin{eqnarray}
\rho=|\psi\rangle \langle \psi| \to d\rho = d|\psi\rangle \langle \psi| +  |\psi\rangle d \langle \psi| +  d|\psi\rangle d \langle \psi|,
\label{eq:SME_psi_psi}
\end{eqnarray}
where by substituting $d |\psi\rangle$ from Equation~\eqref{eq:Schro_norm} we arrive at the Stochastic Master Equation SME\footnote{Note, here we have double-commutator.}
\begin{eqnarray}
d \rho= - \frac{k}{2} [ \sigma_z, [\sigma_z,\rho  ]] dt + \sqrt{k} ( \sigma_z \rho + \rho \sigma_z - 2 \langle \sigma_z \rangle \rho ) d\mathcal{W}
\label{eq:SME0_rho}
\end{eqnarray}

\subsection{Inefficient measurement}
In last two sections, we assumed that the measurement is perfectly efficient, $\eta=1$. In this section we relax this assumption and account for inefficient detection. Inefficient detection can be modeled by considering two concurrent independent measurements on the system but ignoring the measurement outcome of one of them. For that, we consider two measurement apparatuses performing measurements on the system. The measurement strength of the first (second) apparatus is $k^{(1)} (k^{(2)} )$ where $k^{(1)}=\eta k$ and $k^{(2)}=(1-\eta) k$ and the measurement outcome is $V^{(1)} (V^{(2)})$, 
\begin{eqnarray}
V^{(m)} = \langle \sigma_z \rangle + \frac{d\mathcal{W}^{(m)}}{\sqrt{4 k^{(m)}} \Delta t},
\label{eq:V_1V_2}
\end{eqnarray}
where $m=1,2$. Considering both measurements, the qubit evolution would be,
\begin{eqnarray}
d \rho &=& - \frac{k^{(1)} }{2}[ \sigma_z, [\sigma_z,\rho  ]] dt + \sqrt{k^{(1)}} ( \sigma_z \rho + \rho \sigma_z - 2 \langle \sigma_z \rangle \rho ) d\mathcal{W}^{(1)} \nonumber \\
&-& \frac{k^{(2)} }{2} [ \sigma_z, [\sigma_z,\rho  ]] dt + \sqrt{ k^{(2)}} ( \sigma_z \rho + \rho \sigma_z - 2 \langle \sigma_z \rangle \rho ) d\mathcal{W}^{(2)} 
\label{eq:SME_rho1_2}
\end{eqnarray}
Now we ignore the second measurement outcome and average over all possible values for $d\mathcal{W}^{(2)}$. Since $d\mathcal{W}^{(2)}$ is a zero mean Gaussian noise increment, the last term in Equation~\eqref{eq:SME_rho1_2} vanishes and we arrive at the SME for inefficient detection\footnote{Similarly, one can model other types of imperfections and sources of decoherence in the system (e.g. relaxation and dephasing) by considering that the environment performs measurements on the system (via a measurement operator $\sqrt{\gamma} \hat{X} $ which depends on the type of decoherence) and we do not have access to the outcomes of these measurements.}, 
\begin{eqnarray}
d \rho= - \dfrac{k}{2} [ \sigma_z, [\sigma_z,\rho  ]] dt + \sqrt{ \eta k} ( \sigma_z \rho + \rho \sigma_z - 2 \langle \sigma_z \rangle \rho ) d\mathcal{W}
\label{eq:SME_rho}
\end{eqnarray}
where we replace $k^{(1)}=\eta k$, and $k^{(1)} + k^{(2)}=k$ and have simply dropped the superscript for $d\mathcal{W}^{(1)}$. For completeness, let's represent the SME in this form,
\begin{eqnarray}
d \rho&=&  k \left[ \sigma_z \rho  \sigma_z -  \frac{1}{2}( \rho  \sigma_z^2 +   \sigma_z^2 \rho )  \right] dt + \sqrt{ \eta k} ( \sigma_z \rho + \rho \sigma_z - 2 \langle \sigma_z \rangle \rho ) d\mathcal{W} \nonumber\\
\dot{\rho}&=&  k \left[ \sigma_z \rho  \sigma_z -  \rho  \right] + 2 \eta k ( \sigma_z \rho + \rho \sigma_z - 2 \langle \sigma_z \rangle \rho ) (V(t) - \langle \sigma_z \rangle),
\label{eq:SME_rho2}
\end{eqnarray}
where in the last line we substitute $d \mathcal{W}$ in terms of the actual measurement signal $V(t)$ according to Equation~\eqref{eq:nosy_estimate}.

The first term in \ref{eq:SME_rho2} is the Lindbladian term\footnote{ The term $L^{\dagger} \rho L -\frac{1}{2} \{ L^{\dagger} L , \rho \} = \mathcal{D}[L]\rho$ is usually called `dissipation superoperator' term.} $L^{\dagger} \rho L -\frac{1}{2} \{ L^{\dagger} L , \rho \}$ with Lindbladian operator $\hat{L}=\sqrt{k} \sigma_z$ as we introduced in Chapter 2 (See Eq.~\ref{eq:Lindblad_gen}). The second term which includes the measurement record and depends on quantum efficiency is the state update due to the measurement (which is referred to as ``unraveling" in the literature\footnote{In the literature you might find the argument that ``unraveling is not unique" \cite{brun2000continuous,drummond2013quantum}. It is true that there are many ways to unravel SME so that the average of many trajectories converge to the same Lindbladian evolution. But once you choose your efficient detector then the unraveling is unique. What about inefficient detector?}).

In last three subsections, we specifically discussed generalized measurements corresponding to the measurement operator $\sqrt{k}\sigma_z$, and found the the resulting SME. This SME has a general form and simply can be extended to any relevant measurement operator $\sqrt{k}\hat{c}$,
\begin{eqnarray}
\dot{\rho}=  k \left[ \hat{c} \rho  \hat{c}^{\dagger} -  \frac{1}{2}( \rho  \hat{c}^{\dagger}\hat{c} +   \hat{c}^{\dagger} \hat{c} \rho )  \right] + 2\eta k ( \hat{c} \rho + \rho \hat{c}^{\dagger} - \langle \hat{c} + \hat{c}^{\dagger} \rangle \rho )(V(t)- \langle \frac{\hat{c} + \hat{c}^{\dagger}}{2} \rangle ), \nonumber\\
\label{eq:SME_c}
\end{eqnarray}
where $k$ still represents the measurement strength. We will see that the measurement operator can even be non-Hermitian.

We can also add other type of imperfections to the dynamics. For example, the qubit decoherence due to the dephasing can be modeled by considering that environment also measures the system with measurement operator $\sqrt{\frac{\gamma_2}{2}} \hat{\sigma_z}$ where the $\gamma_2$ is the dephasing rate\footnote{This effect is significant if the total measurement time is comparable to the dephasing time or the measurement strength $k$ is comparable to the dephasing rate $\gamma_2$. The dephasing rate is ideally $\gamma_2=\frac{\gamma_1}{2}=\frac{1}{2 T_1}$ where the $T_1$ is the relaxation time for the qubit.}. However we do not have access to that measurement record. Therefore we sum over all possible outcomes for the environment (as we did for inefficient detection treatment) and obtain,
\begin{eqnarray}
\dot{\rho}&=&  (k+\frac{\gamma_2}{2}) \left[ \sigma_z \rho  \sigma_z -  \rho  \right] + 2 \eta k ( \sigma_z \rho + \rho \sigma_z - 2 \langle \sigma_z \rangle \rho ) (V(t) - \langle \sigma_z \rangle)\nonumber \\
\label{eq:SME_rho3}
\end{eqnarray}

One can show that Equation~(\ref{eq:SME_rho3}) has following representation in terms of Bloch components $x\equiv\langle \sigma_x \rangle , z\equiv\langle \sigma_z \rangle $,
\bse\label{eq:SME_zx}
\begin{eqnarray}
\dot{z}&=&4 \eta k (1-z^2)(V(t)-z) \label{eq:SME_z}\\
\dot{x}&=&-2( k+\frac{\gamma_2}{2} )x -4 \eta k x z (V(t)-z) \label{eq:SME_x}.
\end{eqnarray}
\ese
It is worth discussing the ensemble behavior of these equations, which occurs when we average over all possible measurement signals (that means we measure the system but disregard or don't have access to the measurement results). Let's consider the special case where the qubit is prepared in the superposition state $z=0,x=1$. It is apparent that in this case that $\dot{z}=0$ but the quantum coherence $x$ decays by the rate $2\kappa +\gamma_2$,
\begin{eqnarray}
x&=&e^{-(2 k+\gamma_2) t}.
\end{eqnarray}
Apart from the natural dephasing rate $\gamma_2$ which is ideally negligible, the qubit also dephases due to the unmonitored measurement photons,
\begin{eqnarray}
\Gamma=2k=\frac{8 \chi^2 \bar{n}}{\kappa},
\label{eq:MID}
\end{eqnarray}
which is called \emph{measurement induced dephasing}\footnote{Later we will utilize this equation for calibration.}.

One can also add unitary evolution to the SME~(\ref{eq:SME_rho2}) to account for a coherent drive on the qubit and obtain the full version of SME,
\begin{eqnarray}
\dot{\rho}=  -i [H_R, \rho ]  &+& k \left[ \sigma_z \rho  \sigma_z -  \rho  \right] \nonumber\\
&+& 2 \eta k( \sigma_z \rho + \rho \sigma_z - 2 \langle \sigma_z \rangle \rho ) (V(t) - \langle \sigma_z \rangle),
\label{eq:SME_rh0_full}
\end{eqnarray}
where $H_R$ represents the Hamiltonian for a drive on the qubit\footnote{Normally we consider $H_R=\frac{\Omega_R}{2}\sigma_x$ or $\frac{\Omega_R}{2}\sigma_y$ where we assume that drive is resonant. In general, any coherent drive, detuned, or along any axis can be added to the SME. The coherent drive's Hamiltonian is conveniently represented in the rotating frame of the drive. Note that this is convenient because the experiment happens in the rotating frame of the drive. The preparation and tomography pulses are from the same generator that that is used for the drive, therefore we pulse the qubit in rotating frame of the drive. This should not be confused with cavity homodyne measurement. Homodyne measurement happens in the rotating frame of the cavity. Therefore, the experiment involves two independent rotating frames for two different purposes.}.
In Section~\ref{section:z-me} and \ref{section:x-me}, we will study the combined unitary and non-unitary evolution of the qubit for both $z$-measurement $\sqrt{k} \sigma_z$, and $\sigma_-$-measurement\footnote{In this document, we interchangeably use `$\sigma_-$-measurement' and `$x$-measurement' for the measurement corresponding to the measurement operator $\sqrt{\gamma_1} \sigma_-$.} $\sqrt{\gamma_1} \sigma_-$ in more details.

\section{Bayesian update }
Although the SME (Eq.~\ref{eq:SME_rho2}) is a formal description for open quantum systems, the fact that it is a nonlinear equation makes it less convenient to work with. There is a fairly straightforward method to reconstruct qubit trajectory which is based on Bayes' theorem,
\begin{equation}
P(A|B)=\frac{P(B|A) P(A)}{P(B)},
\label{eq:bayes}
\end{equation}
where $P(A|B)$ is the probability of event $A$ given that event $ B $ has happened.
In connection with quantum measurement one can assume that:
\begin{eqnarray}
\mbox{event} \ A & \to & \mbox{finding the qubit in ground/excited state} ,  \nonumber \\
\mbox{event} \ B & \to & \mbox{obtaining the measurement signal} \  V . \nonumber
\end{eqnarray}
Therefore, one can use Bayes' rule to infer the qubit evolution conditioned on the measurement signal $V$.  According to Bayes' rule we have,
\bse\label{eq:bayes12}
\begin{eqnarray}
P_{i+1}(0) = P( 0 | V_i )= \frac{P(V_i|0) P_i(0)}{P(V_i)} \label{eq:bayes1}\\
P_{i+1}(1) = P( 1 | V_i )= \frac{P(V_i|1) P_i(1)}{P(V_i)}\label{eq:bayes2},
\end{eqnarray}
\ese
where $P_i(0)$ and $P_i(1)$ are the probabilities for the qubit to be in the ground and the excited state before the measurement---these are our prior knowledge in the  $i$th step of the update. Then we get the updated probabilities for the qubit state, $P_{i+1}(0)$ and $P_{i+1}(0)$ conditioned on measurement outcome $V_i$. The updated probabilities would be our prior knowledge for the next step of state update. The probability $P(V_i)$ is the unconditioned probability for getting signal $V_i$ based on our prior knowledge.

The Bayesian approach is powerful because it connects the unknown conditional probability $P(0|V_i)$ to a well-known conditional probability $P(V_i|0)$. Note that $P(V_i|0)$ and $P(V_i|1)$ are nothing but the Gaussian distributions separated by $\Delta V =2$ as we discussed in Equation~(\ref{eq:sum_up_POVM2}),
\bse
\begin{eqnarray}
P(V_i) &\propto& \rho_{00} e^{-\frac{(V_i-1)^2}{2 \sigma^2}} + \rho_{11} e^{-\frac{(V_i+1)^2}{2 \sigma^2}},\\
P(V_i|0) &\propto& e^{-\frac{(V_i-1)^2}{2 \sigma^2}}, \label{eq:bayes_PV1}\\
P(V_i|1) &\propto& e^{-\frac{(V_i+1)^2}{2 \sigma^2}},
\label{eq:bayes_PV2}
\end{eqnarray}
\ese
where $\sigma^2=1/(4 k \eta \Delta t)$ as we discussed in Section~\ref{section:CWM}.

Now in order to clearly connect Bayes' theorem to the quantum trajectory, we proceed by dividing two conditional probabilities in Equation~(\ref{eq:bayes1}) and (\ref{eq:bayes2}),
\begin{eqnarray}
\frac{P_{i+1}(0)}{P_{i+1}(1)} = \frac{P( 0 | V_i )}{ P( 1 | V_i )}= \frac{P_i(0)}{P_i(1)}\frac{P(V_i|0)}{P(V_i|1)},
\end{eqnarray}
and substitute the last term form Equation~(\ref{eq:bayes_PV1}) and (\ref{eq:bayes_PV2}),
\begin{eqnarray}
\frac{P_{i+1}(0)}{P_{i+1}(1)} = \frac{P_i(0)}{P_i(1)} \exp(+\frac{\Delta V}{\sigma^2}V_i),
\end{eqnarray}
where we prefer to explicitly have $\Delta V$ (which equals 2) in our representation\footnote{Note that the sign in the exponent depends on which way the Gaussian shifts for the ground and excited states. The convenient choice is when the Gaussian shifts in the positive direction for ground state which is consistent with the interpretation in Equation~(\ref{eq:nosy_estimate}).}. Now by considering the fact that $P_j(0)+P_j(1)=1$, one can calculate ${P_{i+1}(0)}$ and ${P_{i+1}(1)}$ given the prior knowledge ${P_{i}(0)}$ and ${P_{i}(1)}$ and measurement outcome $V_i$. 

Before we proceed further, let's switch the notation to the density matrix language which later allows us to make comparison between Bayesian update and SME update. For that we have $P_{i+1}(0) \to \rho_{00}(t+dt)$ and  $P_{i}(0) \to \rho_{00}(t)$ and similarity for $P_{i+1}(1) \to \rho_{11}(t+dt)$ and  $P_{i}(1) \to \rho_{11}(t)$ and obtain,
\begin{eqnarray}
\frac{\rho_{00}(t+dt)}{\rho_{11}(t+dt)} = \frac{\rho_{00}(t)}{\rho_{11}(t)} \exp(+\frac{\Delta V}{\sigma^2}V_i).
\label{eq:Bayesian_diagonal}
\end{eqnarray}
Equation~(\ref{eq:Bayesian_diagonal}) only allows us to calculate the evolution for diagonal elements of the density matrix. In order to account for off-diagonal elements\footnote{Here I follow Korotkov's discussion in \cite{koro11_bayes}}. Let's assume that the qubit at time $t$, before the $i$th-measurement, was in state $|\psi(t)\rangle = \sqrt{\rho_{00}(t)} |0\rangle + e^{i\phi} \sqrt{\rho_{11}(t)} |1\rangle$. After the measurement the state would be $|\psi(t+dt)\rangle = \sqrt{\rho_{00}(t+1)} |0\rangle + e^{i\phi} \sqrt{\rho_{11}(t+1)} |1\rangle$ where we assume that the measurement doesn't change the relative phase\footnote{In the Bloch sphere picture, this is to say that the measurement back-action only kicks the state up or down but not to the sides. In Korotkov's terminology there is only ``spooky" backaction, no ``realistic" backaction.} $\phi$. The density matrix before the measurement would be,
\begin{eqnarray}
\rho(t)=|\psi(t) \rangle \langle\psi(t)| = \rho_{00}(t) |0 \rangle \langle 0| + \rho_{11}(t) |1 \rangle \langle 1| \hspace{2cm} \nonumber \\
+ e^{-i\phi} \sqrt{\rho_{00}(t) \rho_{11}(t)} |0 \rangle \langle 1| +e^{i\phi} \sqrt{\rho_{11}(t) \rho_{00}(t)} |1 \rangle \langle 0|
\end{eqnarray}
and similarly for after measurement $\rho(t+dt)=|\psi(t+dt) \rangle \langle\psi(t+dt)|$. Therefore we arrive at a relation for off-diagonal elements,
\begin{eqnarray}
\frac{\rho_{01}(t+dt)}{\rho_{01}(t)}= \frac{\sqrt{\rho_{00}(t+dt) \rho_{11}(t+dt)}}{\sqrt{\rho_{00}(t) \rho_{11}(t)}}.
\end{eqnarray}
One can add a damping term to this relation to phenomenologically account for additional dephasing (e.g. a finite $T_2^*$ time, finite efficiency),
\begin{eqnarray}
\frac{\rho_{01}(t+dt)}{\rho_{01}(t)}= \frac{\sqrt{\rho_{00}(t+dt) \rho_{11}(t+dt)}}{\sqrt{\rho_{00}(t) \rho_{11}(t)}}e^{-\gamma dt},
\label{eq:Bayesian_off_diagonal}
\end{eqnarray}
where $\gamma =\frac{8 \chi^2 \bar{n}(1-\eta)}{\kappa} +1/T_2^*$ accounts for both depashing due to imperfect detection and finite qubit coherence time. 

\subsection{Bayesian update in terms of the Bloch components}
It is convenient to represent the Bayesian update in terms of $z=\langle \sigma_z\rangle$, $x=\langle \sigma_x\rangle$. By considering that, $z=2\rho_{00}-1$ and $x=2\rho_{01}$ and the fact that $\rho_{00}+\rho_{11}=1$, one can show that the Equation~\ref{eq:Bayesian_diagonal} and \ref{eq:Bayesian_off_diagonal} can be represented in the following form\footnote{We define $z=\langle \sigma_z\rangle=\mathrm{Tr}(\rho \sigma_z)=\rho_{00}-\rho_{11}=2\rho_{00}-1$. Note for off-diagonal elements we have $\rho_{01}=\rho_{10}^*$ and here we assumed that $\rho_{01}$ is real. Therefore $x=\langle \sigma_x\rangle=\mathrm{Tr}(\rho \sigma_x)=2\rho_{01}$ and $y=0$},
\bse \label{baysf12}
\begin{eqnarray}
z(t+dt)&=&  \frac{1 + z(t)  +(z(t)-1)e^{-V(t) S/\Delta V}}{1 + z(t)-(z(t)-1) e^{-V(t)S/\Delta V}} \label{baysf1}\\
x(t+dt)&=& x(t) \frac{\sqrt{1-z(t+dt)^2}}{\sqrt{1-z(t)^2}} e^{-\gamma dt}\label{baysf2}
\end{eqnarray}
\ese
where $S=(\Delta V/\sigma)^2$ is the signal-to-noise ratio. Theses equations, similar to the SME (\ref{eq:SME_rho3}), can be use to update the qubit trajectory for continuous $z$-measurement.

Note that, unlike the SME, here we have not made any assumption about $dt$ or the measurement strength\footnote{We need to make that assumption if we add unitary dynamics to the Bayesian update.}. So the $dt$ can in general be any duration, $dt \to \Delta t =T$, and in that case $V(t) \to V(T)= 1/T \int_0^T V(t) dt$.
Therefore, the Equation~(\ref{baysf12}) can be used to obtain final Bloch coordinate positions $z(T)$ and $x(T)$ without integration. For example, in a simple situation where the qubit starts in a superposition of the measurement operator eigenstates $x(0)=1,z(0)=0$ we have,
\begin{eqnarray}
z(T)&=&  \frac{1 - e^{-V(T) S/\Delta V}}{1 + e^{-V(T)S/\Delta V}}=\tanh(\frac{S}{2 \Delta V} V(T)) \label{baysf1T}\\
x(T)&=& \sqrt{1-z(T)^2} e^{-\gamma dt}=\mathrm{sech}(\frac{S}{2 \Delta V} V(T))\label{bays2T}
\end{eqnarray}
Therefore the final state is determined only by the averaged signal $V(T)$. This is because all measurements commute with one another and commute with the Hamiltonian\footnote{Note, if we add a Rabi drive then the Hamiltonian would not commute with the measurement operator. So we have to do step-wise integration similar to the SME.}. 

\section{Bayesian vs SME \label{section:Bayes_vs_SME}}
We have introduced two approaches for qubit state update. The SME approach (Eqs.~\ref{eq:SME_zx}) and the Bayesian update approach (Eqs.~\ref{baysf12}). Now the question is ``What is the difference? And what are the pros and cons of each approach? More importantly, do they even agree?''
We know that in order to arrive at the SME, we did a bunch of expansions and approximations regarding the weak measurement limit. However, for the Bayesian update we did not make any assumption (except the assumption that Bayes's rule applies). So, in principle one should arrive at the SME by expanding the Bayesian result. For that, let's start off by Equations~(\ref{baysf1}) and substitute $S/\Delta V = 8 \eta k dt$ and calculate $dz=z(t+dt)-z(t)$ (we drop the notation showing time dependence for compactness),
\bse
\begin{eqnarray}
dz &=&  \frac{(1-z^2)\sinh(4 \eta k V dt )}{\cosh(4 \eta k V dt )+ z  \sinh(4 \eta k V dt )} \label{bays_prove1}\\
&=& 4 \eta k V (1-z^2)dt -(4 \eta k)^2 V^2 (z+z^3) dt^2+\mathcal{O}[dt]^{3/2},
\label{bays_prove2}
\end{eqnarray}
\ese
where we expand\footnote{We keep terms up to the second order of $dt$, but remember $V$ includes a term which is effectively in the order of $1/\sqrt{dt}$, Equation~(\ref{eq:nosy_estimate}).} sinh and cosh to arrive at Equation~(\ref{bays_prove2}). Now we substitute $V= z + \frac{d\mathcal{W}}{\sqrt{4 \eta k} dt}$ only in the second term in \ref{bays_prove2} and keep terms up to the first order of $dt$ (remember $(d\mathcal{W})^2=dt$), therefore we have,
\begin{eqnarray}
dz&=& 4 \eta k V (1-z^2)dt - 4 \eta k (z-z^3) dt+\mathcal{O}[dt]^{3/2}\\
\to \dot{z}&=& 4 \eta k (1-z^2)(V -z) +\mathcal{O}[dt]^{1/2},
\label{bays_prove3}
\end{eqnarray}
where $\dot{z}=dz/dt$. Equation~\eqref{bays_prove3} is in agreement with the SME (\ref{eq:SME_z}).\\

\noindent\fbox{\parbox{\textwidth}{
\textbf{Exercise~2:} By a similar procedure as we did in this subsection, show that the Bayesian equation for $\dot{x}$, from Equations~(\ref{baysf2}) is also in agreement with the SME Equation~(\ref{eq:SME_x}) in the limit of weak measurement.
}} \vspace{0.25cm}

Now the question is why we bother considering SME while we have the exact equations from the Bayesian update. The answer is that SME has greater flexibility and can be used for any measurement operator. We will see in the next section that for $x$-measurement\footnote{By $x$-measurement, we mean the measurement with measurement operator $\sigma_-$. We refer to it as $x$-measurement because the measurement signal in that measurement is related to the $\mathcal{R}e[\sigma_-]=\sigma_x$.} there is no Bayesian update equation.

\section{Generalized measurement in the $\sigma_x$ basis\label{section:generalized_mx}}
In this section, we study continuous measurement with the measurement operator $\sqrt{\gamma} \sigma_-$. This measurement operator occurs in homodyne detection of qubit emission. We may refer to this measurement as $x$-measurement since, as we will see later, we normally set the measurement phase so that the homodyne signal in related to $\mathcal{R}e[\sigma_-]=\sigma_x$.

\subsection{POVM\label{subsection:povm_sigmax}}
We follow a phenomenological approach to formulate the corresponding POVM. Consider a qubit which decays into a transmission line by rate of $\gamma_1$ as depicted in Figure~\ref{fig:x_measurement_simple}. This configuration can be described by the interaction Hamiltonian\footnote{The interaction Hamiltonian before taking the RWA is $H_{\mathrm{int}}= - \gamma_1  ( \sigma_-  + \sigma_+)( \hat{a} +  \hat{a}^{\dagger})$.}
\begin{eqnarray}
H_{\mathrm{int}}= - \gamma_1  ( \sigma_- \hat{a}^{\dagger}  + \sigma_+ \hat{a} ),
\label{eq:JCH_ch4}
\end{eqnarray}
\begin{figure}[ht]
\centering
\includegraphics[width = 0.48\textwidth]{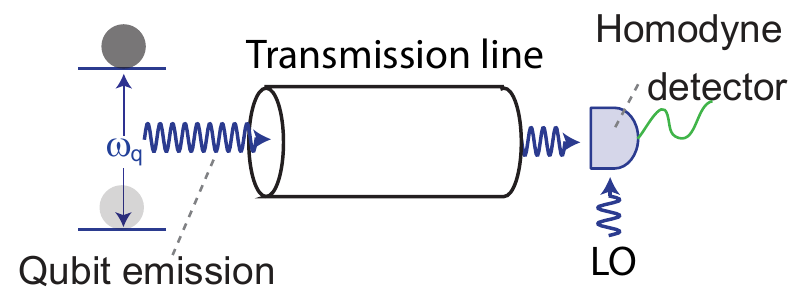}
\caption[$x$-measurement schematics]{ {\footnotesize \textbf{$x$-measurement schematic:} A qubit decays into a modeof the transmission line where we perform homodyne measurement.}} 
\label{fig:x_measurement_simple}
\end{figure}
where $\gamma_1$ is the relaxation rate for the qubit and $\hat{a}^{\dagger} (\hat{a})$ is creation (annihilation) operator for the corresponding electromagnetic mode of the transmission line\footnote{What happens to the cavity in this interpretation? One can think of that the cavity mediates the qubit emission. In this interpretation, the qubit has faster relaxation into the transmission line when the qubit and cavity are closer in frequency. However, a more realistic interpretation considers that the qubit and cavity hybridize. Therefore, the first two eigenstates of the combined qubit-cavity system act as a effective qubit as discussed in Chapter~2. This interpretation is more accurate in the limit of strong hybridization, where the qubit state is a polariton state.}.
Now assume that the qubit is initially is in state
\begin{equation}
\psi=\alpha_0|g\rangle + \beta_0|e\rangle, \label{eq:qu_init}
\end{equation}
and the transmission line is in the vacuum state $|\Phi\rangle=|0\rangle_{\mathrm{tr}}$ where we use superscript $\cdot_{tr}$ for the transmission line. After time $dt$, the unnormalized state of total system would be in an entangled state,
\begin{eqnarray}
\Psi_{\mathrm{tot}} = \alpha_0 |0\rangle |0\rangle_{\mathrm{tr}} + \beta_0 \sqrt{1-\gamma_1 dt}|1\rangle |0\rangle_{\mathrm{tr}} + \beta_0\sqrt{\gamma_1 dt} |0\rangle |1\rangle_{\mathrm{tr}}. \label{eq:psi_tot_relax}
\end{eqnarray}
If we perform photon detection on transmission line, the (unnormalized) state of the qubit would be,
\bse
\begin{eqnarray}
\mathrm{detecting  \ no  \ photon \ |0\rangle_{\mathrm{tr}}}  &\to& \psi = \alpha_0 |g\rangle + \beta_0 \sqrt{1-\gamma_1 dt}|e\rangle \label{eq:psi_tot_relax1}\\
\mathrm{detecting  \ a  \ photon \ |1\rangle_{\mathrm{tr}}}  &\to& \psi =  |g\rangle, \label{eq:psi_tot_relax2}
\end{eqnarray}
\ese
where $\gamma_1 dt$ is the probability of a relaxation event when the qubit is excited.

However, if we perform homodyne measurement instead of photon detection, then the field of the transmission line collapses to a coherent state $|\alpha\rangle_{\mathrm{tr}}$ and we will obtain a measurement outcome $\alpha$,
\begin{eqnarray}
\Psi_{\mathrm{tot}} &=& \left(\alpha_0 |g\rangle +  \beta_0 \sqrt{1-\gamma_1 dt}|e\rangle \right) |\alpha\rangle_{\mathrm{tr}}  \langle \alpha |0\rangle_{\mathrm{tr}}  + \beta_0\sqrt{\gamma_1 dt} |g\rangle |\alpha\rangle_{\mathrm{tr}}  \langle \alpha |1\rangle_{\mathrm{tr}} \nonumber\\
&=& e^{-|\alpha|^2/2} \left(\alpha_0 |g\rangle +  \beta_0 \sqrt{1-\gamma_1 dt}|e\rangle    + \alpha^* \beta_0\sqrt{\gamma_1 dt} |g\rangle  \right) |\alpha\rangle_{\mathrm{tr}},
\end{eqnarray}
where we use $ \langle \alpha | 0 \rangle_{\mathrm{tr}}=e^{-|\alpha|^2/2}$ and  $ \langle \alpha | 1 \rangle_{\mathrm{tr}}=\alpha^* e^{-|\alpha|^2/2}$ and we absorb constants in the normalization factor\footnote{Note, $|\alpha\rangle = e^{-|\alpha|^2/2} \sum_n \frac{\alpha^n}{\sqrt{n!} } |n\rangle$  therefore, $ \langle n |\alpha\rangle = \frac{\alpha^n}{\sqrt{n!}} \langle 0 |\alpha\rangle$ where $ \langle 0 |\alpha\rangle=e^{-|\alpha|^2/2}$.}. We assume that $\alpha$ is real\footnote{This choice makes the measurement to be $\mathcal{R}e[\sigma_-]$ so we call it $x$-measurement. Experimentally, the paramp phase ( the phase of our phase sensitive parametric amplifier) can be set so that all information is encoded only in the ``real" quadrature.} and define $V=\alpha \sqrt{ \gamma_1 dt}$ where $V$ is the homodyne signal. Therefore the qubit state after the measurement will be 
\begin{eqnarray}
\psi &=& e^{-\frac{V^2}{2 \gamma_1 dt}} \left(\alpha_0 |g\rangle +  \beta_0 \sqrt{1-\gamma_1 dt}|e\rangle    + V \beta_0  |g\rangle  \right). \label{eq:qu_after}
\end{eqnarray}
Note that $\psi$ is not normalized yet. One can show that the corresponding POVM connecting the qubit state before the measurement (Equation~\ref{eq:qu_init}) to qubit state after the measurement (Equation~\ref{eq:qu_after}) has this form (up to a normalization factor),
\begin{eqnarray}
\Omega_V &=&e^{-\frac{V^2}{2 \gamma_1 dt}} \left(  |g \rangle \langle g| + \sqrt{1-\gamma_1 dt } |e \rangle \langle e|  + V |g \rangle \langle e| \right),\label{eq:POVM_xm_not_expanded}\\
&=& e^{-\frac{V^2}{2 \gamma_1 dt}} \left( 1- \frac{\gamma_1 dt }{2} \sigma_+\sigma_-  + V  \sigma_- \right), \label{eq:POVM_xm}
\end{eqnarray}
where we find Eq.~\ref{eq:POVM_xm} by expanding Eq.~\ref{eq:POVM_xm_not_expanded} up to the first order in $dt$~\cite{tan2017homodyne}.\\

\noindent\fbox{\parbox{\textwidth}{
\textbf{Exercise~3:} Show that $\Omega_V$ is a POVM by verifying $\int \Omega_V^\dagger  \Omega_V dV  =\mathbb{1}$ and obtain the missing normalization factor in Equation~\eqref{eq:POVM_xm} (for answer see Ref.~\cite{tan2017homodyne}).
}} \vspace{0.25cm}

Now let's look at the probability of getting a measurement signal $V$,
\begin{eqnarray}
P(V)  &=& | \Omega_V |\psi \rangle |^2= \langle \psi | \Omega_V^{\dagger} \Omega_V | \psi\rangle \\
&=& e^{-\frac{V^2}{ \gamma_1 dt}} \left( 1- \gamma_1 (dt-V^2) \langle  \sigma_+\sigma_-\rangle +  V \langle \sigma_+ + \sigma_-\rangle  \right),\\
&=& e^{-\frac{V^2}{ \gamma_1 dt}} \left( 1- \frac{\gamma_1 dt }{2}(1-V^2)(1+z)+ V x   \right),
\label{eq:POVM_xm_P}
\end{eqnarray}
where $z=\langle \sigma_z \rangle$ and $x=\langle \sigma_x \rangle$. In the limit of continuous measurement $dt \to 0$ we have,
\begin{eqnarray}
P(V) &\simeq& e^{-\frac{V^2}{  \gamma_1 dt}} \left( 1 + V x   \right), \\
&\simeq& \exp \left[ -\frac{1 }{ \gamma_1 dt} (V^2 - 2 \gamma_1 dt V x )\right]\\
&\simeq& \exp \left[ -\frac{(V - \gamma_1 x dt/2)^2}{ \gamma_1 dt }\right],
\label{eq:POVM_xm_Pv}
\end{eqnarray}
It is convenient to rescale the signal to have variance $\sigma^2 =\gamma_1 dt$ therefore we arrive at\footnote{We could do this rescaling right at the beginning by defining the homodyne signal as $V=\sqrt{\gamma_1 dt/2}$. This scaling may have to do with the fact that with homodyne measurement we only collect half of the signal on average.},
\begin{eqnarray}
P(V) &\simeq & \frac{1}{\sqrt{2 \pi \gamma_1 dt}} \exp \left[ -\frac{(V - \gamma_1 x dt)^2}{ 2 \gamma_1 dt }\right],
\label{eq:POVM_x_scaled}
\end{eqnarray}
where we also added the normalization factor. Equation~\eqref{eq:POVM_x_scaled} is analogous to Equation~\eqref{eq:POVM_sigz}, However this time the measurement signal distribution is shifted by $\gamma_1 \langle x \rangle$ and has variance of $\gamma_1 dt$ as depicted in Figure~\ref{fig:overlap_PVx}.
\begin{figure}[ht]
\centering
\includegraphics[width = 0.44\textwidth]{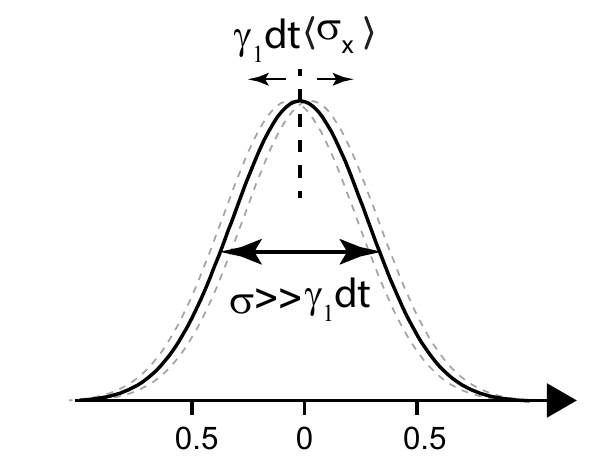}
\captionsetup{font=footnotesize}
\caption[Homodyne measurement signal distribution $x$-measurement]{\textbf{Homodyne measurement signal distribution in the weak limit, $x$-measurement:} Note that $\gamma_1 dt \ll 1$ therefore $\sqrt{\gamma_1 dt} \gg \gamma_1 dt$} 
\label{fig:overlap_PVx}
\end{figure}
Therefore, the homodyne signal can be described in the form\footnote{It worth mentioning that, in case of inefficient detection, the signal would be $V =  \sqrt{\eta} \gamma_1 \langle x \rangle dt + \sqrt{\gamma_1} d\mathcal{W}$. This intuitively makes sense because we always rescale the signal to have variance $\gamma_1 dt$ regardless of the efficiency $\eta$. Still, inefficient measurement decreases the SNR since the greatest mean separation of the homodyne signal conditioned on $\langle x\rangle$ scales linearly in $\eta$.},
\begin{eqnarray}
V &= & \gamma_1 \langle x \rangle dt + \sqrt{\gamma_1} d\mathcal{W}= \gamma_1 \langle x \rangle dt + \sqrt{\gamma_1} d\xi dt,
\label{eq:noisy_estimate_x}
\end{eqnarray}
where $d\mathcal{W}$ and $d\xi$ are zero-mean Gaussian distributions with variance of $dt$ and $dt^{-1}$ respectively (therefore $\sqrt{\gamma_1} d\mathcal{W}$ has variance of $\gamma_1 dt$).

\subsection{SME\label{Subsection:SME_xm_xz}}
Now we turn to state evolution and calculating the SSE and SME. For the SSE we simply need to calculate the change in state $|\psi\rangle$ during the measurement time $dt$,
\begin{eqnarray}
d|\psi \rangle &=& |\psi(t+dt)\rangle  - |\psi(t)\rangle = (\Omega_V -1)\psi \nonumber \\
&=&e^{-\frac{V^2}{2 \gamma_1 dt}} \left( - \frac{\gamma_1 dt }{2} \sigma_+\sigma_-  + V  \sigma_- \right) |\psi\rangle.
\label{eq:SSE_x_m}
\end{eqnarray}
In order to obtain the SME we calculate $d\rho = d|\psi\rangle \langle \psi| +  |\psi\rangle d \langle \psi| +  d|\psi\rangle d \langle \psi|$, 
\begin{eqnarray}
d\rho &=& \left( - \frac{\gamma_1 dt }{2} \sigma_+\sigma_-  + V  \sigma_- \right) \rho + \rho  \left( - \frac{\gamma_1 dt }{2} \sigma_+\sigma_-  + V  \sigma_+ \right)  \nonumber \\
 &+& \left( - \frac{\gamma_1 dt }{2} \sigma_+\sigma_-  + V  \sigma_- \right) \rho  \left( - \frac{\gamma_1 dt }{2} \sigma_+\sigma_-  + V  \sigma_+ \right),
\label{eq:SME_x_m1}
\end{eqnarray}
where we used Equation~(\ref{eq:SSE_x_m}) and ignored normalization constants. Again, we only keep terms up to the first order of $dt$\footnote{Remember $V$ has a term of order $\sqrt{t}$ according to Equation~(\ref{eq:noisy_estimate_x}).},
\begin{eqnarray}
d\rho &=&  - \frac{\gamma_1 dt }{2} ( \sigma_+\sigma_-  \rho +  \rho \sigma_+\sigma_- ) + V ( \sigma_-\rho  + \rho \sigma_+ )\nonumber \\
&\ & + \gamma_1  \sigma_-  \rho \sigma_+  dt,
\label{eq:SME_x_m2}
\end{eqnarray}
where in the last term, we substitute $V$ from Equation~(\ref{eq:noisy_estimate_x}) and keep terms up to $dt$ and use the It\^o rule $(d\mathcal{W})^2 =dt$. By rearranging terms we have,
\begin{eqnarray}
d\rho = \gamma_1 dt \left( \sigma_-  \rho \sigma_+ - \frac{1}{2} (\sigma_+\sigma_-  \rho +  \rho \sigma_+\sigma_- ) \right) + V ( \sigma_-\rho  + \rho \sigma_+ ).
\label{eq:SME_x_m3}
\end{eqnarray}
Since we have ignored the normalization constants, now we need to normalize the result. One can show that the normalized SME has the form
\begin{eqnarray}
d\rho &=& \gamma_1 dt \left( \sigma_-  \rho \sigma_+ - \frac{1}{2} (\sigma_+\sigma_-  \rho +  \rho \sigma_+\sigma_- ) \right) \label{eq:SME_x_m4} \\
 &+& \left(V -\gamma_1 \mathrm{Tr}[ (\sigma_-+\sigma_+)\rho]  dt \right)  ( \sigma_-\rho  + \rho \sigma_+ - \mathrm{Tr}[ (\sigma_-+\sigma_+)\rho]\rho).\nonumber
\end{eqnarray}\\

\noindent\fbox{\parbox{\textwidth}{
\textbf{Exercise~4:} Convince yourself about the normalization step, which is the transition from Equation~(\ref{eq:SME_x_m3})~$\to$~(\ref{eq:SME_x_m4}).
}} \vspace{0.25cm}

Equation~(\ref{eq:SME_x_m4}) can be represented in terms of the dissipation superoperator $\mathcal{D}[L]\rho=L^{\dagger} \rho L - \frac{1}{2}\{L^{\dagger} L, \rho\}$ and jump
superopertator $\mathcal{H}[L]\rho= L \rho +  \rho L^{\dagger} - \mathrm{Tr}[ (L+L^{\dagger})\rho] \rho$ in a more compact form,
\bse
\begin{eqnarray}
d\rho &=&  \gamma_1 dt \mathcal{D}[\sigma_-]\rho + (V -\gamma_1 x dt ) \mathcal{H}[\sigma_-]\rho\\
&=&  \gamma_1 dt \mathcal{D}[\sigma_-]\rho + \sqrt{\gamma_1} d\mathcal{W} \mathcal{H}[\sigma_-]\rho\\
\to \dot{\rho}=\frac{d\rho}{dt}&=&\gamma_1 \mathcal{D}[\sigma_-]\rho + \sqrt{\gamma_1} d\xi \mathcal{H}[\sigma_-]\rho
\label{eq:SME_xm}
\end{eqnarray}
\ese
where we substitute $\mathcal{W}$ and $d\xi$ as defined in Equation~(\ref{eq:noisy_estimate_x}).

Equation~(\ref{eq:SME_xm}) describes the evolution of the qubit under radiative decay with rate $\gamma_1$ and continuous perfect monitoring of that radiation with homodyne detection. By comparing to the general form of the SME (Equation \ref{eq:SME_rho3}), we understand that the homodyne measurement of the qubit radiation is corresponding to the measurement operator $\sigma_-$ and the measurement strength $k=\gamma_1$ is the rate in which the detector receives the emission.

In order to account for imperfect detection we can again use the technique of multiple detectors. Assume that actual detector receives proportion $\eta$ of the total emission at rate $\gamma_1$, thus the measurement strength of this detector is $\eta \gamma_1$. The rest of the emission is then measured by a fictitious detector (the environment) with measurement strength $(1-\eta) \gamma_1$. Both measurement detectors impose their own backaction on the qubit evolution,
\begin{eqnarray}
d\rho &=& \gamma_1 \mathcal{D}[\sigma_-]\rho + \sqrt{\gamma_1} d\xi \mathcal{H}[\sigma_-]\rho + \sqrt{\gamma_1} d\xi^{(f)} \mathcal{H}[\sigma_-]\rho,
\label{eq:SME_xm_eta}
\end{eqnarray}
where $d\xi$ and $d\xi^{(f)}$ represents the collected homdyne signal by actual detector and the fictitious detector respectively. By averaging over all the fictitious detector outcomes we arrive at the SME for inefficient detection of the qubit emission,
\begin{eqnarray}
d\rho &=& \gamma_1 \mathcal{D}[\sigma_-]\rho + \sqrt{\eta \gamma_1} d\xi \mathcal{H}[\sigma_-]\rho,
\label{eq:SME_xm_eta2}
\end{eqnarray}
where that the corresponding inefficient homodyne signal can be described by,
\begin{eqnarray}
V &= & \sqrt{\eta}\gamma_1 \langle x \rangle dt + \sqrt{\gamma_1} d\mathcal{W}= \sqrt{\eta}\gamma_1 \langle x \rangle dt + \sqrt{\gamma_1} d\xi dt.
\label{eq:noisy_estimate_x2}
\end{eqnarray}
Similar to the discussion we had for the SME~(\ref{eq:SME_rho3}), one can also add unitary evolution to the SME~(\ref{eq:SME_xm_eta2}) to account for a coherent drive on the qubit and obtain a full version of the SME,
\begin{eqnarray}
d\rho &=& -i[H_R, \rho] + \gamma_1 \mathcal{D}[\sigma_-]\rho + \sqrt{\eta \gamma_1} d\xi \mathcal{H}[\sigma_-]\rho.
\label{eq:SME_xm_eta3}
\end{eqnarray}

To sum up the discussion in this section, we may recast this stochastic master equation terms of Bloch vector components,
\begin{subequations}\label{SME_xm_eta_xz}
\begin{eqnarray}
 \dot{z} =& +\Omega x  +  \gamma_1 (1-z)  + \sqrt{\eta \gamma_1} x (1-z) d\xi,\label{SME_xm_eta_z},  \\
 \dot{x} =& - \Omega z  - \frac{\gamma_1}{2} x  + \sqrt{\eta} ( 1-z - x^2  )d\xi,\label{SME_xm_eta_x},
\end{eqnarray}
\end{subequations}
where we assume $H_R=\frac{\Omega}{2} \sigma_y$.

\section{z-measurement procedure\label{section:z-me}}
In this section, we are going to utilize the basic techniques mentioned in Chapter 3 to discuss how to actually perform weak measurement and analyze the data to obtain quantum trajectories. A typical $z$-measurement includes:
\begin{itemize}
\item Qubit calibration and characterization as discussed in Chapter 3.
\item Paramp calibration, dumb-signal cancellation, readout calibration, as discussed in Chapter 3.
\item Calibration for $ \chi , \eta, \bar{n}, k$.
\item Calibration for preparation and tomography pulses, Rabi tomography.
\item Data acquisition for quantum state tomography and the actual experiment.
\item Post-processing, verifying the measurement trajectory update method by quantum state tomography.
\end{itemize}
In the following subsections, we discuss each of these steps in greater detail.
\subsection{Basic characterization}
As discussed in Chapter 3 we first need to characterize the qubit-cavity system. The information we need to obtain in this step is the cavity frequency $\omega_q$, cavity linewidth $\kappa$, qubit frequency $\omega_q$, qubit relaxation time $T_1$, and qubit dephasing time $T_2^*$.

In this stage we also find an initial calibration for $\pi$ and $\pi/2$ pulses (usually $T_\pi=20$ ns ,  $T_{\pi/2}=10$ ns for certain amplitude in arbitrary waveform generator (AWG). More careful calibration should be performed after paramp calibration and dumb-signal cancellation. See Chapter 3 for more details on basic experiment characterization.

\subsection{Paramp calibration}
As discussed in Chapter 3, we set up the paramp (preferably in double-pump operation mode) at the cavity frequency (more precisely at $\omega_c -\chi$ so we have an optimum and symmetric response for the states $|g\rangle$ and $|e\rangle$).

\emph{\textbf{``Dumb-signal" cancellation---}} Beside the basic paramp setup and obtaining a proper gain profile, here we need also consider some practical techniques to optimize the low-power readout fidelity.
The point is that in weak measurement the paramp should be adjusted to have best performance for weak signal detection\footnote{Moreover the paramp normally works efficiently in the weak signal limit.}. However, during the readout we use a much stronger signal to project the qubit (basically the readout is a very strong measurement). Having calibrated the paramp for weak measurement, it may not have the best performance for readout where we send a large number of photons during the readout.

The trick to go around this issue is called ``dumb-signal cancellation". The idea is following: although we need a high number of photons during the readout inside the cavity, after passing the cavity we can coherently cancel the unnecessary part of the signal and only net phase shifts are amplified by the paramp as demonstrated in Figure~\ref{fig:dum_sig_cancelation}.
\begin{figure}[ht]
\centering
\includegraphics[width = 0.98\textwidth]{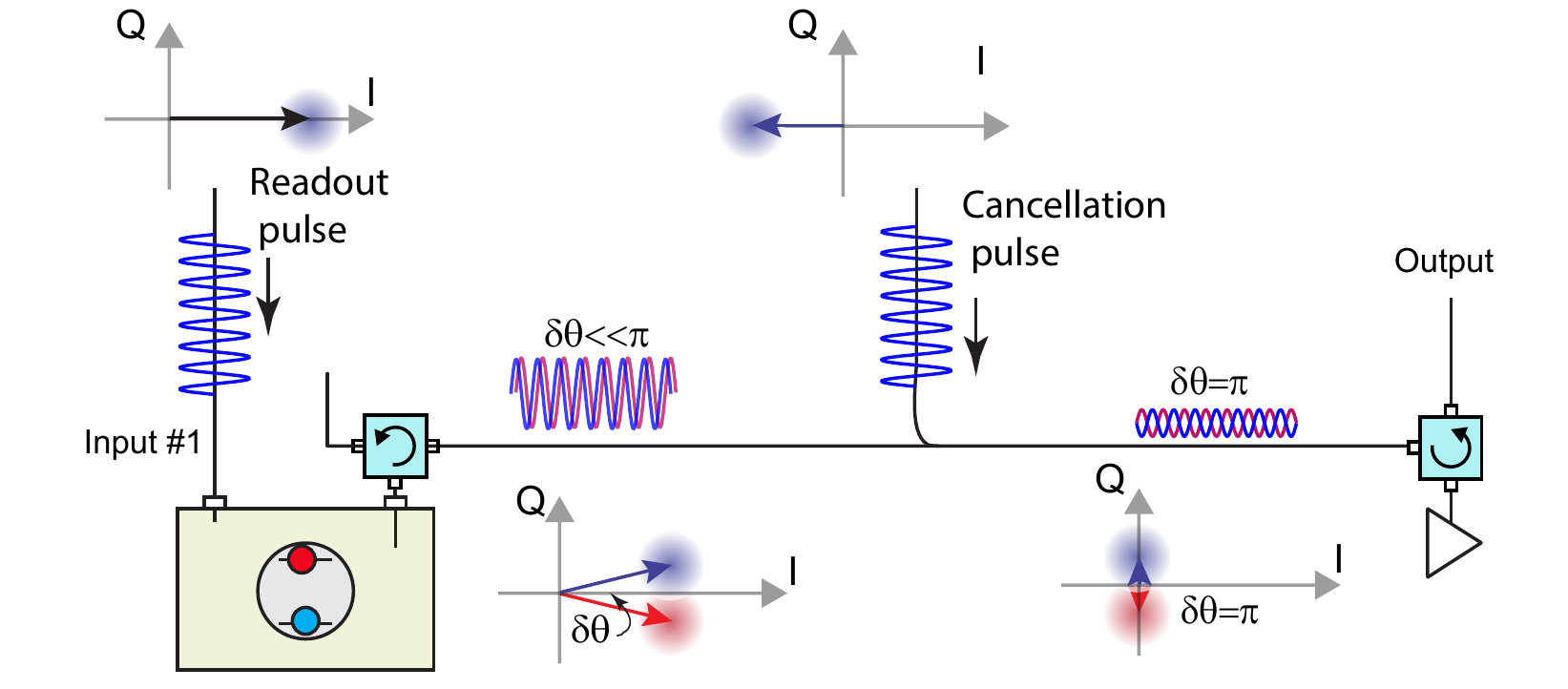}
\caption[Dumb-signal cancellation]{ {\footnotesize \textbf{Dumb-signal cancellation:} A copy of the readout pulse with proper amplitude and phase cancels the readout signal in $I$ quadrature while maintaining the information along the $Q$ quadrature (separation $\mathrm{Amp}\times \delta \theta$ is preserved) since the paramp works efficiently in the weak signal regime.} }
\label{fig:dum_sig_cancelation}
\end{figure}
The dumb-signal cancellation is basically a copy of the readout pulse with the right amount of attenuation\footnote{One way to estimate the proper amplitude for the dumb signal in the experiment is to send a continuous readout pulse to the cavity (e.g. run a long readout sequence in continuous mode, or set the readout pulse always high at mixer) and look at the output signal power with spectrum analyzer. Now disconnect the readout input but this time send the dumb signal cancellation through the pump port and adjust to amplitude to have the same output power signal as we had for continuous readout. By this calibration the amplitude is roughly calibrated and you can find the optimal phase by looking at average homodyne signals in IQ plane and comparing the output signal before readout and during the readout.} and proper phase to cancel the readout signal before reaching to the paramp while maintaining the qubit information.
The room temperature circuitry for dumb-signal cancellation is depicted in Figure~\ref{fig:chi_nbar1}.

\subsection{Quantum efficiency calibration}
After the paramp is set up for optimal readout performance, we are ready to calibrate the quantum efficiency $\eta$. This includes calibration of the dispersive shift $\chi$, the average photon number $\bar{n}$, the measurement strength $k$, and finally measurement of the quantum efficiency.

In order to obtain values for $\chi$ and $\bar{n}$, we use a Ramsey measurement\footnote{Note that we might have a crude estimation of $\chi$ from the punch-out experiment, but that is not accurate enough for the quantum efficiency calibration.}. As discussed in Chapter 2 (Eq.~\ref{eq:H_dis_rearange2}), the qubit frequency is shifted by the average number of photons in the cavity, $\Delta \omega_q= 2 \chi \bar{n}$. Moreover as discussed earlier in this chapter (Equation~\ref{eq:MID}), photons in the cavity also induce dephasing of the qubit coherence by a rate  $\Gamma=8\chi^2 \bar{n}/\kappa$. We can observe these two effects by performing a Ramsey measurement over a range of average photon number occupation in the cavity. Fortunately, the ratio $\Gamma/\Delta \omega_q=4 \chi/\kappa$ is independent of $\bar{n}$, which means we just need to sweep the average number of photons in the cavity (without knowing the actual $\bar{n}$ values) and calculate the ratio to obtain $\chi$ (the value for the cavity linewidth $\kappa$ is independently known from the basic characterization).

For that, we start by running the Ramsey experiment (typically a 5 $\mu$s Ramsey sequence). We set the frequency to be slightly off-resonant\footnote{It is more convenient to avoid being on-resonance with qubit so there are always oscillations which makes an easier fitting procedure. Therefore we prefer to be slightly above the actual qubit frequency (0.4 MHz for 5 $\mu$s Ramsey sequence) and by increasing the $\bar{n}$, qubit will be pushed down (remember $\chi$ is typically negative) and we never Stark shift the qubit into an on-resonance situation. Typically, we set the qubit drive frequency so that we have $\sim$ one oscillation in the limit $\bar{n}\to0$. This usually ensures we will sample enough to resolve Ramsey oscillations at higher $\bar{n}$ in the cavity.} as illustrated in Figure~\ref{fig:chi_nbar1}. The qubit signal generator (BNC2) is set to be $0.4$ MHz above the qubit resonance frequency. By changing the DC offset values at the I/Q inputs of the cavity mixer, we let photons to leak into the cavity and shift the qubit frequency and also dephase the qubit, which is measured by the Ramsey measurement.
\begin{figure}[ht]
\centering
\includegraphics[width = 0.98\textwidth]{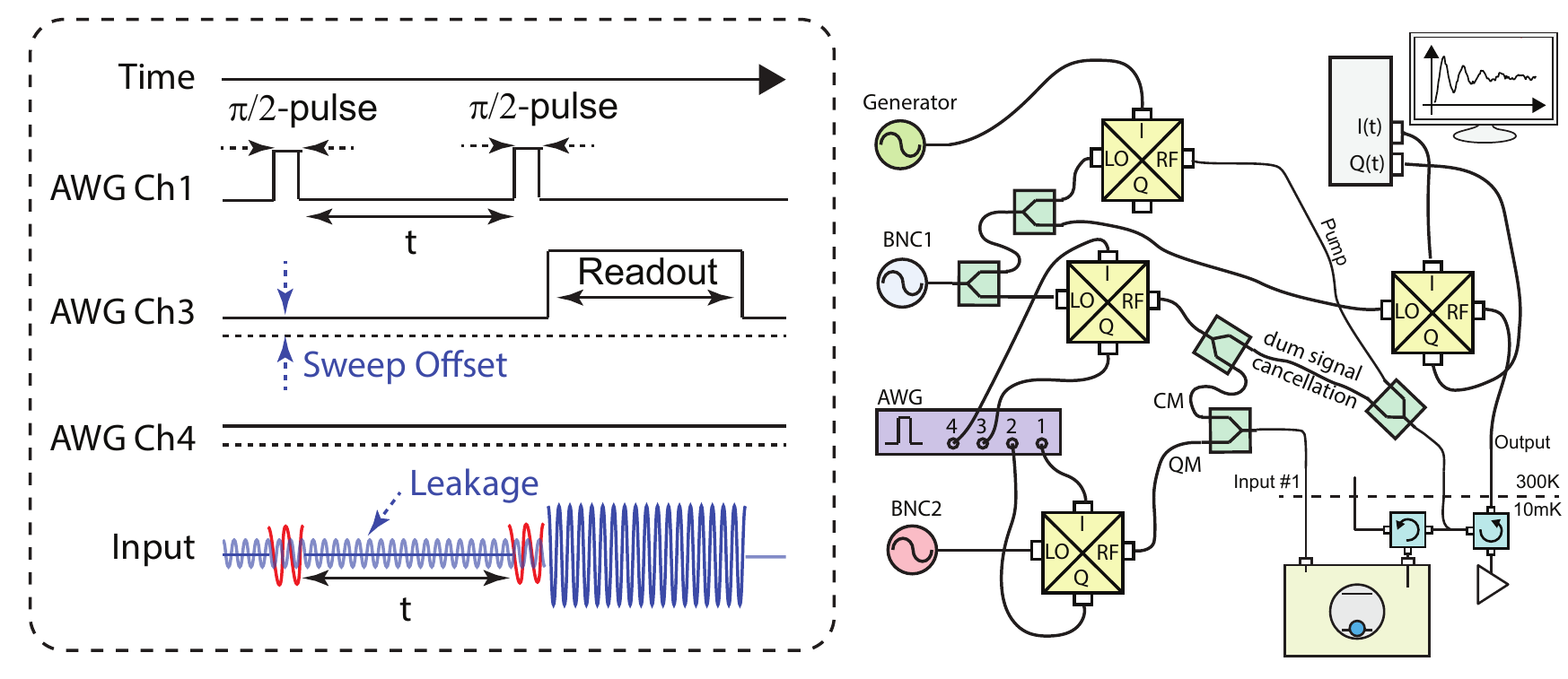}
\caption[Quantum efficiency calibration setup]{ {\footnotesize \textbf{Quantum efficiency calibration setup}. }} 
\label{fig:chi_nbar1}
\end{figure}
The result of this sweep labeled as the Ch3 offset (the applied DC offset voltage to the input Q for the cavity mixer) has been shown in Figure~\ref{fig:chi_nbar2}a. The minimum oscillation, which is $f_{min}=0.4$ MHz, happens somewhere around $55$ mV for the Ch3 offset. As the Ch3 offset deviates from a minimum leakage value, the mixer lets photons populate the cavity and the oscillation frequency increases by $2\chi \bar{n}$. Moreover, the oscillations decay faster as the average number of photons increases in the cavity as expected by the relation $\Gamma=8\chi^2 \bar{n}/\kappa$. Figure~\ref{fig:chi_nbar2}b shows Ramsey oscillations from data both near and far from minimum leakage. By fitting a decaying sinusoid to the data we obtain the oscillation frequency $f$ and the Ramsey decay time $1/\Gamma$ as a function of the Ch3 offset value as depicted in Figure~\ref{fig:chi_nbar2}c. In order to obtain $\chi$ we plot $\Gamma$ versus $f$ and fit a line to the data as depicted in Figure~\ref{fig:chi_nbar2}d. The slope would be $4\chi/\kappa$ (the value of the cavity linewidth $\kappa$ is known from the low-power cavity transmission measurement).
\begin{figure}[ht]
\centering
\includegraphics[width = 0.8\textwidth]{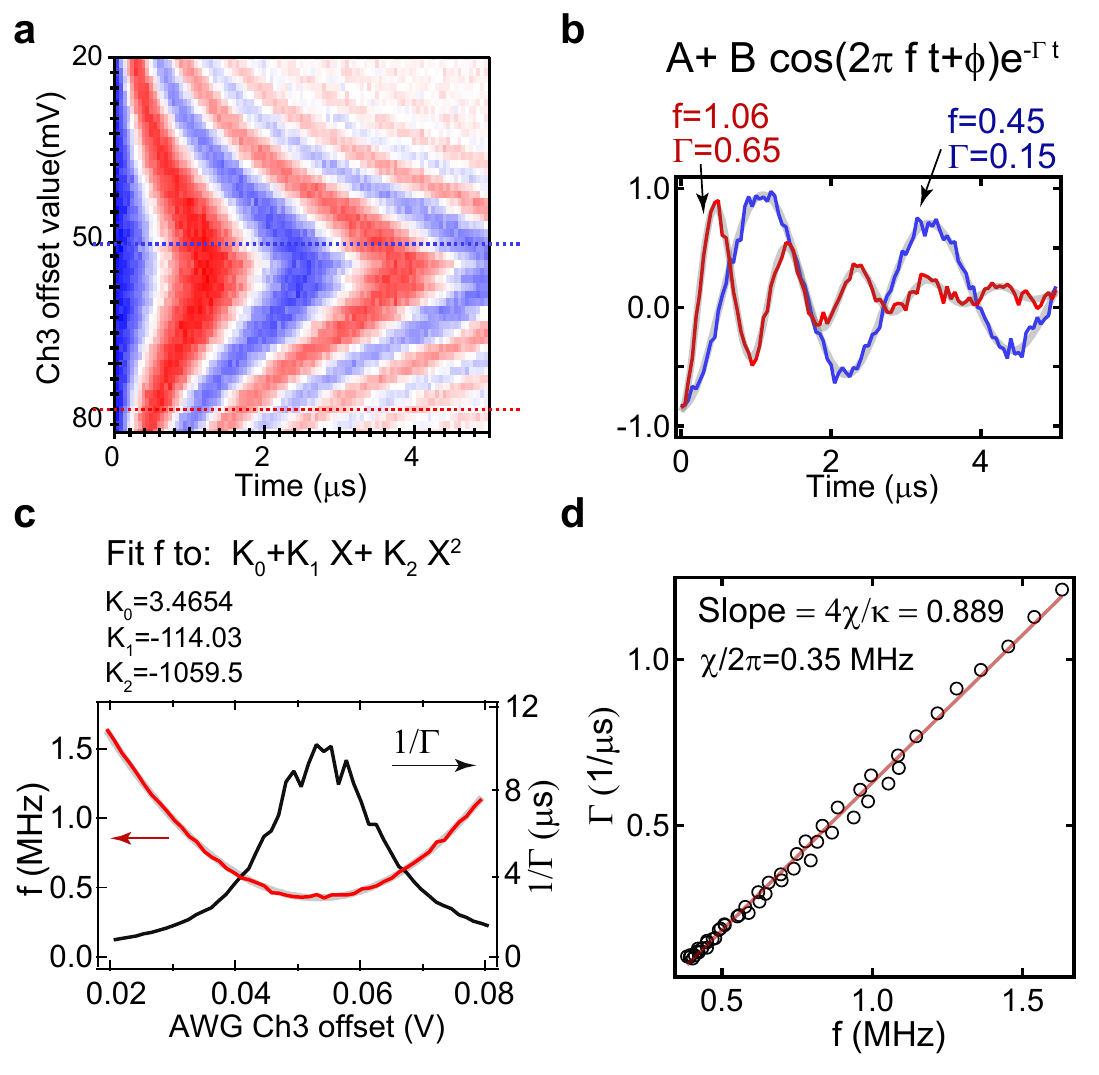}
\caption[$\chi$ calibration result]{ {\footnotesize \textbf{The $\chi$ calibration result:} \textbf{a}, The Ramsey experiment result for different offset values of Ch3. \textbf{b}, Two cuts from the sweep data in \textbf{a} indicated by dashed lines. \textbf{c}, The frequency and the damping rate for the Ramsey data of panel (a) versus the DC offset of Ch3. By fitting the red curve to a parabola we get coefficients $K_0$, and $K_1$, and $K_2$, which will be used to find the optimal quadrature for the measurement \textbf{d}, The damping rate versus the frequency is ideally a line with a slope of $4\chi/\kappa$.}}
\label{fig:chi_nbar2}
\end{figure}
We need one more piece of information from these data. We fit the curve $f$ (frequency versus Ch3 offset) to a polynomial (parabola is enough) and record the fit parameters as depicted in Figure~\ref{fig:chi_nbar2}. Later, this data will be used to find the optimal quadrature for the measurement\footnote{The relative phase of the cavity photons and the paramp pump is important for optimizing amplification and hence quantum efficiency.}.

We repeat the experiment and apply the same analysis for the DC offset of Ch4 while keeping the offset of Ch3 fixed at the minimum leakage value\footnote{To be more accurate, after finishing the Ch4 offset sweep one can redo the Ch3 offset sweep with Ch4 offset fixed at the corresponding minimum leakage value.}. 

Now, we use the parabolic fit to parametrize the mixer output power in terms of the Ramsey oscillation frequency $f$. Here we briefly discuss what this means. Ideally, the mixer output power can be represented by,
\begin{eqnarray}
f_k = K_2^{\mathrm{(Ch3)}} (\mathrm{Ch3}-\mathrm{Ch3}_{\mathrm{min}})^2 + K_2^{\mathrm{(Ch4)}}(\mathrm{Ch4}-\mathrm{Ch4}_{\mathrm{min}})^2
\end{eqnarray}
Where we use the fact that Ch3 and Ch4 are orthogonal and $\mathrm{Ch3}_{\mathrm{min}}=-K_1^{\mathrm{(Ch3)}}/2K_2^{\mathrm{(Ch3)}}$ . The phase of the output signal also can be represented as,
\begin{eqnarray}
\theta = \mathrm{atan} \left[ \sqrt{\frac{K_2^{\mathrm{(Ch4)}}}{K_2^{\mathrm{(Ch3)}}}} \frac{ (Ch4-\mathrm{Ch4}_{\mathrm{min}}) }{(Ch3-\mathrm{Ch3}_{\mathrm{min}})} \right]
\end{eqnarray}
as depicted in Figure~\ref{fig:Mixer_anngle}, the parameter $\theta$ sets the angle of the output signal in the $IQ$ plane (phasor) and $f_k$ parametrizes the length of the phasor which has to do with the number of photons $\bar{n}$ but we usually keep it in terms of frequency $2\pi f_k=2\pi (f-f_{\mathrm{min}})=2\chi \bar{n}$. In fact, $k=4 \chi \cdot 2\pi (f-f_{\mathrm{min}}) /\kappa$ where $k$ is the measurement strength\footnote{Note that, we explicitly express $f$ in MHz (Figure~\ref{fig:chi_nbar2}b). One needs to be careful about this factor of $2\pi$}.
\begin{figure}[ht]
\centering
\includegraphics[width = 0.35\textwidth]{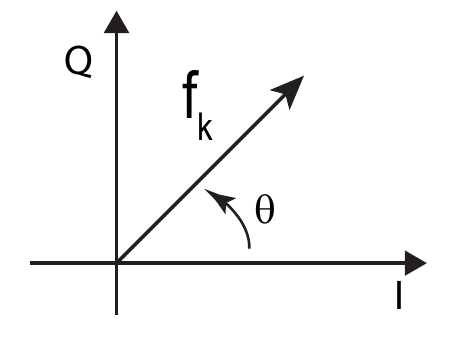}
\caption[Mixer output]{ {\footnotesize \textbf{ Mixer output:} The output of the mixer is a coherent signal whose phase and amplitude depend on Ch3/Ch4 amplitude and offset. The parameter $f_k$ quantifies the strength of measurement and $\theta$ is the measurement quadrature.}} 
\label{fig:Mixer_anngle}
\end{figure}
One can show that for a given mixer output power $f_k$ and angle $\theta$ the value for Ch3 and Ch4 should be,
\bse\label{eq:ch3_theta12}
\begin{eqnarray}
\mathrm{Ch3}(f_k, \theta) = \sqrt{ \frac{f_k}{K_2^{\mathrm{(Ch3)}}}} \cos(\frac{\pi}{180} \theta) - \frac{K_1^{\mathrm{(Ch3)}}}{2 K_2^{\mathrm{(Ch3)}}} \label{eq:ch3_theta},\\
\mathrm{Ch4}(f_k, \theta) = \sqrt{ \frac{f_k}{K_2^{\mathrm{(Ch4)}}}} \sin(\frac{\pi}{180} \theta) - \frac{K_1^{\mathrm{(Ch4)}}}{2 K_2^{\mathrm{(Ch4)}}}.\label{eq:ch4_theta},
\end{eqnarray}
\ese
where we represent $\theta$ in degrees for convenience. Now we use Equation~(\ref{eq:ch3_theta12}) to once again sweep the Ramsey measurement, but this time we sweep the angle $\theta$ and while keeping the frequency $f_k$ fixed at a certain value\footnote{It is convenient to set $f_k$ at or close to the value that you will be performing the actual experiment. Usually $f_k=0.1$ is a very weak measurement and $f_k=1$ is a relatively ``strong" weak measurement.}. Figure~\ref{fig:sweep_theta_Ramsey} shows Ramsey oscillation measurements for different values of $\theta$ value at $f_k=0.5$.
\begin{figure}[ht]
\centering
\includegraphics[width = 0.9\textwidth]{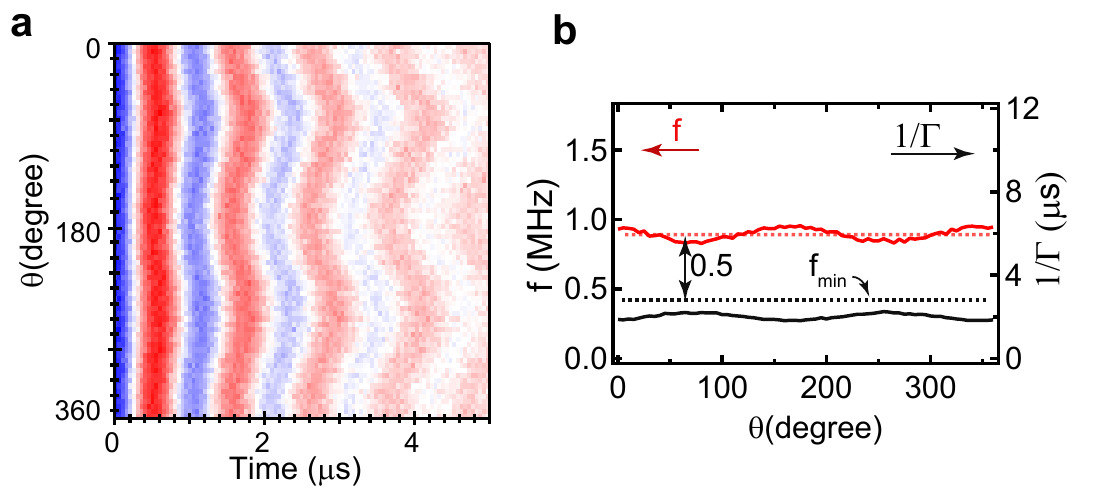}
\caption[Ramsey measurements for a sweep of different angles]{ {\footnotesize  \textbf{Ramsey measurements for a sweep of different angles.}}} 
\label{fig:sweep_theta_Ramsey}
\end{figure}
Ideally the Ramsey oscillation frequency should be fixed, but in practice the mixer may have some imperfections and the Equation~(\ref{eq:ch3_theta12}) does not perfectly predict the mixer output. But this is not be a problem for calibration for a reason that will be clear shortly\footnote{Eventually, for quantum efficiency calibration, we compare the result of two $\theta$ sweeps so imperfections do not contribute to the final result.}.

Now we arrive at the last step of the quantum efficiency calibration. In this step we want to find the optimal angle which gives the best signal-to-noise ratio for measurement of the qubit state. For that, we compare the weak measurement signal for a certain time $T \sim 100$ ns after preparing the qubit in the ground or excited state \footnote{Note there is no readout pulse needed in this step.}. We repeat this measurement for different angles and compare the separation of the two readout histograms to find which angles gives the optimal SNR as depicted in Figure~(\ref{fig:sweep_theta_histo}).
\begin{figure}[ht]
\centering
\includegraphics[width = 0.98\textwidth]{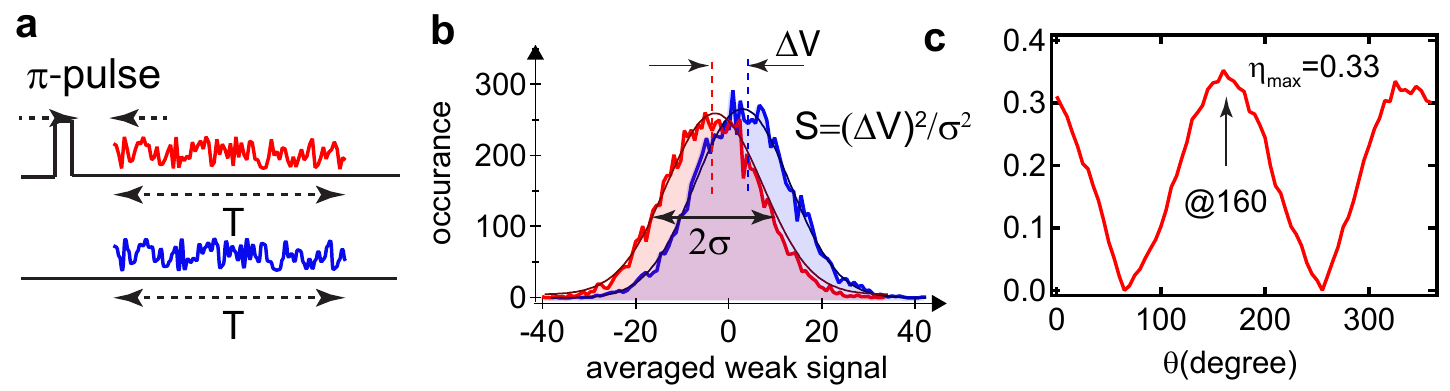}
\captionsetup{font=normalsize}
\caption[Calibration of $\eta$]{ {\footnotesize \textbf{Calibration of $\eta$:} Comparison of the measurement histograms for $|g\rangle$ and $|e\rangle$ for different $\theta$.}} 
\label{fig:sweep_theta_histo}
\end{figure}
Once we find the SNR for different angles $\theta$, we have all the pieces we need to calculate the quantum efficiency,
\begin{eqnarray}
\eta= \frac{S \kappa} {64 \chi^2 \bar{n} T} = \frac{S \kappa} {64 \pi \chi (f - f_{\mathrm{min}}) T}. \label{eq:eta}
\end{eqnarray}
As depicted in Figure~\ref{fig:sweep_theta_histo}d, the quantum efficiency is maximum at a certain angle which is ideally aligned with the paramp amplification quadrature.
\subsection{Tomography pulse calibration}
Before we collect data, it is good to fine-tune the preparation/tomographic pulses. A short Rabi (100ns) sequence with all three types of tomographic readout for $x,y,z$ (as discussed in Chapter 3) is a simple test to verify the preparation and tomographic pulses\footnote{We may have already calibrated $\pi,\pi/2$-pulses in a ``basic qubit characterization" step but note that we are now pumping the paramp and may need to revisit the qubit calibration. Moreover, we might need a more complicated preparation for the actual experiment. So it makes sense to specifically check the preparation pulses before starting the actual experiment.}.
Figure~\ref{fig:rabi_tomog_diagnosis}a shows a Rabi tomography result corresponding to a perfect calibration for preparation and tomography pulses. The fact that the oscillations for both $x$ and $y$ start from the zero and that $y$ remains always zero means that, for most part, the pulses are calibrated\footnote{One can use a longer Rabi sequence with lower amplitude, $T_1$, or Ramsey sequence to further tune the calibration.} Figures~\ref{fig:rabi_tomog_diagnosis}b,c,d,e show some common imperfect calibrations.
\begin{figure}[ht]
\centering
\includegraphics[width = 0.98\textwidth]{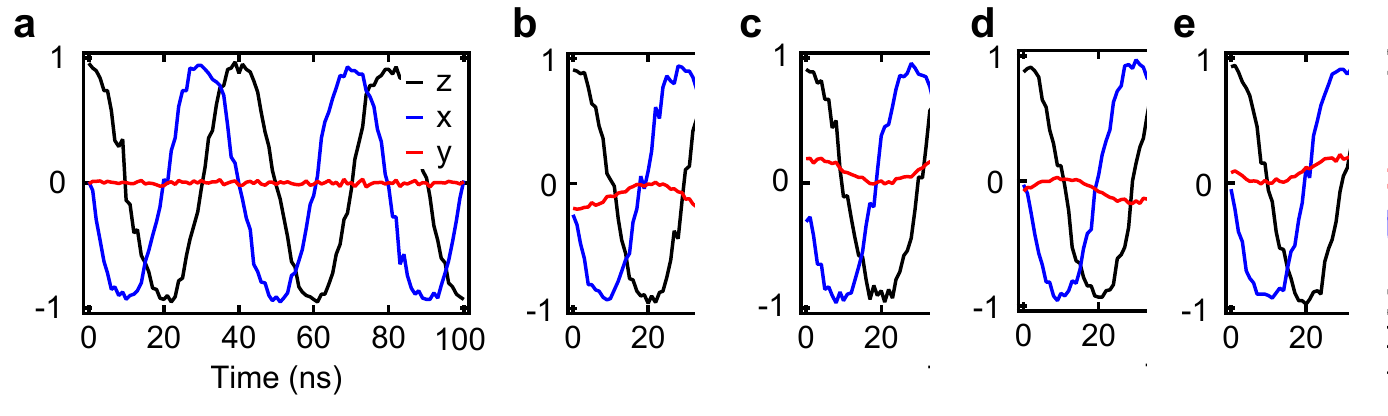}
\caption[Rabi tomography diagnosis]{ {\footnotesize  \textbf{Rabi tomography diagnosis:}  \textbf{a}, A perfect calibration. \textbf{b} The $\pi/2$ pulses needed to be weaker, either shorter pulses or lower amplitude. \textbf{c} The $\pi/2$ pulse for $x\  (y)$ needed lower (higher) amp/duration. \textbf{d}, Mixer orthogonality is slightly higher that 90 degrees. \textbf{d}, Mixer orthogonality is slightly lower that 90 degrees.  } }
\label{fig:rabi_tomog_diagnosis}
\end{figure}

\subsection{Data acquisition}
After recalibrating the preparation and tomographic pulses. We are ready to run experimental sequences (including noise calibration and state tomography sequences).

The noise calibration measurement (depicted in Figure~\ref{fig:sweep_theta_histo}a,b) needs to be collected as a reference to scale the collected digitized weak signal\footnote{The noise calibration sequence doesn't have drive or readout, only ground and excited state preparations and weak measurement for a certain time $\sim$ 1 $\mu$s.}.

For example, the sequence for continuous monitoring of a driven qubit has been depicted in Figure~\ref{fig:driven_z_measur_seq}a which includes pulses for heralding, preparation, weak measurement, and readout. The obtained data is depicted as a color plot in Figure~\ref{fig:driven_z_measur_seq}b. Note that we perform the experiment for different times $t$ (in this case we vary the measurement time from 0 to $2\  \mu$s)\footnote{One may think that only repeating the longest trajectory is enough because then you can update trajectories as long as you wish. However, in order to verify the validity of trajectory update, you will need to have trajectories which have different lengths, which provides you with readout measurement at different times. Later we will discuss how to use the trajectory measurements of different times to tomographically validate the trajectory update method.}. 
\begin{figure}[ht]
\centering
\includegraphics[width = 0.98\textwidth]{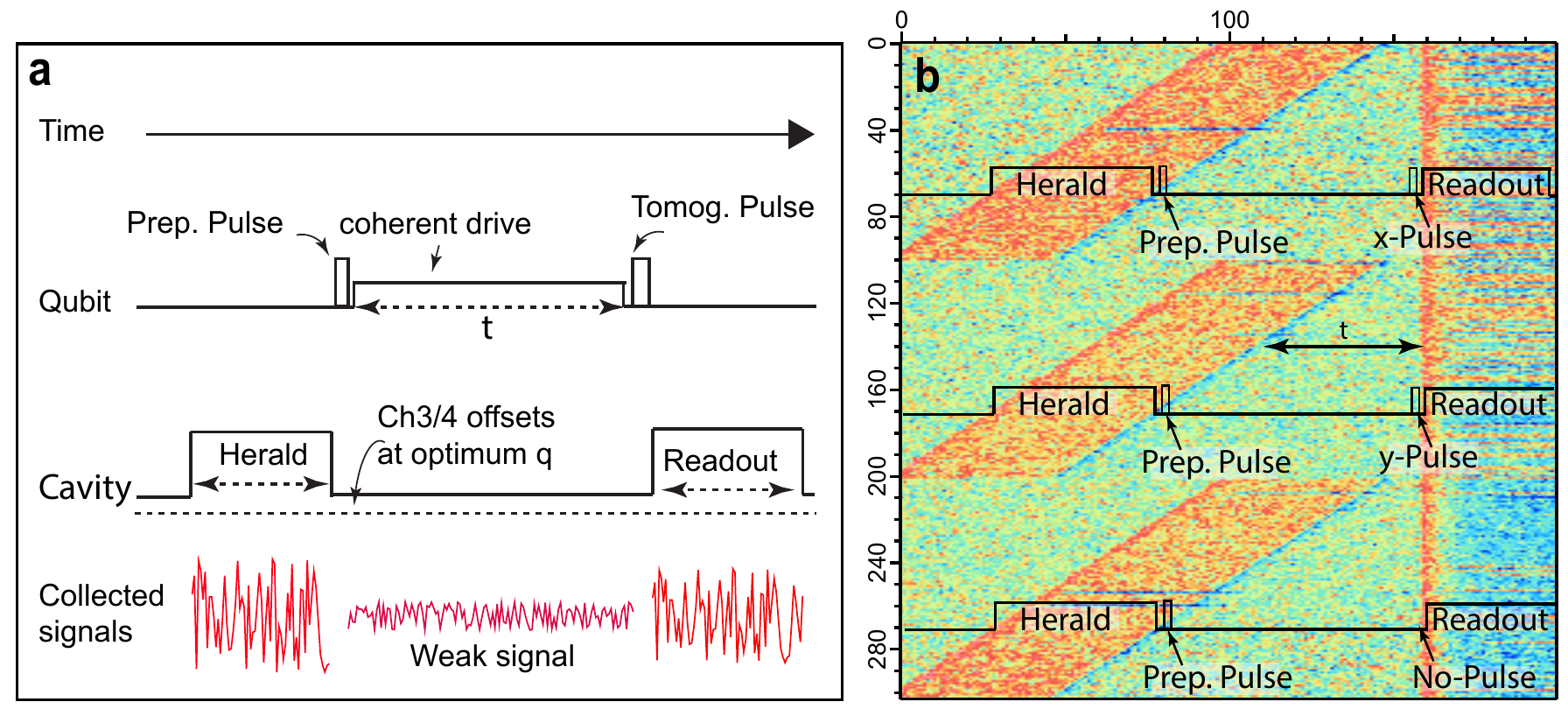}
\caption[Driven $z$-measurement sequence]{ {\footnotesize  \textbf{Driven $z$-measurement sequence}. }} 
\label{fig:driven_z_measur_seq}
\end{figure}

\subsection{Post-processing: Quantum trajectory update\label{subsection:pst_pross}}
In this step we use the SME (Equations~\ref{eq:SME_zx}) or Bayesian update (Equations~\ref{baysf12}) to reconstruct quantum trajectories. First, we need to properly scale the digitized measurement signal to obtain $V(t)$. For that we use the noise calibration data and subtract the overall offset\footnote{Note that the overall offset is determined by averaging both signals regardless of the preparation.}. Then we scale the signal so that the separation between measurement signal histograms of the ground and excited state preparations are equal to two. Moreover, the sign for the scaling factor is chosen so that the histogram corresponding to the ground state preparation is centered at $V=+1$ as depicted in Figure~\ref{fig:scaled_v}b which is consistent with our convention (for example see Equation~\ref{eq:POVM4}).
\begin{figure}[ht]
\centering
\includegraphics[width = .9\textwidth]{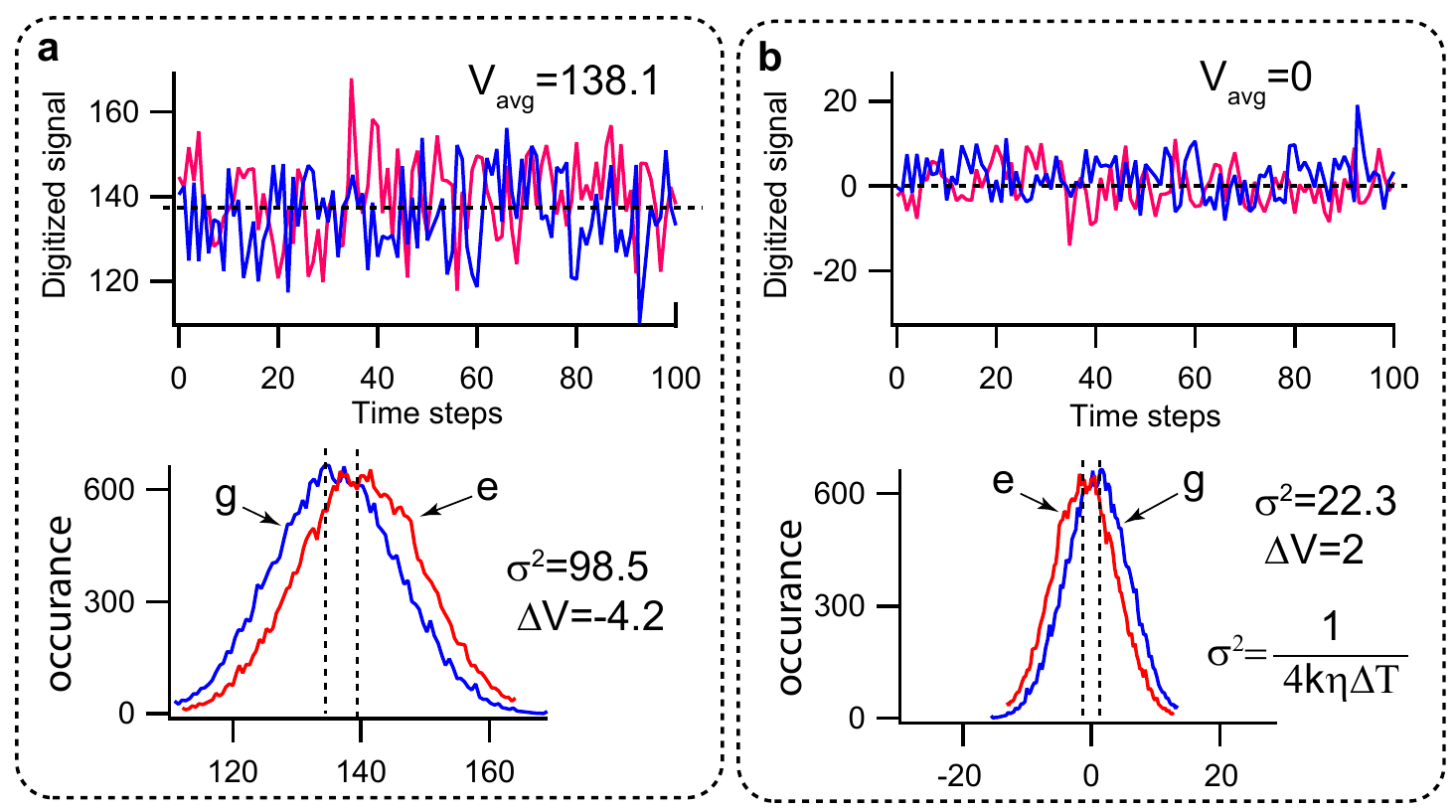}
\caption[Digitized weak measurement signal scaling]{ {\footnotesize  \textbf{Digitized weak measurement signal scaling}. }} 
\label{fig:scaled_v}
\end{figure}
One can check at this point to make sure that the variance of the signal is consistent with the calibrated quantum efficiency.

\subsubsection{SME update}
For quantum trajectories, we use the scaled signal in the SME. In order to account for a coherent drive $H_R=-\Omega_R \sigma_y/2$, we use the full version of the SME (Equation~\ref{eq:SME_rh0_full}). We represent this in terms of Bloch components as,
\bse\label{eq:SME_zx_i}
\begin{eqnarray}
z[i+1]&=&z[i] + \Omega_R x[i]dt + 4 \eta k (1-z[i]^2)(V[i]-z[i]) dt \label{eq:SME_z_i}\\
x[i+1]&=&x[i] - \Omega_R z[i+1]dt   -( 2k+\gamma_2)x[i]dt -4 \eta k x[i] z[i] (V[i]-z[i]) dt \nonumber\\ \label{eq:SME_x_i},
\end{eqnarray}
\ese
where we also discretized\footnote{Note that, $x[i+1]$ uses $z[i+1]$ for the rotation terms. Why?} the equations to be consistent with the digitized measurement signal with timestep $dt \sim 20$ ns. Equation~(\ref{eq:SME_zx_i}) may not be numerically stable or accurate when the timestep $dt$ in the experiment is not small enough. There is an alternative way to update the SME which involves two steps. In this method the unitary evolution separately implemented by a geometric rotation,
\bse\label{eq:SME_zx_ii}
\begin{eqnarray}
z_{\mathrm{d}}&=&z[i]\cos(\Omega_R dt) + x[i] \sin(\Omega_R dt)\\
x_{\mathrm{d}}&=&x[i]\cos(\Omega_R dt) - z[i] \sin(\Omega_R dt)\\
z[i+1]&=&z_{\mathrm{d}} + 4 \eta k (1-z_{\mathrm{d}}^2)(V[i]-z_{\mathrm{d}}) dt \label{eq:SME_z_ii}\\
x[i+1]&=& x_{\mathrm{d}}  -( 2k+\gamma_2)x_{\mathrm{d}}dt -4 \eta k x_{\mathrm{d}}z_{\mathrm{d}} (V[i]-z_{\mathrm{d}}) dt \label{eq:SME_x_ii},
\end{eqnarray}
\ese
where $z_d$ and $x_d$ are dummy variables connecting the two steps. The two-step update has better performance when $dt$ is not small enough to ensure the stability in single-step update. In most of practical situations $dt \Omega_R\ll1$ and the two methods are almost the same (see Figure~\ref{fig:SME_traj_update}).
\begin{figure}[ht]
\centering
\includegraphics[width = 0.9\textwidth]{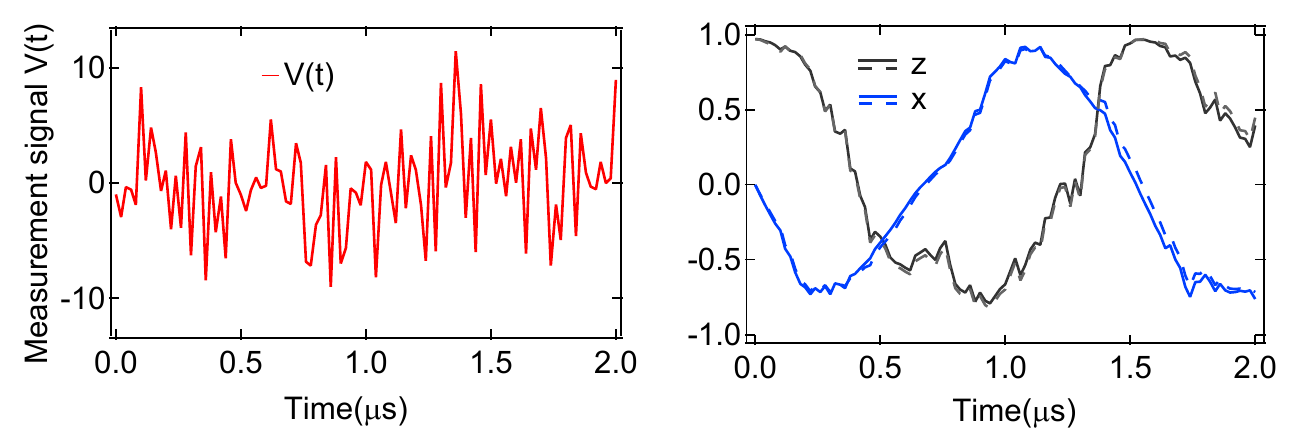}
\captionsetup{font=footnotesize}
\caption[Quantum trajectory updated by SME]{ { \textbf{Quantum trajectory updated by SME:} \textbf{a.} A typical homodyne measurement signal $V(t)$ corresponding to $z$-measurement of a driven qubit, $\Omega_R/2\pi=0.6, k=1,\eta=0.35,dt=20$ ns. \textbf{b.} The corresponding updated quantum trajectory using Equations~(\ref{eq:SME_zx_i}) for solid lines and Equations~(\ref{eq:SME_zx_ii}) for dashed line. As you can see, for these drive parameters both methods are practically the same.}} 
\label{fig:SME_traj_update}
\end{figure}
\subsubsection{Tomographic validation}
In order to verify that the updated trajectories accurately predict the state evolution of the qubit, we show the qubit state predicted by the trajectory is consistent with measurement from quantum state tomography. The idea is to compare the expectation values for $x$, $y$, and $z$ predicted by the quantum trajectory to the expectation value obtained by the result of projective measurements (readouts). Of course the readout is a destructive measurement with binary outcome. Therefore in order to obtain the expectation values one need to repeat the readout measurement on the same state many times. But it is not possible to perform many readouts on a single trajectory hence it is impossible to obtain expectation value for a single trajectory from a projective measurement.

However, instead of using single trajectory, we can use many different trajectories as long as all that trajectories have the same prediction for $\langle x \rangle , \langle y \rangle , \langle z \rangle$ at a given verification time $t_{\mathrm{v}}$. Therefore the tomographic verification at any given time $t_{\mathrm{v}}$ involves post-selection of trajectories that agree at that time. 

A nice way to do this is by choosing a random trajectory as a reference, and for each time step we post-select trajectories that have same prediction as the reference trajectory. Therefore we can reconstruct the reference trajectory by using the readout outcome of post-selected trajectories. Figure~\ref{fig:tomog_verification}a shows a reference $z$-trajectory in (black line) and a few post-selected trajectories that have the same prediction for $\langle z \rangle$ at $t_{\mathrm{v}}=0.8\  \mu$s within some tolerance indicated by a red window. Note these post-selected trajectories are from the experiment time $t=t_{\mathrm{v}}$ so their readout outcomes at $t=t_{\mathrm{v}}$ are available. The average of the readout outcomes from post-selected trajectories reconstruct the reference trajectory at that time step which is indicated by a green circular marker in the zoomed-inset. The agreement between the green circle and the reference trajectory indicates that quantum trajectories truly predict the state of the qubit at that time step.
\begin{figure}[ht]
\centering
\includegraphics[width = .9\textwidth]{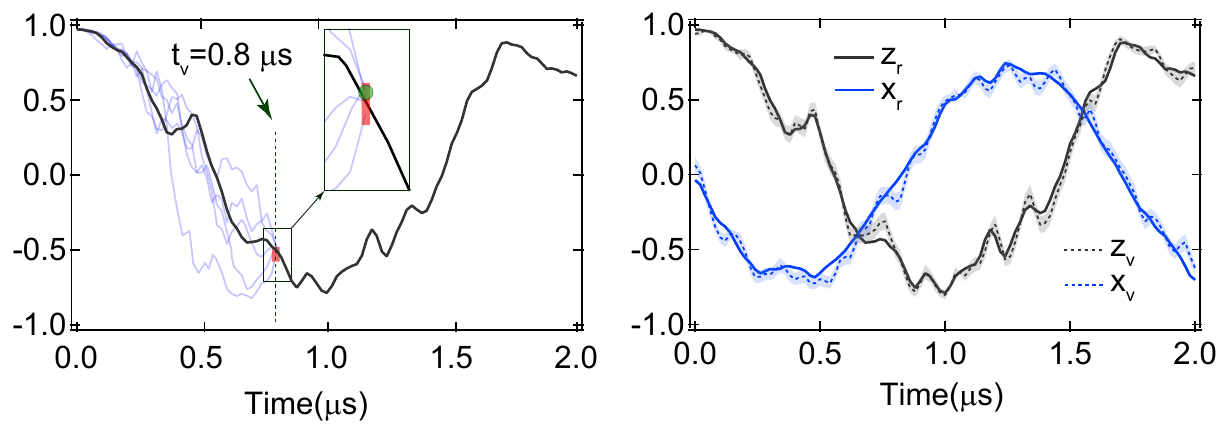}
\caption[Tomographic reconstruction]{ {\footnotesize  \textbf{Tomographic reconstruction}.}} 
\label{fig:tomog_verification}
\end{figure}
By repeating this process for both $z$, and $x$ for all time steps, one can reconstruct the reference trajectory and validate the state update as depicted in Figure~\ref{fig:tomog_verification}b. The shaded area indicates the binomial error from the readout outcomes of post-selected trajectories at each time step. The binomial error can be calculated as,
\begin{eqnarray}
\mathrm{Bionamial \ Error} = \sqrt{\frac{p \cdot q}{N}} = \sqrt{\frac{N_+ N_-}{(N_+ +  N_-)^3}}
 \label{eq:bionamial_error}.
\end{eqnarray}
Where $p=N_+/N$ and $q=N_-/N$ are the probabilities for two possible outcomes $(p=1-q)$ and $N=N_+ + N_-$ is the total number of outcomes. In this case, the total number of outcomes is equal to the total number of post-selected trajectories for each verification time.

\section{$\sigma_x$ measurement procedure\label{section:x-me}}
In this section we discuss the experimental procedure for $x$-measurement. This section follows the theoretical discussion of Section~\ref{section:generalized_mx}. For most part, the procedure for $x$-measurement is similar to $z$-measurement which is discussed in Section~\ref{section:z-me}. Here we discuss only two steps that are slightly different: the quantum efficiency calibration and the quantum trajectory update.

\subsection{Quantum efficiency calibration} 
The paramp setup is slightly different from the $z$-measurement. Here, the paramp pump is similar to the qubit frequency. Practically, the qubit pulses and paramp pump have to be from the same generator. In an $x$-measurement the paramp is only used for state tracking but not readout. Therefore there is no dumb-signal cancellation and no readout fidelity optimization. High power readout is used and often the fidelity can be improved by transferring population to the higher excited states prior to the readout pulse.

The quantum efficiency calibration is relatively easier for $x$-measurement than $z$-measurement. For this, we only need to run the noise calibration sequence with $\pm x$ state preparations and plot the histogram of the weak signal after a certain time of integration and scale the variance to be $\gamma_1 dt$. Then the separation would be $\Delta V = 2 \sqrt{\eta} \gamma_1$.

\subsection{State update and quantum trajectory}
As discussed in Subsection~\ref{Subsection:SME_xm_xz}, the SME for $x$-measurement in terms of Bloch components is described by Equation~(\ref{SME_xm_eta_xz}). In order to calculate quantum trajectories, the digitized homodyne signal needs to be properly scaled. For that we first subtract the offset (the offset can be determined by taking the average of the signal from the noise calibration sequence regardless of preparation). Then we scale the signal so that the variance of the histograms is $\gamma_1 dt$. The signal is then ready to be used in the discretized SME,
\bse
\begin{eqnarray}
\resizebox{.85\hsize}{!}{$ z[i+1] = z[i] +\Omega_R x[i]  +  \gamma_1 (1-z[i])  + \sqrt{\eta \gamma_1} x (1-z[i]) (V[i]-\sqrt{\eta} \gamma_1 x[i] dt),\label{SME_xm_eta_z2}$} \\
\resizebox{.85\hsize}{!}{$x[i+1] = x[i] - \Omega_R z[i+1]  - \frac{\gamma_1}{2} x[i]  + \sqrt{\eta} ( 1-z[i] - x[i]^2  )(V[i]-\sqrt{\eta} \gamma_1 x[i] dt).\label{SME_xm_eta_x2}$}
\end{eqnarray}
\ese
\chapter{Monitoring Spontaneous Emission of a Quantum Emitter \label{ch5}} 
In this chapter, I discuss the experimental study of a continuously monitored quantum system. We focus on the dynamics of a decaying emitter under homodyne detection of its radiation. The aim of this chapter is to connect the this experiment with discussions provided in the previous chapters.

Unlike classical mechanics, measurement has an inevitable disturbance on quantum systems. This disturbance which is known as measurement backaction and depends on the type of detector that we use for measurement. Therefore, it is natural to ask how the same quantum system, with the same interaction Hamiltonian to the environment, behaves differently under different detection schemes on the environment. Although this doesn't make much sense in a classical framework, it is understandable in the quantum case, owing to the entanglement between the detector and the emitter as we have already seen in the simple model in Chapter~4 (Section~\ref{section:generalized_m}).

A prime example is the detection of spontaneous emission of an excited emitter. How does the emitter decay under continuous monitoring? Does the decay dynamics depend on the type of the detector? In other words, does an atom decay regardless of the detection or it does decay because of the detection? Exploring these questions underpin the topic of our study in this chapter. 

\section{Spontaneous emission}
Spontaneously emission is ubiquitous in nature and accounts for most of the light that we see around us~\cite{milonni1984spontaneous}. It is often an undesirable effect but also essential for diverse applications ranging from fluorescence imaging to quantum encryption using single photons. 
 
In the spontaneous emission process, an excited emitter (excited atom) releases its energy in form of photons into one of the available electromagnetic modes of the environment\footnote{Therefore the spontaneous emission rate can be altered by manipulating the electromagnetic modes that are available to the emitter via engineering the environment~\cite{houck2008controlling, gambetta2011superconducting}.}.
From the quantum measurement point of view, spontaneous emission is due to the light-matter interaction and entanglement of the state of the emitter to its electromagnetic environment~\cite{blinov2004observation,eichler2012observation}. In this picture, measurements on the environment (e.g. photon detection, homodyne detection) collapse the entangled wavefunction in a specific basis and convey information about the state of the emitter and consequently cause backaction \cite{wiseman2009quantum}. Therefore, the choice of measurement may change the quantum evolution of the emitter~\cite{wiseman2012dynamical,bolund2014stochastic,jordan2016anatomy, campagne2016observing}. 

A goal in this chapter is to study the dynamics of spontaneous emission under continuous homodyne measurement. But before discussing homodyne measurement, it would be illuminating to discuss photon detection. This will be helpful to draw a connection between these two types of detection.

\section{Photon Detection}
Consider a qubit (as a quantum emitter) interacting with an electromagnetic mode of the environment. Assume we use a photon detector to monitor the existence of photon in that mode of the environment\footnote{In general, one can assume that the emitter is interacting with many modes. Then for our discussion, we should also assume that the detector is sensitive to all of the modes.} as depicted in Figure~\ref{fig:photon_detection}a. 
\begin{figure}[ht]
\centering
\includegraphics[width = 0.98\textwidth]{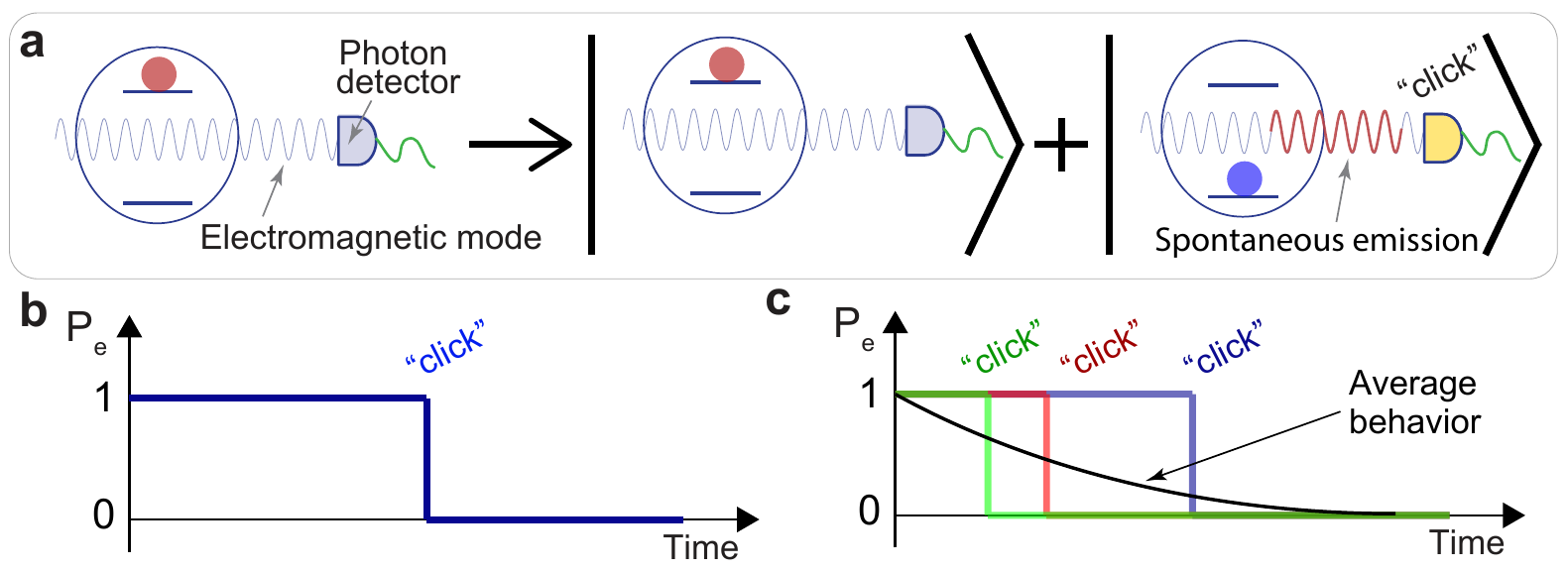}
\caption[Photon detection]{ { \footnotesize \textbf{Photon detection:} \textbf{a}, The qubit is initially prepared in the excited state and interacts with an electromagnetic mode. The qubit state and its emission to the mode are entangled via the interaction Hamiltonian~\eqref{eq:int_hamch5}. \textbf{b}, The detection of a photon results is a sudden jump for the emitter state. \textbf{c}, The average of many jump detections results in an exponential decay for the state of the qubit.} }
\label{fig:photon_detection} 
\end{figure}

The emitter which is initially prepared in the excited state interacts with the electromagnetic mode of the environment via the interaction Hamiltonian,
\begin{equation}
H_\mathrm{int} = \gamma (a^\dagger \sigma_- + a \sigma_+),
\label{eq:int_hamch5}
\end{equation}
where $a$ and $a^\dagger$ correspond to the creation and annihilation of a photon in that mode. The parameter $\gamma$ quantifies the interaction strength which, in this case, is related to the decay rate of the emitter to the environment\footnote{As discussed earlier, $\gamma$ is proportional to the available density of state for the emitter to decay.}.

The interaction Hamiltonian entangles the state of the emitter and the electromagnetic mode which can be represented as\footnote{Note that this is similar to our discussion in Chapter~4 and Equation~\eqref{eq:psi_tot_relax}, except here the emitter initially is in the excited state $\alpha_0=0, \beta_0=1$. This means that if we do not detect a photon, the emitter is still in the excited state with certainty, which is not the case when the emitter is prepared in a superposition state as we discussed in Equation~\eqref{eq:psi_tot_relax}.} (also depicted in Figure~\ref{fig:photon_detection}a),
\begin{eqnarray}
\Psi_{\mathrm{tot}}(0) = |e\rangle |0\rangle \to \Psi_{\mathrm{tot}}(t) = \beta |e\rangle |0\rangle + \alpha |g\rangle |1\rangle.\\
\end{eqnarray}

The photon detector monitors the state of the environment by performing measurements in the photon number basis. If we detect a ``click" we learn that the wave-function of the environment has been collapsed to the state $|1 \rangle$. This means the state of the qubit must be in ground state (measurement backaction). If we do not detect a click, then the emitter is still in the excited state. Therefore the detection of the spontaneous emission in the form of photons (energy quanta), results in an instantaneous jump of the emitter from the excited state to the ground state as depicted in Figure~\ref{fig:photon_detection}b~\cite{dalibard1992wave, blocher2017many}. If we average over many jump detections (or equivalently, if we disregard the detection results) the state of the qubit would exponentially decay from the excited to the ground state (Fig.~\ref{fig:photon_detection}c).

Before we conclude this subsection, it is worth mentioning a key point. You may notice that in the quantum measurement interpretation of the spontaneous emission, the atom decays because a detector collapses the wave function. In other words `the atom decays because the detector clicks'. This is so counterintuitive with our classical understanding of detection where we would say that the detector clicks because the atom has decayed\footnote{The argument `the atom decays because the detector clicks' is true when there is an entanglement between the emitter and the photon.}. We will return to this point again in the discussion on homodyne measurement.\\

\noindent\fbox{\parbox{\textwidth}{
\textbf{Exercise~1:} Consider detecting a photon from a star lightyears away. Does this mean that our detection of that photon causes that atom decay years ago? Explain this in terms of the quantum measurement interpretation.
}} \vspace{0.25cm}

\section{Homodyne detection of spontaneous emission}

In the previous section, we discussed a situation where spontaneous emission is measured by a photon detector. Now the question is, ``What if the emission is measured with a detector that is not sensitive to quanta, but rather to the amplitude of the field?'' In other words, what if we use a detector that addresses the wave notion of light as opposed to a photon detector which addresses the particle notion of light. What would the backaction be in this case? How are measurement outcomes correlated with the state of the emitter?'' In this section, we experimentally explore these questions by performing homodyne measurement of the spontaneous emission of a qubit.

As we discussed in Chapter~4 (Subsection~\ref{subsection:povm_sigmax}), Homodyne measurement can be thought of as projections to the coherent basis $|\alpha\rangle$, where $\alpha=|\alpha|e^{i\phi}$~\cite{jordan2016anatomy}. The measurement outcome $\alpha$ corresponds to the amplitude of the field in the quadrature $a^\dagger e^{i \phi} + a e^{-i \phi}$ which also contains fluctuations in that quadrature. 

In practice, when we perform homodyne measurement along a certain quadrature, we basically squeeze the outgoing emission along that quadrature as depicted in Figure~\ref{fig:homodyne_detection}b. This means we amplify the signal along the $\phi$-axis and de-amplify along the orthogonal axis. Therefore the measurement (or the collapse) happens only along the quadrature\footnote{Because we do not obtain any information along the other quadrature.} $\phi$. Returning to our discussion of `the atom decays because the detector clicks', this means that the emitter only ``decays''\footnote{The word decay is in quotes because, unlike the photon detection, homodyne detection does not necessarily fully collapse the emitter state.} along the $\phi$-quadrature\footnote{The fact that collapse happens in a certain quadrature, results in a certain type of backaction on the qubit which may confine the qubit evolution in a certain subspace.}.
\begin{figure}
\centering
\includegraphics[width = 0.88\textwidth]{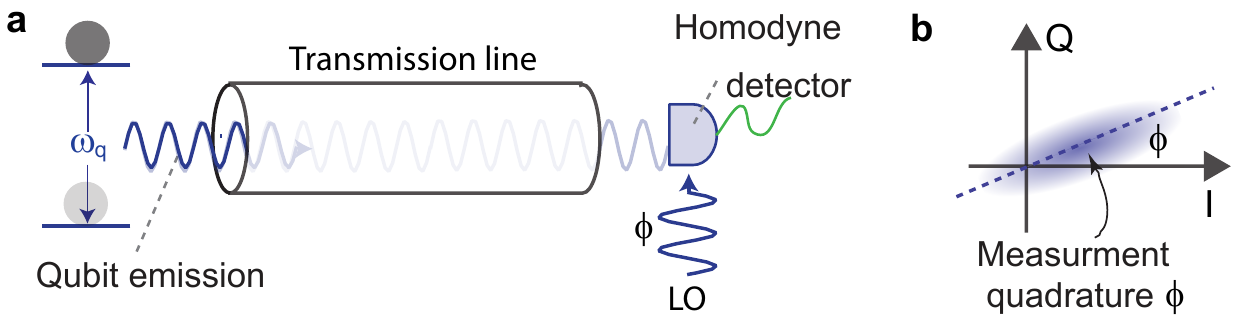}
\caption[Homodyne detection]{ {\footnotesize \textbf{Homodyne detection:} \textbf{a}, The spontaneous emission of the emitter is detected by homodyne measurement. The local oscillator has a well defined relative phase $\phi$ with respect to the qubit rotating frame which determines the amplification quadrature shown in \textbf{b}. The measurement happens only in one quadrature along $\phi$-axis. The fluctuations in the orthogonal quadrature are de-amplified, which means we do not learn about fluctuations in that quadrature. This allows for noiseless amplification for the $\phi$-quadrature~\cite{caves1982quantum}}} 
\label{fig:homodyne_detection}
\end{figure}

Therefore the idea for the experiment is 1) to study the spontaneous emission dynamics of a qubit by performing a homodyne measurement along the $\phi$ quadrature and 2) to explore the dynamics for different homodyne quadrature measurements.

Note that the interaction Hamiltonian (Eq.~\ref{eq:int_hamch5}) connects the quadrature $a^\dagger e^{i \phi} + a e^{-i \phi}$ to the corresponding dipole moment of the qubit (emitter), $ \sigma_-e^{i \phi} + \sigma_+e^{-i \phi}$. For example if we set the phase $\phi=0$, we showed in Chapter~4 that the homodyne measurement is actually a noisy estimate of $\langle \sigma_x \rangle$ which can be described by (see Equation~\ref{eq:noisy_estimate_x}),
\begin{eqnarray}
dV_t =\sqrt{\eta}\gamma\langle\sigma_x\rangle dt + \sqrt{\gamma} dW_t.
\label{eq:noisy_estimate_x_ch5}
\end{eqnarray}

We are interested to know what a detection of the homodyne signal $dV_t$ tells us about the state of the decaying qubit. For that, we use the experimental setup to perform the sequence depicted in Figure \ref{seq_spon}.
For the experimental setup, note that the qubit pulse and paramp pump share a same generator\footnote{This is a practical way to ensure that the paramp pump and the qubit pulse have a well defined and stable relative phase.} (BNC2) and the paramp is operated in a double-pump mode. Moreover, regarding the homodyne measurement of the emitter's emission, the demodulation should be happen at the qubit frequency but high-power readout demodulation should be at the bare cavity frequency. Therefore we use an RF switch to toggle between two frequencies for demodulation purposes. 

For the experimental sequence, we prepare the qubit in an initial state (in this case we prepared the qubit in the excited state, $+x$, and $+y$) then start collecting the homodyne signal for a variable time $t$ ($t$=40 ns, 80 ns,...). Finally, we perform a projective measurement to determine the final state of the qubit at that time. 
\begin{figure}
\centering
\includegraphics[width=0.98\textwidth]{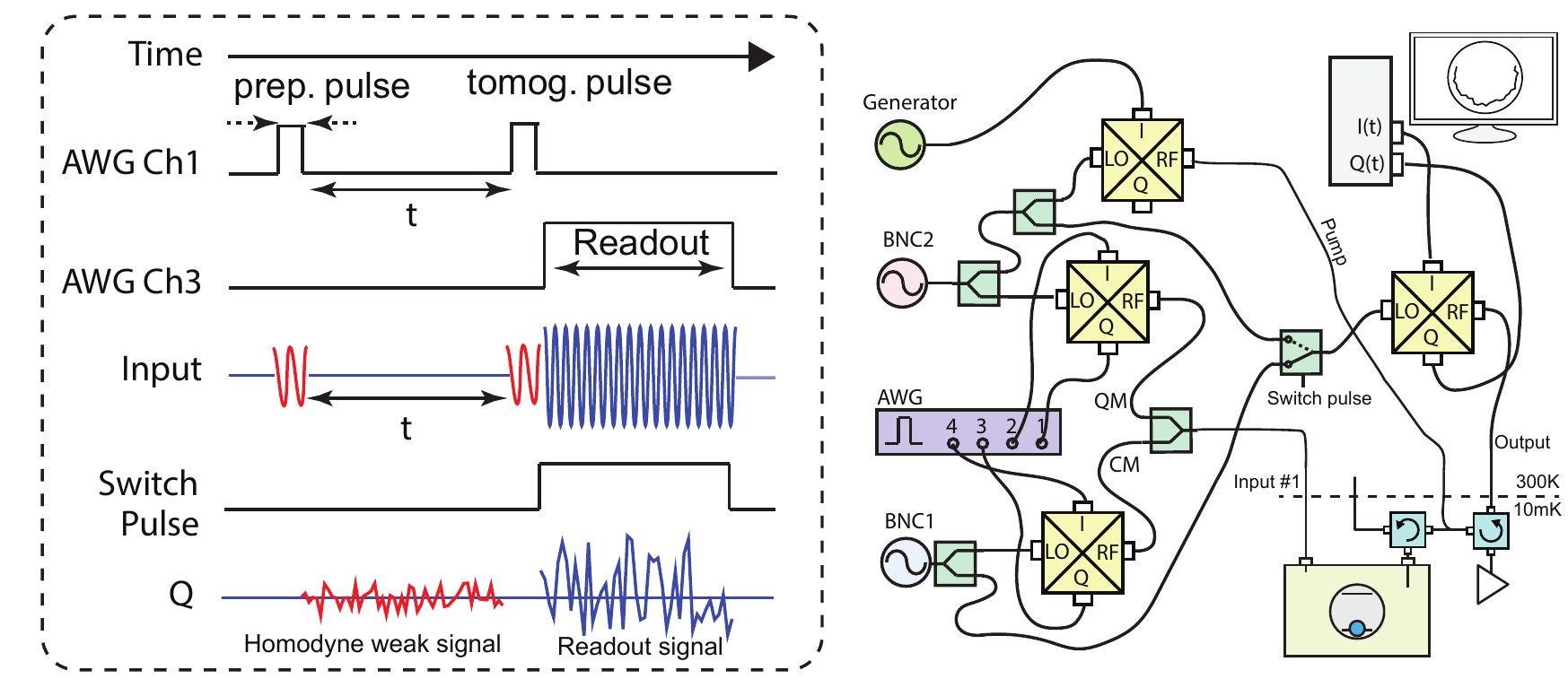}
\caption[The experimental setup (spontaneous emission experiment)]{\footnotesize {\bf The experimental setup and sequence:} The emitter is initialized in the excited state or in a superposition state by a preparation pulse. Right after the preparation, the homodyne signal is collected. Finally, we apply a tomographic rotation pulses along different axes followed by a high-power readout pulse to projectively measure the final state of the emitter.} \label{seq_spon}
\end{figure}

We characterize the correlation between the average of the collected homodyne signal and the final state (at time $t$). For that, we average the projective result conditioned at the average homodyne signal $\bar{V}$. Therefore we obtain the conditional expectation values, $\langle \sigma_x \rangle |_{\bar{V}}$, $\langle \sigma_y \rangle |_{\bar{V}}$, $\langle \sigma_z \rangle |_{\bar{V}}$.  In Figure \ref{fig2_spon}a-c we plot  $\langle \sigma_z \rangle |_{\bar{V}}$ and $\langle \sigma_x \rangle |_{\bar{V}}$ parametrically on the $X$--$Z$ plane of the Bloch sphere for different integration times. 
\begin{figure}
\begin{center}
\includegraphics[width=0.98\textwidth]{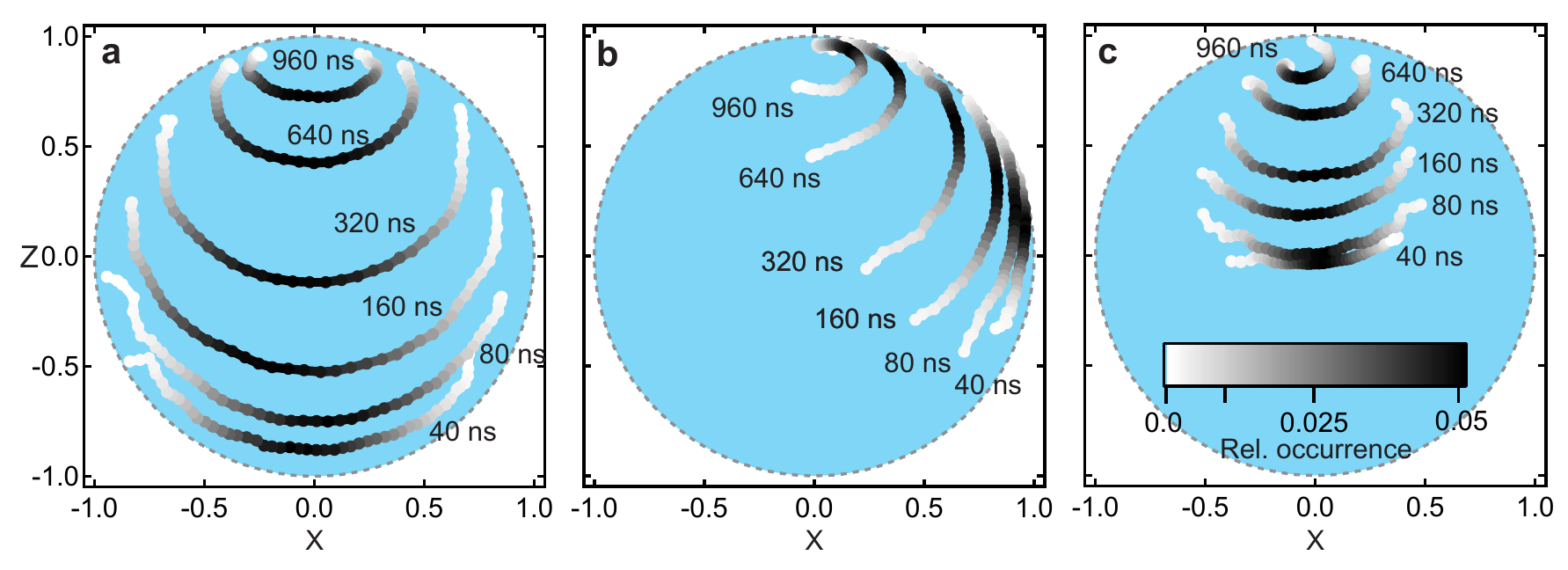}
\caption[Conditional dynamics of spontaneous decay]{\footnotesize {\bf Conditional dynamics of spontaneous decay:}  The tomographic result is the averaged tomographic readout conditioned on the outcome of the homodyne measurement to determine $x \equiv \langle \sigma_x \rangle |_{\bar{V}}$, and  $z\equiv \langle \sigma_z \rangle |_{\bar{V}}$ . These correlated tomography results are displayed on the $X$-$Z$ plane of the Bloch sphere for three different initial states: $-z$ ({\bf a}), $+x$ ({\bf b}), and $+y$ ({\bf c}). The gray scale indicates the relative occurrence of each measurement value. Note that different backaction between ({\bf b}), and $+y$ ({\bf c}) is the result of the phase-sensitive amplification on different quadratures of the homodyne signal.
} \label{fig2_spon}
\end{center}
\end{figure}

Looking at the experimental result in Figure~\ref{fig2_spon}, a few points are noticeable;
\begin{itemize}
\item When the qubit is prepared in the excited state, we see that the $x$-component of the state develops a correlation with the averaged homodyne signal.
\item The emitter state evolves in a deterministic curve inside the Bloch sphere. Therefore one can use these ``smiley" curves for heralding the system in a nearly arbitrary point in the Bloch sphere.
\item When the emitter is prepared as the $+x$ state, the qubit state sometimes gets more excited during the decay~\cite{bolund2014stochastic}. This stochastic excitation happens only in the amplitude measurements of the field and such excitations are not possible in the case of photodetection~\cite{jordan2016anatomy}.
\item If we rotate the amplification phase by 90 degrees\footnote{or equivalently prepare the system in $+y$.},  As depicted in Figure~\ref{fig2_spon}c, the state evolution for qubit is totally different. This is because the backaction happens in a different quadrature. This demonstrates how the choice of homodyne measurement phase can be used to control the evolution of the emitter.
\end{itemize}

We can take advantage of the deterministic ``smiley" evolution of the qubit to characterize the backaction for the qubit at different points in the Bloch sphere. For that, we let the system evolve from the excited state to a nearly arbitrary place inside the Bloch sphere on a smiley curve $(x_i, z_i)$. This acts as heralding of the qubit state to a specific point in the Bloch sphere, $(x_i, z_i)$. Then we collect the homodyne signal for an additional 40 ns. We use results from tomography to calculate the final position of the qubit ($x_f, z_f$). Therefore, for each point on the smiley curve, we obtain the conditioned evolution of the qubit based on the sign of the additional collected homodyne signal. This method tells us about the measurement backaction for positive and negative homodyne signals at each point on the Bloch sphere.
\begin{figure}[ht]
  \begin{center}
    \includegraphics[width=0.8\textwidth]{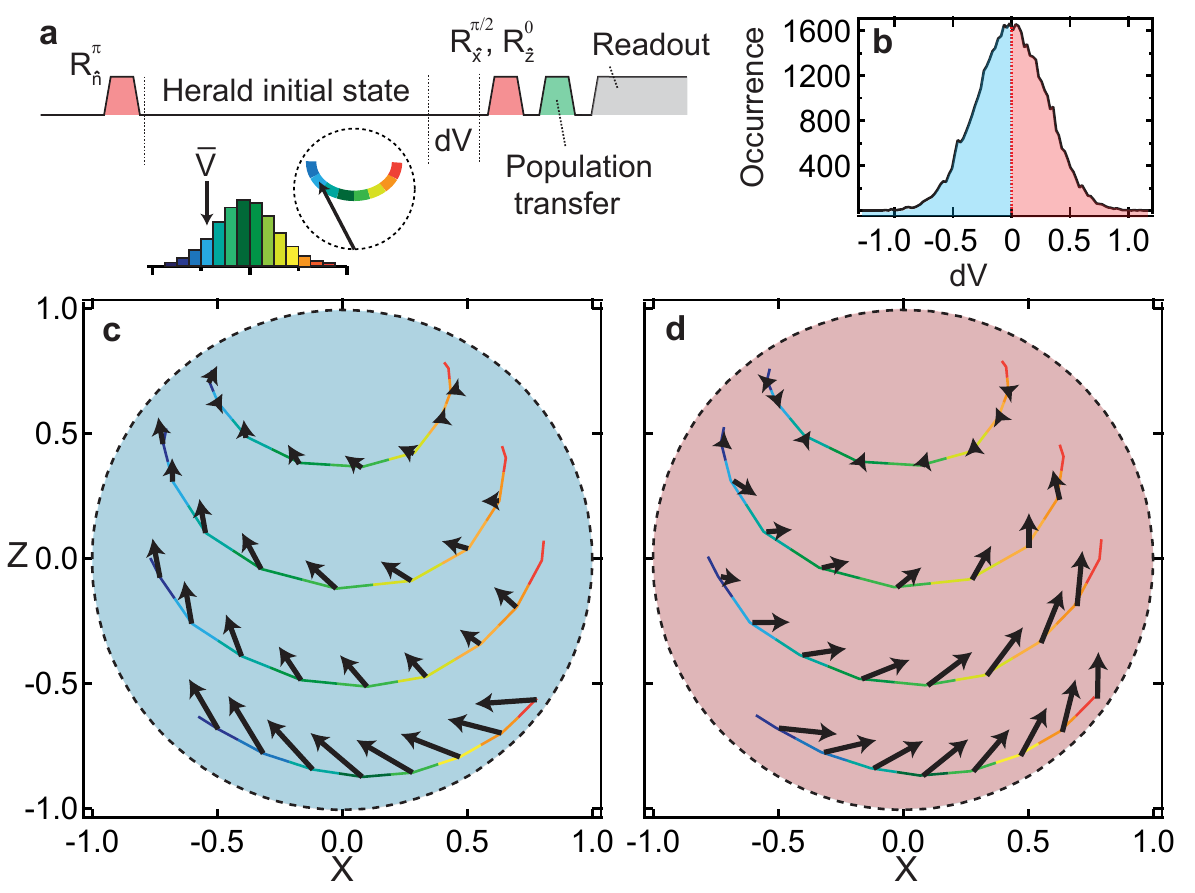}
  \caption[Backaction vector maps]{\footnotesize {\bf Backaction vector maps~\cite{naghiloo2016mapping}.} {\bf a}, We use the deterministic relation between the average homodyne signal $\bar{V}$ and the emitter's state to herald a nearly arbitrary initial state in the $X$-$Z$ plane of the Bloch sphere. The conditional backaction is obtained by quantum state tomography based on a small portion of the signal $dV$. {\bf b}, Histogram of the signals $dV$ which we separate into positive or negative $dV$. The corresponding backaction imparted on the emitter for negative ({\bf c}) or positive ({\bf d}) values of $dV$ are depicted by arrows at different locations in the $X$-$Z$ plane of the Bloch sphere.}\label{fig3_spon}
\end{center}
\end{figure}
The results are summarized in Figure~\ref{fig3_spon}. The backaction at a specific location in state space, associated with the detection of a given value of $dV$, is demonstrated by the vector connecting $(x_i, z_i)$ and  ($x_f, z_f$).  The backaction vector maps demonstrate how positive (negative) measurement results push the state toward $+x$ $(-x)$. Furthermore, the maps suggest that measurement backaction is stronger near the state $-z$ suggesting that the measurement strength is proportional to the emitter's excitation.

Finally, we can look at the individual quantum trajectories of this process. As we discussed in Chapter~4, all we need is to properly scale the homodyne signal and use that in the SME~\eqref{SME_xm_eta_xz}, 
\begin{eqnarray}
dx &=& - \frac{\gamma}{2} x dt + \sqrt{\eta} ( 1-z - x^2  )(dV_t - \gamma \sqrt{\eta} x dt), \label{smex}\\
dz &=&\gamma (1-z) dt + \sqrt{\eta} x(1-z)(dV_t - \gamma \sqrt{\eta}  x dt ), \label{smez}\\
dy &=& - \frac{\gamma}{2} y dt - \sqrt{\eta} xy (dV_t - \gamma \sqrt{\eta} x dt). \label{smey} 
\end{eqnarray}
Figure~\ref{fig4_map_spon} shows the result for state update from the exited state and the $+x$ state for 2 $\mu$s of continuous measurement. As we see, the evolution of the qubit during the decay process is no longer jumpy as opposed to the case of photon detection. However, the average of many trajectories would recover the same exponentially damped behavior as we discussed in the previous section\footnote{Regardless of the type of the detector, we will recover the Lindbladian evolution for the system if we average over many detection outcomes (this is equivalent to disregarding all measurement outcomes).}. Moreover, in Figure~\ref{fig4_map_spon}b stochastic excitations of individual trajectories toward the excited state is clearly apparent.
\begin{figure}[ht]
  \begin{center}
    \includegraphics[width=0.7\textwidth]{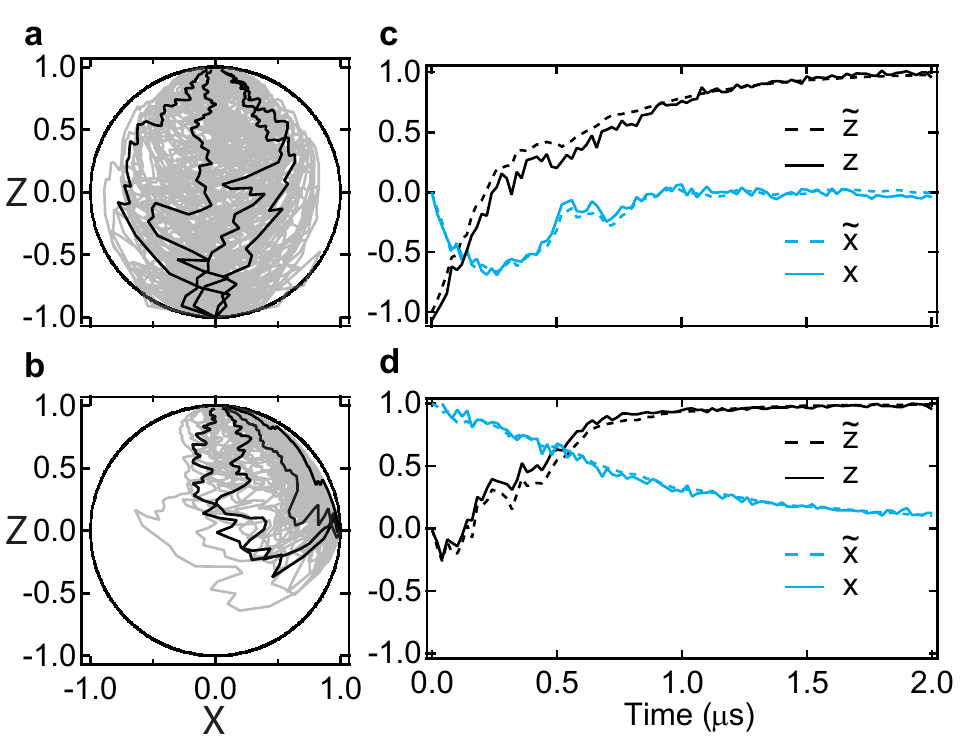}
\caption[Quantum trajectories for a decaying atom]{ \footnotesize \textbf{Quantum trajectories for decaying atom:} {\bf a,b}, Quantum trajectories of spontaneous decay calculated by the stochastic master equation, initiated from $-z$ ({\bf a}) and $+x$ ({\bf b}).  Several trajectories are depicted in gray, and a few individual trajectories are highlighted in black. {\bf c,d}  Individual trajectories ($\tilde{x}, \tilde{z}$) that originate from  $-z$ ({\bf c}) and $+x$ ({\bf d}) are shown as dashed lines and the tomographic reconstruction (see Chapter~4) based on projective measurements are shown as solid lines.}\label{fig4_map_spon}
\end{center}
\end{figure}
One can quantify the stochastic excitation by extracting the probability of excitation above a certain threshold at different times. Looking at the measurement term (proportional to $\sqrt{\eta}$) in Equation~\eqref{smez} it is clear that the state at $+x$ will be stochastically excited if the Weiner increment $dW_t$, obtained from the detected signal $dV_t$, is less than $- \sqrt{\gamma/\eta} dt$, predicting that $\sim 35\%$ of the trajectories should be excited in the first time step~\cite{naghiloo2016mapping}.


Having access to the stochastic trajectories of a quantum system opens new doors to investigate the dynamics of open quantum systems. In particular, the stochastic and non-unitary dynamics of quantum systems combined with a unitary evolution exhibits a rich dynamics which can be utilized for studying fundamental question in quantum physics~\cite{jordan2016anatomy,foroozani2016correlations,naghiloo2017quantum,lewalle2017prediction}.\\

\chapter{Quantum Thermodynamics: Quantum Maxwell's Demon \label{ch6}} 
In this chapter we explore quantum thermodynamics at the extreme level of a single atom interacting with a bath. The atom is a two level quantum system in contact with a detector which acts as the atom's environment. In this chapter we attempt to put our understanding about quantum dynamics into the language of quantum thermodynamics. In particular, we study the information-energy connection in quantum thermodynamics in the context of Maxwell's demon.

\section{Fluctuation theorems: thermodynamics at the microscope scale}
Thermodynamics is normally considered as a theory which describes systems in the limit of a large number of particles, $N \to \infty$.  In this limit, often known as the thermodynamic limit, fluctuations of energy are absolutely negligible compared to the total energy in the system. Therefore, it makes sense to describe the state of the system by a few macroscopic parameters regardless of fluctuations in individual degrees of freedom. For example, we define an equilibrium state and characterize the total energy in terms of heat and work by only a few thermodynamic parameters (e.g volume, pressure, temperature) for a gas inside a piston regardless of the position and the velocity of individual gas molecules. As depicted in Figure~\ref{fig:fluctu}a, the work fluctuations in a thermodynamic process are negligible in thermodynamic limit so that the work distribution is effectively a delta function. 

However, for microscopic systems which have a finite number of degrees of freedom, the fluctuations are no longer negligible. In this limit, fluctuations basically drive the systems in a stochastic manner during the process\footnote{Similar to the quantum trajectories which are stochastic due to quantum fluctuations.} as depicted in Figure~\ref{fig:fluctu}b. Therefore, the traditional thermodynamics laws need to be revisited for microscopic systems where thermal fluctuations are significant.
\begin{figure}[ht]
\centering
\includegraphics[width = 0.78\textwidth]{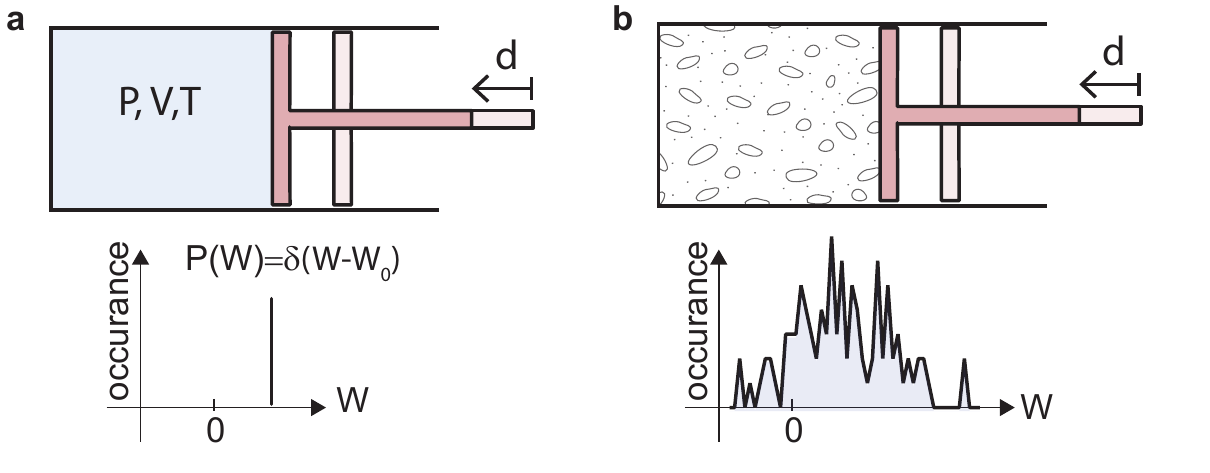}
\caption[Work fluctuations in the macroscopic and microscopic limit]{ { \footnotesize \textbf{Work fluctuations in the macroscopic and microscopic limit:} \textbf{a}. The work distribution for a system (gas inside a cylinder) in the thermodynamic limit (number of particles $N \to \infty$). The fluctuations are absolutely negligible $\propto N^{-\frac{1}{2}} \sim 0$ in this limit. \textbf{b}. The corresponding system in the limit of a finite number of particles ($N \to 1$). The work distribution fluctuates substantially $\propto N^{-\frac{1}{2}} \sim N$ due to thermal fluctuations.} }
\label{fig:fluctu} 
\end{figure}

In the past decades, thermodynamics has been successfully extended to nonequilibrium microscopic systems to account for thermal fluctuations. In particular, the generalized second law of the thermodynamics, in terms of a fluctuation theorem, has been experimentally verified for classical systems~\cite{collin2005verification}. For example, it has been shown that work fluctuations in a nonequilibrium process follow a fairly strong rule known as the Jarsynski equality (JE),
\begin{equation}\label{eq:jar0}
\langle e^{- \beta W} \rangle = e^{- \beta \Delta F},
\end{equation}
which connects the work distribution $W$ from a nonequilibrium process to the equilibrium free energy difference $\Delta F$~\cite{jarzynski1997nonequilibrium}. One can recover the second law of thermodynamics from JE by using Jensen's inequality, 
\begin{equation}\label{eq:jar1}
\langle e^{- \beta W} \rangle = e^{- \beta \Delta F} \xrightarrow{\langle e^{x}\rangle \geq e^{\langle x \rangle}} \langle W \rangle \geq \Delta F.
\end{equation}

Therefore, the JE is considered as the 2nd law of thermodynamics for microscopic systems. This equality has been verified experimentally for classical systems (see for example Ref.~\cite{liphardt2002equilibrium}).
However, the extension of thermodynamics to include quantum fluctuations faces
unique challenges because quantum fluctuations and coherence do not have a clear role in thermodynamics.
The newfound experimental capability to track single quantum trajectories adds to an intense endeavor to study and define thermodynamic quantities for individual quantum systems.

\section{Maxwell's demon and the 2nd law}
Consider the schematic in Figure~\ref{fig:fluctu}b in the limit of a few particles in the cylinder. If we are able to track the particles and react fast enough, we can basically displace the piston without doing any work! In this case, the work distribution is ideally a delta function at zero, but we have displacement in the piston $\Delta F \neq 0$. Thus the JE is no longer valid. In fact, Maxwell came up with a similar idea which was in apparent violation of the 2nd law soon after the establishment of thermodynamics. Maxwell considered a box full of air molecules and an intelligent being who has access to the velocity and position of individual molecules. The demon can sort the hot and cold particle to either side of the box without doing any work as depicted in Figure~\ref{fig:demon}.
\begin{figure}[ht]
\centering
\includegraphics[width = 0.8\textwidth]{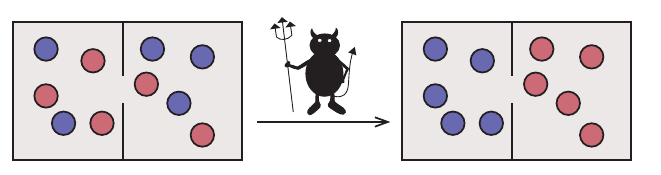}
\caption[Maxwell's demon]{ { \footnotesize \textbf{Maxwell's demon:} By knowing the position and the velocity of the particles, a demon sorts hot and cold particle in a box in apparent violation of the 2nd law.} }
\label{fig:demon} 
\end{figure}
The question of, ``How the demon can make a oven next to a fridge without doing any work and violate the 2nd law?'', reveals a profound connection between the energy and information in thermodynamics\footnote{This question of how the demon actually violates the 2nd law was unsolved for decades.}.

Owing to the dominant contribution of fluctuations in the dynamics of microscopic systems, a lot of effort has been directed toward the understanding of the connection between energy and information in microscopic systems. In particular, the Jarzynski equality (2nd law) has been generalized to account for the demon's information,
\begin{equation}\label{eq:jar2}
\langle e^{- \beta W - I} \rangle = e^{- \beta \Delta F},
\end{equation}
where $I$ is the mutual information between the demon's measurement outcome and the state of the system. 

The generalized Jarzynski equality (GJE) has been studied and verified for classical microscopic systems in which the demon is realized by measuring the thermal fluctuations and by applying subsequent feedback on the system~\cite{serreli2007molecular,raizen2009comprehensive,toyabe2010experimental,koski2014experimental}.

The recent advances in fabrication and control over quantum systems allow for unprecedented study of the concept of Maxwell's demon in quantum systems where instead of thermal fluctuations, the quantum fluctuations are dominant. For example, in the minimal quantum situation of a two level quantum system, the generalized Jarzynski equality is verified in the experiment by considering the mutual information between projective measurement outcomes and the state of the qubit~\cite{camati2016experimental,cottet2017observing,ciampini2017experimental,masuyama2018information}. Although these experiments use quantum systems, their result can be interpreted as a classical mixture either because the dynamics doesn't include quantum coherence or because the projective measurement destroys the quantum coherence.
However, in an actual quantum situation, the demon can also gain information about the quantum coherences; the off-diagonal elements in the density matrix\footnote{One can think of it in this way that; the classical demon is able to identify the particles as either hot or cold. But the quantum demon in general can also identify particles that are in superposition of hot and cold.} (Fig.~\ref{fig:demon_c_q}).
\begin{figure}[ht]
\centering
\includegraphics[width = 0.98\textwidth]{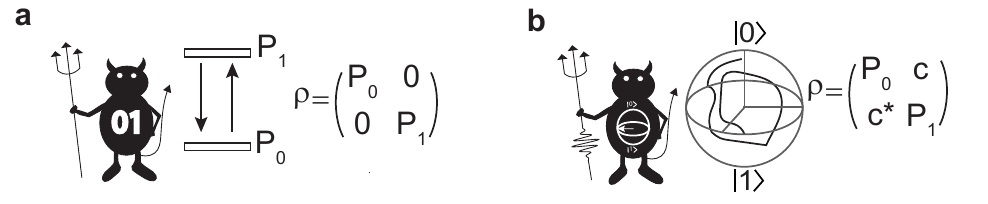}
\caption[Classical demon vs. quantum demon]{ { \footnotesize \textbf{Classical demon vs. quantum demon:} \textbf{a}, A classical demon's knowledge about a quantum system is limited to the populations in the definite states. \textbf{b}, A quantum demon also has knowledge about the quantum coherence in the system.} }
\label{fig:demon_c_q} 
\end{figure}

In previous chapters, we studied continuous monitoring and weak measurement of a quantum systems. Through weak measurement we can learn about the quantum state and quantum coherences without destroying them (completely). Therefore in this chapter, we attempt to utilize continuous monitoring to study Maxwell's demon in the context of quantum measurement.

\section{Continuous monitoring: a quantum Maxwell's demon}
The idea is to use our ability of tracking and manipulating the quantum state to realize a truly quantum Maxwell's demon. For that, consider the $z$-measurement setup (discussed in Chapter~4) and the experimental sequence demonstrated in Figure~\ref{fig:seq_demon}. The experimental protocol consists of five steps:
\begin{figure}[ht]
\centering
\includegraphics[width = 0.98\textwidth]{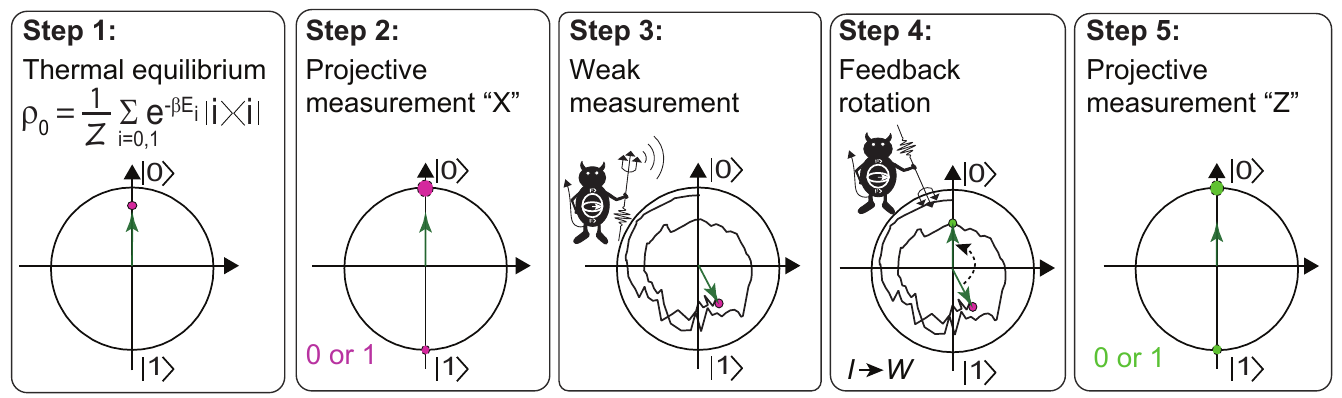}
\caption[Maxwell's demon experimental sequence]{ { \footnotesize \textbf{Experimental sequence}.} }
\label{fig:seq_demon} 
\end{figure}

\begin{itemize}
\item In Step~1, the qubit is prepared in a thermal state characterized by an inverse temperate\footnote{Here we represent $\beta$ in the qubit energy scale so that, initially for the qubit populations we have $P_1/P_0=e^{-\beta}$.} $\beta$. Practically this can be done by a proper rotation pulse followed by a projective measurement and by then disregarding the measurement outcome.
\item In Step~2, a projective measurement is performed so that the qubit is projected to one of its eigenstates. The binary measurement outcome $X \in \{0,1\}$ is recorded. This result, along with the projective result in Step~5, will be used to calculate the transition probabilities and characterize the work distribution for the experiment.
\item In Step~3, the demon, without knowing about the projective measurement result $X$, starts monitoring the qubit state while an external drive also acts on the system. Note the effective Hamiltonian for a resonantly driven qubit in the rotating frame is $H_t=- \Omega_R \sigma_y/2$ where $\Omega_R$ quantifies the drive strength as discussed in Chapter~2.
\item In Step~4, at a certain time, the demon uses his knowledge about the state of the system to rotate the system back to the ground state and extract work\footnote{In the actual experiment, in order to avoid feedback delay, we perform a random rotation pulse and the correct pulses are post-selected in the data analysis.}.
\item In Step~5, the experiment is finished by a second projective measurement which results in a binary measurement outcome $Z \in\{ 0, 1\}$.
\end{itemize} 

We repeat this experimental protocol and gather measurement statistics to experimentally study the 2nd law of thermodynamics. For example, Figure~\ref{fig:scatter_demon} shows the scatter plot of final states of the qubit before and after the rotation feedback in Step~4 for 200 experiment runs.

\begin{figure}[ht]
\centering
\includegraphics[width = 0.28\textwidth]{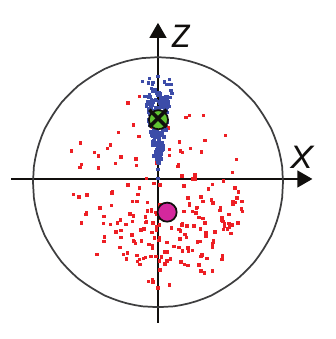}
\caption[The act of the demon in Step~4]{ { \footnotesize \textbf{The act of the demon in Step~4:} The red (blue) dots are the state of the qubit before (after) rotation feedback. The circular markers shows the average of the states before and after feedback. All data are from weak measurement and quantum trajectory reconstructions except the black cross which comes from the projective measurement after feedback. The agreement between the cross and green circular markers indicates that the trajectory update and feedback rotations are faithfully executed.} }
\label{fig:scatter_demon} 
\end{figure}

\subsection{Examining the Jarzynski equality}

Now, we examine the Jarsynski equality~\ref{eq:jar1} in the following form,
 \begin{eqnarray}\label{eq:jar0_exp}
\langle e^{-\beta W } \rangle  &=& \int P(W) e^{- \beta W} dW
\end{eqnarray}
where we set $\Delta F=0$ since the initial and final Hamiltonian are practically the same in our experiment.
In order to obtain the work distribution $P(W)$ we only use the projective measurement result and calculate the transition probabilities $P_{m,n}$ as demonstrated in Figure~\ref{fig:trans_demon}. The work distribution then can be calculated in this form\footnote{Note, because quantum systems do not necessarily occupy states with well defined energy (only eigenstates of the Hamiltonian have a well defined energy), the work distribution is described in terms of transition probabilities between energy eigenstates \cite{talkner2007fluctuation}.},
\begin{equation}\label{eq:pu}
P(W)  =  \sum_{m,n} P_{m,n}^{\tau} P_{n}^{0} \delta(\Delta U - (E^\tau_m-E^0_n)),
\end{equation}
where the $P_{n}^{0}$ denote the initial occupation probabilities, $P_{m,n}^{\tau}$ are the transition probabilities between initial and final eigenvalues $E^0_n$ and $E^\tau_m$ of the Hamiltonian $H_t$,  and $\tau$ is the duration of the protocol.

\begin{figure}[ht]
\centering
\includegraphics[width = 0.48\textwidth]{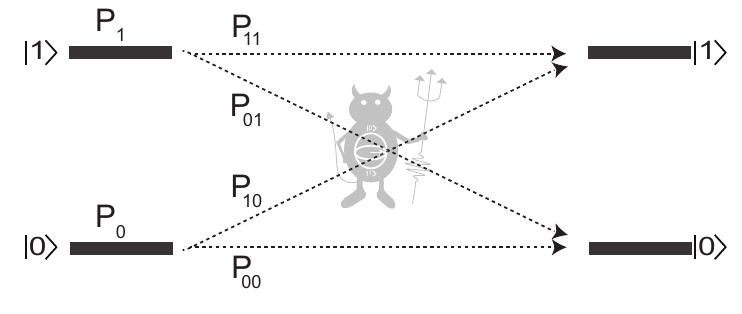}
\caption[Transition probabilities]{ { \footnotesize \textbf{Transition probabilities:} The projective measurement results are used to calculate the transition probabilities. The initial probabilities $P_0(0)$ and $P_1(0)$  are simply calculated based on the relative occurrence of outcome 0 or 1 in the first projective measurement. For the transition probabilities $P_{nm}(\tau)$, we calculate the relative occurrence of the result $n \in \{0,1\}$ in the second projective measurement conditioned that the result $m \in \{0,1\}$ is obtained in the first projective measurement.} }
\label{fig:trans_demon} 
\end{figure} 
Therefore, we examine the Jarzynski equality by using transition probabilities as follows,
\begin{eqnarray}\label{eq:jar1_exp}
\int P(W) e^{- \beta W} dW &=& P_0(0)P_{00}(\tau)+  P_1(0)P_{11}(\tau)\nonumber \\
&\ & +  P_0(0) P_{10}(\tau) e^{+\beta}  + P_1(0) P_{01}(\tau)  e^{-\beta} \nonumber\\
&=& 1.
\end{eqnarray}

Figure~\ref{fig:jar_demon} (square markers) shows the experimental result for the left-hand side of the Equation~\eqref{eq:jar1_exp} for five different duration times. There is no surprise that the result deviates from unity, because in Equation~\eqref{eq:jar1_exp}, we have ignored the act of the demon on the system. In other words, the demon violates the second law unless we account for the information of the demon.

\begin{figure}[ht]
\centering
\includegraphics[width = 0.48\textwidth]{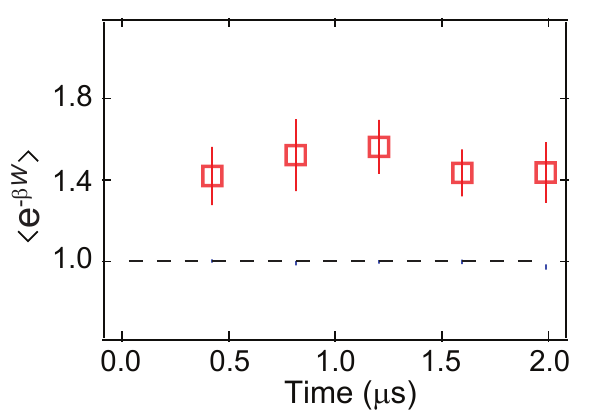}
\caption[Violation of the 2nd law]{ { \footnotesize \textbf{Violation of the 2nd law:} The experimental results violates the Jarzynski equality. This violation is because we have ignored the demon's information.} }
\label{fig:jar_demon} 
\end{figure}

\subsection{The demon's information}
In this section, we quantify the information that the demon obtains during the measurement. But what is the information? One way to quantify the information is to measure how much you learned that you didn't already know~\cite{lutz2015maxwell}. If we already know that the qubit state is $\rho= 0.99 |0\rangle \langle 0 | + 0.01 |1\rangle \langle 1 | $ and someone measures the qubit and lets us know that the qubit is in the ground state, we would not learn much. But it turns out that if the qubit is in the excited state our state of knowledge about the qubit would be substantially changed. Therefore the amount of information can be quantified by how unexpected the outcome is. For that, consider $I_z(\rho)=- \ln P_z(\rho)$ as the information content of $\rho$ along $z-$basis which quantifies how much we learn if we obtain result $z=-1,1$ along the $z-$basis. Now we define information exchange for the demon as the difference between initial and final information content,
\begin{eqnarray} 
I_{z',z}(t)&=& \ln P_{z'}(\rho_{t|r}) - \ln P_z(\rho_0),
\label{Mut_inf1}
\end{eqnarray}
where, $P_{z'}$ represents the probability of getting the result $z'=-1,1$ in the $z'$-basis where the system is diagonal\footnote{The initial state is always a thermal state so the diagonal basis initially is $z$ basis.}~\cite{funo2013integral}.
We calculate the probabilities in the diagonal basis to account for the information encoded in the populations (diagonal elements in density matrix) as well as coherences (off-diagonal elements) as depicted in Figure~\ref{fig:info_demon}a.
\begin{figure}[ht]
\centering
\includegraphics[width = 0.78\textwidth]{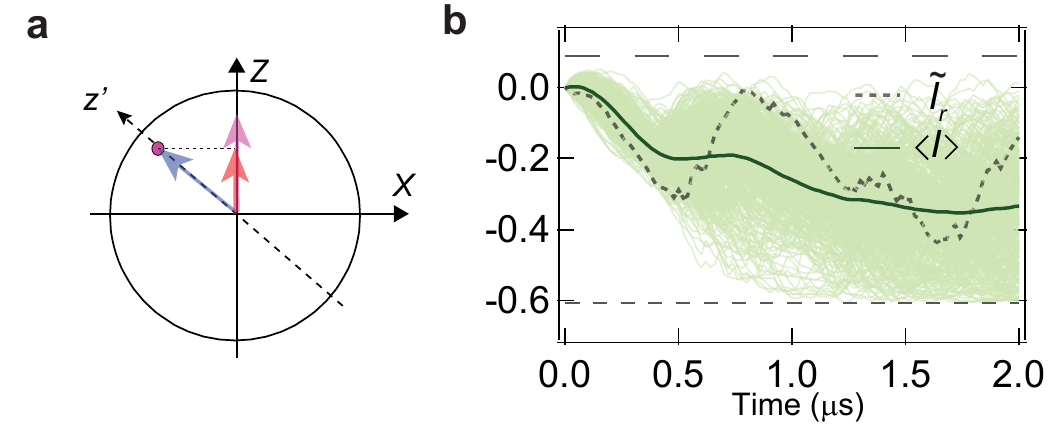}
\caption[Information dynamics for the quantum Maxwell's demon]{ { \footnotesize \textbf{Information dynamics for the quantum Maxwell's demon:} \textbf{a}. The information is quantified by considering the probabilities in the diagonal basis to account for information encoded in the coherences. \textbf{b}. The information exchange dynamics along quantum trajectories; the dashed line shows a typical trace from the ensemble of data (shaded background color). The black solid line is the average of the information exchange.} }
\label{fig:info_demon} 
\end{figure} 
For example, the state of the qubit is indicated by the blue arrow in Figure~\ref{fig:info_demon}a has the same amount of quantum information as the magenta arrow has provided that the probabilities are calculated in the diagonal basis for each state.

The expectation value for the information exchange along a quantum trajectory would be,
\begin{eqnarray}
\tilde{I}_r  = \sum_{z,z'=\pm 1} [P_{z'}(\rho_{t|r}) \ln P_{z'}(\rho_{t|r}) - P_z(\rho_0)\ln P_z(\rho_0)],
\label{Mut_inf2}
\end{eqnarray}
where the conditional probabilities come from a single quantum trajectory. The subscript $\cdot_r$ indicated by $\rho_t$ is the conditional evolution found by averaging over many trajectories. We obtain this average value for the information exchange as,
\begin{eqnarray}
\langle I \rangle = \sum_r \tilde{I}_r = \sum_{z,z'=\pm1,r} P_{z'}(\rho_t) \ln P_{z'}(\rho_t) - P_z(\rho_0)\ln P_z(\rho_0).
\label{Mut_inf3}
\end{eqnarray}

\subsection{Test of the generalized Jarzynski equality}

Now we attempt to verify the generalized Jarzynski equality which includes the information term. For that, we represent Equation~\eqref{eq:jar2} as\footnote{The sign for $W$ in GJE depends on our definition of the work; the work done by the system, or the work done on the system $e^{\pm\beta W - I + \Delta F}$.},
\begin{equation}
\begin{split}
\langle e^{-\beta W - I + \Delta F} \rangle =& P_0(0)P_{00}(t)e^{-I_{00}} +  P_1(0)P_{11}(t)e^{-I_{11}} \\
+&  P_0(0) P_{10}(t) e^{-\beta-I_{10}}  + P_1(0) P_{01}(t)  e^{+\beta-I_{01}},
\end{split}
\label{jarcq}
\end{equation}
where $I_{ij}= \ln P_{i}(\rho_t) - \ln P_j(\rho_0)$ as we discussed in the previous Subsection. Figure~\ref{fig:gen_jar_demon} shows the experimental result for Equation~\eqref{jarcq} which indicates that the generalized Jarzynski is indeed verified.

\begin{figure}[ht]
\centering
\includegraphics[width = 0.48\textwidth]{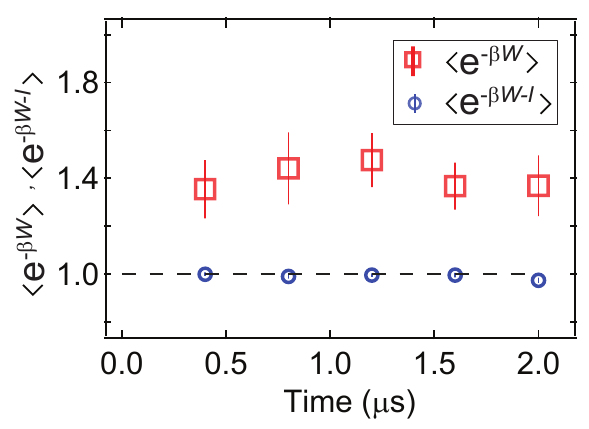}
\caption[Generalized Jarzynski equality for the quantum Maxwell's demon]{ { \footnotesize \textbf{Generalized Jarzynski equality for the quantum Maxwell's demon:} The blue round markers are the experimental result for the generalized Jarzynski equality for different time durations. The agreement between the dashed line and markers indicates that the GJE is verified, as opposed to the JE (red square markers).} }
\label{fig:gen_jar_demon} 
\end{figure}

\section{Information gain and loss}
Now, we study the information dynamics at the ensemble level. In Figure \ref{fig:info_demon}d (solid curve) we showed the average information change from many trajectories. You may notice that the average information is negative. This loss of information is due to decoherence which is a uniquely quantum feature and only appears if coherences contribute to the dynamics, and is thus not possible in a classical situation (e.g. See reference \cite{masuyama2018information}).

One may categorize the information change in two parts and distinguish the contribution of information gain through measurement and information loss due to imperfect detection \cite{funo2013integral}. 
In principle, imperfect detection arises because the state evolution of the detector is not exactly known and we must average over possible configurations of the detector as illustrated in Figure~\ref{fig:info_transition_demon}a. If we consider the detector uncertainty as an average over inaccessible degrees of freedom, parameterized by a stochastic variable $a$, the exchanged information \eqref{Mut_inf3} can be written as a sum of information gain and information loss $\langle I \rangle =I_{\rm gain} - I_{\rm loss}$ where~\cite{funo2013integral},
\begin{eqnarray}
I_{\rm gain} &=&S(\rho_0)- \sum_a p(a,r) S(\rho_{t|r,a}) \geqslant 0\\
I_{\rm loss} &=&\sum_r S(\rho_{t|r})- \sum_a p(a,r) S(\rho_{t|r,a}) \geqslant 0
\end{eqnarray}

\begin{figure}[ht]
\centering
\includegraphics[width = 0.78\textwidth]{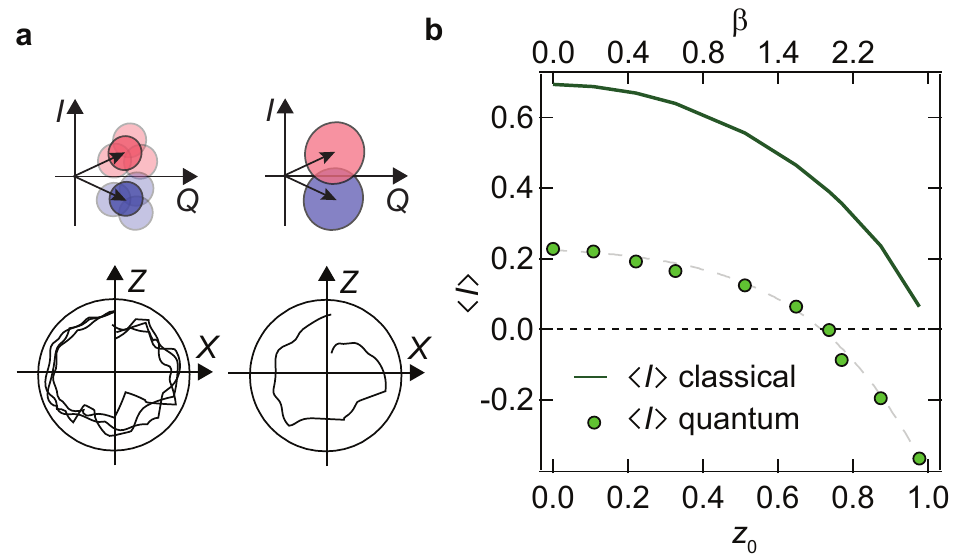}
\caption[Information gain and loss for the quantum Maxwell's demon]{ { \footnotesize \textbf{Information gain and loss for the quantum Maxwell's demon:} \textbf{a}, The inefficient detection can be modeled by averaging over unknown degrees of freedom for the detector. This basically lowers the signal-to-noise ratio. \textbf{b}, By adjusting the initial preparation for the qubit, we change the effective temperature for the system. The average of information exchange is negative for lower temperature (initially purer states) but it is positive for higher temperatures (initially more mixed states). Considering that only information encoded in coherences are susceptible to the loss, the more coherences involved in the dynamics, the more information will be lost.}}
\label{fig:info_transition_demon} 
\end{figure}

However, we do not have access to $a$ in this experiment. But still, we can explore the regimes where quantum coherence has different contributions to the dynamics meaning that the loss has different contributions to the total information change. To do this, we prepare the system in different initial thermal states and calculate the average information exchange for different initial temperatures. In Figure~\ref{fig:info_transition_demon}b  we plot the final information exchange (at $2\mu$s) versus different initial thermal states ( characterized by $z_{in}= \langle z\rangle|_{t=0}$). The total information change is positive for higher temperature (for more mixed initial states) but it is negative for lower temperature (more pure initial states). This transition of information gain to information loss can be understood by considering the fact that loss comes from decoherence of lower temperature (more pure) states. In our case, initially, more pure states will acquire more coherence through the unitary drive which turns the initial populations into coherences, these coherences are then lost due to inefficient detection, ultimately leading to a loss of information.






\begin{thebibliography}{100}

\bibitem{gisin2007quantum}
Nicolas Gisin and Rob Thew.
\newblock Quantum communication.
\newblock {\em Nature photonics}, 1(3):165, 2007.

\bibitem{degen2017quantum}
Christian~L Degen, F~Reinhard, and P~Cappellaro.
\newblock Quantum sensing.
\newblock {\em Reviews of modern physics}, 89(3):035002, 2017.

\bibitem{wendin2017quantum}
G~Wendin.
\newblock Quantum information processing with superconducting circuits: a
  review.
\newblock {\em Reports on Progress in Physics}, 80(10):106001, 2017.

\bibitem{rotter2015review}
Ingrid Rotter and JP~Bird.
\newblock A review of progress in the physics of open quantum systems: theory
  and experiment.
\newblock {\em Reports on Progress in Physics}, 78(11):114001, 2015.

\bibitem{zurek1992environment}
Wojciech~H Zurek.
\newblock The environment, decoherence, and the transition from quantum to
  classical.
\newblock In {\em Quantum Gravity And Cosmology-Proceedings Of The Xxii Gift
  International Seminar On Theoretical Physics}, page 117. World Scientific,
  1992.

\bibitem{modi2012classical}
Kavan Modi, Aharon Brodutch, Hugo Cable, Tomasz Paterek, and Vlatko Vedral.
\newblock The classical-quantum boundary for correlations: discord and related
  measures.
\newblock {\em Reviews of Modern Physics}, 84(4):1655, 2012.

\bibitem{tan2016quantum}
Dian Tan, Mahdi Naghiloo, Klaus M{\"o}lmer, and KW~Murch.
\newblock Quantum smoothing for classical mixtures.
\newblock {\em Physical Review A}, 94(5):050102, 2016.

\bibitem{dressel2017arrow}
Justin Dressel, Areeya Chantasri, Andrew~N Jordan, and Alexander~N Korotkov.
\newblock Arrow of time for continuous quantum measurement.
\newblock {\em Physical review letters}, 119(22):220507, 2017.

\bibitem{manikandan2018fluctuation}
Sreenath~K Manikandan, Cyril Elouard, and Andrew~N Jordan.
\newblock Fluctuation theorems for continuous quantum measurement and absolute
  irreversibility.
\newblock {\em arXiv preprint arXiv:1807.05575}, 2018.

\bibitem{harrington2018characterizing}
PM~Harrington, D~Tan, M~Naghiloo, and KW~Murch.
\newblock Characterizing a statistical arrow of time in quantum measurement
  dynamics.
\newblock {\em arXiv preprint arXiv:1811.07708}, 2018.

\bibitem{vinjanampathy2016quantum}
Sai Vinjanampathy and Janet Anders.
\newblock Quantum thermodynamics.
\newblock {\em Contemporary Physics}, 57(4):545--579, 2016.

\bibitem{houck2008controlling}
AA~Houck, JA~Schreier, BR~Johnson, JM~Chow, Jens Koch, JM~Gambetta,
  DI~Schuster, L~Frunzio, MH~Devoret, SM~Girvin, et~al.
\newblock Controlling the spontaneous emission of a superconducting transmon
  qubit.
\newblock {\em Physical review letters}, 101(8):080502, 2008.

\bibitem{gladchenko2009superconducting}
Sergey Gladchenko, David Olaya, Eva Dupont-Ferrier, Benoit Dou{\c{c}}ot, Lev~B
  Ioffe, and Michael~E Gershenson.
\newblock Superconducting nanocircuits for topologically protected qubits.
\newblock {\em Nature Physics}, 5(1):48, 2009.

\bibitem{gambetta2011superconducting}
JM~Gambetta, AA~Houck, and Alexandre Blais.
\newblock Superconducting qubit with purcell protection and tunable coupling.
\newblock {\em Physical review letters}, 106(3):030502, 2011.

\bibitem{kockum2018decoherence}
Anton~Frisk Kockum, G{\"o}ran Johansson, and Franco Nori.
\newblock Decoherence-free interaction between giant atoms in waveguide quantum
  electrodynamics.
\newblock {\em Physical review letters}, 120(14):140404, 2018.

\bibitem{ningyuan2015time}
Jia Ningyuan, Clai Owens, Ariel Sommer, David Schuster, and Jonathan Simon.
\newblock Time-and site-resolved dynamics in a topological circuit.
\newblock {\em Physical Review X}, 5(2):021031, 2015.

\bibitem{dempster2014understanding}
Joshua~M Dempster, Bo~Fu, David~G Ferguson, DI~Schuster, and Jens Koch.
\newblock Understanding degenerate ground states of a protected quantum circuit
  in the presence of disorder.
\newblock {\em Physical Review B}, 90(9):094518, 2014.

\bibitem{kelly2015state}
Julian Kelly, Rami Barends, Austin~G Fowler, Anthony Megrant, Evan Jeffrey,
  Theodore~C White, Daniel Sank, Josh~Y Mutus, Brooks Campbell, Yu~Chen, et~al.
\newblock State preservation by repetitive error detection in a superconducting
  quantum circuit.
\newblock {\em Nature}, 519(7541):66, 2015.

\bibitem{ofek2016extending}
Nissim Ofek, Andrei Petrenko, Reinier Heeres, Philip Reinhold, Zaki Leghtas,
  Brian Vlastakis, Yehan Liu, Luigi Frunzio, SM~Girvin, L~Jiang, et~al.
\newblock Extending the lifetime of a quantum bit with error correction in
  superconducting circuits.
\newblock {\em Nature}, 536(7617):441, 2016.

\bibitem{corcoles2015demonstration}
Antonio~D C{\'o}rcoles, Easwar Magesan, Srikanth~J Srinivasan, Andrew~W Cross,
  Matthias Steffen, Jay~M Gambetta, and Jerry~M Chow.
\newblock Demonstration of a quantum error detection code using a square
  lattice of four superconducting qubits.
\newblock {\em Nature communications}, 6:6979, 2015.

\bibitem{reed2012realization}
Matthew~D Reed, Leonardo DiCarlo, Simon~E Nigg, Luyan Sun, Luigi Frunzio,
  Steven~M Girvin, and Robert~J Schoelkopf.
\newblock Realization of three-qubit quantum error correction with
  superconducting circuits.
\newblock {\em Nature}, 482(7385):382, 2012.

\bibitem{korotkov2010decoherence}
Alexander~N Korotkov and Kyle Keane.
\newblock Decoherence suppression by quantum measurement reversal.
\newblock {\em Physical Review A}, 81(4):040103, 2010.

\bibitem{kim2012protecting}
Yong-Su Kim, Jong-Chan Lee, Osung Kwon, and Yoon-Ho Kim.
\newblock Protecting entanglement from decoherence using weak measurement and
  quantum measurement reversal.
\newblock {\em Nature Physics}, 8(2):117, 2012.

\bibitem{gillett2010experimental}
GG~Gillett, RB~Dalton, BP~Lanyon, MP~Almeida, Marco Barbieri, Geoff~J Pryde,
  JL~O’brien, KJ~Resch, SD~Bartlett, and AG~White.
\newblock Experimental feedback control of quantum systems using weak
  measurements.
\newblock {\em Physical review letters}, 104(8):080503, 2010.

\bibitem{vijay2012stabilizing}
R~Vijay, Chris Macklin, DH~Slichter, SJ~Weber, KW~Murch, Ravi Naik, Alexander~N
  Korotkov, and Irfan Siddiqi.
\newblock Stabilizing rabi oscillations in a superconducting qubit using
  quantum feedback.
\newblock {\em Nature}, 490(7418):77, 2012.

\bibitem{sayrin2011real}
Cl{\'e}ment Sayrin, Igor Dotsenko, Xingxing Zhou, Bruno Peaudecerf, Th{\'e}o
  Rybarczyk, S{\'e}bastien Gleyzes, Pierre Rouchon, Mazyar Mirrahimi, Hadis
  Amini, Michel Brune, et~al.
\newblock Real-time quantum feedback prepares and stabilizes photon number
  states.
\newblock {\em Nature}, 477(7362):73, 2011.

\bibitem{sorensen2003measurement}
Anders~S S{\o}rensen and Klaus M{\o}lmer.
\newblock Measurement induced entanglement and quantum computation with atoms
  in optical cavities.
\newblock {\em Physical review letters}, 91(9):097905, 2003.

\bibitem{ruskov2003entanglement}
Rusko Ruskov and Alexander~N Korotkov.
\newblock Entanglement of solid-state qubits by measurement.
\newblock {\em Physical Review B}, 67(24):241305, 2003.

\bibitem{roch2014observation}
Nicolas Roch, Mollie~E Schwartz, Felix Motzoi, Christopher Macklin, Rajamani
  Vijay, Andrew~W Eddins, Alexander~N Korotkov, K~Birgitta Whaley, Mohan
  Sarovar, and Irfan Siddiqi.
\newblock Observation of measurement-induced entanglement and quantum
  trajectories of remote superconducting qubits.
\newblock {\em Physical review letters}, 112(17):170501, 2014.

\bibitem{murch2013observing}
KW~Murch, SJ~Weber, Christopher Macklin, and Irfan Siddiqi.
\newblock Observing single quantum trajectories of a superconducting quantum
  bit.
\newblock {\em Nature}, 502(7470):211, 2013.

\bibitem{hacohen2016quantum}
Shay Hacohen-Gourgy, Leigh~S Martin, Emmanuel Flurin, Vinay~V Ramasesh,
  K~Birgitta Whaley, and Irfan Siddiqi.
\newblock Quantum dynamics of simultaneously measured non-commuting
  observables.
\newblock {\em Nature}, 538(7626):491, 2016.

\bibitem{foroozani2016correlations}
N~Foroozani, M~Naghiloo, D~Tan, K~M{\o}lmer, and KW~Murch.
\newblock Correlations of the time dependent signal and the state of a
  continuously monitored quantum system.
\newblock {\em Physical review letters}, 116(11):110401, 2016.

\bibitem{naghiloo2016mapping}
Mahdi Naghiloo, N~Foroozani, Dian Tan, A~Jadbabaie, and KW~Murch.
\newblock Mapping quantum state dynamics in spontaneous emission.
\newblock {\em Nature communications}, 7:11527, 2016.

\bibitem{weber2014mapping}
SJ~Weber, Areeya Chantasri, Justin Dressel, Andrew~N Jordan, KW~Murch, and
  Irfan Siddiqi.
\newblock Mapping the optimal route between two quantum states.
\newblock {\em Nature}, 511(7511):570, 2014.

\bibitem{naghiloo2017quantum}
M~Naghiloo, D~Tan, PM~Harrington, P~Lewalle, AN~Jordan, and KW~Murch.
\newblock Quantum caustics in resonance-fluorescence trajectories.
\newblock {\em Physical Review A}, 96(5):053807, 2017.

\bibitem{cujia2018watching}
KS~Cujia, JM~Boss, J~Zopes, and CL~Degen.
\newblock Watching the precession of a single nuclear spin by weak
  measurements.
\newblock {\em arXiv preprint arXiv:1806.08243}, 2018.

\bibitem{naghiloo2017achieving}
M~Naghiloo, AN~Jordan, and KW~Murch.
\newblock Achieving optimal quantum acceleration of frequency estimation using
  adaptive coherent control.
\newblock {\em Physical review letters}, 119(18):180801, 2017.

\bibitem{kiilerich2016bayesian}
Alexander~Holm Kiilerich and Klaus M{\o}lmer.
\newblock Bayesian parameter estimation by continuous homodyne detection.
\newblock {\em Physical Review A}, 94(3):032103, 2016.

\bibitem{naghiloo2017thermodynamics}
M~Naghiloo, D~Tan, PM~Harrington, JJ~Alonso, E~Lutz, A~Romito, and KW~Murch.
\newblock Thermodynamics along individual trajectories of a quantum bit.
\newblock {\em arXiv preprint arXiv:1703.05885}, 2017.

\bibitem{brandao2015second}
Fernando Brandao, Michal Horodecki, Nelly Ng, Jonathan Oppenheim, and Stephanie
  Wehner.
\newblock The second laws of quantum thermodynamics.
\newblock {\em Proceedings of the National Academy of Sciences},
  112(11):3275--3279, 2015.

\bibitem{toyabe2010experimental}
Shoichi Toyabe, Takahiro Sagawa, Masahito Ueda, Eiro Muneyuki, and Masaki Sano.
\newblock Experimental demonstration of information-to-energy conversion and
  validation of the generalized jarzynski equality.
\newblock {\em Nature physics}, 6(12):988, 2010.

\bibitem{gogolin2016equilibration}
Christian Gogolin and Jens Eisert.
\newblock Equilibration, thermalisation, and the emergence of statistical
  mechanics in closed quantum systems.
\newblock {\em Reports on Progress in Physics}, 79(5):056001, 2016.

\bibitem{parrondo2015thermodynamics}
Juan~MR Parrondo, Jordan~M Horowitz, and Takahiro Sagawa.
\newblock Thermodynamics of information.
\newblock {\em Nature physics}, 11(2):131, 2015.

\bibitem{naghiloo2018information}
M~Naghiloo, JJ~Alonso, A~Romito, E~Lutz, and KW~Murch.
\newblock Information gain and loss for a quantum maxwell's demon.
\newblock {\em arXiv preprint arXiv:1802.07205}, 2018.

\bibitem{kurizki2015quantum}
Gershon Kurizki, Patrice Bertet, Yuimaru Kubo, Klaus M{\o}lmer, David
  Petrosyan, Peter Rabl, and J{\"o}rg Schmiedmayer.
\newblock Quantum technologies with hybrid systems.
\newblock {\em Proceedings of the National Academy of Sciences}, page
  201419326, 2015.

\bibitem{murch2012cavity}
KW~Murch, U~Vool, D~Zhou, SJ~Weber, SM~Girvin, and I~Siddiqi.
\newblock Cavity-assisted quantum bath engineering.
\newblock {\em Physical review letters}, 109(18):183602, 2012.

\bibitem{harrington2018bath}
PM~Harrington, Mahdi Naghiloo, D~Tan, and KW~Murch.
\newblock Bath engineering of a fluorescing artificial atom with a photonic
  crystal.
\newblock {\em arXiv preprint arXiv:1812.04205}, 2018.

\bibitem{el2018non}
Ramy El-Ganainy, Konstantinos~G Makris, Mercedeh Khajavikhan, Ziad~H
  Musslimani, Stefan Rotter, and Demetrios~N Christodoulides.
\newblock Non-hermitian physics and pt symmetry.
\newblock {\em Nature Physics}, 14(1):11, 2018.

\bibitem{peng2014parity}
Bo~Peng, {\c{S}}ahin~Kaya {\"O}zdemir, Fuchuan Lei, Faraz Monifi, Mariagiovanna
  Gianfreda, Gui~Lu Long, Shanhui Fan, Franco Nori, Carl~M Bender, and Lan
  Yang.
\newblock Parity--time-symmetric whispering-gallery microcavities.
\newblock {\em Nature Physics}, 10(5):394, 2014.

\bibitem{chen2017exceptional}
Weijian Chen, {\c{S}}ahin~Kaya {\"O}zdemir, Guangming Zhao, Jan Wiersig, and
  Lan Yang.
\newblock Exceptional points enhance sensing in an optical microcavity.
\newblock {\em Nature}, 548(7666):192, 2017.

\bibitem{bender2016pt}
Carl~M Bender.
\newblock Pt symmetry in quantum physics: From a mathematical curiosity to
  optical experiments.
\newblock {\em Europhysics News}, 47(2):17--20, 2016.

\bibitem{bender2017behavior}
Carl~M Bender, Nima Hassanpour, Daniel~W Hook, SP~Klevansky, Christoph
  S{\"u}nderhauf, and Zichao Wen.
\newblock Behavior of eigenvalues in a region of broken pt symmetry.
\newblock {\em Physical Review A}, 95(5):052113, 2017.

\bibitem{bender2016comment}
Carl~M Bender, Mariagiovanna Gianfreda, Nima Hassanpour, and Hugh~F Jones.
\newblock Comment on “on the lagrangian and hamiltonian description of the
  damped linear harmonic oscillator”[j. math. phys. 48, 032701 (2007)].
\newblock {\em Journal of Mathematical Physics}, 57(8):084101, 2016.

\bibitem{bender2018series}
Carl~M Bender, C~Ford, Nima Hassanpour, and B~Xia.
\newblock Series solutions of pt-symmetric schr{\"o}dinger equations.
\newblock {\em Journal of Physics Communications}, 2(2), 2018.

\bibitem{bender2016analytic}
Carl~M Bender, Alexander Felski, Nima Hassanpour, SP~Klevansky, and Alireza
  Beygi.
\newblock Analytic structure of eigenvalues of coupled quantum systems.
\newblock {\em Physica Scripta}, 92(1):015201, 2016.

\bibitem{bender2018p}
Carl~M Bender, Nima Hassanpour, SP~Klevansky, and Sarben Sarkar.
\newblock P t-symmetric quantum field theory in d dimensions.
\newblock {\em Physical Review D}, 98(12):125003, 2018.

\bibitem{bender2018pt}
Carl~M Bender.
\newblock {\em PT Symmetry: In Quantum and Classical Physics}.
\newblock World Scientific Publishing, 2018.

\bibitem{naghiloo2019quantum}
M~Naghiloo, M~Abbasi, Yogesh~N Joglekar, and KW~Murch.
\newblock Quantum state tomography across the exceptional point in a single
  dissipative qubit.
\newblock {\em arXiv preprint arXiv:1901.07968}, 2019.

\bibitem{malek17cutoff}
Moein Malekakhlagh, Alexandru Petrescu, and Hakan~E T{\"u}reci.
\newblock Cutoff-free circuit quantum electrodynamics.
\newblock {\em Physical review letters}, 119(7):073601, 2017.

\bibitem{gely17converg}
Mario~F Gely, Adrian Parra-Rodriguez, Daniel Bothner, Ya~M Blanter, Sal~J
  Bosman, Enrique Solano, and Gary~A Steele.
\newblock Convergence of the multimode quantum rabi model of circuit quantum
  electrodynamics.
\newblock {\em Physical Review B}, 95(24):245115, 2017.

\bibitem{gerr05}
Christopher Gerry, Peter Knight, and Peter~L Knight.
\newblock {\em Introductory quantum optics}.
\newblock Cambridge university press, 2005.

\bibitem{walls2007quantum}
Daniel~F Walls and Gerard~J Milburn.
\newblock {\em Quantum optics}.
\newblock Springer Science \& Business Media, 2007.

\bibitem{houck2007generating}
AA~Houck, DI~Schuster, JM~Gambetta, JA~Schreier, BR~Johnson, JM~Chow,
  L~Frunzio, J~Majer, MH~Devoret, SM~Girvin, et~al.
\newblock Generating single microwave photons in a circuit.
\newblock {\em Nature}, 449(7160):328, 2007.

\bibitem{schu07thesis}
David~Isaac Schuster.
\newblock {\em Circuit quantum electrodynamics}.
\newblock Yale University, 2007.

\bibitem{tinkham2004introduction}
Michael Tinkham.
\newblock {\em Introduction to superconductivity}.
\newblock Courier Corporation, 2004.

\bibitem{devoret2004shortreviwe}
Michel~H Devoret, Andreas Wallraff, and John~M Martinis.
\newblock Superconducting qubits: A short review.
\newblock {\em arXiv preprint cond-mat/0411174}, 2004.

\bibitem{khezri26beyand}
Mostafa Khezri, Eric Mlinar, Justin Dressel, and Alexander~N. Korotkov.
\newblock Measuring a transmon qubit in circuit qed: Dressed squeezed states.
\newblock {\em Phys. Rev. A}, 94:012347, Jul 2016.

\bibitem{sank16beyond}
Daniel Sank, Zijun Chen, Mostafa Khezri, J.~Kelly, R.~Barends, B.~Campbell,
  Y.~Chen, B.~Chiaro, A.~Dunsworth, A.~Fowler, E.~Jeffrey, E.~Lucero,
  A.~Megrant, J.~Mutus, M.~Neeley, C.~Neill, P.~J.~J. O'Malley, C.~Quintana,
  P.~Roushan, A.~Vainsencher, T.~White, J.~Wenner, Alexander~N. Korotkov, and
  John~M. Martinis.
\newblock Measurement-induced state transitions in a superconducting qubit:
  Beyond the rotating wave approximation.
\newblock {\em Phys. Rev. Lett.}, 117:190503, Nov 2016.

\bibitem{niemczyk2010circuit}
Thomas Niemczyk, F~Deppe, H~Huebl, EP~Menzel, F~Hocke, MJ~Schwarz,
  JJ~Garcia-Ripoll, D~Zueco, T~H{\"u}mmer, E~Solano, et~al.
\newblock Circuit quantum electrodynamics in the ultrastrong-coupling regime.
\newblock {\em Nature Physics}, 6(10):772, 2010.

\bibitem{bosman2017multi}
Sal~J Bosman, Mario~F Gely, Vibhor Singh, Alessandro Bruno, Daniel Bothner, and
  Gary~A Steele.
\newblock Multi-mode ultra-strong coupling in circuit quantum electrodynamics.
\newblock {\em npj Quantum Information}, 3(1):46, 2017.

\bibitem{koch2007charge}
Jens Koch, M~Yu Terri, Jay Gambetta, Andrew~A Houck, DI~Schuster, J~Majer,
  Alexandre Blais, Michel~H Devoret, Steven~M Girvin, and Robert~J Schoelkopf.
\newblock Charge-insensitive qubit design derived from the cooper pair box.
\newblock {\em Physical Review A}, 76(4):042319, 2007.

\bibitem{pozar}
David~M Pozar.
\newblock {\em Microwave engineering}.
\newblock John Wiley \& Sons, 2009.

\bibitem{megrant2012planar}
Anthony Megrant, Charles Neill, Rami Barends, Ben Chiaro, Yu~Chen, Ludwig
  Feigl, Julian Kelly, Erik Lucero, Matteo Mariantoni, Peter~JJ O’Malley,
  et~al.
\newblock Planar superconducting resonators with internal quality factors above
  one million.
\newblock {\em Applied Physics Letters}, 100(11):113510, 2012.

\bibitem{khalil2012analysis}
MS~Khalil, MJA Stoutimore, FC~Wellstood, and KD~Osborn.
\newblock An analysis method for asymmetric resonator transmission applied to
  superconducting devices.
\newblock {\em Journal of Applied Physics}, 111(5):054510, 2012.

\bibitem{reed2010}
M.~D. Reed, L.~DiCarlo, B.~R. Johnson, L.~Sun, D.~I. Schuster, L.~Frunzio, and
  R.~J. Schoelkopf.
\newblock High-fidelity readout in circuit quantum electrodynamics using the
  jaynes-cummings nonlinearity.
\newblock {\em Phys. Rev. Lett.}, 105:173601, Oct 2010.

\bibitem{slichterthesis}
Daniel~Huber Slichter.
\newblock Quantum jumps and measurement backaction in a superconducting qubit.
\newblock {\em University of California, Berkeley}, 2011.

\bibitem{drummond2013quantum}
Peter~D Drummond and Zbigniew Ficek.
\newblock {\em Quantum squeezing}, volume~27.
\newblock Springer Science \& Business Media, 2013.

\bibitem{gambetta2008quantum}
Jay Gambetta, Alexandre Blais, Maxime Boissonneault, Andrew~A Houck,
  DI~Schuster, and Steven~M Girvin.
\newblock Quantum trajectory approach to circuit qed: Quantum jumps and the
  zeno effect.
\newblock {\em Physical Review A}, 77(1):012112, 2008.

\bibitem{jaco06}
Kurt Jacobs and Daniel~A Steck.
\newblock A straightforward introduction to continuous quantum measurement.
\newblock {\em Contemporary Physics}, 47(5):279--303, 2006.

\bibitem{brun2000continuous}
Todd~A Brun.
\newblock Continuous measurements, quantum trajectories, and decoherent
  histories.
\newblock {\em Physical Review A}, 61(4):042107, 2000.

\bibitem{koro11_bayes}
Alexander~N Korotkov.
\newblock Quantum bayesian approach to circuit qed measurement.
\newblock {\em arXiv preprint arXiv:1111.4016}, 2011.

\bibitem{tan2017homodyne}
Dian Tan, Neda Foroozani, Mahdi Naghiloo, AH~Kiilerich, K~M{\o}lmer, and
  KW~Murch.
\newblock Homodyne monitoring of postselected decay.
\newblock {\em Physical Review A}, 96(2):022104, 2017.

\bibitem{milonni1984spontaneous}
PW~Milonni.
\newblock Why spontaneous emission?
\newblock {\em American Journal of Physics}, 52(4):340--343, 1984.

\bibitem{blinov2004observation}
BB~Blinov, DL~Moehring, L-M Duan, and Chris Monroe.
\newblock Observation of entanglement between a single trapped atom and a
  single photon.
\newblock {\em Nature}, 428(6979):153, 2004.

\bibitem{eichler2012observation}
C~Eichler, C~Lang, JM~Fink, J~Govenius, S~Filipp, and A~Wallraff.
\newblock Observation of entanglement between itinerant microwave photons and a
  superconducting qubit.
\newblock {\em Physical review letters}, 109(24):240501, 2012.

\bibitem{wiseman2009quantum}
Howard~M Wiseman and Gerard~J Milburn.
\newblock {\em Quantum measurement and control}.
\newblock Cambridge university press, 2009.

\bibitem{wiseman2012dynamical}
Howard~M Wiseman and Jay~M Gambetta.
\newblock Are dynamical quantum jumps detector dependent?
\newblock {\em Physical review letters}, 108(22):220402, 2012.

\bibitem{bolund2014stochastic}
Anders Bolund and Klaus M{\o}lmer.
\newblock Stochastic excitation during the decay of a two-level emitter subject
  to homodyne and heterodyne detection.
\newblock {\em Physical Review A}, 89(2):023827, 2014.

\bibitem{jordan2016anatomy}
Andrew~N Jordan, Areeya Chantasri, Pierre Rouchon, and Benjamin Huard.
\newblock Anatomy of fluorescence: quantum trajectory statistics from
  continuously measuring spontaneous emission.
\newblock {\em Quantum Studies: Mathematics and Foundations}, 3(3):237--263,
  2016.

\bibitem{campagne2016observing}
Philippe Campagne-Ibarcq, Pierre Six, Landry Bretheau, Alain Sarlette, Mazyar
  Mirrahimi, Pierre Rouchon, and Benjamin Huard.
\newblock Observing quantum state diffusion by heterodyne detection of
  fluorescence.
\newblock {\em Physical Review X}, 6(1):011002, 2016.

\bibitem{dalibard1992wave}
Jean Dalibard, Yvan Castin, and Klaus M{\o}lmer.
\newblock Wave-function approach to dissipative processes in quantum optics.
\newblock {\em Physical review letters}, 68(5):580, 1992.

\bibitem{blocher2017many}
Philip~Daniel Blocher and Klaus M{\o}lmer.
\newblock How many atoms get excited when they decay?
\newblock {\em Quantum Science and Technology}, 2(3):034011, 2017.

\bibitem{caves1982quantum}
Carlton~M Caves.
\newblock Quantum limits on noise in linear amplifiers.
\newblock {\em Physical Review D}, 26(8):1817, 1982.

\bibitem{lewalle2017prediction}
Philippe Lewalle, Areeya Chantasri, and Andrew~N Jordan.
\newblock Prediction and characterization of multiple extremal paths in
  continuously monitored qubits.
\newblock {\em Physical Review A}, 95(4):042126, 2017.

\bibitem{collin2005verification}
Delphine Collin, Felix Ritort, Christopher Jarzynski, Steven~B Smith, Ignacio
  Tinoco~Jr, and Carlos Bustamante.
\newblock Verification of the crooks fluctuation theorem and recovery of rna
  folding free energies.
\newblock {\em Nature}, 437(7056):231, 2005.

\bibitem{jarzynski1997nonequilibrium}
Christopher Jarzynski.
\newblock Nonequilibrium equality for free energy differences.
\newblock {\em Physical Review Letters}, 78(14):2690, 1997.

\bibitem{liphardt2002equilibrium}
Jan Liphardt, Sophie Dumont, Steven~B Smith, Ignacio Tinoco, and Carlos
  Bustamante.
\newblock Equilibrium information from nonequilibrium measurements in an
  experimental test of jarzynski's equality.
\newblock {\em Science}, 296(5574):1832--1835, 2002.

\bibitem{serreli2007molecular}
Viviana Serreli, Chin-Fa Lee, Euan~R Kay, and David~A Leigh.
\newblock A molecular information ratchet.
\newblock {\em Nature}, 445(7127):523, 2007.

\bibitem{raizen2009comprehensive}
Mark~G Raizen.
\newblock Comprehensive control of atomic motion.
\newblock {\em Science}, 324(5933):1403--1406, 2009.

\bibitem{koski2014experimental}
Jonne~V Koski, Ville~F Maisi, Takahiro Sagawa, and Jukka~P Pekola.
\newblock Experimental observation of the role of mutual information in the
  nonequilibrium dynamics of a maxwell demon.
\newblock {\em Physical review letters}, 113(3):030601, 2014.

\bibitem{camati2016experimental}
Patrice~A Camati, John~PS Peterson, Tiago~B Batalhao, Kaonan Micadei,
  Alexandre~M Souza, Roberto~S Sarthour, Ivan~S Oliveira, and Roberto~M Serra.
\newblock Experimental rectification of entropy production by maxwell’s demon
  in a quantum system.
\newblock {\em Physical review letters}, 117(24):240502, 2016.

\bibitem{cottet2017observing}
Nathana{\"e}l Cottet, Sebastien Jezouin, Landry Bretheau, Philippe
  Campagne-Ibarcq, Quentin Ficheux, Janet Anders, Alexia Auff{\`e}ves, R{\'e}mi
  Azouit, Pierre Rouchon, and Benjamin Huard.
\newblock Observing a quantum maxwell demon at work.
\newblock {\em Proceedings of the National Academy of Sciences},
  114(29):7561--7564, 2017.

\bibitem{ciampini2017experimental}
Mario~A Ciampini, Luca Mancino, Adeline Orieux, Caterina Vigliar, Paolo
  Mataloni, Mauro Paternostro, and Marco Barbieri.
\newblock Experimental extractable work-based multipartite separability
  criteria.
\newblock {\em NPJ Quantum Information}, 3(1):10, 2017.

\bibitem{masuyama2018information}
Y~Masuyama, K~Funo, Y~Murashita, A~Noguchi, S~Kono, Y~Tabuchi, R~Yamazaki,
  M~Ueda, and Y~Nakamura.
\newblock Information-to-work conversion by maxwell’s demon in a
  superconducting circuit quantum electrodynamical system.
\newblock {\em Nature communications}, 9(1):1291, 2018.

\bibitem{talkner2007fluctuation}
Peter Talkner, Eric Lutz, and Peter H{\"a}nggi.
\newblock Fluctuation theorems: Work is not an observable.
\newblock {\em Physical Review E}, 75(5):050102, 2007.

\bibitem{lutz2015maxwell}
Eric Lutz and Sergio Ciliberto.
\newblock From maxwell’s demon to landauer’s eraser.
\newblock {\em Physics Today}, 68(9):30, 2015.

\bibitem{funo2013integral}
Ken Funo, Yu~Watanabe, and Masahito Ueda.
\newblock Integral quantum fluctuation theorems under measurement and feedback
  control.
\newblock {\em Physical Review E}, 88(5):052121, 2013.

\end{thebibliography}

\addcontentsline{toc}{chapter}{Bibliography}

\end{main}



\end{document}